\documentclass[12pt, a4paper]{report}

\usepackage[english]{babel}
\usepackage{amsthm,thmtools}
\usepackage{cleveref}

\newtheoremstyle{break}{\topsep}{\topsep}{\itshape}{}{\bfseries}{.}{\newline}{}

\newcommand*{\dupcntr}[2]{%
	\expandafter\let\csname c@#1\expandafter\endcsname\csname c@#2\endcsname
}

\newcommand{\Sref}[1]{\textbf{\textsection\ref{#1}}\xspace}

\theoremstyle{definition}
\newtheorem{example}{Example}[chapter]
\theoremstyle{plain}
\newtheorem{definition}{Definition}[chapter]
\theoremstyle{break}
\newtheorem{definition-break}{Definition}[chapter]
\crefname{definition-break}{Definition}{Definitions}
\dupcntr{definition-break}{definition}

\theoremstyle{definition}
\newtheorem{remark}{Remark}[chapter]
\theoremstyle{definition}
\newtheorem{theorem}{Theorem}[chapter]

\addto\captionsenglish{%
}

\newcommand{\req}{\actN{req}}
\newcommand{\ans}{\actN{ans}}
\newcommand{\cls}{\actN{cls}}

\newcommand{\patReqA}{\recv{\dvV}{\req}}
\newcommand{\patReqB}{\recv{\dvV}{\req}}
\newcommand{\patReqC}{\recv{\dvV}{\req}}
\newcommand{\patReqD}{\recv{\dvVV}{\req}}

\newcommand{\patAns}{\send{\dvV}{\ans}}
\newcommand{\patAnsC}{\send{\dvV}{\ans}}

\newcommand{\symReqA}{\actSN{\patReqA}{\dvV{\neq}j}}
\newcommand{\symReqB}{\actSN{\patReqB}{\ctru}}

\newcommand{\dvVA}{\ensuremath{\dvV^1}}
\newcommand{\dvVB}{\ensuremath{\dvV^2}}
\newcommand{\dvVC}{\ensuremath{\dvV^3}}
\newcommand{\dvVD}{\ensuremath{\dvV^4}}
\newcommand{\dvVE}{\ensuremath{\dvV^5}}
\newcommand{\dvVF}{\ensuremath{\dvV^6}}
\newcommand{\dvVG}{\ensuremath{\dvV^7}}
\newcommand{\dvVH}{\ensuremath{\dvV^8}}

\newcommand{\symReqC}{\actSN{\patReqC}{\dvV{\neq}h}}
\newcommand{\symReqD}{\actSN{\patReqD}{\dvVV{\neq}j}}
\newcommand{\symReqE}{\actSN{\patReqD}{\ctru}}

\newcommand{\symAns}{\actSN{\patAns}{\ctru}}
\newcommand{\symAnsC}{\actSN{\patAnsC}{\ctru}}

\newcommand{\actReq}{\recv{i}{\req}}
\newcommand{\actAns}{\send{i}{\ans}}
\newcommand{\actCls}{\recv{i}{\cls}}

\newcommand{\trnsReqO}{\actSTN{\patReqA}{\dvV{\,=\,}i}{\patReqA}}
\newcommand{\trnsReqA}{\actSTN{\patReqA}{\dvV\neq j}{\patReqA}}
\newcommand{\trnsReqB}{\actSTN{\patReqB}{\ctru}{\actt}}
\newcommand{\trnsReqC}{\actSTN{\patReqA}{\dvV\neq j}{\actt}}
\newcommand{\trnsReqD}{\actSTN{\patReqA}{\ctru}{\patReqA}}

\newcommand{\trnsAns}{\actSTN{\send{\dvV}{\ans}}{\ctru}{\send{\dvV}{\ans}}}

\newcommand{\actReqV}{\recv{\vV}{\req}}
\newcommand{\actReqVV}{\actReqV}
\newcommand{\actAnsV}{\send{\vV}{\ans}}

\newcommand{\procPVdef}{\rec{\mx}{\bigl(\esel{\prf{\actReq}{\prf{\actAns}\mx}}{\prf{\actCls}{\nil}}\bigr)}}
\newcommand{\procPVVdef}{\rec{\mx}{\bigl(\esel{\prf{\actReq}{\prf{\actAns}\mx}{\esel{\,}{\,\prf{\actReq}\mx}}}{\prf{\actCls}{\nil}}\bigr)}}

\newcommand{\hVSF}{\ensuremath{\hV^{\textsf{sf}}}\xspace}
\newcommand{\hVNF}{\ensuremath{\hV^{\textsf{nf}}}\xspace}
\newcommand{\sysREC}{\ensuremath{\sys^\textsf{comb}}\xspace}
\newcommand{\eqqREC}{\ensuremath{\eqq^\textsf{comb}}\xspace}

\newcommand{\eqqUNI}{\ensuremath{\eqq^\textsf{uni}}\xspace}
\newcommand{\eqqUNIP}{\ensuremath{\eqq^\textsf{uni'}}\xspace}

\newcommand{\sysUNI}{\ensuremath{\sys^\textsf{\;uni}}\xspace}

\newcommand{\hVBdef}{\hmax{\hVarX}{(\hnec{\symReqC}\hnec{\symAnsC}\hVarX)\!\hand\!(\hnec{\symReqD}\hnec{\symReqE}\hfls) }}
\newcommand{\hVBdefConc}{\hmax{\hVarX}{(\hnec{\actReq}\hnec{\actAns}\hVarX)\!\hand\!(\hnec{\actReq}\hnec{\actReq}\hfls) }}

\newcommand{\hVdef}{\hmax{\hVarX}{\hnec{\symReqA}(\hnec{\symAns}\hVarX\!\hand\!\hnec{\symReqB}\hfls)}}
\newcommand{\hVdefConc}{\hmax{\hVarX}{\hVdefConcAP}}

\newcommand{\hVdefConcAP}{\hnec{\actReq}\;(\hnec{\actAns}\hVarX\!\hand\!\hnec{\actReq}\hfls)}

\newcommand{\mVdef}{\mrec{\mx}{\bigl(\prf{\patReqA}{\bigl(\mch{\prf{\patAns}{\mx}}{\prf{\patReqB}{\mno}}\bigr)}\bigr)}}

\newcommand{\eVOdef}{\mrec{\mx}{\bigl(\mact{\trnsReqO}{\mrec{\my}{\bigl( \mch{\mact{\trnsReqB}{\my}}{\mact{\trnsAns}{\mx}} \bigr)}}\bigr)}}
\newcommand{\eVAdef}{\mrec{\mx}{\bigl( \mch{\mact{\trnsReqC}{\mx}}{\mact{\trnsAns}{\mx}} \bigr)}}
\newcommand{\eVBdef}{\mrec{\mx}{\bigl(\mact{\trnsReqA}{\mrec{\my}{\bigl( \mch{\mact{\trnsReqB}{\my}}{\mact{\trnsAns}{\mx}} \bigr)}}\bigr)}}
\newcommand{\eVDdef}{\mrec{\mx}{\begin{xbrackets}{c}\mch{\;\;\;\bigl(\mact{\trnsReqD}{\mact{\trnsAns}{{\mx}}} \bigr) \\[2mm]\;}{\;\bigl(\mact{\trnsReqO}{\mact{\trnsReqB}{\mx}} \bigr)} \end{xbrackets}} }

\newcommand{\pVVAdef}{\rec{\mx}{\bigl(\esel{\prf{\actReq}{\prf{\actAns}\mx}{\esel{\,}{\,\underline{\prf{\actReq}\mx}}}}{\prf{\actCls}{\nil}}\bigr)}}

\input{model.sty}

\usepackage{enumerate}
\usepackage{jointphd}
\usepackage{colornames}
\usepackage{graphicx,mathpartir}
\usepackage{amsmath}
\usepackage{amsthm}
\usepackage{tikz, amssymb,bm,color,stmaryrd}
\usepackage{semantic}
\usepackage{mathpartir}
\usepackage{colortbl}
\usepackage{scrextend}
\usepackage{setspace}
\usepackage{subfig}

\usetikzlibrary{calc}
\usetikzlibrary{shapes,arrows,positioning}
\usetikzlibrary{decorations.markings}
\usetikzlibrary{shapes.geometric}
\pgfdeclarelayer{edgelayer}
\pgfdeclarelayer{nodelayer}
\pgfsetlayers{edgelayer,nodelayer,main}

\tikzstyle{simple}=[-, line width=2.000]
\tikzstyle{arrow}=[->,line width=2.000]
\tikzstyle{tick}=[-,postaction={decorate},decoration={markings,mark=at position .5 with {\draw (0,-0.1) -- (0,0.1);}},line width=2.000]
\tikzstyle{none}=[inner sep=0pt]
\tikzstyle{rect}=[rectangle,rounded corners,minimum width=40mm, draw=black,thick, fill=white, minimum height=8mm]
\tikzstyle{shadowRect}=[rectangle,draw=white , rounded corners, fill=darkgray, minimum height=8mm]

\makeatletter
\g@addto@macro \normalsize {%
	\setlength\abovedisplayskip{5pt plus 2pt minus 2pt}%
	\setlength\belowdisplayskip{5pt plus 2pt minus 2pt}%
}

\makeatletter
\renewcommand{\@chapapp}{}

\makeatother

\begin{document}

\title{Developing Theoretical Foundations for Runtime Enforcement}
\firstuniversity{University of Malta}
\seconduniversity{Reykjavik University}
\firstuniversitycrestpath{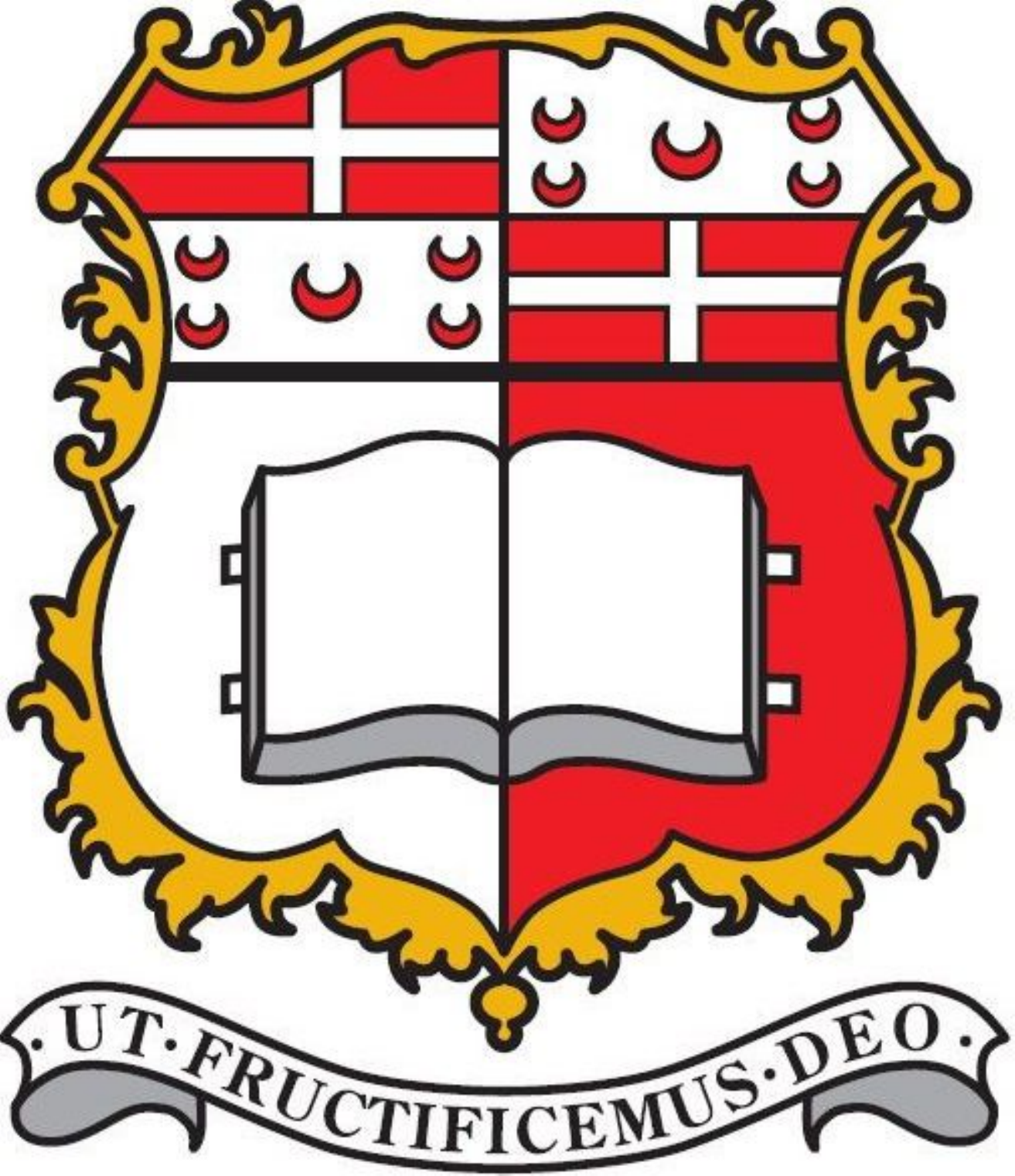}
\seconduniversitycrestpath{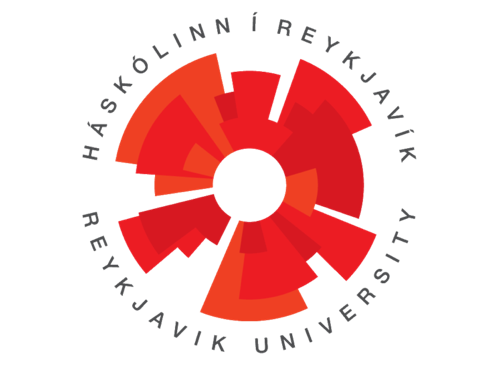}
\reporttype{Transfer Report}
\author{Ian Cassar}
\date{\today}
\supervisors{\!Prof.~Adrian Francalanza, \\
		Prof.~Luca Aceto,\, and \;\; \\
		Prof.~Anna Ing\'{o}lfsd\'{o}ttir}
\submitdate{\today} 
\include{}

\begin{acknowledgements}
	The research work disclosed in this publication is partially supported by the project ``Developing Theoretical Foundations for Runtime Enforcement'' (184776-051) of the Icelandic Research Fund.
\end{acknowledgements}

\begin{abstract}
The ubiquitous reliance on software systems increases the need for ensuring that systems behave correctly and are well protected against security risks. Runtime enforcement is a dynamic analysis technique that utilizes software monitors to check the runtime behaviour of a software system with respect to a correctness specification. Whenever the runtime behaviour of the monitored system is about to deviate from the specification (either due to a programming bug or a security hijack attack), the monitors apply enforcement techniques to prevent this deviation. 

Current Runtime Enforcement techniques require that the correctness specification defines the behaviour of the enforcement monitor itself. This burdens the specifier with not only having to define property that needs to be enforced, but also with having to specify \emph{how} this should be enforced at runtime; we thus relieve the specifier from this burden by resorting to a highly expressive logic. Using a logic we allow the specifier to define the correctness specification as a logic formula from which we can automatically synthesise the appropriate enforcement monitor. 

Highly expressive logics can, however, permit for defining a wide variety of formulae, some of which cannot actually be enforced correctly at runtime. We thus study the enforceability of Hennessy Milner Logic with Recursion (\uhml) for which we identify a subset that allows for defining enforceable formulae. This allows us to define a synthesis function that translates enforceable formulae into enforcement monitors. As our monitors are meant to ensure the correct behaviour of the monitored system, it is imperative that they work correctly themselves. We thus study formal definitions that allow us to ensure that our enforcement monitors behave correctly. 
\end{abstract}

\tableofcontents

\listoffigures


\mainmatter

\chapter{Introduction}
\label{sec:intro}
Modern society is becoming more dependent on software solutions, thus increasing the need for software systems to \emph{behave correctly}. In an ideal world, the correctness of software systems should be entirely verified \emph{pre-deployment} using \emph{static verification} techniques (\eg Theorem Proving or Model Checking). These techniques can statically determine whether a system is \emph{well-behaved}, or not, as specified by a \emph{correctness property}, which is often expressed in terms of an abstract \emph{logic}.

However, as software systems become increasingly larger and more complex, pre-deployment verification becomes exponentially harder due to the \emph{state explosion} problems inherent to static verification techniques \cite{Clarke1999Book,Aceto2007Book}. As a compromise, correctness properties can be \emph{decomposed} \cite{Martinelli2005,Lang2012TACAS,Andersen1995} into smaller parts and verified using a combination of static (pre-deployment) and dynamic (post-deployment) verification techniques. \emph{Runtime Monitoring} is a lightweight, dynamic verification technique in which the correctness of a program is assessed by only analysing the \emph{current execution} \wrt some correctness property. 

In most monitoring settings \cite{Bauer2011,Cassar2017RV,Cassar2015,Francalanza2015Mon}, the correctness property is generally specified as a formula in a \emph{logic} with precise formal semantics, from which a \emph{monitor} is then automatically synthesised. This monitor is essentially the executable software which observes and analyses the runtime execution of a program in relation to the given property, and reacts accordingly. Monitoring is currently gaining interest in verification \cite{Ligatti2005,schneider2000} since it provides a mechanism for verifying and (in certain cases) ensuring correct system behaviour after a system is deployed.

Runtime Enforcement (RE) \cite{Falcone2011,Ligatti2005,Ligatti2010} is a monitoring technique which ensures that the system behaviour is \emph{always} in agreement with the correctness specification. The monitor (\aka the \emph{enforcer}) should therefore be capable of anticipating incorrect behaviour and countering it before it actually happens.
Hence, enforcers are typically designed to act as an \emph{intermediary} which wraps around the system and scrutinises its interactions. During analysis, the enforcers are thus able to \emph{transform} incorrect executions into correct ones by either \emph{suppressing} incorrect events (interactions) exhibited by the system, or by \emph{inserting} events by executing actions on behalf of the system \cite{Ligatti2005,Ligatti2010}.

The execution transformation capabilities of action suppression and insertion were first introduced in \cite{Ligatti2005} in terms of special finite state automata called \emph{Edit-Automata}. However, specifying correctness properties directly in terms of Edit-Automata, burdens the specifier with having to \emph{manually identify} the points in which the enforcer must suppress or insert a specific system action. As identified in earlier work by Bielova \etal \cite{Bielova2011Predictability,Bielova2011PhD}, in most RE approaches there does \emph{not} exist a distinction between the specification and the enforcer, which means that the specifiers must not only reason about \emph{what} property they want to enforce, but also about \emph{how} they should enforce it. Hence, an \emph{algorithmic} enforcer should ideally be automatically derived from a \emph{declarative} correctness property that is expressed as a \emph{logic formula}.

Being able to derive enforcers from logic formulae allows for integrating Runtime Enforcement within a multi-pronged verification approach that combines static and dynamic verification. Such an approach can therefore benefit from the dynamic nature of runtime enforcement to completely ensure correct system behaviour, while minimizing the possibilities of incurring state explosion when applying pre-deployment verification in cases where enforcement might not be possible. This, however, requires understanding the boundaries of \emph{enforceability}, \ie determining which types of properties can actually be enforced at runtime or not.

When used in conjunction to other verification techniques, Runtime Enforcement raises a number of issues related to:
\begin{enumerate}[$(i)$]
	\item the \emph{expressiveness} of the logic used for defining correctness specifications,
	\item the \emph{correctness} of the enforcers themselves, and 
	\item the \emph{implementability} and \emph{feasibility} of enforcing a property at runtime.
\end{enumerate}

In the case of $(i)$, making use of a \emph{highly expressive logic} is important since the more expressive the logic is, the more types of correctness properties one can express. Although some parts of the logic might not be enforceable, when used in combination with other verification techniques, one can employ standard techniques \cite{Martinelli2005,Andersen1995,Lang2012TACAS} to decompose a large (possibly non-enforceable) property into a collection of smaller properties, such that the enforceable ones can be enforced at runtime, while the others are statically verified. Identifying which parts of the logic are \emph{enforceable} is therefore crucial.

The second issue, \ie $(ii)$, stipulates that ensuring a degree of \emph{correctness} about the enforcers is essential, especially since enforcers are often treated as part of the \emph{trusted computing base}. Prior work \cite{Ligatti2005,Bielova2011,Falcone2012,Pinisetty2016} suggests that enforcers must at least be both \emph{Sound} and \emph{Transparent}. A sound enforcer is one which \emph{always} manages to enforce the given property, while a transparent one only applies enforcement when necessary \ie if a system is well-behaved the enforcers should not modify its runtime behaviour, and if they do, the modified behaviour should be somehow equivalent to the original one.

Finally, issue $(iii)$ concerns the fact that most work carried out so far on runtime enforcement has either focussed entirely on its theoretical aspects (\eg \cite{Bielova2011,Falcone2012,Ligatti2005,Ligatti2010,schneider2000}), or else on the implementation aspect (\eg \cite{Bruening2011,Bruening2004PhdDynamoRIO,Charafeddine2014,Pinisetty2017}). To our knowledge, little to no work has been conducted to study the enforceability of a logic, assess whether it is possible to synthesise sound and transparent enforcers that are also implementable, and if so develop an actual implementation that is based on these provably correct enforcers. Assessing whether these enforcement mechanisms can actually be implemented thus enables the understanding of the potential of runtime enforcement \wrt to real world constraints.

Despite these issues, software development and maintenance can however benefit from Runtime Enforcement in various ways. For instance, enforcement techniques provide an excellent way of \emph{ensuring} the correct functionality of \emph{critical systems}, \ie systems which do not afford to misbehave. Such techniques can also be used as a means of \emph{protecting} systems against \emph{security attacks} that attempt to hijack the control flow of the enforced system, \eg enforcers can shield the system by suppressing harmful external stimuli \cite{Bielova2011}, or steer the execution of the system to a more stable state from where the system can be controlled using safe and well-understood procedures \cite{Chen2005}.

\section{Aims and Objectives}\label{sec:intro-aims}
In this report we investigate ways of enabling correctness properties to first be \emph{specified} using a \emph{highly-expressive logic}, which is independent of the verification or enforcement technique, and then be \emph{automatically converted} into an enforcer capable of enforcing the specified property by inserting or suppressing specific system actions as necessary. 

We thus follow the line of research investigated in \cite{Francalanza2015Mon,Francalanza2017FMSD,FraCini2015,Achilleos2018FSTTCS,Achilleos2018Fossacs} where they establish a \emph{correspondence} between a \emph{declarative} model of correctness, \ie the logic, and an \emph{operational} model of correctness, \ie the monitors. We aim to apply this methodology to runtime enforcement, and in turn develop a notion of enforceability, \ie a relation between the meaning of a property expressed as a logic formula, and the ability to enforce it at runtime. Based on this notion, we identify a \emph{maximally expressive subset} of our logic that allows for defining \emph{enforceable properties}.

More concretely, we aim to study the enforceability of properties defined in terms of \emph{Hennessy Milner Logic with recursion} (\uhml) \cite{Aceto2007Book,Larsen1990,Francalanza2015Mon} $-$ a well studied branching time logic. Due to its high expressivity, \uhml can allow for defining a wide variety of properties, some of which might not be enforceable. It is, however, an ideal logic to use in a multi-pronged verification approach, by which a large property can be rewritten into smaller parts and verified (or enforced) using multiple techniques. Moreover, since \uhml is one of the most expressive logics, it embeds other widely used logics and formalisms, such as LTL \cite{Pnueli1977,Bauer2011,Bauer2010}, CTL and CTL* \cite{Clarke2008}. By conducting this study in the context of highly expressive logics, we permit for less expressive logics to also benefit from some results obtained for this logic.  

To our knowledge, the enforceability of properties expressible via a highly expressive logic such as \uhml, has never been studied in depth since no one has yet presented a formal relation between a logic and the existing enforcement mechanisms. To fulfil our aims we therefore subdivide our work in the following 4 objectives:
\begin{enumerate}[\bf O1.]
	\item \textbf{Defining Enforceability and Abstract Enforcers:} To address issue $(i)$ (expressivity), we investigate how \uhml specifications can be synthesised into enforcers. Enforcers will first be defined using automata-based abstractions as this permits the study of runtime enforcement without having to deal with the complexities of a full implementation. This should lead to defining a notion of enforceability \wrt which we can identify the maximally expressive enforceable subset of \uhml specifications.
	\item  \textbf{A Formal Evaluation for Abstract Enforcers:} We aim to prove a number of correctness guarantees about our enforcement enforcers, such as \emph{soundness} and \emph{transparency} \cite{Ligatti2005}, in order to address issue $(ii)$ (correctness), \ie that of guaranteeing that the abstract enforcers \emph{O1} exhibit a level of correctness.
	\item \textbf{Defining Implementable Enforcers:} In preparation for tackling issue $(iii)$ (implementability and feasibility), we intend to develop another synthesis function that converts enforceable \uhml properties into implementable enforcers. To completely address research problem $(ii)$ (\ie correctness) we must ensure that these enforcers follow the formal guarantees proven for the abstract enforcers in \emph{O2}. We thus intend to prove correspondence, ascertaining that the behaviour of our implementable enforcers is equivalent to that described by the abstract enforcers of \emph{O2.} This allows the former to inherit every guarantee proven for the latter.
	\item \textbf{Tool Development and Evaluation:} Finally, we intend to completely address feasibility	by developing a \emph{RE prototype tool} which implements the synthesis of implementable enforcers	introduced in $O3$. Having an actual implementation will permit us to analyse and assess the	performance overheads that the enforcers impose upon the enforced system during runtime.
\end{enumerate} 
%

\section{Report Structure} In this report we discuss the initial investigations addressing our first two objectives $O1$ and $O2$; this is presented in \Cref{sec:enf-model,sec:enf-synthesis}. We outline the rest of the objectives as part of our future work in \Cref{sec:conc}. We structure our document as follows:

\begin{itemize}
	\item In \Cref{sec:preliminaries} we provide the necessary preliminary material required for understanding our novel contributions; in this chapter we thus explain the chosen logic (\uhml), labelled transition systems, detection monitors and monitorability. 
	\item In \Cref{sec:enf-model} we define a formal \emph{runtime enforcement model} capable of \emph{transforming} system events, with the aim of converting invalid system executions into valid ones. We also present novel definitions by which we formally define the meaning of \emph{enforceability}, \ie we define the criteria required for a \uhml formula to be enforceable.	
	\item In \Cref{sec:enf-synthesis} we identify a subset of \uhml formulae that are enforceable via suppressions and establish a \emph{synthesis function} that converts formulae from the identified enforceable subset into the \resp suppression enforcers. To ensure that the synthesised enforcers behave deterministically, we first apply a \emph{normalization algorithm} that converts the given formula into a semantically equivalent normalized formula from which we generate the required enforcer. As means to assess the correctness of our synthesis function, we prove that the synthesised enforcers are deterministic, sound and transparent.	
	\item Finally, in \Cref{sec:background} we compare other related research, and then we conclude in \Cref{sec:conc} with a summary of our contributions and future work.
	\item The Appendix \Cref{sec:app:supporting-material,sec:app:proofs-new-determinization,sec:app:correctness-proofs} \resp provide: additional background material for better understanding the concept of bisimilarity; proofs for lemmas required for the normalization algorithm presented in \Cref{sec:enf-synthesis}; and further proofs for lemmas required for proving that the synthesised enforcers are deterministic, sound and transparent as specified in \Cref{sec:enf-synthesis}.
\end{itemize}

\chapter{Preliminaries}
\label{sec:preliminaries}
In this section we overview preliminary material that is required for understanding the novel work that we will be presenting in the forthcoming chapters.

\section{Runtime Monitoring} \label{sec:runtimeMonitoring}
In general, monitoring can be seen as the empirical observation of the behaviour of some dynamic entity; when applied to software verification, the dynamic entity is the software system being verified, while its behaviour is its runtime execution. In software verification, monitoring is therefore a lightweight compromise for automatically assessing the correctness of a system by observing and analysing its current execution. Runtime monitoring constitutes the basis of several other techniques including Runtime Verification, Adaptation and Enforcement.

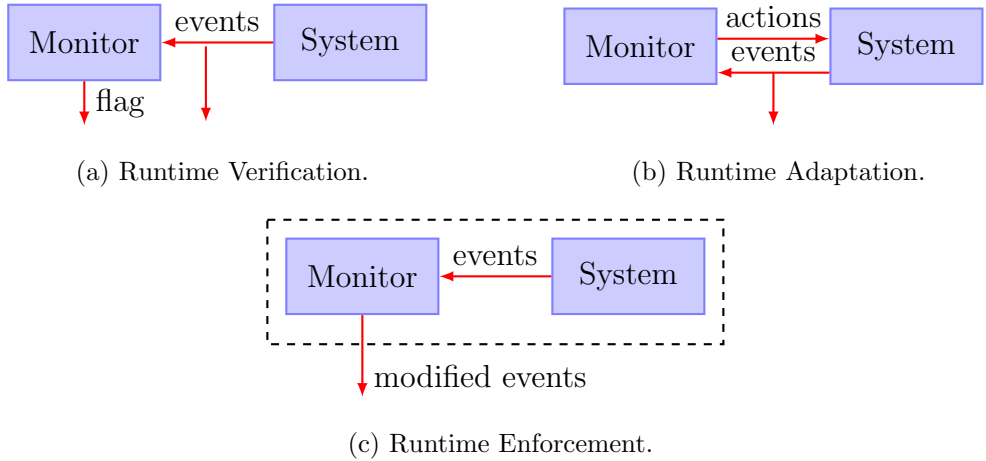
\begin{figure}[t]
	\centering
	\subfloat[Runtime Verification.\label{fig:back:rv}]{

\begin{tikzpicture}[>=latex,auto,thick]
		\begin{scope}[draw=blue!50,fill=blue!20,minimum width=2cm,minimum height=1cm]
		\node (monitor) at ( 0,1) [shape=rectangle,draw,fill] {Monitor};
		\node (system) at (3.5,1) [shape=rectangle,draw,fill] {System};
		\end{scope}
		\node (supervisor) at (0,-0.25) {};
		\node (env) at (1.6,-0.2) {};
		\node (envsrc1) at (1.6,1.1) {};
		\begin{scope}[draw=red,fill=red]
		\draw[->] (system) to node[above]{events} (monitor);
		\draw[->] (envsrc1) -- (env);
		\draw [->] (monitor) to  node {flag} (supervisor);
		\end{scope}		
\end{tikzpicture}} \qquad\qquad
	\subfloat[Runtime Adaptation.\label{fig:back:ra}]{

\begin{tikzpicture}[>=latex,auto,thick]
	\begin{scope}[draw=blue!50,fill=blue!20,minimum width=2cm,minimum height=1cm]
		\node (monitor) at ( 0,1) [shape=rectangle,draw,fill] {Monitor};
		\node (system) at (3.5,1) [shape=rectangle,draw,fill] {System};
	\end{scope}
	\node (env) at (1.75,-0.2) {};
	\begin{scope}[draw=red,fill=red]
		\draw [->] ($(system.west)+(0,-0.35)$) to   node[above] (envsrc2) {events} ($(monitor.east)+(0,-0.35)$);
		\draw [->] ($(monitor.east)+(0,0.1)$) to  node {actions} ($(system.west)+(0,0.1)$);
		\draw[->] (envsrc2) to (env);
	\end{scope}
\end{tikzpicture}}\\
	\subfloat[Runtime Enforcement.\label{fig:back:re}]{
\begin{tikzpicture}[>=latex,auto,thick]
	\begin{scope}[draw=blue!50,fill=blue!20,minimum width=2cm,minimum height=1cm]
		\node (monitor) at ( 0,1) [shape=rectangle,draw,fill] {Monitor};
		\node (system) at ( 3.5,1) [shape=rectangle,draw,fill] {System};
	\end{scope}
	\node (supervisor) at (0,-0.75) {};
	\begin{scope}[draw=red,fill=red]
		\draw[->] (system) to node[above]{events} (monitor);
		\draw [->] (monitor) to  node[near end] {modified events} (supervisor);
	\end{scope}
	\begin{scope}[draw=black, dashed]
		\draw (-1.25,1.75) --(4.75,1.75) -- (4.75,0.1) -- (-1.25,0.1) -- (-1.25,1.75);
	\end{scope}
\end{tikzpicture}}
	\caption{Distinguishing between Runtime Verification, Adaptation and Enforcement}
\end{figure}

In Runtime verification (RV) \cite{Leucker2009,Cassar2017RV} monitors adopt a \emph{passive} role \cite{Bauer2011,Cassar2017Betty} and are exclusively concerned with receiving system events, analysing them, and \emph{detecting} (flagging) violations (or satisfactions) of their respective correctness properties; this is illustrated in \Cref{fig:back:rv}. Hence, RV monitors are capable of \emph{recognising} a (valid or invalid) execution and produce a verdict accordingly, while refraining from directly modifying the system's behaviour in any way.

By contrast, monitors in Runtime Adaptation (RA) \cite{Cassar2016IFM,Kell2008Survey,Cassar2015,JacquesSilva2012} break this passivity by executing adaptation actions after analysing a particular sequence of system events. As shown in \Cref{fig:back:ra}, rather than flagging violations, RA monitors can execute adaptation actions upon \emph{recognising} a specific execution sequence, \eg one which denotes incorrect behaviour. The adaptation actions executed by the RA monitor do not necessarily correct or revert the detected misbehaviour \cite{Rinard2012,Kell2008Survey}; instead they attempt to mitigate its effect by changing certain aspects of the system as it executes, with the aim of preventing either future occurrences of the same error, or of other errors that may potentially occur as a side-effect of the detected violation. RA may also be used to optimise \cite{ibm2005,Kell2008Survey} the system's behaviour based on the information collected by the monitor, \eg switching off redundant processes when under a small load, or increasing processes and load balancing when under a heavy load. 

In Runtime Enforcement (RE) \cite{Falcone2011,Ligatti2005,Ligatti2010} the system behaviour is kept in line with the correctness requirement by anticipating incorrect behaviour and countering it before it actually happens. In RE the monitor (\aka the \emph{enforcer}) 
is typically designed to act as a \emph{proxy} which wraps around the system and analyses its external interactions (see the dotted-line in \Cref{fig:back:re}). The allows the enforcer to \emph{transform} incorrect executions into correct ones by either \emph{suppressing} incorrect events exhibited by the system, or by \emph{inserting} events by executing actions on behalf of the system \cite{Ligatti2005,Ligatti2010}. This contrasts with runtime adaptation, where the monitors may allow violations to occur but then execute remedial actions to mitigate the effects of the violation.

\section{Concrete Events and System Actions} \label{sec:symevents}
Monitors have the task of observing and analysing the behaviour of a given system. System behaviour is generally represented as a stream of observable discrete operations that can be performed by the system. We thus represent these operations as atomic \emph{concrete events}.
Concrete events, $\acta,\actb\in\CEvt$, are used to explicitly represent and identify a single, specific system operation, \eg \recv{i}{3} denotes an atomic \emph{input} operation where a process with identifier $i$ inputs the value $3$, while \send{i}{4} denotes an \emph{output} operation where process $i$ outputs the value $4$. Given that concrete events describe only actual values (\eg \recv{i}{3} describes an actual process id $i$ and value $3$), these events can \emph{easily be distinguished} depending on the \emph{type of operation} they describe, and the \emph{concrete data values} they specify. We thus say that two concrete events are \emph{disjoint} from one another whenever they are \emph{not syntactically equal}, \ie as formally defined below.

\begin{definition}[Disjoint Concrete Events] \label{def:conc-distinct}
	Two Concrete Events \acta and \actb are \emph{disjoint} (\distinctSym) whenever they are \emph{distinct}, \ie
	\begin{displaymath}
		\distinctBi{\acta}{\actb} \defeq \acta\neq\actb \qedhere
	\end{displaymath} 
	We abuse notation and define a list of disjoint concrete events as follows:
	\begin{displaymath}
		 \bigdistinct{i\in\IndSet}\acta_i \defeq \forall i,j\in\IndSet\cdot\acta_i\neq\acta_j
	\end{displaymath} 
	where \IndSet is a set of indices, \ie $\IndSet=\set{1,\ldots, n}$. \bqed
\end{definition}

\begin{example}[Disjoint Concrete Events]
	Consider the following input events \recv{i}{3}, \recv{i}{4} and output event \send{i}{3}. Even though the data specified by events \recv{i}{3} and \send{i}{3} is the same, namely id $i$ and value $3$, these two events differ since they describe a different operation, \ie input (\recv{i}{3}) \vs output (\send{i}{3}). 
	The input events \recv{i}{3} and \recv{i}{4} differ since they define different values, \ie since $3\neq 4$. 
\end{example}
\paragraph{System Actions.} Systems in general can act either in a \emph{verbose} manner by executing operations that are perceivable by external entities, or else, \emph{silently} by executing internal unobservable actions. We thus represent the set of actions that a system can perform as $\actu\!\in\!\Act\!=\!\CEvt\cup\set{\actt}$ which include all the system's \emph{observable} concrete events, $\acta,\actb\!\in\!\CEvt$, along with a distinguished \emph{silent action}, $\actt\notin\CEvt$, denoting \emph{unobservable} internal system operations.

\section{Pattern Matching, Data-binding and Symbolic Events} \label{sec:symevent-match}
Pattern matching allows for \emph{open values} \aka \emph{patterns}, $\pate,\patee\!\in\!\Pat$, defining \emph{data variables}, to be compared to \emph{closed values} such as concrete events.  As defined below pattern \pate serves to define the \emph{operation type} (\eg input or output operation), along with \emph{concrete values} (including process identifiers, $i,j\!\in\!\Pid$, or generic data, $\vV,\vVV\!\in\!\Val$), or \emph{variables}, $\dvV,\dvVV,\dvVVV\!\in\!\Var$. 
\begin{definition}[Patterns] \normalfont 
	\[
	\begin{array}{rclcl} 
		\pate,\patee&\in&\Pat &\bnfdef& \recv{\delta\,}{\,\gamma} \;\; \text{(Input)} \quad \vert \quad \send{\delta\,}{\,\gamma} \;\; \text{(Output)}\\[-1mm]
		\delta&\in&\SymId &\bnfdef& \Var \quad\vert\quad \Pid \\[-1mm]
		\gamma&\in&\Data &\bnfdef& \Var \quad\vert\quad \Val 
	\end{array} 	
	\]\\[-10mm]\bqed
\end{definition}

For instance, pattern $\pate=\recv{i}{\dvV}$ describes the set of all input operations that can be performed by a process $i$, \ie pattern $\recv{i}{\dvV}$ describes the following set of concrete events $\set{\ldots,\recv{i}{1},\recv{i}{2},\ldots}$; hence this pattern specifies that a concrete process $i$ may input any value $\dvV$, where $\dvV$ is a variable. When referring to a symbolic pattern, \pate, defining an arbitrary system operation (\ie input or output) ranging over variables $x_0\ldots x_n$, we abuse notation and use $\pate(x_0\ldots x_n)$.

\subsection{Pattern Matching} \label{sec:ptrn-match}
We follow the standard way of representing pattern matching functionality in terms of the function \mtch{\pate}{\,\acta\!\!\!}. As stated in \Cref{def:ptrn-match}, this function matches a (possibly) open pattern (\ie defining data variables) with a concrete event \acta, and upon a \emph{successful match} returns \emph{substitution environment} \s, where \s defines a bijective function which maps the variables, $\dvV,\dvVV\in\Vars$, defined in the pattern \pate to the respective values, $\vV,\vVV\in\Vals$, defined in the matching concrete event \acta.

\begin{definition}[Pattern Matching]\label{def:ptrn-match} 
	Given a pattern \pate and a concrete event \acta,
	\renewcommand{\qedsymbol}{\ensuremath{\blacksquare}}
	\begin{displaymath}
	\pushQED{\bqed}
		\mtch{\pate}{\acta}=\s \;\text{ such that }\; \pate\s=\acta \qedhere
	\popQED
	\end{displaymath} 
\end{definition}

\begin{example}[Pattern Matching] Consider the following pattern matching applications:
	\begin{align}
	\label{eq:1}
	\mtch{\,\recv{\dvV}{\dvVV}\,}{ \recv{i}{3} \,} &\; =\; \set{\dvV\mapsto i, \dvVV\mapsto 3} \\
	\label{eq:2}
	\mtch{\,\recv{i}{3}\,}{ \recv{i}{3}\,} &\; =\; \set{} \\
	\label{eq:3}
	\mtch{\, \recv{i}{\dvVV}\,}{ \recv{j}{3} \,}  & \; =\; \sundef \\
	\label{eq:4}
	\mtch{\,\recv{\dvV}{\dvVV}\,}{ \send{i}{3} \,}  & \; =\; \sundef
	\end{align}
	In (\ref{eq:1}) the input pattern $\recv{\dvV}{\dvVV}$ is successfully matched with the concrete input event \recv{i}{3}, where $\dvV$ and $\dvVV$ are pattern matched with the values $i$ and $3$ \resp In (\ref{eq:2}) the two concrete events are matched (exactly), returning the empty substitution. The mismatch in (\ref{eq:3})  is due to mismatching identifiers of the input events \ie $i$ \vs $j$, whereas the mismatch in (\ref{eq:4}) is because the input pattern $\recv{\dvV}{\dvVV}$ cannot be matched with actions defining a different operation, \eg output event \send{i}{3}. \bqed
\end{example}

\paragraph{Pattern Variants.} A pattern is said to be \emph{fully closed} if it only defines \emph{concrete values}, \eg \recv{i}{3}, while a \emph{fully open} pattern \emph{does not define any concrete value}, \ie defines data variables only, \eg \recv{\dvV}{\dvVV}.

\paragraph{Pattern Equivalence.} We identify \emph{patterns} up to the consistent renaming of their variables, \ie two patterns $\actS_1$ and $\actS_2$ are \emph{equivalent} if there exists a bijective relation $\s:\Vars\mapsto\Vars$, such that $\actS_1=\s\actS_2$, \eg  $\pate(\dvV_0,\ldots,\dvV_n)$ and $\pate(\dvVV_0,\ldots,\dvVV_n)$ are equivalent since  $\pate(\dvVV_0,\ldots,\dvVV_n)[\subE{\dvV_0}{\dvVV_0},\ldots,\subE{\dvV_n}{\dvVV_n}]$ becomes \emph{syntactically equal} to $\pate(\dvV_0,\ldots,\dvV_n)$. Hence, two equivalent patterns can match the \emph{same} set of concrete events and can thus define the exact same set of concrete events as shown in the definition below.

\begin{definition}[Equivalence] Patterns $\pate_1$ and $\pate_2$ are equivalent ($\equiv$) whenever
	\renewcommand\qedsymbol{\ensuremath{\blacksquare}}
	\[	
	\pushQED{\qed}
	\pate_1\equiv\pate_2 \;\defeq\; \syn{\pate_1}\!=\!\syn{\pate_2} \qedhere
	\popQED
	\]  
\end{definition}

\subsection{Symbolic Events}
A generalization of concrete events can be attained through \emph{Symbolic Events}, $\actS\!\in\!\SEvt$, denoting a \emph{set of concrete events} specified by a \emph{fully open pattern}, $\pate\!\in\!\Pat$, (as defined earlier) and a \emph{filtering condition} $\predc\!\in\!\Cond$; a symbolic event \actS is thus defined as $\actS\!=\!\actSN{\pate}{\predc}$. 

Filtering conditions specified in symbolic events represent a \emph{decidable predicate}, $\predc$, ranging over the variables, $\dvV_0,\ldots,\dvV_n$, defined in the respective pattern $\pate(\dvV_0,\ldots,\dvV_n)$. Once again we abuse notation and use $\predc(\dvV_0\ldots \dvV_n)$ to denote any condition that analyses the values bound to variables $\dvV_0\ldots \dvV_n$. Finally, we define condition evaluation as follows
\begin{definition}[Condition Evaluation] Given a \emph{closed, decidable condition} \predc,
	\normalfont
	\begin{itemize}
		\item \ceval{\predc}{\ctru} \quad \textsl{ iff } \predc evaluates to \emph{true}.
		\item \ceval{\predc}{\cfls} \quad \textsl{ iff } \predc evaluates to \emph{false}.\bqed
	\end{itemize}
\end{definition}
For example, by using these filtering conditions we can restrict the range of input operations described by $\pate\!=\!\recv{\dvV}{\dvVV}$ via condition $\predc\!=\!(\dvV{=}i\land\dvVV{\ge}10\land \dvVV{\le}15)$ such that the resultant symbolic event \actSN{\pate}{\predc}, defines a set containing every concrete input event (as stated by pattern \recv{\dvV}{\dvVV}) that is performed by a process with identifier $i$ in which the input value is between 10 and 15 (as specified by conditions $\dvV{=}i$ and $\dvVV{\ge}10\land \dvVV{\le}15$ \resp), such that the resultant set is $\set{\recv{i}{10},\ldots,\recv{i}{15}}$. 

As specified by the denotational semantics given in \Cref{def:sym-sem} (below), a symbolic event $\actS\!=\!\actSN{\pate}{\predc}$ thus defines a set containing every concrete event which matches pattern \pate and satisfies condition \predc as a result of the pattern match.

\begin{definition}[Denotational Semantics for Symbolic Events] \label{def:sym-sem}
	For an arbitrary symbolic event $\actS=\actSN{\pate}{\predc}$, 
	\begin{displaymath} \normalfont
	\syn{\actSN{\pate}{\predc}} \defeq \Setdef{\,\acta\,}{\forall\acta\cdot\mtch{\pate}{\acta}\!=\!\s \textsl{ and } \ceval{\predc\s}{\ctru}}
	\end{displaymath} 
	We therefore say that a concrete event \acta is \emph{an element of} a symbolic event $\actS\!=\!\actSN{\pate}{\predc}$ whenever  $\acta\!\in\!\syn{\actS}$, \ie when \acta \emph{pattern matches} \pate  creating substitution \s as a result (\ie $\mtch{\pate}{\acta}=\s$), such that when \s is applied to the filtering condition \predc, this evaluates to \emph{true} (\ie $\ceval{\predc\s}{\ctru}$). 
	
	Alternatively, we use notation $\mtchS{\actS}{\acta}$ (as defined below) whenever we need to be aware about the substitution environment, \s, that is obtained when \acta matches the pattern \pate of $\actS$ and satisfies the associated filtering condition \predc. 
	\begin{align*}
	\mtchS{\actS}{\acta} & \defeq 
	\begin{xbrace}{cl}
	\s & \qquad \textsl{when } \mtch{\pate}{\acta}\!=\!\s\land\ceval{\predc\s}{\ctru} \\
	\sundef & \qquad \textsf{otherwise}
	\end{xbrace} 
	\end{align*}  \\[-15mm]\bqed 
\end{definition}

\paragraph{Shorthand Notations} Although we assume that symbolic events define fully opened patterns, \ie $\pate(\dvV_0,\ldots,\dvV_n)$, we adopt a shorthand notation and write \actSN{\pate(\vV_0,\ldots,\vV_n)}{\ctru} in lieu of \actSN{\pate(\dvV_0,\ldots,\dvV_n)}{\dvV_0{=}\vV_0\land\ldots\dvV_n{=}\vV_n}.
\begin{example} \label{ex:sym-events-shorthand} Consider event $\actSN{\recv{\dvV}{\dvVV}}{\dvV{=}i\land\dvVV{=}\req}$, as a shorthand we can represent this as \actSN{\recv{i}{\req}}{\ctru}, since conditions $\dvV{=}i$ and $\dvVV{=}\req$ evaluate to false in case \dvV and \dvVV are matched with some other value. \bqed
\end{example}
Moreover, whenever the filtering condition \predc evaluates to true, we simply write \pate instead of \actSN{\pate}{\ctru}. We will be using these shorthand notations interchangeably throughout this report.

\paragraph{Singleton Symbolic Events} As specified in \Cref{def:sym-sem}, a symbolic event \actSN{\pate}{\predc} denotes a set of 0 or more concrete events. A symbolic event is said to be \emph{singleton} whenever it denotes a set containing a \emph{single concrete event}.
\begin{example}[Singleton Symbolic Event] Recall from \Cref{ex:sym-events-shorthand} that \actReq is a shorthand for symbolic event \actSN{\recv{\dvV}{\dvVV}}{\dvV{=}i\land\dvVV{=}\req}. This symbolic event is a \emph{singleton event} since \syn{\actSN{\recv{\dvV}{\dvVV}}{\dvV{=}i\land\dvVV{=}\req}}{=}\set{\actReq} where the resultant set of the denotation only contains concrete event \actReq. \bqed
\end{example}

\paragraph{Distinguishing between Symbolic Events.} \label{sec:symevent-equiv}
Contrary to concrete events, distinguishing between symbolic events is, however, not quite straight forward. As an example, consider events \actSN{\recv{\dvV}{3}}{\dvV\!\neq\!j} and \actSN{\recv{i}{\dvVV}}{\dvVV{>}2}, although these events are syntactically different, concrete events such as \recv{i}{3} is an element of both \actSN{\recv{\dvV}{3}}{\dvV\!\neq\!j} and \actSN{\recv{i}{\dvVV}}{\dvVV{>}2}, since it defines an output action \recv{(}{)} and the concrete values $i$ and $3$ which match and satisfy the patterns and conditions of both symbolic events. By \Cref{def:sym-sem} we thus know that the sets of concrete events denoted by these two symbolic events intersect with one another. Hence, as defined by \Cref{def:sym-distinct}, symbolic events are said to be \emph{disjoint} whenever the sets of concrete events they denote are \emph{disjoint}. This happens whenever they define, either:
\begin{enumerate}[$(i)$]
	\item a \emph{different operation}, \eg \actSN{\recvB{\dvV_1}{\dvVV_1}}{\predc_1} and \actSN{\sendB{\dvV_2}{\dvVV_2}}{\predc_2};
	\item \emph{conflicting concrete data}, \eg \actSN{\recv{\pmb{i}}{\dvVV_1}}{\predc_1} and \actSN{\recv{\pmb{j}}{\dvVV_1}}{\predc_1}, where $i\neq j$; or
	\item conditions which \emph{contradict} each other, eg \actSN{\recv{i}{\dvVV}}{\pmb{\dvVV{>}10}} and \actSN{\recv{i}{\dvVV}}{\pmb{y\!\leq\!10}}.
\end{enumerate}

\begin{definition}[Disjoint Symbolic Events] \label{def:sym-distinct}
	Two Symbolic Events $\actS_1$ and $\actS_2$ are \emph{disjoint} whenever:
	\[	
		\distinctBi{\actS_{1}}{\actS_{2}} \defeq \syn{\actS_1}\cap\syn{\actS_2}=\emptyset
	\]
	Once more we abuse this notation to define a list of disjoint symbolic events as follows:
	\[	
		\bigdistinct{i\in\IndSet}\actS_i \defeq \bigcap_{i\in\IndSet}\syn{\actS_i}=\emptyset
	\]  
	where \IndSet is a set of indices, \ie $\IndSet{\,=\,}\set{1,\ldots,n}$. \bqed
\end{definition}

\begin{example}[Disjoint Symbolic Events]
	\newcommand{\act}{\recv{i}{3}}
	\newcommand{\patA}{\recv{\dvV}{3}}
	\newcommand{\patB}{\recv{i}{\dvVV}}
	\newcommand{\patC}{\recv{i}{\dvVVV}}	
	\newcommand{\predA}{\dvV\!\neq\!j}
	\newcommand{\predB}{\dvVV\!>\!2}
	\newcommand{\predC}{\dvVVV\!\leq\!2}
	\newcommand{\actA}{\actSN{\patA}{\predA}}
	\newcommand{\actB}{\actSN{\patB}{\predB}}
	\newcommand{\actC}{\actSN{\patC}{\predC}}
	
	Consider events \actA, \actB and \actC. Notice that the sets denoted by the first two events, \actA and \actB, intersect with each other on event \recv{i}{3} as shown by \eqref{ex:sym-distinct-eq1}, meaning that they are \emph{not} disjoint.
	\begin{align}
		\syn{\actA}\cap\syn{\actB}&=\set{\act} \label{ex:sym-distinct-eq1} \\
		\syn{\actA}\cap\syn{\actC}&=\emptyset \label{ex:sym-distinct-eq2} \\
		\syn{\actB}\cap\syn{\actC}&=\emptyset \label{ex:sym-distinct-eq3}
	\end{align}
	By contrast, \eqref{ex:sym-distinct-eq2} shows that even though patterns \patA and \patB can match the \emph{same} concrete event \act, symbolic event \actA is \emph{disjoint} from \actC (and vice-versa) since this concrete event does not satisfy the filtering condition of \actC, \ie $(\predC)\sub{3}{\dvVVV} \,\equiv\,3\!\leq\!2=\textsl{false}$. Similarly, \eqref{ex:sym-distinct-eq3} states that events \actB and \actC are disjoint from each other even though their patterns can match the \emph{exact same} set of concrete events. They are, however, guaranteed to be disjoint as they define \emph{contradicting} filtering conditions, \ie $\nexists n\cdot n\!>\!2\land n\!\leq\!2$. \bqed
\end{example}



\section{Labelled Transition Systems} \label{sec:lts}
Labelled Transition Systems (LTSs) provide a convenient framework for defining an operational description of the behaviour of a system. An LTS, is a triple $\langle\Proc,\Act,\rightarrow\rangle$ which is composed from:
\begin{enumerate}[$(i)$]
	\item a set of \emph{states}, $\pV,\pVV,\pVVV\in\Proc$, corresponding to processes;
	\item a set of \emph{actions}, $\actu,\actg\in\Act$, which include all the system's \emph{observable} ($\acta,\actb\in\CEvt$) and \emph{internal} $\actt\notin\Evt$ actions; and
	\item a \emph{transition} relation, $\reduc\;\subseteq(\Proc\times\Act\times\Proc)$ that relates the state on the RHS, to the other state on the LHS of the transition via a unidirectional reduction from left to right over a specific action, \eg $(\pV,\acta,\pV')\in\reduc$ describes a unidirectional reduction from state \pV to $\pV'$ over action \acta.
\end{enumerate} 

\begin{figure}[t]
	\textbf{Syntax}
	\begin{align*}
	\pV,\pVV,\pVVV\in\Proc & \bnfdef \nil && (\text{inaction}) &&
	 \bnfsepp \prf{\actu}{\pV} && (\text{prefixing}) &&
	 \bnfsepp \mCh\,\pV_i && (\text{choice}) \\ 
	& \bnfsepp \rec{x}{\pV} && (\text{recursion}) && 
	 \bnfsepp x && (\text{rec. variable})
	\end{align*}
	\textbf{Dynamics}
	\begin{mathpar}
		\inference[\rtit{Act}]{}{\prf{\actu}{\pV}  \traSS{\actu} \pV} 
		\and
		\inference[\rtit{Sel}]{\pV_j \traSS{\actu} \pVV_j}{\mCh\,\pV_i \traSS{\actu} \pVV_j}[$j\in\IndSet$] 
		\and
		\inference[\rtit{Rec}]{\pV\sub{\rec{\mx}{\pV}}{\mx} \traSS{\actu} \pV'}{\rec{\mx}{\pV} \traSS{\actu} \pV'}  
	\end{mathpar}	
	\caption{A Model for describing Systems}
	\label{fig:lang}
\end{figure}

For convenience, we define processes, \Proc, using the \emph{regular fragment} of CCS \cite{Milner1992CCS} as defined by the syntax in \Cref{fig:lang}.   Assuming a specific set of (visible) concrete events, $\acta$ and a denumerable set of (recursion) variables $\mx,\my,\mz\in\Vars$, processes are defined as either the \emph{inactive process} \nil, an \emph{action-prefixed process} \prf{\actu}{\pV} \ie prefixed by an action $\actu$ where \actu is either a visible (\acta) or a silent (\actt) action, a \emph{mutually-exclusive choice} amongst processes where $\mCh\,\pV_i$ sums up the processes identified by the unique indices in \IndSet (such that $\mCh\,\pV_i$ represents $\mch{\pV_1}{\mch{\ldots}{\pV_n}}$, where $1,\ldots,n\in\IndSet$), or a \emph{recursive process} where \rec{x}{\pV} acts as a \emph{binder} for $\mx$ in $\pV$. We work up to alpha-conversion of bound recursion variables and assume that all recursive processes are guarded, meaning that all occurrences of bound recursion variables occur under an action prefix (either directly or indirectly). Closed terms are processes where all occurrences of recursion variables are bound.


When describing the dynamic behaviour of processes, we use the more intuitive notation $\pV\traS{\actu}\pV'$ in lieu of $(\pV,\actu,\pV')\in\reduc$, along with notation $\pV\ntraS{\actu}$ to denote $\lnot(\exists\pV'\cdot\pV\traS{\actu}\pV')$. The rules in \Cref{fig:lang} are standard. Rule \rtit{Act} allows a \actu-prefixed process $\prf{\actu}{\pV}$ to reduce over action \actu to the derivative \pV, \ie $\prf{\actu}{\pV}\traS{\actu}\pV$, while by rule \rtit{Sel}, a choice process, $\mCh\,\pV_i$ reduces to $\pVV_j$ whenever there exists a process identified by some index $j\in\IndSet$ in this summation, \ie $\pV_j$ which performs a \actu-transition, \ie $\pV_j\traS{\actu}\pVV_j$ (\resp $\pVV\traS{\actu}\pVV'$). By rule \rtit{Rec}, a recursive process $\rec{\mx}{\pV}$ can reduce to $\pV'$ over action \actu, \ie $\rec{\mx}{\pV} \traSS{\actu} \pV'$, when its unfolded version, $\pV\sub{\rec{\mx}{\pV}}{\mx}$, reduces to $\pV'$ over \actu, \ie $\pV\sub{\rec{\mx}{\pV}}{\mx} \traSS{\actu} \pV'$.

We also employ the usual notation $\pV \wreduc \pV'$ and $\pV \wtraS{\acta} \pV'$ to denote weak transitions representing $\pV (\traS{\actt})^{\ast} \pV'$ and $\pV \wreduc\cdot\traS{\acta}\cdot\wreduc \pV'$ \resp, referring to $\pV'$ as a \actu-derivative of $\pV$. We also employ notation $\pV\wtraS{\hat{\actu}}\pV'$ to collectively refer to $\pV \wreduc \pV'$ and $\pV \wtraS{\acta} \pV'$. Sequences of visible actions are expressed as traces $\tr,\trr \in \CEvt^\ast$, such that sequences of transitions are defined as $\pV\wtra{\acta_1}\ldots\wtra{\acta_n} p_n$ as $\pV \wtraS{\tr} p_n$, where a trace $\tr = \acta_1,\ldots,\acta_n$. For more details, consult standard texts such as \cite{Milner1992CCS,Aceto2007Book}.

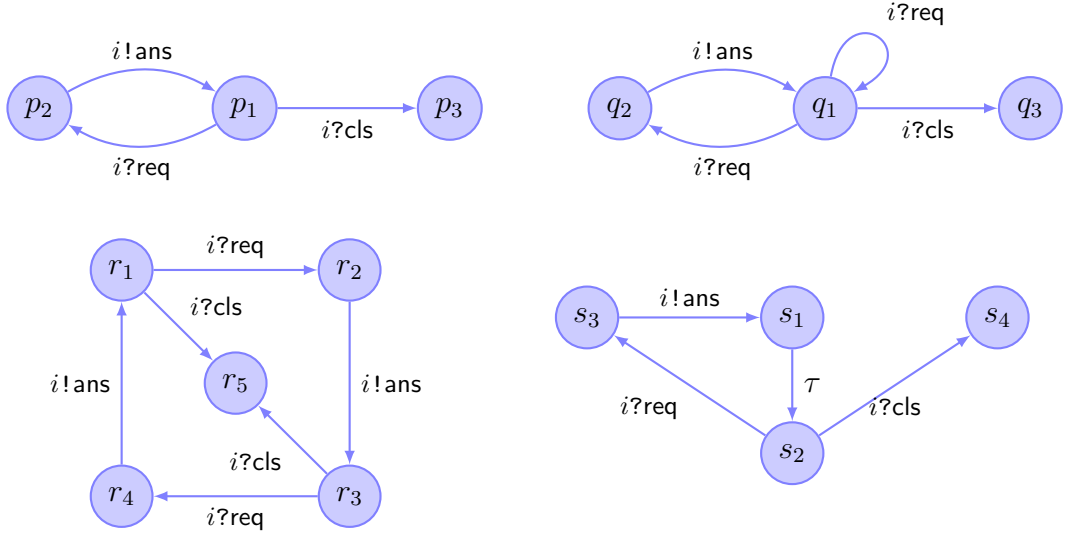
\begin{figure}[t]
	\begin{center}
		\begin{tikzpicture}[>=latex,auto,thick, scale=0.9]
			\begin{scope}[draw=blue!50,fill=blue!20,minimum size=0.5cm]
				\node (p1) at ( 0,0) [shape=circle,draw,fill] {$\pV_1$};
				\node (p2) at (-3,0) [shape=circle,draw,fill] {$\pV_2$};
				\node (p3) at ( 3,0) [shape=circle,draw,fill] {$\pV_3$};
				\node (p4) at ( 4.5,0)  {\phantom{$\pV_4$}};
				\end{scope}
				\begin{scope}[draw=blue!50,fill=blue!50]
				\draw[->] (p1) [bend left=30] to node {\footnotesize\actReq} (p2); 
				\draw[->] (p2) [bend left=30] to node {\footnotesize\actAns} (p1); 
				\draw[->] (p1) to node[below] {\footnotesize\actCls} (p3);
			\end{scope}
		\end{tikzpicture}	
		\begin{tikzpicture}[>=latex,auto,thick, scale=0.9]
			\begin{scope}[draw=blue!50,fill=blue!20,minimum size=0.5cm]
			\node (q1) at ( 0,0) [shape=circle,draw,fill] {$\pVV_1$};
			\node (q2) at (-3,0) [shape=circle,draw,fill] {$\pVV_2$};
			\node (q3) at ( 3,0) [shape=circle,draw,fill] {$\pVV_3$};
			\end{scope}
			\begin{scope}[draw=blue!50,fill=blue!50]
			\draw[->] (q1) [in=30,out=80,loop] to node {\footnotesize\actReq} (q1); 
			\draw[->] (q1) [bend left=30] to node {\footnotesize\actReq} (q2); 
			\draw[->] (q2) [bend left=30] to node {\footnotesize\actAns} (q1); 
			\draw[->] (q1) to node[below] {\footnotesize\actCls} (q3);
			\end{scope}
		\end{tikzpicture}
		\begin{tikzpicture}[>=latex,auto,thick, scale=0.75]
			\begin{scope}[draw=blue!50,fill=blue!20,minimum size=0.5cm]
			\node (r1) at ( 0,2) [shape=circle,draw,fill] {$\pVVV_1$};
			\node (r2) at ( 4,2) [shape=circle,draw,fill] {$\pVVV_2$};
			\node (r3) at ( 4,-2) [shape=circle,draw,fill] {$\pVVV_3$};
			\node (r4) at ( 0,-2) [shape=circle,draw,fill] {$\pVVV_4$};
			\node (r5) at ( 2,0) [shape=circle,draw,fill] {$\pVVV_5$};
			\node (r6) at ( 7,3)  {\phantom{$\pVV_7$}};
			\end{scope}
			\begin{scope}[draw=blue!50,fill=blue!50]
			\draw[->] (r1) to node {\footnotesize\actReq} (r2); 
			\draw[->] (r2) to node {\footnotesize\actAns} (r3); 
			\draw[->] (r3) to node {\footnotesize\actCls} (r5);
			
			\draw[->] (r3) to node {\footnotesize\actReq} (r4); 
			\draw[->] (r4) to node {\footnotesize\actAns} (r1); 
			\draw[->] (r1) to node {\footnotesize\actCls} (r5);
			\end{scope}
		\end{tikzpicture}
		\begin{tikzpicture}[>=latex,auto,thick, scale=0.9]
		\begin{scope}[draw=blue!50,fill=blue!20,minimum size=0.5cm]
		\node (n0) at ( 0,0) [shape=circle,draw,fill] {$\pVVVV_1$};
		\node (n1) at ( 0,-2) [shape=circle,draw,fill] {$\pVVVV_2$};
		\node (n2) at (-3,0) [shape=circle,draw,fill] {$\pVVVV_3$};
		\node (n3) at ( 3,0) [shape=circle,draw,fill] {$\pVVVV_4$};
		\node (n4) at ( 0,-3)  {\phantom{$\pVVVV_3$}};
		\end{scope}
		\begin{scope}[draw=blue!50,fill=blue!50]
		\draw[->] (n0)  to node [right] {\actt} (n1); 
		\draw[->] (n1)  to node {\footnotesize\actReq} (n2); 
		\draw[->] (n2)  to node {\footnotesize\actAns} (n0); 
		\draw[->] (n1) to node[below] {\footnotesize\actCls} (n3);
		\end{scope}
		\end{tikzpicture}	
	\end{center}	
	\caption{A depiction of the system in \Cref{ex:lts}}
	\label{fig:ex:lts}
\end{figure}


\begin{example}
	\label{ex:lts}
	Consider a (reactive) system that acts as a server that is identified with process id $i$, and which repeatedly accepts \emph{requests} and subsequently responds by outputting an \emph{answer}, with the possibility of terminating through the special \emph{close} request (\actN{cls}). Such a system may be expressed as the following process, $\pV_1$.
	\begin{displaymath}
		\pV_1=\procPVdef
	\end{displaymath}
	We can outright notice that process $\pV_1$ is designed to output an answer (\actAns) for every input request (\actReq). The \emph{same} behaviour can also be represented by processes $\pVVV_1$ and $\pVVVV_1$ (below), which differ from $\pV_1$ since $\pVVV_1$ is an \emph{unfolded} version of $\pVVV_1$, while $\pVVVV_1$ performs an internal \actt action before inputting a request.
	\begin{align*}
		\pVVV_1\defeq&\;\rec{\mx}{\bigl(\esel{\prf{\actReq}{ \prf{\actAns}{\bigl(\underline{\esel{\prf{\actReq}{\prf{\actAns}{\mx}}}
						{\prf{\actCls}{\nil}}}\bigr) }}}{\prf{\actCls}{\nil}}\bigr)} \\
		\pVVVV_1\defeq&\;\rec{\mx}{\bigl(\esel{\prf{\underline{\,\actt\,}}{ \prf{\actReq}{ \prf{\actAns}\mx}}}{\prf{\actCls}{\nil}}\bigr)}
	\end{align*}
	By contrast, process $\pVV_1$ denotes a server that, although similar to $\pV_1$, can also \emph{non-deterministically} refuse to answer for a given request.
	\begin{displaymath}
		\pVV_1=\pVVAdef
	\end{displaymath}
	Pictorially, the \resp LTSs denoted by processes $\pV_1$, $\pVV_1$, $\pVVV_1$ and $\pVVVV_1$ may be represented by the graphs in \Cref{fig:ex:lts}, where the nodes correspond to processes and the arcs correspond to transitions, $\traS{\actu}$. 
\end{example}

\paragraph{Bisimilarity of LTSs}
Intuitively, a \emph{bisimulation} \cite{Aceto2007Book,Milner1992CCS} is a binary relation that associates the behaviour of two Labelled Transition Systems that \emph{exhibit the same behaviour}, \ie in a way that one system \emph{simulates} the other and vice versa. Variants of bisimulation includes \emph{Strong and Weak Bisimulation} (see \Cref{def:strong-bisim,def:weak-bisim}, below). 

\begin{definition}[Strong Bisimulation]\label{def:strong-bisim}
	\renewcommand{\qedsymbol}{ \;}
	A binary relation \R is a \emph{Strong Bisimulation relation} whenever $(\pV,\pVV)\in\R$, such that
	\begin{enumerate}[(a)]
		\item if $\pV\traS{\actu}\pV'$ then there exists a \emph{strong transition} $\pVV\traS{\actu}\pVV'$ such that $(\pV',\pVV')\in\R$  \qed
		\item if $\pVV\traS{\actu}\pVV'$ then there exists a \emph{strong transition} $\pV\traS{\actu}\pV'$ such that $(\pV',\pVV')\in\R$ \bqed
	\end{enumerate}
\end{definition}

Hence, for a pair of processes, $(\pV,\pVV)$, to be in a \emph{Strong Bisimulation relation}, it is required that each process is able to simulate both visible (\acta) and internal (\actt) actions of the other process, \eg if $\pV\traS{\acta}\pV'$ then $\pVV\traS{\acta}\pVV'$ and if $\pVV\traS{\actt}\pVV'$ then $\pV\traS{\actt}\pV'$. By contrast, \emph{Weak Bisimulation} abstracts away from internal (\actt) actions, \ie each process is only required to simulate the visible actions of the other process, and can thus simulate visible actions after performing a zero or more internal (\actt) transitions, \eg if $\pV\traS{\acta}\pV'$ then $\pVV\traS{\actt}^{\ast}\traS{\acta}\traS{\actt}^{\ast}\pVV'\,\equiv\,\pVV\wtraS{\acta}\pVV'$ and if $\pVV\traS{\actt}\pVV'$ then $\pV\traS{\actt}^{\!0}\!\!\pV'$.

\begin{definition}[Weak Bisimulation]\label{def:weak-bisim}
	\renewcommand{\qedsymbol}{ \;}
	A binary relation \R is a \emph{bisimulation relation} whenever $(\pV,\pVV)\in\R$, such that
	\begin{enumerate}[(a)]
		\item if $\pV\traS{\actu}\pV'$ then there exists a \emph{weak transition} $\pVV\wtraS{\hat{\actu}}\pVV'$ such that $(\pV',\pVV')\in\R$ \qed
		\item if $\pVV\traS{\actu}\pVV'$ then there exists a \emph{weak transition} $\pV\wtraS{\hat{\actu}}\pV'$ such that $(\pV',\pVV')\in\R$ \bqed
	\end{enumerate} 
\end{definition}

\paragraph{Proving Bisimilarity of LTS processes} 
Since \emph{Strong (}\resp\emph{Weak) Bisimilarity} is the \emph{largest} Strong (\resp Weak) Bisimulation relation, proving that two LTS processes \pV and \pVV are Strong (\resp Weak) Bisimilar, \ie $\pV\bisim\pVV$ (\resp $\pV\wbisim\pVV$), only requires showing that there exists a Strong (\resp Weak) bisimulation relation \R that can relate them as stated by \Cref{def:strong-bisim,def:weak-bisim}. Note that since Strong Bisimilarity is \emph{stricter} than its Weak counterpart, two Strong bisimilar processes are inherently Weak bisimilar as well.

However, proving the contrary \ie $\pV{\not\bisim}\pVV$ (\resp $\pV{\not\wbisim}\pVV$) is not as straight forward. In fact, to show that a process \pV is \emph{not} Strong (\resp Weak) bisimilar to \pVV, we must show that \emph{every} binary relation that exists between \pV and \pVV \emph{do not satisfy} \Cref{def:strong-bisim} (\resp \Cref{def:weak-bisim}), and hence do not constitute towards valid bisimulation relations. Given the lack of practicality of this exhaustive approach, a \emph{game characterization for Strong (\resp Weak) bisimulation} \cite{Aceto2007Book,Stirling1996} is generally employed to prove that two LTS processes are either \emph{bisimilar} ($\pV\bisim\pVV$) or \emph{not} ($\pV{\not\bisim}\pVV$); in the former case, a relation \R can be deduced from the game derivation. The definition for a bisimulation game is given below. For more information regarding the bisimulation game characterisation one may consult Appendix \Cref{sec:app:bisim-games} or standard texts such as \cite{Aceto2007Book,Stirling1996}.

\begin{example}[Proving Strong Bisimilarity]\label{ex:strong-bisim-p-r}
	To prove that $\pV_1\bisim\pVVV_1$, consider the following relation \R,
	\[\R\defeq\Set{(\pV_1,\pVVV_1),(\pV_2,\pVVV_2),(\pV_1,\pVVV_3), (\pV_2,\pVVV_4), (\pV_3,\pVVV_5)}\]
	Since $\pV_1\bisim\pVVV_1$ is the largest strong bisimulation relation relating processes $\pV_1$ and $\pVVV_1$, it thus suffices showing that relation \R is a Strong Bisimulation Relation as stated by \Cref{def:strong-bisim}; we prove that this is true as follows:
	\begin{proofn}[$\;\;\pmb{\pV_1\bisim\pVVV_1}$]
		To prove that the pair $(\pV_1,\pVVV_1)\in\R$ satisfies \Cref{def:strong-bisim}, we show that $\pV_1\traS{\actReq}\pV_2$ can be \emph{strongly simulated} by $\pVVV_1\traS{\actReq}\pVVV_2$ and vice-versa, such that the resultant process pair $(\pV_2,\pVVV_2)\in\R$ is \emph{true}. In addition, we show that similarly reduction $\pV_1\traS{\actCls}\pV_3$ can be simulated by $\pVVV_1\traS{\actCls}\pVVV_5$ and vice-versa, such that $(\pV_3,\pVVV_5)\in\R$ is also \emph{true}.
		
		The same argument applies for pairs $(\pV_2,\pVVV_2)$, $(\pV_1,\pVVV_3)$ and $(\pV_2,\pVVV_4)$, \ie
		\[
		\begin{array}{rl}
			(\pV_2,\pVVV_2): & \pV_2\traS{\actAns}\pV_1 \emph{ bisimulates } \pVVV_2\traS{\actAns}\pVVV_3 \text{ such that } (\pV_1,\pVVV_3)\in\R \text{ is \emph{true}.}\\
			(\pV_1,\pVVV_3): & \pV_1\traS{\actReq}\pV_2 \emph{ bisimulates } \pVVV_3\traS{\actReq}\pVVV_4 \text{ such that } (\pV_2,\pVVV_4)\in\R \text{ is \emph{true}.}\\
			\hfill \text{ and} & \pV_1\traS{\actCls}\pV_3 \emph{ bisimulates } \pVVV_3\traS{\actAns}\pVVV_5 \text{ such that } (\pV_3,\pVVV_5)\in\R \text{ is \emph{true}.}\\			 
			(\pV_2,\pVVV_4): & \pV_2\traS{\actAns}\pV_1 \emph{ bisimulates } \pVVV_4\traS{\actAns}\pVVV_1 \text{ such that } (\pV_1,\pVVV_1)\in\R \text{ is \emph{true}.}
		\end{array}
		\]
		
		\noindent Finally, since both processes in the pair $(\pV_3,\pVVV_5)$ are unable to perform any further reductions, \ie $\pV_3\ntraS{\actu}$ and $\pVVV_5\ntraS{\actu}$, they still satisfy the constraints of \Cref{def:strong-bisim}.
		
	\end{proofn}$\\$
	\noindent Hence, with the above proof we can conclude that relation \R is a \emph{Strong Bisimulation relation}, such that since $(\pV_1,\pVVV_1)\in\R$, we can also conclude that $\pV_1\bisim\pVVV_1$.
	 
	Similarly, to prove that $\pV_1\bisim\pVVVV_1$, we must once again find a binary relation and prove that it is a Strong Bisimulation Relation as defined by \Cref{def:strong-bisim}. Hence, we consider relation $\R'$,
 	\[\R'\defeq\Set{(\pV_1,\pVVVV_1),(\pV_1,\pVVVV_2),(\pV_2,\pVVVV_3), (\pV_3,\pVVVV_4)}\]
 	From the case of $(\pV_1,\pVVVV_1)\in\R'$, we can immediately notice that relation $\R'$ is \emph{not} a Strong Bisimulation relation since reduction $\pV_1\traS{\actReq}\pV_2$ cannot be strongly simulated by process $\pVVVV_1$, \ie $\pVVVV_1\ntraS{\actReq}$. Hence, from relation $\R'$ we are \emph{unable to conclude} whether $\pV_1\bisim\pVVVV_1$ or not. 
 	
 	However, as proven below, by abstracting over internal (\actt) transitions, by \Cref{def:weak-bisim} we can conclude that relation $\R'$ is a \emph{Weak Bisimulation} relation.
 	\begin{proofn}[$\;\;\pmb{\pV_1\wbisim\pVVVV_1}$]
 		To prove that the pair $(\pV_1,\pVVVV_1)\in\R'$ satisfies \Cref{def:weak-bisim}, we show that transitions $\pV_1\traS{\actReq}\pV_2$ and $\pV_1\traS{\actCls}\pV_3$ can be \emph{weakly simulated} as $\pVVVV_1\traS{\actt}\traS{\actReq}\pVVVV_3$ and $\pVVVV_1\traS{\actt}\traS{\actCls}\pVVVV_4$ \resp, such that $(\pV_2,\pVVVV_3)\in\R'$ and $(\pV_3,\pVVVV_4)\in\R'$ are both true; while transition $\pVVVV_1\traS{\actt}\pVVVV_2$ can also be weakly simulated as $\pV_1\traS{\actt}^{0}\pV_1$, where $(\pV_1,\pVVVV_2)\in\R'$ is \emph{true}.
 	 	
 	 	In the case of $(\pV_1,\pVVVV_2)\in\R'$, we show that transitions $\pV_1\traS{\actReq}\pV_2$ and $\pV_1\traS{\actCls}\pV_3$ can be simulated by $\pVVVV_2\traS{\actReq}\pVVVV_3$ and \resp by $\pVVVV_2\traS{\actCls}\pVVVV_4$, such that $(\pV_2,\pVVVV_3)\in\R'$ is true and $(\pV_3,\pVVVV_4)\in\R'$ is true as well; dually, $\pVVVV_2\traS{\actReq}\pV_3$ and $\pVVVV_2\traS{\actCls}\pVVVV_4$ can also be simulated by $\pV_1\traS{\actReq}\pV_2$ and \resp by $\pV_1\traS{\actCls}\pV_3$. 
 	 	
 	 	Finally, we know that the pair $(\pV_3,\pVVVV_4)\R'$ satisfies the constraints of \Cref{def:weak-bisim} since neither of the two processes can perform a \actu-reduction.
 	 	
 	\end{proofn}$\\$
	\noindent Since $\R'$ is a Weak Bisimulation relating processes $\pV_1$ and $\pVVVV_1$, we can therefore conclude that $\pV_1\wbisim\pVVVV_1$. \bqed
\end{example}

One may already notice that processes $\pV_1$ and $\pVV_1$ from \Cref{fig:ex:lts} appear to behave differently even when observed as a black box from an external perspective; this is because $\pV_1$ is \emph{always} obliged to provide an answer to an external request, while $\pVV_1$ may occasionally refuse to do so. 

In general, proving that $\pV_1{\not\bisim}\pVV_1$ is an exhaustive technique that requires proving that \emph{all binary relations} \R that can relate processes $\pV_1$ and $\pV_2$ are \emph{not} a Strong Bisimulation; this inherently proves that there \emph{does not exist} some Strong Bisimulation relation \R, such that $(\pV_1,\pVV_1)\in\R$. Alternatively, the bisimulation game characterisation provides a more practical alternative to formally prove that $\pV_1\not\bisim\pV_2$ (see Appendix \Cref{sec:app:bisim-games} for more details).

\section{Linear vs Branching Time Logics}
As advocated by \cite{Francalanza2015Mon,Bauer2010,Bauer2011}, in runtime monitoring and enforcement, correctness properties can be defined in a wide variety of \emph{logics}. These logics can be categorized into two major classes \cite{Clarke1999Book,Nain2007}, namely, \emph{Linear Time} and \emph{Branching Time} Logics. 

\emph{Linear Time Logics}, \cite{Clarke1999Book,Nain2007} treat time as if each moment has a unique possible future. Hence, formulas expressed in linear time logics are regarded as specifying the behaviour of a \emph{single program computation}, as they are interpreted over \emph{linear sequences} of system actions. By contrast, in \emph{Branching Time Logics} \cite{Clarke1999Book,Nain2007}, each moment in time may split into several possible futures. Hence, branching time formulas describe properties of both \emph{finite} and \emph{infinite computation trees}, each of which describes the behaviour of the possible computations of a non-deterministic program. 

Several researchers \cite{Nain2007,Emerson1980,Clarke1989,BenAri1983,Lamport1980,Emerson1986,Emerson1985} have been discussing the relative advantages of linear versus branching time logics \wrt system specification and verification, since the 1980s. Certain discussions \cite{Clarke1989,Lamport1980,Emerson1986} led to the conclusion that linear and branching time logics are expressively incomparable. However in \cite{Nain2007}, the authors discuss that branching time logics are more expressive in the context of \emph{algorithmic verification}. 

Due to their high expressiveness, branching time logics such as \uhml \cite{Aceto2007Book} and CTL*, thus allow for specifying a wide variety of properties which can be verified using various verification techniques. However, since dynamic verification techniques, such as monitoring, are incapable of verifying every expressible property \cite{Francalanza2017FMSD}, these logics are instead often favoured to be used with static analysis techniques such as model checking \cite{Aceto2007Book}. 

Multi-pronged verification approaches that combine static and dynamic verification \cite{Azzopardi2016}, provide a practical compromise for verifying such properties by employing dynamic verification as much as possible and only use static verification for those parts that cannot be verified statically. This helps minimize the state explosion problems that are inherent to classic static verification techniques such as model checking \cite{Clarke1999Book,Aceto2007Book}.

%

\section{The Logic}

Hennessy-Milner Logic with recursion (\uhml) \cite{Larsen1990,Aceto2007Book} is a highly expressive, branching-time logic that allows for defining correctness properties as logical formulae. \uhml assumes a countable set of logical variables $\hVarX,\hVarY\!\in\!\LVars$, and is defined as  the set of \emph{closed} formulae 
generated by the grammar of \Cref{fig:recHML}. The logic is equipped with the standard constructs for \emph{truth}, \emph{falsehood}, \emph{conjunction} and \emph{disjunction}, where $\hAnd{\hV_i}$ (\resp $\hOr{\hV_i}$) describes a \emph{compound} conjunction, $\hV_1{\hand}\ldots{\hand}\hV_n$, (\resp disjunction, $\hV_1{\hor}\ldots{\hor}\hV_n$) where $1\ldots n\in\IndSet$. 

\uhml also provides the \emph{possibility} and \emph{necessity modal operators}, together with recursive formulae expressing \emph{least or greatest fixpoints} denoted by formulae \hmin{\hVarX}{\hV} and  \hmax{\hVarX}{\hV} \resp Fixpoints bind free instances of the logical variable \hVarX in \hV, inducing the usual notions of open/closed formulae and formula equality up to alpha-conversion. Modal operators allow for defining \emph{symbolic events} $\actS{=}\actSN{\pate}{\predc}$, consisting in a pattern \pate and filtering condition \predc. As defined in \Cref{sec:symevents}, the pattern may contain \emph{data variables} $\dvV,\dvVV,\dvVVV\in\Var$ that \emph{bind to system data} from a \emph{matching} concrete system event and can be used to evaluate the associated filtering condition. 

\begin{figure}[t]
	\textbf{Syntax}
	\begin{align*}
	\hV,\hVV \in \uhml &\bnfdef  \htru && (\text{truth})& 
	&\bnfsepp  \hfls && (\text{falsehood})\\[1mm]
	& \bnfsepp \hOr{\hV_i}  && (\text{disjunction}) &
	& \bnfsepp \hAnd{\hV_i}  && (\text{conjunction}) \\[1mm]
	&\bnfsepp \hpos{\actS}{\hV} && (\text{possibility}) &                                                                        
	&\bnfsepp \hnec{\actS}{\hV} && (\text{necessity}) \\[1mm]
	& \bnfsepp \hmin{\hVarX}{\hV} && (\text{min. fixpoint}) &
	& \bnfsepp \hmax{\hVarX}{\hV} && (\text{max. fixpoint}) \\[1mm]
	& \bnfsepp\; \hVarX && (\text{rec. variable})
	\end{align*}
	\textbf{Semantics}
	\[\begin{array}{l@{\qquad\qquad}c@{\qquad\qquad}r} 
	\syn{\htru,\rho}  \defEquals \Proc &
	\syn{\hfls,\rho}  \defEquals \emptyset &
	\syn{\hVarX,\rho} \defEquals \rho(\hVarX)
	\end{array}\]
	\[\begin{array}{rl} 
%
	\syn{\hAnd{\hV_i},\rho} & \defEquals \bigintersect{i\in\IndSet}\syn{\hV_i,\rho} 
	\\
	\syn{\hOr{\hV_i},\rho} & \defEquals \bigunion{i\in\IndSet}\syn{\hV_i,\rho}\\[2mm]
	\!\syn{\hmin{\!\hVarX}{\hV},\rho} & \defEquals \bigcap \Set{S \;|\;  \syn{\hV,\rho[\hVarX\mapsto S]} \subseteq S}
	\\
	\!\syn{\hmax{\!\hVarX}{\hV},\rho} & \defEquals \bigcup \Set{S \;|\;  S \subseteq \syn{\hV,\rho[\hVarX\mapsto S]}} 
	\end{array}\]
	\[\begin{array}{rl} 
	\syn{\hnec{\actS}{\hV},\rho}  & \defEquals \Set{\pV \;|\;  (\forall\acta,\pVV\cdot\pV  \tra{\acta} \pVV  \text{ and } \mtchS{\actS}{\acta}\!=\!\s) \text{ implies } q \in \syn{\hV\s,\rho}\!}	\\
	\syn{\hpos{\actS}{\hV},\rho}  & \defEquals \Set{\pV \;|\; \exists\acta,\pVV\cdot\pV  \tra{\acta} \pVV   \text{ and } \mtchS{\actS}{\acta}\!=\!\s \text{ and } q \in \syn{\hV\s,\rho} \!}
	\end{array}\]
	\caption{\uhml Syntax and Semantics}
	\label{fig:recHML}
\end{figure}

Formulae are interpreted over 
the process powerset domain where $S\in\pset{\Proc}$. The semantic definition of \Cref{fig:recHML} is given for  \emph{both} open and closed formulae and employs a valuation (\ie a map) from  logical variables to sets of processes, $\rho\in(\LVars \rightarrow \pset{\Proc})$, where $\rho'=\rho[\hVarX\mapsto S]$ denotes a valuation such that $\rho'(\hVarX) = S$ and  $\rho'(\hVarY) = \rho(\hVarY)$ for all other $\hVarY\neq \hVarX$. This permits an inductive definition for \syn{\hV,\rho}, the set of processes satisfying the formula \hV \wrt an environment $\rho$, based on the structure of the formula.  

For instance, in \Cref{fig:recHML}, the semantic meaning of a variable \hVarX in relation to a map $\rho$ is the mapping $\rho(\hVarX)$.  The semantics of truth, falsehood, conjunction and disjunction are standard, \ie \!\hor and \!\hand are interpreted as set-theoretic union and intersection. Possibility formulae \hpos{\actS}{\hV} describe processes that can perform an action $\acta$, where ${\mtchS{\actS}{\acta}=\s}$ (see \Cref{def:ptrn-match}), such that \emph{at least one} $\acta$-derivative satisfies $\hV\s$. By contrast, necessity formulae \hnec{\actS}{\hV} describe processes capable of performing a compliant action $\acta$, where ${\mtchS{\actS}{\acta}=\s}$, such that \emph{all} of their $\acta$-derivatives (possibly none) satisfy $\hV\s$. 

The powerset domain \pset{\Proc} is a complete lattice \wrt set-inclusion, $\subseteq$, which guarantees the existence of least and largest solutions for the recursive formulae of the logic --- these are defined \resp as the intersection of all the pre-fixpoint solutions and the union of all post-fixpoint solutions \cite{Aceto2007Book}. Since the interpretation of  closed formulae is independent of the environment $\rho$,  we write \syn{\hV} in lieu of  \syn{\hV,\rho}. 

\begin{example} \label{ex:uhml-formula} Recall processes $\pV_1$ and $\pVV_1$ from \Cref{ex:lts}, using \uhml we can formally define that two consecutive requests indicate invalid behaviour, as the desired behaviour entails that every request must be provided with an answer. This safety property can be defined as formula $\hV_0$:
	$$ \hV_0\defeq\hVdefConc $$ 
Formula $\hV_0$ describes a recursive (\hmaxB{\hVarX}{\ldots}) property requiring that whenever a process identified by process id $i$, inputs a first request (\actReq), then it cannot input a subsequent request (\ie \hnec{\actReq}\hfls), unless it outputs an answer beforehand, in which case the formula recurses (\ie \hnec{\actAns}\hVarX).  

With formula $\hV_1$ (given below), we generalize formula $\hV_0$ to range over \emph{any} process id \dvV, as opposed to a specific process id $i$. We however assume that this formula should not apply for the process identified by id $j$, and hence we add this restriction as the filtering condition $\dvV\neq j$ of the symbolic event defined in the first necessity of $\hV_1$. 
	$$ \hV_1\defeq\hVdef $$ 
Although $\hV_0$ and $\hV_1$ both describe the same system behaviour, they differ in terms of the LTS they define. For instance, formula $\hV_0$ specifies a \emph{finite} LTS describing the finite set of actions of a server process that is specifically identified by id $i$, namely actions \actReq and \actAns. By contrast, since $\hV_1$ specifies an \emph{infinite} LTS since it describes the actions that can be preformed by an infinite number of server processes each identified by some process id $\dvV\neq j$. \bqed
\end{example}

We say that a formula \hV is \emph{satisfiable} (\ie $\hV\in\Sat$) whenever there exists some process \pV such that $\pV\in\syn{\hV}$, \ie as formally defined below. Hence, in order to find whether $\hV_1$ is satisfiable or not it suffices finding a single process \pV which satisfies the formula.
\begin{definition}[Satisfiable Formulae] \label{def:sat}
	\renewcommand{\qedsymbol}{\bqed}
	\begin{displaymath}
		\pushQED{\bqed}
		\hV\in\Sat \;\textsl{ iff }\;\exists\pV\cdot\pV\in\syn{\hV} \qedhere
		\popQED
	\end{displaymath}
\end{definition}

Satisfiable formulae can also be used to differentiate between processes. As formally specified in \Cref{thm:HMT}, the Hennessy-Milner theorem \cite{Aceto2007Book,Hennessy1985} dictates that given an image-finite LTS (\ie an LTS in which all of its states have a finite number of outgoing transitions), two LTS states (processes) are \emph{bisimilar} if they both \emph{satisfy the same (non-recursive) \uhmlsl formulae} (and vice-versa).

\begin{theorem}[The Hennessy-Milner Theorem] \label{thm:HMT}
	\renewcommand{\qedsymbol}{\ensuremath{\blacksquare}}
	Given an image finite Labelled Transition System, $\langle\Proc,\Act,\rightarrow\rangle$, assuming two states $\pV,\pVV\in\Proc$,
	\[
	 \pushQED{\bqed}		
	 \pV \bisim\pVV \quad \textsl{  iff  } \quad \forall\hV\cdot \pV \!\in\!\syn{\hV}\Leftrightarrow\pVV \!\in\!\syn{\hV} \qedhere
	\]
\end{theorem}

\begin{example}\label{ex:uhml-formula-derivation}
	Using formula $\hV_1$ we can thus distinguish between processes $\pV_1$ and $\pVV_1$ by showing that process $\pV_1$ satisfies $\hV_1$ (\ie $\pV_1\!\in\!\syn{\hV_1}$) and that $\pVV_1$ does not (\ie $\pVV_1\!\notin\!\syn{\hV_1}$). We prove result using \emph{Tarski's Fixpoint Algorithm} \cite{Aceto2007Book,Tarski1955} in the Appendix \Cref{sec:app:tarski-proof}. 
	
	Hence, since process $\pVV_1$ is unable to satisfy all \uhml properties that can be satisfied by $\pV_1$ (this is proven to be true by $\hV_1$), by the Hennessy-Milner theorem \cite{Aceto2007Book,Hennessy1985} (recited in \Cref{thm:HMT}), we can also conclude that $\pV_1{\not\bisim}\pVV_1$. \bqed
\end{example}

\section{A model for Detection Monitors}\label{sec:det-mon-model}

Runtime Verification provides an alternative mechanism for checking whether a program exhibits the expected behaviour or not. In RV, this is achieved via detection monitors, which analyse the current execution trace of the system so as to determine whether this behaves correctly or not as specified by some correctness property. Detection monitors are said to \emph{recognise} a good (\resp bad) trace whenever they are able to conclude that the program \emph{satisfies} (\resp \emph{violates}) the given correctness property, by just analysing the trace of events generated by the executing program.

\begin{figure}[t]
	\textbf{Syntax}
	\begin{align*}
	\mV,\mVV\in\Mon\ &\bnfdef\  \vV && \bnfsepp \prf{\actS}{\mV}  && \bnfsepp \mCh{\mV_i} && \bnfsepp \rec{\mx}{\mV} &&\bnfsepp \mx \\
	\vV,\vVV \in \Verd & \bnfdef\ \mend &&  \bnfsepp \mno && \bnfsepp \myes
	\end{align*}
	
	\textbf{Dynamics}
	\begin{mathpar}
		\inference[\rtit{mAct}]{\mtchS{\actS}{\acta}\!=\!\s & \ceval{c\s}{\ctru}}{\prf{\actS}{\mV}  \traSS{\acta} \mV} \and
		\inference[\rtit{mRec}]{\mV\sub{\rec{\mx}{\mV}}{\mx} \traSS{\acta} \mVV}{\rec{\mx}{\mV} \traSS{\acta} \mVV} 
		\\
		\inference[\rtit{mSel}]{\mV_j \traSS{\actu} \mVV_j}{\mCh\,\mV_i \traSS{\actu} \mVV_j}[$j\in\IndSet$] 
		\and
		\inference[\rtit{mVer}]{}{\vV \traSS{\acta} \vV}\\
	\end{mathpar}
	
	\textbf{Instrumentation}
	\begin{mathpar}
		\inference[\rtit{iMon}]{\pV \traSS{\acta} \pV' & \mV \traSS{\acta} \mV'}{\iRV{\mV}{\pV} \traSS{\acta} \iRV{\mV'}{\pV'}} 
		\and
		\inference[\rtit{iAsyP}]{\pV \traSS{\actt} \pV'}{\iRV{\mV}{\pV} \traSS{\tau} \iRV{\mV}{\pV'}}
		\and
		\inference[\rtit{iTer}]{\pV \traSS{\acta} \pV' & \mV \ntraSS{\acta}  
		 }{\iRV{\mV}{\pV} \traSS{\acta} \iRV{\mend}{\pV'}}
	\end{mathpar}
	\caption{Monitors and Instrumentation}
	\label{fig:monit-instr}
\end{figure}

In \cite{Francalanza2015Mon,Francalanza2016TheoMon}, Francalanza \etal defined the structure and dynamic behaviour of detection monitors in terms of the LTS syntax and semantics provided in \Cref{fig:monit-instr}. Assuming a specific set of symbolic events, $\SEvt$, and a denumerable set of (recursion) variables $\mx,\my\in\Vars$, detection monitors are defined as either a \emph{verdict} $v$, a prefixed process by a symbolic action $\actS\!=\!\actSN{\pate}{\predc}$, a mutually-exclusive choice amongst two monitors, or a recursive monitor.

A monitor can issue one of three verdicts, namely, \myes, \mno\ and \mend, \resp denoting acceptance, rejection and termination (\ie an inconclusive outcome). Verdicts in detection monitors are irrevocable as specified by rule \rtit{mVer}, which states that a verdict may transition with \emph{any} system action $\acta\in\Act$ and go back to the same state. The monitor $\rec{\hVarX}{\mV}$ acts as a \emph{binder} for recursion variable $\mx$ in $\mV$ where, by rule \rtit{mRec} a recursive monitor can reduce over \acta, \ie $\mrec{\mx}{\mV}\traS{\acta}\mVV$, whenever its unfolded version reduces over \acta, \ie $\mV\sub{\mrec{\mx}{\mV}}{\mx}\traS{\acta}\mVV$.  All recursive monitors are assumed to be guarded, meaning that all occurrences of bound recursive variables occur under an action prefix (either directly or indirectly). 

As specified by rule \rtit{mAct}, a prefix monitor $\prf{\actSN{\pate}{\predc}}{\mV}$ binds in \mV any concrete event \acta that is within the constraints of the symbolic event prefix, \ie $\prf{\actSN{\pate}{\predc}}{\mV}\traS{\acta}\mV\s$ whenever \acta matches the pattern \pate and satisfies condition \predc via $\mtchS{\actSN{\pate}{\predc}}{\acta}\!=\!\s$. Substitution environment \s is used to bind in the derived monitor \mV, any data variable, $\dvV\in\Var$, (defined in pattern \pate), to the \resp concrete values provided by the concrete system event \acta. For example, if $\mtchS{\actSN{\recv{i}{\dvV}}{\dvV\!>\!5}}{\recv{i}{6}}\!=\!\sub{6}{\dvV}$ then $\prf{\actSN{\recv{i}{\dvV}}{\dvV\!>\!5}}{\prf{\actSN{\recv{i}{\dvVV}}{\dvVV\!>\!\dvV}}{\mV}} \traS{\recv{i}{6}} \bigl((\prf{\actSN{\recv{i}{\dvVV}}{\dvVV\!>\!\dvV}}{\mV})\sub{6}{\dvV} \,\equiv\, \prf{\actSN{\recv{i}{\dvVV}}{\dvVV\!>\!6}}{(\mV\sub{6}{\dvV})}\bigl) $.
The behaviour of a mutually exclusive choice is specified by rule \rtit{mSel}, which states that $\mCh{\mV_i}\traS{\acta}\mVV_j$ if there exists an index $j\in\IndSet$ such that $\mV_j\traS{\acta}\mVV_j$.

%

The rules we have seen so far, specify the runtime behaviour of monitors in isolation, without describing any notion of interaction between the monitor and the process under scrutiny. Hence, \Cref{fig:monit-instr} also describes an instrumentation relation, connecting the behaviour of a process $\pV$ with that of a monitor \mV such that the configuration \iRV{\mV}{\pV} denotes a \emph{monitored system}.  

In an instrumentation, the process leads the (visible) behaviour of a monitored system (\ie if the process cannot \acta-transition, then the monitored system will not either) while the monitor passively follows, transitioning accordingly.  Specifically, rule \rtit{iMon} states that if a process can transition with action \acta and the assigned monitor can follow this by transitioning with the same action, then in an instrumented monitored system they transition in lockstep. 

However, if the monitor is unable to perform such a transition, \ie $\mV\!\ntraS{\acta}$, even after any number of internal actions, \ie $\mV\!\ntraS{\tau}$, the instrumentation rule \rtit{iTer} forces it to terminate with an inconclusive verdict, \mend, while the process is allowed to proceed unhindered. Also, note that since \mend\ can follow \emph{any} visible system event, future transitions by $\pV$ are still allowed while the terminated monitor maintains its state, using the  rule \rtit{iMon}. Finally, rule \rtit{iAsyP} allows processes to transition independently from the monitor \wrt internal moves, thus reducing the coupling between the process and the monitor.

\begin{example}  \label{ex:monitor-instrum}
	By using detection monitors we can analyse the runtime execution of processes $\pV_1$ and $\pVV_1$ from \Cref{ex:lts} (restated below) in order to recognise witness traces that testify for negative behaviour as specified by $\hV_1$ (given in \Cref{ex:uhml-formula}), \ie when monitoring for $\hV_1$, the monitor should be able to detect the cases where processes do not always provide an answer for a given request. 
	\begin{displaymath}
		\begin{array}{rcl}
				\pV_1&\defeq&\procPVdef \\
				\pVV_1&\defeq&\procPVVdef
		\end{array}
	\end{displaymath}
	
	To detect such instances consider monitor $\mV_1$ (defined below) that reaches a \emph{violation} verdict, \mno, after observing that a process that is identified by \emph{any} process id \dvV, has executed two consecutive request actions (\ie $\prf{\symReqA}{\prf{\symReqB}{\mno}}$), but recurses after observing that the request was serviced by an answer output action, \ie $\prf{\symReqA}{\prf{\symAns}{\mx}}$. 
	\begin{displaymath}
	\mV_1\defeq\mVdef
	\end{displaymath}
	When instrumented with process $\pVV_1$ from \Cref{ex:lts}, we observe the following behaviour for the monitored system whereby on line \eqref{eq:mon-loop} the monitor preserves the $\mno$ verdict for all remaining transitions.
	\begin{align*}
	\iRV{\mV_1}{\pVV_1}  \traS{\actReq}\;\; & \iRV{\bigl(\mch{\prf{\symAns}{\mV_1}}{\prf{\symReqB}{\mno}}\bigl)\underline{\sub{i}{\dvV}}}{\pVV_1} \\
			 \equiv\;\;& \iRV{\mch{\prf{\actAns}{\mV_1}}{\prf{\actReq}{\mno}}}{\pVV_1}
		&& \hspace{-8mm} {\small\text{using \rtit{iMon}(\rtit{mRec}+\rtit{mAct})}}\\
	 \traSS{\actReq}\;\; & \iRV{\mno\sE}{\pVV_2} \;\; \equiv \;\; \iRV{\mno}{\pVV_2}		 
		 && \hspace{-8mm} {\small\text{using \rtit{iMon}(\rtit{mSelR}+\rtit{mAct})}}\\
	 \wtraSS{\;\; t'\;\;} \;& \iRV{\mno}{\pVV_1} && \hspace{-8mm} {\small\text{using \rtit{iMon}(\rtit{mVer})}} \tag{$\ast$} \label{eq:mon-loop}
	\end{align*}
	
	Note how the monitor's runtime analyses abstracts over the system's data via variables defined in the prefixing symbolic events, \eg \symReqA and \symAns. This allows the monitor to detect the specified behaviour regardless of the data associated to the system's action, \eg if the process id of $\pVV_1$ changes from $i$ to $j$, monitor $\mV_1$ would still be able to analyse the behaviour this modified version of $\pVV_1$, since variable \dvV can match with \emph{any} process id. 
	
	Also, note that monitor $\mV_1$ was only able to detect a property violation since $\pVV_1$ has actually executed two consecutive request, such that $t\,{=}\,\actReq.\actReq.t'$. However, given the non-deterministic behaviour of $\pVV_1$, it is possible that while executing, this process provides an answer for every request and never executes two consecutive requests. In this way, the monitor would never be able to conclude any verdict about $\pVV_1$, despite the fact that $\pVV_1$ actually violates $\hV_1$, \ie $\pVV_1\notin\syn{\hV_1}$ (see \Cref{ex:uhml-formula-derivation}). Hence, this shows that the monitor's verdicts are limited to the behaviour exhibited by the process at runtime, and can only be issued when the system actually executes the specified behaviour.
	
	Similarly, in the case of process $\pV_1$, the monitors are unable to issue a negative verdict, \mno, since the process is designed to always issue an answer for any given request; they are also incapable of producing a positive verdict, \myes, since the monitor keeps on recursing upon perceiving a request-answer sequence. \bqed
\end{example}

\section{The monitorability of \uhml}\label{sec:monitorability}

One can immediately start to notice the structural resemblance that exists between monitor $\mV_1$, defined in \Cref{ex:monitor-instrum}, and formula $\hV_1$, defined in \Cref{ex:uhml-formula} (both restated below). Intuitively, the maximal fixpoint and fixpoint variable, the conjunction operation and the necessity operations in $\hV_1$, \resp map to the recursive constructs, the choice operation and the action prefixes in $\mV_1$.
\begin{align*}
	\mV_1 &\defeq\,\mVdef
	\\
	\hV_1 &\defeq\,\hVdef
\end{align*}
In \cite{Francalanza2015Mon}, Francalanza \etal proved 
the existence a formal correspondence between monitors and logic formulae. In their work they present a notion of \emph{monitorability} as a property of a correctness specification describing the ability to be adequately analysed at runtime. This definition is fundamentally dependent on the monitoring setup assumed and the conditions that constitute an \emph{adequate runtime analysis}. The authors start by distinguishing between \emph{acceptance} and \emph{rejection} monitors (\ie monitors that yield a \myes and \mno verdicts \resp), by defining the predicates in \Cref{def:acc-rej}.

\begin{definition}[Acceptance and Rejection Predicates]	\label{def:acc-rej} \leavevmode
\begin{enumerate}[\hspace{18pt}1.]
	\item $\monacc{\pV}{\mV} \;\defeq\; \exists t,\pV'\cdot \iRV{\mV}{\pV} \wtraS{t} \iRV{\myes}{\pV'}$
	\item $\monrej{\pV}{\mV} \;\defeq\; \exists t,\pV'\cdot \iRV{\mV}{\pV} \wtraS{t} \iRV{\mno}{\pV'}$ \bqed
\end{enumerate}
\end{definition}
\noindent The acceptance predicate, \monacc{\pV}{\mV}, states that a program \pV is \emph{accepted} by a monitor \mV whenever the process is able to generate a trace of execution $t$ from which the monitor can deduce a positive verdict, \myes. Similarly, the rejection predicate, \monrej{\pV}{\mV}, states that a process \pV is \emph{rejected} by a monitor \mV whenever the process is capable of generating an execution trace $t$ from which the monitor can conclude a negative verdict, \mno.

\begin{example}[Rejecting a Process] \label{ex:proc-rej}
	\newcommand{\mVA}{\prf{\actReq}{\bigl(\mch{\prf{\actAns}{\mV_1}}{\prf{\actReq}{\mno}}\bigr)}}
	\newcommand{\mVB}{\mch{\prf{\actAns}{\mV_1}}{\prf{\actReq}{\mno}}}
	Recall process $\pVV_1$ defined in \Cref{ex:lts}, and monitor $\mV_1$ defined in \Cref{ex:monitor-instrum} (both restated below).
	\begin{align*}
	\pVV_1 &\defeq\,\procPVVdef \\
	\mV_1 &\defeq\, \mVdef
	\end{align*}
	As shown in \Cref{ex:monitor-instrum}, monitor $\mV_1$ issues a \mno verdict when process $\pVV_1$ executes trace $t=\actReq.\actReq.t'$. Hence, by the definition of \monrej{\pV}{\mV}, we can conclude that monitor $\mV_1$ is capable of \emph{rejecting} process $\pVV_1$, \ie $\monrej{\pVV_1}{\mV_1}$. \bqed
\end{example}

Based on the predicates discussed in \Cref{def:acc-rej}, the authors define the criteria expected of a monitor \mV when it \emph{monitors soundly for a property} \hV as \smondef{\mV}{\hV} in \Cref{def:sound-mon}.

\begin{definition}[Sound Monitoring] A monitor \mV monitors soundly for the property represented by the formula \hV, denoted as \smondef{\mV}{\hV}, whenever for \emph{all} processes $\pV\in\Proc$ the following hold:
	\label{def:sound-mon} 
	\begin{enumerate}[\hspace{18pt}(i).]
		\item $\monacc{\pV}{\mV} \,\imp\,\pV\in\syn{\hV}$ \label{item:smon-acc}
		\item $\monrej{\pV}{\mV} \,\imp\,\pV\notin\syn{\hV}$ \label{item:smon-rej}
	\end{enumerate}
\end{definition}
\noindent Sound monitoring thus \emph{relates} a logic formula \hV to a monitor \mV in such a way that whenever the monitor is able to accept (\resp reject) any process $\pV\in\Proc$, then the related formula must be \emph{satisfied} (\resp violated) by the same set of processes, \ie \Proc.

\begin{example}[Sound Monitoring] \label{ex:sound-mon} In general it is very difficult to prove that a monitor \mV soundly monitors for a formula \hV, since as stated by \Cref{def:sound-mon}, this requires proving that predicates \eqref{item:smon-acc} and \eqref{item:smon-rej} hold for \emph{every} possible process $\pV\in\Proc$ (where \Proc can be an infinite set). However, to better explain the concept of soundness, in this example we \emph{limit ourselves} to processes $\pV_1$ and $\pVV_1$ (\ie we assume that \Proc is \emph{restricted to} $\set{\pV_1,\pVV_1}$). Therefore, using the result from \Cref{ex:proc-rej} and the result from \Cref{ex:uhml-formula-derivation}, \ie that $\pVV_1\notin\syn{\hV_1}$, we can conclude
	\begin{gather}
		\monrej{\pVV_1}{\mV_1} \,\imp\,\pVV_1\notin\syn{\hV_1} \label{ex:sound-mon:1}
	\end{gather}
	Since $\mV_1$ is incapable of producing a positive verdict, \myes, none of the processes in \Proc can ever generate a trace that can be accepted by $\mV_1$. Since this premise is false, we can conclude
	\begin{gather}
		\monacc{\pVV_1}{\mV_1} \,\imp\,\pVV_1\in\syn{\hV_1} \label{ex:sound-mon:2} \\
		\monacc{\pV_1}{\mV_1} \,\imp\,\pV_1\in\syn{\hV_1} \label{ex:sound-mon:3} 
	\end{gather}
	On the other hand, from the LTS given in \Cref{fig:ex:lts}, we can deduce that process $\pV_1$ is incapable of producing a trace that can be rejected by monitor $\mV_1$, which allows us to conclude
	\begin{gather}
		\monrej{\pV_1}{\mV_1} \,\imp\,\pV_1\notin\syn{\hV_1} \label{ex:sound-mon:4} 
	\end{gather}
	Hence by \eqref{ex:sound-mon:1}, \eqref{ex:sound-mon:4} and \eqref{ex:sound-mon:2}, \eqref{ex:sound-mon:3}, and since we \emph{assume} that $\Proc$ is \emph{restricted to just} $\set{\pV_1,\pVV_1}$, we can deduce
	\begin{gather}
		\forall\pV\in\Proc\cdot\monrej{\pV}{\mV_1} \,\imp\,\pV\notin\syn{\hV_1} \label{ex:sound-mon:5} \\ 
		\forall\pV\in\Proc\cdot\monacc{\pV}{\mV_1} \,\imp\,\pV\in\syn{\hV_1} \label{ex:sound-mon:6} 
	\end{gather}
	Finally, by \eqref{ex:sound-mon:5}, \eqref{ex:sound-mon:6} and \Cref{def:sound-mon} we can conclude that monitor \mV is able to \emph{soundly monitor} processes $\pV_1$ and $\pVV_1$, \wrt formula \hV, \ie we can conclude that \smondef{\mV_1}{\hV_1}. \bqed	
\end{example} 

In addition, the authors also define relate formula satisfactions and violation to monitor detections in the opposite direction, by defining the notion of \emph{partially complete monitoring} in \Cref{def:comp-mon}. This notion entails that monitoring should be either \emph{satisfaction or violation complete}, \ie if a formula \hV is \emph{satisfiable} (\resp unsatisfiable), the monitor should \emph{accept} (\resp reject) it at runtime.

\begin{definition-break}[Satisfaction, Violation, and Partially-Complete Monitoring] \label{def:comp-mon} \leavevmode
	$\begin{array}{cccc@{\qquad\quad\;}r}
		\scmondef{\mV}{\hV} & \defeq \; \forall \pV\cdot \pV\!\in\!\syn{\hV}\;\imp\;\monacc{\pV}{\mV} && \hfill \text{(satisfaction complete)} &\\
		\vcmondef{\mV}{\hV} & \defeq \; \forall \pV\cdot \pV\!\notin\!\syn{\hV}\;\imp\;\monrej{\pV}{\mV} && \hfill \text{(violation complete)} &\\
		\cmondef{\mV}{\hV} & \defeq \; \scmondef{\mV}{\hV} \text{ or } \vcmondef{\mV}{\hV} && \hfill \text{(partially complete)}  & \hspace{-5mm}\bqed
	\end{array}$	
\end{definition-break}

\begin{example}[Partially-Complete Monitoring] \label{ex:comp-mon}  
	Same as per Sound Monitoring, in general, proving that a monitor \mV monitors for a formula \hV in a Partially Complete manner is a very hard task, since as stated by \Cref{def:comp-mon}, this requires proving satisfaction and violation completeness \emph{for every possible process} $\pV\in\Proc$ (where \Proc can be an infinite set). Once again, in this example we simplify the proof by limiting ourselves to just processes $\pV_1$ and $\pVV_1$ (\ie we assume that \Proc is limited to just $\set{\pV_1,\pVV_1}$). Hence, using the result from \Cref{ex:uhml-formula-derivation}, \ie that $\pVV_1\notin\syn{\hV_1}$, and the result from \Cref{ex:proc-rej} \ie that $\monrej{\pVV_1}{\mV_1}$, we can conclude
	\begin{gather}
		\pVV_1\notin\syn{\hV_1} \,\imp\,\monrej{\pVV_1}{\mV_1} \label{ex:comp-mon:1}
	\end{gather}
	Although from \Cref{ex:uhml-formula-derivation} we know that $\pV_1\in\syn{\hV_1}$, since monitor $\mV_1$ is \emph{unable} to produce positive verdicts, we can conclude
	\begin{gather}
		(\pV_1\in\syn{\hV_1} \,\imp\,\monacc{\pV_1}{\mV_1}) \textsl{  is false} \label{ex:comp-mon:2}
	\end{gather}
	Hence, the result in \eqref{ex:comp-mon:2} prevents us from immediately deducing partial completeness, however, we also know that $\pV_1{\in}\syn{\hV_1}$, which inherently means that $\pV_1{\notin}\syn{\hV_1}$ is a \emph{false} statement. Therefore, given that a \emph{false statement can imply any result}, we can simply conclude
	\begin{gather}
		\pV_1\notin\syn{\hV_1} \,\imp\,\monrej{\pV_1}{\mV_1} \textsl{  is true} \label{ex:comp-mon:3}
	\end{gather}
	Therefore, from \eqref{ex:comp-mon:1}, \eqref{ex:comp-mon:3} and the definition of \vcmondef{\mV}{\hV} we can conclude that monitor \mV can monitor processes $\pV_1$ and $\pVV_1$ \wrt formula \hV in a violation-complete manner, \ie we know \vcmondef{\mV_1}{\hV_1}. Finally, by the definition of \cmondef{\mV}{\hV} we can conclude that there exists a \emph{partially-complete} monitoring relation between monitor $\mV_1$ and formula $\hV_1$, \ie we can conclude \cmondef{\mV_1}{\hV_1}. \bqed
\end{example}

\noindent Finally, using \Cref{def:sound-mon,def:comp-mon}, the authors thus define a \emph{monitor-formula correspondence relation} as stated by \Cref{def:mon-form-corr}. This relation states that a monitor \mV is said to \emph{monitor} for a formula \hV, whenever it can do so in a \emph{sound} and \emph{partially-complete} manner.

\begin{definition-break}[Monitor-Formula Correspondence]
	\label{def:mon-form-corr}  
	\indent\indent$\quad\mondef{\mV}{\hV} \; \defeq\; \smondef{\mV}{\hV} \text{  and  } \cmondef{\mV}{\hV} $ \bqed
\end{definition-break}

\begin{example}[Establishing a Monitor-Formula Correspondence] \label{ex:mon-form-corr}
	The results of \Cref{ex:sound-mon,ex:comp-mon}, \ie \smondef{\mV_1}{\hV_1} and \cmondef{\mV_1}{\hV_1}, along with \Cref{def:mon-form-corr}, allow us to conclude that there exists a \emph{correspondence relation} between monitor $\mV$ and formula $\hV_1$, meaning that $\mV_1$ is able to monitor for$\hV_1$. Hence, we can conclude \mondef{\mV_1}{\hV_1}. \bqed
\end{example}

\noindent Based on the correspondence that was established between monitors and logic formulae in \Cref{def:mon-form-corr}, the authors define the meaning of a \emph{monitorable formula} and a \emph{monitorable language subset} as \emph{monitorability} (stated by \Cref{def:monitrability}, below).

\begin{definition}[Monitorability]
	\label{def:monitrability}  
	Formula \hV is \emph{monitorable} iff there exists a monitor $\mV$  such that $\mondef{\mV}{\hV}$. A logical language $\mathcal{L}\subseteq \uhml$ is monitorable iff  every $\hV \in \mathcal{L}$ is monitorable. \bqed
\end{definition}

\begin{example}[Monitorability] \label{ex:monitrability}
	Using the result obtained in \Cref{ex:mon-form-corr}, \ie $\mondef{\mV_1}{\hV_1}$, we know that since monitor $\mV_1$ is able to monitor for formula $\hV_1$, then this \uhml formula is \emph{monitorable}. \bqed
\end{example}

Francalanza \etal also showed that not all logical formulae are monitorable and as a result, they identified a syntactic subset of \uhml formulae called \mhml, for which they proved it is a \emph{monitorable subset}. The language subset consists of the safe and co-safe syntactic subsets of \uhml, denoted as \shml and \chml \resp in \Cref{def:monit-logic}.
\begin{definition}[Monitorable Logic]\label{def:monit-logic} $\hV,\hVV \in \mhml \defEquals \shml \cup \chml$ where:
	\renewcommand{\qedsymbol}{\ensuremath{\blacksquare}}
		\begin{align*}\normalfont
	\qquad\quad\qquad\shV,\shVV \in\shml & \bnfdef \htru &&\bnfsepp \hfls &&\bnfsepp \hnec{\actS}{\shV} &&\bnfsepp \shV\hand\shVV &&\bnfsepp \hmax{\hVarX}{\shV} &&\bnfsepp \hVarX \\
	\chV,\chVV \in\chml & \bnfdef \htru && \bnfsepp \hfls &&\bnfsepp \hpos{\actS}{\chV} &&\bnfsepp \chV\hor\chVV &&\bnfsepp \hmin{\hVarX}{\chV} &&\bnfsepp \hVarX \qquad \qed
	\end{align*} 
\end{definition}

\begin{example}[Monitorable Formulae]
	Reconsider formula $\hV_1$, from \Cref{ex:uhml-formula} (restated below), along with formulae $\hV_2$ and $\hV_3$ (defined below).
	\begin{align*}
		 \hV_1\;\defeq&\:\hVdef \\
		 \hV_2\;\defeq&\;\hpos{\symReqB}\hpos{\symAns}\htru \\
		 \hV_3\;\defeq&\;\hV_1\hor\hV_2
	\end{align*}
	The syntactic restrictions given in \Cref{def:monit-logic} allow us to conclude that formula $\hV_1\in\shml$, while $\hV_2\in\chml$, thus meaning that both are \emph{monitorable} according to \Cref{def:monitrability,def:monit-logic}. However, since $\hV_3$ can neither be defined in terms of \shml nor \chml, we are unable to draw any conclusion regarding its monitorability, \ie we cannot say whether it is monitorable or not. \bqed 
\end{example}

To show that \mhml is monitorable as implied by \Cref{def:monitrability}, the authors define the synthesis function \gRV{\!-\!} (defined in \Cref{fig:mon-synthesis}) that generates a detection monitor for each $\hV\in\mhml$. They also show that \gRV{\hV} is able to generate the witness monitor required by \Cref{def:monitrability} to demonstrate the monitorability of \hV, \ie they show that the synthesis generates \emph{sound} and \emph{partially-complete} detection monitors.

\begin{figure}[t]\small
	\vspace{-6mm}
	\begin{addmargin}{-5mm}
		\[\begin{array}{rlrl}
		\gRV{\!\hfls}    & \multicolumn{3}{l}{\defeq \mno  \qquad\qquad\qquad \qquad\qquad \gRV{\!\htru}\defeq \myes   \qquad\qquad\qquad\qquad \qquad\gRV{\!\hVarX}  \defeq \mx }  
		\end{array}\]
		\[\begin{array}{@{\hspace{0mm}}r@{\,}l@{\hspace{-2mm}}r@{\,}l}
		\gRV{\!\hnec{\actS}{\hV}} & \!\defeq\!
		\begin{cases}
		\prf{\actS}{\gRV{\!\hV\!}} & \text{if}\; \gRV{\!\hV\!} \neq \myes\\
		\myes & \text{otherwise}
		\end{cases}
		\quad&
		\gRV{\!\hpos{\actS}{\hV}} & \!\defeq\!
		\begin{cases}
		\prf{\actS}{\gRV{\!\hV\!}} & \text{if}\; \gRV{\!\hV\!} \neq \mno\\
		\mno & \text{otherwise}
		\end{cases}\\[1.3em]
		\gRV{\!\hAnd\hV_i} & \!\defeq\!
		\!\begin{cases}
			\mno  & \exists j\in\IndSet\cdot\gRV{\!\hV_j\!}{\neq}\mno \\[2mm]
			\mCH{i\in\IndSet'}\gRV{\hV_i} &  \!\!\!\begin{cases}
								\text{if}\, \IndSet{=}\IndSet'{\cupplus}\IndSet'' \textsl{ st. }\\
								\gRV{\hAND{j\in\IndSet''\!\!\!\!}\hV_i}{=}\mCH{\!\!j\in\IndSet''}\myes
							\end{cases}		
		 \end{cases}
		&
	  \gRV{\!\hOr\hV_i} & \!\defeq\!
		\!\begin{cases}
			\myes  & \exists j\in\IndSet\cdot\gRV{\!\hV_j\!}{\neq}\myes \\[2mm]
			\mCH{i\in\IndSet'}\gRV{\hV_i} &  \!\!\!\begin{cases}
				\text{if}\, \IndSet{=}\IndSet'{\cupplus}\IndSet'' \textsl{ st. }\\
				\gRV{\hOR{j\in\IndSet''\!\!\!\!}\hV_i}{=}\mCH{j\in\IndSet''}\mno
			\end{cases}		
		\end{cases}
		\\[3.4em]
		\gRV{\!\hmax{\!\hVarX}{\hV}} & \!\defeq\!
		\begin{cases}
		\rec{x}{\gRV{\!\hV\!}} & \text{if}\;\gRV{\!\hV\!} \neq \myes\\
		\myes   & \text{otherwise}
		\end{cases}
		&
		\gRV{\!\hmin{\!\hVarX}{\hV}} & \!\defeq\!
		\begin{cases}
		\rec{x}{\gRV{\!\hV\!}} & \text{if}\;\gRV{\!\hV\!} \neq \mno\\
		\mno   & \text{otherwise}
		\end{cases}
		\end{array}\]	
	\end{addmargin}
	\caption{The Synthesis function for Detection Monitors.}
	\label{fig:mon-synthesis}
\end{figure}

Their synthesis (restated in \Cref{fig:mon-synthesis}) converts the logic falsehood $\hfls{\,\in\,}\mhml$ and truth $\htru{\,\in\,}\mhml$ into monitor verdicts $\mno,\myes{\,\in\,}\Verd$ \resp, and logical variables $\hVarX{\,\in\,}\mhml$ into the corresponding recursive variables $\mx{\,\in\,}\Mon$. Logic necessities, $\hnec{\actS}\hV{\,\in\,}\shml$, and possibilities, $\hpos{\actS}\hV{\,\in\,}\chml$, are both mapped to monitor actions, $\prf{\actS}{\gRV{\hV}}{\,\in\,}\Mon$ in the general case, yet in certain cases, the synthesis simplifies the monitor based on logical equivalences, \eg since $\hnec{\actS}\htru\equiv\htru$, \gRV{\hnec{\actS}\htru} synthesises monitor \myes instead of \prf{\actS}{\myes}, similarly since $\hpos{\actS}\hfls\equiv\hfls$, \gRV{\hpos{\actS}\hfls} synthesises monitor \mno.

Conjunctions, $\hAnd\hV_i{\,\in\,}\shml$, and disjunctions, $\hOr\hV_i{\,\in\,}\chml$, are mapped to a monitor summation, $\mCh{\gRV{\hV_i}}$ unless they can be optimized. One optimization for conjunctions and \resp disjunctions consists in synthesising a \emph{single} \mno (\resp \myes) monitor when there exists at least one formula in the conjunction (\resp disjunction) which yields a \mno (\resp \myes) monitor, \eg since $\hfls\hand\hV=\hfls$ and $\htru\hor\hV=\htru$, then $\gRV{\hfls\hand\hV}$ and $\gRV{\htru\hor\hV}$ \resp yield monitors \mno and \myes, rather than \mch{\mno}{\gRV{\hV}} and \mch{\myes}{\gRV{\hV}}. Another optimization serves to yield smaller summations $\mCH{i\in\IndSet'}{\gRV{\hV_i}}$ (where $\IndSet'\subseteq\IndSet$), whenever there exists a \emph{disjoint subset of indices} $\IndSet''$, \ie $\IndSet=\IndSet'\cupplus\IndSet''$, such that every formula identified by the indices in $\IndSet''$ yield \myes for conjunctions, and \mno for disjunctions, \eg since $\hfls\hor\hV\equiv\hV$ and $\htru\hand\hV\equiv\hV$, then $\gRV{\hfls\hor\hV}$ and $\gRV{\htru\hand\hV}$ both yield $\gRV{\hV}$ instead of $\mch{\mno}{\gRV{\hfls\hor\hV}}$ and $\mch{\myes}{\gRV{\hfls\hor\hV}}$ \resp. 

In the general case, the fixpoint binders, $\hmax{\hVarX}{\hV}{\,\in\,}\shml$ and $\hmin{\hVarX}{\hV}{\,\in\,}\chml$, are both mapped to the recursive construct $\mrec{\mx}{\gRV{\hV}}{\,\in\,}\Mon$. However, in certain cases the synthesis also optimizes the synthesised monitors, \eg since $\hmax{\hVarX}{\htru}\equiv\htru$, \gRV{\hmax{\hVarX}{\htru}} synthesises a monitor \myes instead of \mrec{\mx}{\myes}, since $\gRV{\htru}=\myes$; same applies for $\hmin{\hVarX}{\hfls}\equiv\hfls$, where $\gRV{\hmin{\hVarX}{\hfls}}$ synthesises monitor \mno instead of \mrec{\mx}{\mno}.

Although the synthesis covers both \shml and \chml, the syntactic constraints of \Cref{def:monit-logic} implicitly infer that the synthesis for a formula \hV uses at most the first row and then either the first column (in the case of \shml) or the second column (in  case of \chml). 

\begin{example}[Synthesising Detection Monitors] 
	\normalfont
	Recall formula \hV defined in \Cref{ex:uhml-formula}, from this formula we can synthesise a detection monitor using the synthesis function defined in \Cref{fig:mon-synthesis}.
	\begin{align*}
	& \quad \gRV{\hmax{\hVarX}{\hnec{\symReqA}(\hnec{\symReqB}\hfls\,\hand\,\hnec{\symAns}\hVarX)}} \\ 
	= & \quad \mrec{\mx}{ \bigl( \gRV{\hnec{\symReqA}(\hnec{\symReqB}\hfls\,\hand\,\hnec{\symAns}\hVarX)} \bigr) } \\	
	= & \quad \mrec{\mx}{ \bigl( \mact{\symReqA}{(\gRV{\hnec{\symReqB}\hfls\,\hand\,\hnec{\symAns}\hVarX})\!\!}\!\! \bigr) }\\	
	= & \quad \mrec{\mx}{ \bigl( \mact{\symReqA}{(\mch{\gRV{\hnec{\symReqB}\hfls}}{\gRV{\hnec{\symAns}\hVarX}})}\!\! \bigr) }\\	
	= & \quad \mrec{\mx}{ \bigl( \mact{\symReqA}{(\mch{\mact{\symReqB}{\g{\hfls}}}{\mact{\symAns}{\g{\hVarX}}})}\!\! \bigr) }\\[-8mm]
\end{align*}
\noindent Resultant Monitor: 
\begin{displaymath}
	\mVdef
\end{displaymath}
Notice how the resultant monitor is \emph{identical} to monitor $\mV_1$ (defined earlier in \Cref{ex:monitor-instrum}). Hence, this derivation demonstrates an automated way of constructing monitors directly from a formula, such that the derived monitor corresponds (in the sense of \Cref{def:mon-form-corr}) to the formula it was derived from. \bqed
\end{example}

Finally, the authors also proved that \shml and \chml are \emph{maximally expressive} \wrt safety and co-safety properties \ie \emph{every} safety (\resp co-safety) property that can be defined in \uhml can be converted into a \emph{semantically equivalent} property expressed in \shml (\resp \chml). 

\begin{example}[Maximally expressive subsets]
	Consider the following safety property $\syn{\hnec{a}\hfls\hor\hnec{a}\hfls}\notin\shml$. This property can be redefined into the semantically equivalent \shml property $\syn{\hnec{a}\hfls}\in\shml$.  \bqed
\end{example} 

\chapter{A Framework for Runtime Enforcement}
\label{sec:enf-model}
Enforcement monitors (\aka enforcers) implement mechanisms which ensure that the runtime behaviour of some process is kept in line with some correctness property. Hence, unlike detection monitors, enforcers are not only capable of \emph{recognising} execution traces by detecting whether they satisfy or violate some property, but are also able to \emph{transform} invalid executions into valid ones, thereby enforcing the behaviour dictated by the said property upon the process under scrutiny.

\paragraph{Chapter Overview}
We open this chapter by giving a brief background in relation to runtime enforcement in \Cref{sec:re-background}. Following this, in \Cref{sec:symbolic-transducers} we define a formal mechanism for \emph{transforming} concrete system events which are then employed by the \emph{runtime enforcement framework}, presented in \Cref{sec:re-framework}, to transform invalid system executions into valid ones.

In \Cref{sec:enforceability}, we present novel definitions by which we formally define the meaning of \emph{enforceability}, \ie we define the criteria required for a \uhml formula to be enforceable. Finally, we conclude this chapter in \Cref{sec:re-conc} with a summary of the presented content, in which we highlight the main contributions of this chapter.

\section{A brief account of Runtime Enforcement} \label{sec:re-background}
Runtime Enforcement (RE) \cite{Falcone2012,Ligatti2005,Ligatti2010} is a monitoring technique that aims to ensure that a given system always behaves in accordance to a given correctness property. Enforcement monitors are thus used to keep track of the system's behaviour at runtime, and if necessary, modify the system's dynamic behaviour to keep it in line with the given property. 

Runtime Enforcement is used in areas such as software security \cite{schneider2000,Ligatti2005}, since enforcement monitors provide an excellent mechanism to counter malicious attacks that hijack the control flow of the enforced system. Runtime Enforcement, however, raises a number of issues relating to the expressiveness of the logic used for defining specifications, the correctness of the enforcers themselves, and the performance overheads imposed by the enforcers.  

In general, the more expressive the logic is, the more types of correctness properties one can express. Despite the high expressiveness of the logic, it may allow for defining properties that cannot be enforced at runtime by an enforcer $-$ identifying which parts of the logic are enforceable is therefore crucial. Moreover, ensuring that the derived enforcers behave correctly is also essential as they must ensure correct system behaviour, meaning that if they behave erroneously, they might corrupt well-behaved systems. Furthermore, designing the enforcers to be as efficient as possible is necessary since the enforcers need to execute alongside the system under scrutiny. Hence, an enforced system needs to make use of additional hardware resources (CPU time, memory, \etc) when executing, since it must also execute the enforcer's code; this is often seen as an overhead which must be minimized so to avoid rendering the enforced system unusable in practice. 

The notion of Runtime Enforcement was first introduced by Schneider \etal \cite{Schneider1999,schneider2000} as \emph{Security Automata} --- later renamed to \emph{Truncation Automata} by Ligatti \etal in \cite{Ligatti2005}. These automata, however, are limited \wrt the type of properties that they can enforce, since these automata can only prevent the specified bad behaviour from occurring by terminating the system just before the property is violated $-$ this means that Truncation automata can only enforce safety properties. Truncation Automata are \emph{sequence recognizers}, \ie they are only able to read an action from a trace of system events, and transition from one state to another without altering the trace in any way.

Ligatti \etal \cite{Ligatti2002,Ligatti2005} sought to widen the set of enforceable properties by defining \emph{Suppression} and \emph{Insertion automata}. These two enforcement mechanisms differ from Truncation automata as they are based on \emph{sequence transformation}, rather then recognition. Sequence transformers are automata that are not only capable of reading an action from the system's trace and of transitioning from one state to another accordingly (as in a standard automaton), but are also capable of modifying the trace as a result of the applied transition. 

A Suppression automaton is able to enforce properties by \emph{suppressing} specific program actions; this allows for enforcing safety properties by suppressing violating actions rather than by terminating the program outright. Insertion automata seek to enforce properties by \emph{inserting} (executing) a sequence of one or more actions on behalf of the enforced program; this allows for enforcing more expressive types of properties such as co-safety and infinite-renewal properties, amongst others. Automata defining both suppressions and insertions are known as \emph{Edit Automata} \cite{Ligatti2005,Ligatti2010}.

Ligatti \etal realised the importance of providing correctness guarantees about enforcement automata, and thus proposed that enforcement automata should at least guarantee \emph{Soundness} and \emph{Transparency} \cite{Ligatti2005,Ligatti2010}. Soundness requires that a system that is being enforced by a monitor must never violate the enforced property, while \emph{Transparency} states that valid executions should not be altered in any way by the enforcement automaton. 

Bielova \cite{Bielova2011Predictability,Bielova2011PhD} however stressed that until now there is \emph{no distinction between the specification and the enforcement monitor}, which burdens the specifier with having to specify the enforcers themselves in terms of edit-automata. This means that the specifier must \emph{manually} identify the points in which the monitor must suppress or insert a system action. 

\section{From Symbolic Events to Transformations} \label{sec:symbolic-transducers}
As described in \Cref{sec:symevent-match}, symbolic events, $\actS\!=\!\actSN{\pate}{\predc}$, provide a neat way for describing a set of concrete events, thereby enabling monitors to \emph{recognise} a wider range of execution traces. However, by themselves symbolic events are not capable of modifying the events that they describe. We thus introduce \emph{Symbolic Transformations}, $\actSTN{\pate}{\predc}{\pate'}$, which extend symbolic events by allowing for \emph{replacing} concrete events which match pattern \pate and satisfy condition {\predc\s}, with the concrete event $\pate'\s$, where \s is obtained as a result of successful pattern matching. This behaviour is formally defined below.

\begin{definition}[Denotational Semantics for Symbolic Transformations] For an arbitrary \emph{Symbolic Transformation} $\actSTN{\pate}{\predc}{\pate'}$,
	$$	{ \normalfont{\;\normalfont\syn{\actSTN{\pate}{\predc}{\pate'}}}\;\defeq\;
	\fun{x}{\begin{xbrace}{cl}
		(\pate'\s,\s) & \quad \text{when } \mtchS{\actSN{\pate}{\predc}}{x}{\,=\,}\s \\
		\bot & \quad\text{otherwise}
	\end{xbrace}}}$$
	A symbolic transformation, \actSTN{\pate}{\predc}{\pate'}, thus denotes a \emph{function} which accepts a concrete event \acta via argument $x$, and returns a \emph{pair} containing a (possibly) different concrete event $\pate'\s$ along with the substitution environment, whenever the input concrete event \acta satisfies the symbolic event \actSN{\pate}{\predc}, \ie when $\mtchS{\actSN{\pate}{\predc}}{x}{\,=\,}\s$ (see \Cref{def:sym-sem}).
	
	As a result of this function application, \ie {\;\normalfont$\syn{\actSTN{\pate}{\predc}{\pate'}}(\acta)$}, the input concrete event \acta is transformed into another concrete event $\pate'\s=\actb$, where the transformation pattern $\pate'$ is a \emph{closed symbolic pattern}, \ie any data variables present in $\pate'$ must also be defined in \pate, such that when substitution \s is applied on $\pate'$, this yields a concrete event $\actb=\pate'\s$. 
	
	We abuse notation and say that two symbolic transformations are \emph{disjoint} whenever the symbolic events they define are disjoint, \ie {\;\normalfont$\distinctBi{\actSTN{\pate_1}{\predc_1}{\pate'_1}}{\actSTN{\pate_2}{\predc_2}{\pate'_2}}$} whenever {\;\normalfont$\distinctBi{\actSN{\pate_1}{\predc_1}}{\actSN{\pate_2}{\predc_2}}$}. Finally, we adopt the shorthand notation {\;\normalfont$\actSTN{\pate}{\predc}{\pate'}(\acta)$} to denote {\;\normalfont$\syn{\actSTN{\pate}{\predc}{\pate'}}(\acta)$}. \bqed
\end{definition}

\begin{example}[Symbolic Transformations]
	Symbolic Transformations thus allow for \emph{replacing} a concrete event with another. For instance, consider event \send{i}{3} and transformation \actSTN{\send{i}{\dvV}}{\dvV{>}2}{\send{i}{\actN{err}(\dvV)}}, where the latter transformation \emph{replaces} input \send{i}{3} into an error report \send{i}{\actN{err}(3)}, since value $3$ was matched with variable $\dvV$, such that $(\dvV>2)\sub{3}{\dvV}$, \ie $3>2$ evaluates to true.
	
	As \actt-transitions are not perceivable by external observers, the replacement mechanism provided by the Symbolic Transformations, can also be used to replace visible concrete events into unobservable silent \actt-actions. For instance, by using \actSTN{\send{i}{\dvV}}{\dvV{>}2}{\actt} we can suppress events such as \send{i}{3} by transforming them into a \actt-action, thereby making them invisible to external observers.
	
	By omitting transformation, an identity transformation such as \actSTN{\send{i}{\dvV}}{\dvV{>}2}{\send{i}{\dvV}} can simulate action recognition by outputting the original input event, and hence $\actSTN{\send{i}{\dvV}}{\dvV{>}2}{\send{i}{\dvV}}(\send{i}{3})$ produces the pair $(\send{i}{3},\sub{3}{\dvV})$. \bqed
\end{example}


\noindent Symbolic Transformations will therefore serve to provide the bases for action transformation in our enforcement framework.

\section{The Framework} \label{sec:re-framework}
We model enforcers in terms of LTSs, through the syntax of \Cref{fig:mod-re}. The syntax allows for defining: the \emph{identity} enforcer (\miden), enforcers that are \emph{prefixed by symbolic transformations} (\prf{\actSTN{\pate}{\predc}{\pate'}}{\eV}), \emph{recursive} enforcers (\mrec{\mx}{\eV}), and \emph{selections} ($\mCh\eV_i$ where \IndSet is a set of indices such that $\mCh\eV_i$ represents $\mch{\eV_1}{\mch{\ldots}{\eV_n}}$, where $1,\ldots,n\in\IndSet$). The structure of enforcers is thus very similar to that of detection monitors, with the exception that enforcers cannot issue verdicts \myes and \mno, and action recognition via symbolic events, is extended to \emph{action transformation} via symbolic transformations.

The behaviour of the enforcers is also similar to that of detection monitors for the common constructs such as recursion (\rtit{eRec}) and selections (\rtit{eSel}. In fact, for recursion we use a \actt-free semantics via rule \rtit{eRec}, which allows a recursive enforcer, \mrec{\mx}{\eV}, to reduce to some $\eV'$ whenever its unfolded version, $\eV\sub{\mrec{\mx}{\eV}}{\mx}$, reduces to $\eV'$. For selections, rule \rtit{eSel} states that whenever a single enforcer $\eV_j$ (where $\eV_j$ is part of the summation $\mCh\eV_i$, \ie $j\in\IndSet$) is able to transform some action \acta and reduce to $\eV_j'$, \ie $\eV_j\traS{\actau}\eV'_j$, then the entire summation reduces to $\eV_j'$ \ie $\mCh\eV_i\traS{\actau}\eV'_j$.

As described by the transition rules in \Cref{fig:mod-re}, enforcer transitions range over \emph{action transformation} rather than action recognition, \ie the enforcer is able to modify an input action by outputting a different one, $\actau$ (where \acta might differ from $\actu$). Our enforcers thus achieve action transformation as described by rule \rtit{eTrns}. This rule states that at runtime an enforcer prefixed by a symbolic transformation, \prf{\actSTN{\pate}{\predc}{\pate'}}{\eV}, can replace an action \acta into a (possibly) different concrete action \actu, thereby reducing into its derivative $\eV\s$, where \actu and \s are obtained by applying the symbolic transformer to \acta, \ie via \mtchS{\actSTN{\pate}{\predc}{\pate'}}{\acta}{\,=\,}(\actu,\s).

\begin{figure}[t]
	\noindent\textbf{Syntax}
	\begin{align*}
	\eV,\eVV\in\Enf &\quad::=\quad \; \mtrns{\pate}{c}{\pate'}{\eV} & \vert\; & \mCh\eV_i&  \vert\;&  \mrec{\mx}{\eV}&  \vert\; & \mx  &\vert\; &  \miden 
	\end{align*}
	
	\noindent\textbf{Dynamics}
	\begin{mathpar}
		\inference[\rtit{eId}]{ }{\miden \traS{\ioact{\acta}{\acta}} \miden } \and 
		\inference[\rtit{eSel}]{\eV_j \traSS{\actau} \eV_j'}{\mCh\,\eV_i \traSS{\actau} \eV_j'}[$j\in\IndSet$] 
		\and
		\inference[\rtit{eRec}]{\eV\sub{\mrec{\mx}{\eV}}{\mx} \tra{\actau} \eV' }{\mrec{\mx}{\eV} \tra{\actau} \eV'} \and
		\inference[\rtit{eTrns}]{ \mtchS{\actSTN{\pate}{\predc}{\pate'}}{\acta}{\,=\,}(\actu,\s) }{\mtrns{\pate}{c}{\pate'}{\eV} \traS{\ioact{\acta}{\actu}} \eV\s}
		\\
	\end{mathpar}
	
	\noindent\textbf{Instrumentation}
	\begin{mathpar}		
		\inference[\rtit{iTer}]{\pV \tra{\acta}\pV' && \eV\ntra{\acta}}{ \i{\eV}{\pV} \tra{\acta} \i{\miden}{\pV'}} \and 
		\inference[\rtit{iAsyP}]{ \pV \tra{\actt}\pV' }{ \i{\eV}{\pV} \tra{\actt} \i{\eV}{\pV'} }  \and 
		\inference[\rtit{iEnf}]{\pV \tra{\acta}\pV' && \eV\tra{\ioact{\acta}{\actu}}\eV'}{ \i{\eV}{\pV} \tra{\actu} \i{\eV'}{\pV'}} 
	\end{mathpar}	
	\caption{A model for Enforcement monitors}
	\label{fig:mod-re}
\end{figure}

\Cref{fig:mod-re} also describes an instrumentation relation for enforcement monitors which relates the behaviour of an LTS process $\pV$ with that of an enforcer \eV where the resultant LTS process \i{\eV}{\pV} denotes the \emph{enforced system}. As in the case of detection monitors, in an enforcement instrumentation, the process leads the behaviour of an enforced system: whenever the process cannot perform a transition, the enforced system will not either. However, unlike detection monitors, this type of instrumentation also allows the enforcer to determine the visible actions of the enforced system, \ie this instrumentation allows the monitor to change an \acta-transition into a \actu-transition, where \actu can either be the same as the input (\ie remains \acta), be changed into a different concrete event \actb, or else be suppressed into a \actt-transition. 

Specifically, rule \rtit{iEnf} states that if a process can transition with action \acta and the \resp enforcer can \emph{transform} this action into \actu, then in an instrumented system, \i{\eV}{\pV}, the enforcer, \eV, and the system, \pV, transition in lockstep over the enforcer's output action \actu.  However, if the enforcer cannot perform such a reduction, $\eV{\,\ntraSS{\acta}}$, \ie $\nexists\actu,\eV'$ such that $\eV\traS{\actau}\eV'$, the instrumentation forces it to terminate by reducing to the identity enforcer, \miden, while the process is allowed to proceed unaffected as shown by rule \rtit{iTer}. Same as per the instrumentation for detection monitors, rule \rtit{iAsyP} allows processes to transition independently \wrt internal transitions. However, the instrumentation does not allow enforcers to transition independently over \actt-transitions, given that our enforcers are \emph{\actt-free}, \ie do not perform silent actions themselves.

We use the notation $\eV{\wtraSS{\wioact{\tr}{\trr}}}\eV'$, to denote a \emph{sequence of transformations}, \eg $\wioact{\tr}{\trr}=\ioact{\acta_1}{\actu_1},\ldots,\ioact{\acta_n}{\actu_n}$, performed by enforcer \eV, where \tr denotes the \emph{input trace}, \eg $\tr=\acta_1,\ldots,\acta_n$, while \trr denotes the \emph{output trace}, \eg $\tr=\actu_1,\ldots,\actu_n$. Since the instrumented system, \i{\eV}{\pV} is a standard LTS, we use standard notation $\i{\eV}{\pV}\wtra{\trr}\i{\eV'}{\pV'}$ where \trr denotes the sequence of enforced events, \ie \trr is the output trace generated by the enforcer \eV. 

\begin{example} \normalfont \label{ex:enf-instrum} Consider processes $\pV_1$ and $\pVV_1$ (defined in \Cref{ex:lts}). Assume that this time we want to enforce the property stated in \Cref{ex:monitor-instrum}, \ie that every request is followed by an answer, as formally specified by formula $\hV_1$ (see \Cref{ex:uhml-formula}). One drastic way to enforce this behaviour is by suppressing every request sent to the server which can be done by transforming every request action matching pattern \patReqA, into a silent \actt-action, when $\dvV\neq j$, as shown below.
\begin{displaymath}
	\eV_1 \defeq \eVAdef
\end{displaymath}
From an external point of view, by suppressing every request, the server can never execute two requests in a row, and so it can never be the case where a request is not followed by an answer. For instance, given the non-determinism inherent to process $\pVV_1$, a violating trace, $\actReq.\actReq.\actAns$, may be exhibited. As shown in the derivation below, enforcer \eV neutralises this invalid behaviour by suppressing every request via transformation $\trnsReqC$ such that the output trace is $\actt.\actt.\actAns$, which is equivalent to $\actAns$.  
\begin{align*}
	\i{\eV_1}{\pVV_1} & \traS{\;\;\actt\;\;} \i{\eV_1}{\pVV_1}  && \text{\small when } \pVV_1\tra{\actReq}\pVV_1\emph{, since by \rtit{iEnf}}, \eV_1 \tra{\ioact{\actReq}{\actt}}\eV_1 \\ 
	& \traS{\;\;\actt\;\;} \i{\eV_1}{\pVV_2}  && \text{\small when } \pVV_1\tra{\actReq}\pVV_2 \emph{ (same) } \\
	& \traSS{\actAns} \i{\eV_1}{\pVV_1} && \text{\small when } \pVV_2\tra{\actAns}\pVV_1\emph{, since by \rtit{iEnf}}, \eV_1 \tra{\ioact{\actAns}{\actAns}}\eV_1 \\[-2mm]
	& \; \ldots  
\end{align*}
Although $\eV_1$ enforces the desired behaviour as required, it employs a premature enforcement stance by which it may unnecessarily edit correct executions as well. For instance, the derivation below shows how $\eV_1$ needlessly modifies the behaviour of $\pVV_1$ such that when it produces a correct execution trace $\actReq.\actAns.\actReq$, this is transformed into, $\actt.\actAns.\actt$.
\begin{align*}
	\i{\eV_1}{\pVV_1} & \traS{\;\;\actt\;\;} \i{\eV_1}{\pVV_2}  && \text{\small when } \pVV_1\tra{\actReq}\pVV_2\emph{, since by \rtit{iEnf}}, \eV_1 \tra{\ioact{\actReq}{\actt}}\eV_1 \\
	& \traSS{\actAns} \i{\eV_1}{\pVV_1} && \text{\small when } \pVV_2\tra{\actAns}\pVV_1\emph{, since by \rtit{iEnf}}, \eV_1 \tra{\ioact{\actAns}{\actAns}}\eV_1 \\
	& \traS{\;\;\actt\;\;} \i{\eV_1}{\pVV_2}  && \text{\small when } \pVV_1\tra{\actReq}\pVV_2\emph{, since by \rtit{iEnf}}, \eV_1 \tra{\ioact{\actReq}{\actt}}\eV_1 \\[-2mm]
	& \; \ldots  
\end{align*}

An alternative enforcer can be defined such that enforcement is delayed to the very end, \ie until the enforcer is sure that the desired behaviour will definitely be compromised; for instance consider $\eV_2$ (defined below).
\begin{align*}
	\eV_2 &\defeq \mrec{\mx}{\bigl(\mact{\trnsReqA}{\eV_2'}\bigr)}\\
	\eV_2' &\defeq \mrec{\my}{\bigl( \mch{\mact{\trnsAns}{\mx}}{\mact{\trnsReqB}{\my}} \bigr)}
\end{align*}
Enforcer, $\eV_2$, prevents the violation by using an additional recursive construct, \mrec{\my}{(\ldots)}, to continuously suppress every request matching pattern \patReqA such that $\dvV\neq j$, succeeding the primary occurrence of the request action. In this way, whenever the enforced process executes two (or more) consecutive requests matching \patReqA such that $\dvV\neq j$, the enforcer keeps suppressing requests succeeding an unanswered request until an answer matching \patAns is produced by the enforced process. 

The following derivation thus shows how enforcer $\eV_2$ modifies the invalid execution trace $\actReq.\actReq.\actAns$, produced by $\pVV_1$, into the corresponding valid trace $\actReq.\actt.\actAns$, \ie $\actReq.\actAns$.
\begin{align*}
	\i{\eV_2}{\pVV_1} & \traSS{\actReq} \i{\eV_2'\sub{\eV_2}{\mx}}{\pVV_1}  && \text{\small when } \pVV_1\tra{\actReq}\pVV_1\emph{, since by \rtit{iEnf}}, \eV_2 \tra{\ioact{\actReq}{\actReq}}\eV_2'\sub{\eV_2}{\mx} \\
	& \traS{\;\;\actt\;\;} \i{\eV_2'\sub{\eV_2}{\mx}}{\pVV_2}  && \text{\small when } \pVV_1\tra{\actReq}\pVV_2\emph{, since by \rtit{iEnf}}, \eV_2'\sub{\eV_2}{\mx} \tra{\ioact{\actReq}{\actt}}\my\sub{\eV_2'\sub{\eV_2}{\mx}}{\my} \\
	& \traSS{\actAns} \i{\eV_2}{\pVV_1} && \text{\small when } \pVV_2\tra{\actAns}\pVV_1\emph{, since by \rtit{iEnf}}, \eV_2'\sub{\eV_2}{\mx} \tra{\ioact{\actAns}{\actAns}}\mx\sub{\eV_2}{\mx} \\[-2mm]
	& \; \ldots  
\end{align*}
In contrast to enforcer $\eV_1$, the derivation below demonstrates how the late enforcement methodology adopted by $\eV_2$ helps to preserve correct execution traces such as $\actReq.\actAns.\actReq$ by leaving them unchanged.
\begin{align*}
	\i{\eV_2}{\pVV_1} & \traSS{\actReq} \i{\eV_2'\sub{\eV_2}{\mx}}{\pVV_2}  && \text{\small when } \pVV_1\tra{\actReq}\pVV_2\emph{, since by \rtit{iEnf}}, \eV_2\tra{\ioact{\actReq}{\actReq}}\eV_2'\sub{\eV_2}{\mx} \\
	& \traSS{\actAns} \i{\eV_2}{\pVV_1} && \text{\small when } \pVV_2\tra{\actAns}\pVV_1\emph{, since by \rtit{iEnf}}, \eV_2'\sub{\eV_2}{\mx} \tra{\ioact{\actAns}{\actAns}}\mx\sub{\eV_2}{\mx} \\
	& \traSS{\actReq} \i{\eV_2'\sub{\eV_2}{\mx}}{\pVV_2}  && \text{\small when } \pVV_1\tra{\actReq}\pVV_2\emph{, since by \rtit{iEnf}}, \eV_2\tra{\ioact{\actReq}{\actReq}}\eV_2'\sub{\eV_2}{\mx}\\[-2mm]
	& \; \ldots  
\end{align*} \\[-15mm] \bqed
\end{example}

\section{Defining Enforceability of the Logic} \label{sec:enforceability}
In this section we shift back to the logic and investigate what it means for an enforcer to enforce a property. As specified below, we define \emph{enforceability} of a logic as being the relationship between the meaning of a property, expressed in the said logic, and the ability to enforce it at runtime upon a specific system. 

\begin{definition}[Enforceability] A \uhml formula $\hV\in\Sat$ is \emph{enforceable} whenever
	\renewcommand{\qedsymbol}{\bqed}	
	\[
		\pushQED{\bqed}
			\exists\eV\in\Enf\cdot\eV \emph{ enforces }\hV \qedhere
		\popQED
	\]
\end{definition}
Intuitively, for a property \hV to be enforceable, there must exist an enforcer \eV which is capable of modifying the dynamic behaviour of \emph{any process} \pV in order to keep it in line with the behaviour specified in \hV. However, as discussed in \Cref{ex:enf-instrum},an enforcer may adopt different approaches in order to enforce the behaviour dictated by the given property, \eg in \Cref{ex:enf-instrum} although $\eV_1$ actually prevents the violation of $\hV_1$, it unnecessarily modifies correct behaviours, whereas $\eV_2$ does not.

As stated by Ligatti \etal in \cite{Ligatti2005,Falcone2012,Bielova2011PhD} (see \Cref{sec:re-background}), for an enforcer to adequately enforce a property, it must at least ensure that the enforced process always executes correctly \wrt \hV, \ie by either preventing it from executing violating runtime behaviour, or by ensuring the execution of runtime behaviour satisfying \hV. An enforcer capable of doing so is said to be \emph{Sound}. In our case, since we represent the enforced system, \i{\eV}{\pV}, as an LTS in itself, we can formally specify \emph{Enforcement Soundness} as follows:
\begin{definition}[Sound Enforcement] \label{def:senf} We say that enforcer \eV \emph{soundly enforces} a formula \hV, denoted as \senfdef{\eV}{\hV}, iff \; $\forall \pV\in\Proc\cdot \i{\eV}{\pV}\in\syn{\hV}$.\bqed
\end{definition}
\noindent The above definition specifies that enforcer \eV \emph{soundly enforces} a formula \hV if \eV can enforce any process $\pV$ such that the resultant enforced LTS, \i{\eV}{\pV}, \emph{always} satisfies \hV. 

Enforcement \emph{soundness} on its own is, however, a relatively \emph{weak constraint} as it does not regulate the extent of the applied enforcement. For instance, \Cref{ex:enf-instrum} demonstrates that enforcer although $\eV_1$ manages to keep the execution of process $\pVV_1$ in line with property $\hV_1$, it however adopts a conservative enforcement approach which needlessly modifies the correct behaviour of $\pVV_1$ as well. Intuitively, this example indicates\footnote{In general proving enforcement soundness is very hard as it requires proving that $\i{\eV}{\pV}\in\syn{\hV}$, for \emph{every possible process} \pV. However, \Cref{ex:enf-instrum} only shows that $\eV_1$ soundly enforces property $\hV_1$ on $\pVV_1$ (it might not be the case other processes).} that enforcer $\eV_1$ \emph{soundly enforces} property $\hV_1$, yet lacks an element of \emph{Transparency} \cite{Ligatti2005,Falcone2012,Bielova2011}, since the enforcer can unnecessarily modify valid system behaviour. Similarly, to enforce a simple property $\hV=\hnec{a}\hnec{b}\hfls$ on a process $\pV$, we can use an enforcer that \emph{suppresses every action} produced by $\pV$ thereby suppressing the entire execution: in doing so $\pV$ will surely never violate \hV, but will neither exhibit any kind of valid behaviour.

\emph{Transparency} \cite{Ligatti2005,Falcone2012,Bielova2011}, thus dictates that whenever a process $\pV$ already satisfies the property \hV, the assigned enforcement monitor $e$ must refrain from altering the runtime behaviour of $\pV$ as this would not be necessary. Transparency therefore aims to preserve the original behaviour of the process as much as possible by imposing the least number of enforcement actions, \ie enforcement is only applied when necessary. For instance, enforcer $\eV_2$ in \Cref{ex:enf-instrum}, applies a lazy enforcement approach whereby the behaviour of process $\pVV_1$ is only modified when the runtime behaviour of $\pVV_1$ violates $\hV_1$. This ensures that valid behaviour is never modified unnecessarily. Hence, once again \Cref{ex:enf-instrum} demonstrates that enforcer $\eV_2$ is more transparent then $\eV_1$ when enforcing $\hV_1$ on $\pVV_1$, as unlike $\eV_1$, it does not affect the valid behaviour of $\pVV_1$. Hence, we formally define \emph{enforcement transparency} in \Cref{def:tenf} below.

\begin{definition}[Transparent Enforcement] \label{def:tenf} An enforcer \eV is \emph{transparent} when enforcing a formula \hV, denoted as \tenfdef{\eV}{\hV}, iff 
	\renewcommand{\qedsymbol}{\bqed}	
	\[
	\pushQED{\bqed}
		\forall \pV\in\Proc\cdot \pV\in\syn{\hV} \imp \i{\eV}{\pV}\sim \pV  \qedhere
	\popQED
	\]
\end{definition}

\noindent This definition states that an enforcer \eV enforces a formula \hV in a \emph{transparent} manner, if for every LTS process $\pV$ that \emph{satisfies} \hV, the enforced LTS \i{\eV}{\pV} is \emph{bisimilar} to the original process $\pV$. This ensures that although the resultant enforced LTS, \i{\eV}{\pV}, is structurally different from the original LTS $\pV$, its behaviour is still \emph{perceived to be the same} as that of $\pV$. Based on \Cref{def:senf,def:tenf} we can thus define a \emph{stronger notion of enforcement} as defined below.

\begin{definition}[Strong Enforcement] \label{def:strong-enf} We say that an enforcer \eV \emph{strongly enforces} formula \hV, denoted as \enfdef{\eV}{\hV}, when
	\begin{itemize}
		\item \senfdef{\eV}{\hV}, \ie \eV \emph{soundly enforces} \hV; and
		\item \tenfdef{\eV}{\hV}, \ie \eV is \emph{transparent} when enforcing \hV. \bqed
	\end{itemize} 
\end{definition}
\noindent Strong enforcement thus requires an enforcer \eV to \emph{soundly} and \emph{transparently} enforce a property \hV.

\begin{remark}[Novelty]
	It is important to note that unlike existing work \cite{Ligatti2005,Ligatti2010,Falcone2009EM,Falcone2012,Bielova2011PhD}, we define the resultant enforced system as the LTS \i{\eV}{\pV}. This allows is to give \emph{stronger definitions} for enforcement \emph{soundness} and \emph{transparency} (\ie \Cref{def:senf,def:tenf}), as these are given in terms of the original LTS process $\pV$ and the enforced LTS process $\i{\eV}{\pV}$, as opposed to the more classical definitions that present them in terms of an \emph{input and output (enforced) trace}.
	
	In fact, the classic definitions for soundness (\eg \cite{Ligatti2005,Ligatti2010,Falcone2009EM,Falcone2012,Bielova2011PhD}) require that every enforced \emph{execution trace} that can be produced by an enforcement automaton (monitor), \eV, should satisfy the enforced property \hV. By contrast, in our definition of soundness, we require that for \emph{every process} \pV, the resultant enforced LTS process \i{\eV}{\pV} must always satisfy property \hV.
	
	Similarly, unlike the classic definitions of transparency (\eg \cite{Ligatti2005,Ligatti2010,Falcone2009EM,Falcone2012,Bielova2011PhD}), our definition does not only require trace equivalence between \i{\eV}{\pV} and \pV (as per classic definitions), but instead imposes a \emph{stronger equivalence} criterion, \ie that $\i{\eV}{\pV}\bisim\pV$ (see \cite{Aceto2007Book} for Trace Equivalence vs Bisimilarity). \bqed
\end{remark}

\subsection{Concluding Remarks}\label{sec:re-conc} In this chapter we have presented a novel framework describing the runtime behaviour of enforcers. The main novel contributions of this chapter include:
\begin{description}
	\item[Symbolic Transformations\!\!], which formally define a mapping mechanism for transforming a concrete system event into a (possibly) different one as specified the transformation pattern (see \Cref{sec:symbolic-transducers}).
	\item[An LTS semantics for Enforcers\!\!], which formalise the structure and dynamic behaviour of enforcers, along with the interaction between the enforcer, \eV and the process under scrutiny, \pV, in the form of the instrumented LTS, \i{\eV}{\pV} (see \Cref{sec:re-framework}); and 
	\item[A Formal Definition for Enforceability\!\!], defining the relationship between the meaning of a \uhml property and its ability to be adequately enforced at runtime by an enforcer. With this definition we establish that an enforcer \eV \emph{strongly enforces} a formula \hV whenever it is able to do it in a \emph{sound} and \emph{transparent} manner. By defining the enforced system as an LTS, \i{\eV}{\pV}, we were able to provide novel definitions Soundness and Transparency which are \emph{stronger} than the classic definitions (see \Cref{sec:enforceability}).
\end{description}

\paragraph{Discussion.} So far we have defined the meaning of enforceability in relation to our logic, however, we still have to explore which of the properties, expressible via \uhml, can actually be enforced. Several works in RE \cite{Ligatti2005,Bielova2008,Falcone2012} have already established that suppression enforcers are ideal for detecting potentially violating executions and suppressing parts of them to \emph{prevent} the violation of \emph{safety properties}. 
%
%
Similarly, it was established that insertion enforcers can also be used to enforce other types of properties such as \emph{co-safety}. 
%

We thus explore the enforceability of \uhml properties in an incremental manner. We will first start by exploring the enforceability of \uhml properties,\wrt suppression enforcement, and later on we aim to explore it \wrt insertion enforcement.

\chapter{Enforcing Safety Properties via Suppressions}
\label{sec:enf-synthesis}
As stated in other work \cite{Ligatti2005,Ligatti2010,Bielova2011PhD,Falcone2012}, suppression enforcement is ideal to prevent the violation of \emph{safety properties} by stopping erroneous events from occurring. In this chapter we limit ourselves to identifying a subset of \uhml formulae that are enforceable via suppressions and establish a \emph{synthesis function} that converts formulae from the identified enforceable subset into the \resp suppression enforcers. 

Particularly, we investigate the \emph{enforceability} of safety properties expressed in terms of \shml since, in \cite{Francalanza2015Mon}, this syntactic subset was proven to be \emph{maximally expressive} \wrt safety properties, \ie any safety property that can be defined in \uhml can be expressed in terms of a \emph{semantically equivalent} \shml formula.

\begin{figure}[t]
	$$ \hV,\hVV\in\shml\; ::=\; \htru\;\vert\;\hfls\;\vert\;\hVarX \;\vert\;\hmax{\hVarX}{\hV} \;\vert\;
	\hV\!\hand\!\hVV \;\vert\; \hnec{\actS}{\hV} $$ 
	\caption{The Syntax for the \shml subset.}
	\label{fig:shml-syn}
\end{figure}

In \Cref{fig:shml-syn} we recall the syntax for \shml. The logic is restricted to \emph{truth} and \emph{falsehood} (\htru and \hfls), conjunctions (\hV{\,\hand\,}\hVV), and necessity modalities (\hnec{\actS}{\hV}), while recursion may only be expressed through maximal fixpoints (\hmax{\hVarX}{\hV}). The semantics for these constructs follows from that of \Cref{fig:recHML}.

\begin{example}[Enforcing \shml formulae] \label{ex:enf-shml} Consider the following recursive formula $\hV_2$,
	$$ \hV_2 \defeq \hVBdefConc $$
	\noindent Formula $\hV_2$ defines the same invariant property as $\hV_0$ (defined in \Cref{ex:uhml-formula}), which holds when a request is not immediately followed by a subsequent request after an arbitrary number of answered requests. The formula thus specifies that a process is incorrect when it performs two consecutive requests, \ie $\hnec{\actReq}\hnec{\actReq}\hfls$, but recurses whenever an answer is produced following a request, \ie $\hnec{\actReq}\hnec{\actAns}\hVarX$.
	
	One way how to enforce $\hV_2$ is by generating a suppression enforcer such as $\eV_0$ that prevents a process from performing two or more subsequent requests.
	$$ \eV_0 \defeq \eVOdef $$
	Enforcer $\eV_0$ enforces formula $\hV_0$ by suppressing every request action, $\actReq$, that is performed  after an unanswered request, until an answer \actAns is produced, in which case the enforcer recurses. This ensures that the invariant property is enforceable after an arbitrary number of requests as defined by the property.
	
	In \Cref{ex:enf-instrum} we had also shown that this invariant property (that is formalized by both $\hV_0$ and $\hV_2$) can also be enforced via enforcers $\eV_1$ and $\eV_2$ (restated below).
	\begin{align*}
			\eV_1 \defeq&\; \eVAdef\\
			\eV_2 \defeq&\; \eVBdef
	\end{align*} 
	Notice how enforcer $\eV_0$ is very similar to $\eV_2$. Their main difference is that $\eV_0$ only enforces the property when the requesting process is identified by $i$ (as defined by $\hV_0$). Enforcer $\eV_2$ is, however, more generic as it applies enforcement for any requesting process that is identified by any identifier (including $i$) except for the process identified by $j$. \bqed
\end{example} 

In general, a property can be enforced either \emph{deterministically} or \emph{non-deterministically} as defined by \Cref{def:det-enf}.
\begin{definition}[Deterministic Enforcement] \label{def:det-enf}
	An enforcer \eV behaves \emph{deterministically} whenever
	$$ \eV\wtraS{\wioact{\tr}{\tr'}}\eV' \textsl{  and  } \eV\wtraS{\wioact{\tr}{\tr''}}\eV''\;\imp\;\eV'{\,=\,}\eV'' \textsl{  and  } \tr'{\,=\,}\tr''$$
	where \wioact{\tr}{\tr'} and \wioact{\tr}{\tr''} represent \emph{sequences of transformations} performed by enforcer \eV over the \emph{same} input trace \tr, that may result in different output (enforced) traces. \bqed
\end{definition}
Informally, an enforcer \eV is deterministic whenever it is \emph{unable to react differently} for the \emph{same input trace} of events \tr, \ie it \emph{always} reduces to the \emph{same state}, \ie $\eV'{\,=\,}\eV''$, and \emph{always} produces the same output (enforced) trace, \ie $\tr'{\,=\,}\tr''$. In general, an enforcer behaves non-deterministically whenever it defines a selection of two (or more) \emph{non-disjoint symbolic transformations}, \eg enforcer $\mch{\mact{\trnsReqA}{\eV'}}{\mact{\trnsReqO}{\eV''}}$ is \emph{non-deterministic} since the symbolic transformations that are guarding the summation branches are \emph{not disjoint}, \ie $\distinctBi{\trnsReqA}{\trnsReqO}$ is \emph{false}.

Determining statically whether an enforcer is deterministic or not, is not a straightforward task. In fact, concluding that two (or more) symbolic transformations are disjoint cannot be done via a simple syntactic check, instead it requires making sure that there \emph{does not exist} some concrete event \acta that can be transformed by \emph{multiple transformations} guarding different branches in the summation.

\begin{example}[Deterministic Enforcement]
	In \Cref{ex:enf-shml} we claimed that enforcers $\eV_1$ and $\eV_2$ behave deterministically. Intuitively, this is due to the fact that the selections (\mch{\!}{\!}) in both $\eV_1$ and $\eV_2$ are guarded by \emph{disjoint} symbolic transformations, \ie in both cases \distinctBi{\trnsReqC}{\trnsAns}. This ensures that when instrumented with a process \pV, the enforcers \emph{consistently make the same selections} upon analysing the \emph{same concrete events} generated by \pV, and thus the enforcers always behave in the same way.
	
	In fact, enforcer $\eV_1$ is bound to always choose: the \emph{left branch}, \ie \mact{\trnsReqC}{\mx}, upon analysing a \emph{request event} \ie \actReq, the \emph{right branch}, \ie \mact{\trnsAns}{\mx}, upon an \emph{answer event} \ie \actAns, and reduces to the \emph{identity enforcer}, \miden, upon a \emph{close event}, \ie \actCls; a similar argument applies for the selection applied in $\eV_2$.
	
	In this way, $\eV_1$ and $\eV_2$ can \emph{never reach} a point in which they are able to (non-deterministically) choose between two or more branches, and so they are \emph{always bound to react in the same way} for the same input events, thereby producing the same output (enforced) events. \bqed
\end{example}

Deterministic enforcers are appealing due to their \emph{predictable} runtime behaviour, \ie it is easier to predict how a deterministic enforcer will transform a given concrete input event. By contrast, non-determinism introduces subtleties that can lead to harmful unpredictable behaviour, in which the enforcer may sometimes (non-deterministically) select a branch which does not adequately enforce the given property.

\begin{example}[Harmful Unpredictable Enforcement] \label{ex:enf-inconsistent}
	Consider the non-deterministic enforcer $\eV_3$ (defined below) obtained by applying a ``naive'' synthesis function on $\hV_2$, which informally converts the \emph{maximal fixpoint} in $\hV_2$ into a \emph{recursive construct}, the \emph{modal necessities} that are immediately followed by falsehood, into a \emph{suppression transformation}, \eg $\hmaxB{\hVarX}{(\,\ldots\,\hnec{\actReq}\hfls)}$ into $\mrec{\mx}{(\,\ldots\,\mact{\trnsReqB}{\mx})}$, and other necessities into the \emph{identity transformations}, \eg $\hmaxB{\hVarX}{(\,\ldots\,\hnec{\actAns}\hVarX)}$ into $\mrec{\mx}{(\,\ldots\,\mact{\trnsAns}{\mx})}$. 
	\begin{align*}
		\hV_2 &\defeq\; \hVBdefConc \\
		\eV_3 &\defeq\; \eVDdef
	\end{align*}
	Now, consider a very simple process \pVVV which only issues two consecutive request actions and terminates.
	\begin{align*}
		\pVVV  &\defeq\; \prf{\actReq}{\pVVV'} \\
		\pVVV' &\defeq\; \prf{\actReq}{\nil}
	\end{align*}
	Since the transformations guarding the selections in $\eV_3$, \ie \trnsReqD and \trnsReqO, are \emph{not disjoint} (since $\actReq{\,\in\,}\syn{\actSN{\patReqA}{\ctru}}{\cap\,}\syn{\actSN{\patReqA}{\dvV{=}i}}$), enforcer $\eV_3$ can make a \emph{non-deterministic selection} (using rule \rtit{eSel}) whenever process $\pVVV$ makes an initial request. However, based on this choice, $\eV_3$ might not always enforce the required behaviour. In the derivation below we can see that $\eV_3$ manages to enforce the required behaviour by non-deterministically choosing the second branch.
	\begin{align*}
	\i{\eV_3}{\pVVV} & \traSS{\actReq} \i{\mact{\trnsReqB}{\eV_3}}{\pVVV'}  && \text{\small By \rtit{iEnf}, since } \pVVV\tra{\actReq}\pVVV' \text{ and } 
	\eV_3 \tra{\ioact{\actReq}{\actReq}}{\mact{\trnsReqB}{\eV_3}} \\[2mm]
	& \traS{\;\;\actt\;\;} \i{\eV_3}{\nil}  && \text{\small By \rtit{iEnf}, since } \pVVV'\tra{\actReq}\nil \text{ and } 
	 \mact{\trnsReqB}{\eV_3} \tra{\ioact{\actReq}{\actt}}\eV_3 
	\end{align*}
	However, being non-deterministic, $\eV_3$ can also choose the other branch which allows for the occurrence of violating runtime behaviour as shown below.
	\begin{align*}
	\i{\eV_3}{\pVVV} & \traSS{\actReq} \i{\mact{\trnsAns}{\eV_3}}{\pVVV'}  && \text{\small By \rtit{iEnf}, since } \pVVV\tra{\actReq}\pVVV' \text{ and } \\[1mm] &&& \qquad 
	\eV_3 \tra{\ioact{\actReq}{\actReq}}{\mact{\trnsAns}{\eV_3}} \\[2mm]
	& \traSS{\actReq} \i{\miden}{\nil}  && \text{\small Since } \pVV_1\tra{\actReq}\pVV_2\emph{, but by \rtit{iTer}}, \mact{\trnsReqB}{\eV_3}\!\!\ntra{\actReq}
	\end{align*} 
	Notice how in the above derivation, the monitored process still manages to issue two subsequent request actions.\bqed
\end{example}

If we want to adhere to the notion of enforceability from \Cref{sec:enforceability} (\Cref{def:strong-enf}), our enforcers must, first and foremost, guarantee soundness, \ie that a process $\pV$ monitored by a sound enforcer \eV \emph{must always satisfy} a given \uhml formula \hV. This implies that the outcome of the applied transformations must always prevent the property from being violated.

From \Cref{ex:enf-inconsistent} we can thus notice that enforcer $\eV_3$ is not always capable of is enforcing $\hV_2$, and is therefore \emph{unsound}. In general, however, despite exhibiting unpredictable behaviour, non-deterministic enforcers are not necessarily unsound.

\begin{example}[Harmless Non-Deterministic Enforcement] \label{ex:enf-consistent}
	Recall process $\pVV_1$, from \Cref{ex:lts}, and consider the non-deterministic enforcer $\eV_4$ defined below.
	\begin{align*}
		\pVV_1 &\defeq\; \pVVAdef \\
		\eV_4 &\defeq\; \mch{\eV_1}{\eV_2}
	\end{align*}
	Enforcer $\eV_4$ enforces property $\hV_2$ by first making a non-deterministic selection (using rule \rtit{eSel}) upon the occurrence of an initial request action \actReq. Based on this non-deterministic choice, it either enforces the property by \emph{eagerly} suppressing each and every request action via $\eV_1$ or else on a \emph{by-need-basis}, by suppressing requests that occur after an initial unanswered request following an initial request via $\eV_2$ (see \Cref{ex:enf-instrum} for the \resp derivations). Despite being non-deterministic, enforcer $\eV_4$ still \emph{soundly enforces} $\hV_2$, as it always prevent $\pVV_1$ from violating $\hV_2$, even though it does not always apply the same enforcement strategy. \bqed
\end{example}

However, recall from \Cref{def:strong-enf} that soundness is not the only criterion that an enforcer must abide by in order to \emph{strongly enforce} a property; an enforcer is also required to be \emph{transparent}. Non-determinism can, once again, introduce intricate behaviour that may breach the transparency criterion.

\begin{example}[Non-Determinism and Transparency] Despite being sound, enforcer $\eV_4$ is \emph{not transparent} as it may occasionally select enforcer $\eV_1$ which employs an eager enforcement strategy causing it to unnecessarily modify valid system behaviour (as shown in \Cref{ex:enf-instrum}). Hence, enforcer $\eV_4$ \emph{fails} to strongly enforce formula $\hV_2$. \bqed 
\end{example} 

A correct synthesis function must therefore be aware of the subtleties introduced by non-deterministic behaviour. Since the runtime behaviour of deterministic enforcers is more \emph{predictable} compared to non-deterministic ones, they are generally less subtle and thus easier to understand and debug. For instance, if a deterministic enforcer makes a mistake while enforcing a property, it is easier (or rather more intuitive) to backtrack to the point where the mistake was made, since the enforcer is always forced to react in the same way for the same input. By contrast, understanding the behaviour of a non-deterministic enforcer is harder as one needs to take into consideration the selections that the enforcer chose during its execution, thereby making it harder to understand and debug.

We thus develop a synthesis function which only yields \emph{deterministic enforcers} (as defined by \Cref{def:det-enf}) from a given \shml formula. In \Cref{sec:detenf-norm} we present a normalization algorithm for converting a given \shml formula into a \emph{semantically equivalent formula} that is in a \emph{normal form}, from which we can easily \emph{synthesise deterministic enforcers}. In \Cref{sec:enf-syn}, we then present a novel synthesis algorithm which converts normalized \shml formulae into deterministic enforcers for which we prove that the synthesised enforcers always \emph{behave deterministically} and are guaranteed to \emph{strongly enforce} the formula they were derived from, \ie the synthesised enforcers are guaranteed to adhere to soundness and transparency (see \Cref{def:strong-enf}).

\section{Towards Synthesising Deterministic Enforcers through Normalization} \label{sec:detenf-norm}

In order to obtain deterministic monitors from a given logic formula, we follow the approach presented in \cite{Aceto2016Determinization} and adapt it to our setting; the approach works in two phases. The first phase converts the given \shml (\resp \chml) formula into a \emph{semantically equivalent} formula that is in an intermediary format which can then be easily converted (by the second phase) into the required monitor. Since this phase works at the level of the logic, namely \wrt the \shml and \chml subsets, it is thus independent of the type of monitor (detection or enforcement) being synthesised by the second phase. 

The second phase thus employs a synthesis function 
that converts the result obtained from the previous phase, into the required deterministic monitors; different synthesis algorithms may be created depending on the desired type of resultant monitor, \eg the synthesis function in \cite{Francalanza2015Mon} can be used to obtain deterministic \emph{detection} monitors, while in our case, we must define a synthesis function that converts the output of the first phase into a deterministic \emph{enforcement} monitor.



In this section we focus on explaining the first phase of this approach \wrt \shml formulae, as applying the same approach for \chml formula only requires minimal (syntactic) changes. We thus subdivide this section as follows: in \Cref{sec:prelim-norm} we provide some preliminary material for understanding normalization, then in \Cref{sec:normalization} we present a normalization algorithm which only works for \shml formulae defining \emph{singleton symbolic events}. We then \emph{extend} this algorithm in \Cref{sec:new-determinization} which enables for normalizing \shml (\resp \chml) formulae defining symbolic events which are \emph{not necessarily singleton}.

\subsection{Preliminaries for Normalization} \label{sec:prelim-norm}
Normalization represents a conversion process which translates a given \shml (\resp \chml) formula into an intermediary form known as the \emph{normal form}\footnote{In \cite{Aceto2016Determinization} this is also referred to as the deterministic form.}.

\begin{definition}[Normal Form]\label{def:normal-form}
	A formula is in \emph{normal form} when every conjunction branch is \emph{guarded} by a \emph{disjoint} necessity modality, which denotes a set of concrete events that does not intersect with the set denoted by any other symbolic event defined in the necessities guarding the other branches, \ie a concrete event can only match \emph{one} of the symbolic events. \bqed
\end{definition}

\begin{example}[Normal Form Formulae] Recall formulae $\hV_0$ from \Cref{ex:uhml-formula}, and $\hV_2$ from \Cref{ex:enf-shml} (both restated below).
	\begin{align*}
		\hV_0 &\defeq \hVdefConc \\
		\hV_2 &\defeq \hVBdefConc
	\end{align*}
	Notice how the conjunct branches in $\hV_0$ are guarded by necessities defining \emph{disjoint} concrete events \actReq and \actAns, while the branches in $\hV_2$ are both guarded by the \emph{same} concrete event \actReq. Hence, by \Cref{def:normal-form}, we can deduce that $\hV_0$ is in normal form, while $\hV_2$ is not. \bqed
\end{example}

\begin{figure}[t]
	\noindent\textbf{The Normalized Syntax}
	$$ \hV,\hVV\in\shmlwf\; ::=\; \htru\;\vert\;\hfls\;\vert\;\hVarX \;\vert\;\hmax{\hVarX}{\hV} \;\vert\;
	\hAnd\hnec{\actS_i}{\hV_{i}} \;\;\text{ where } \bigdistinct{i\in\IndSet}{\actS_{i}}$$
	where $i{\,\in\,}\IndSet$ is an index that identifies a branch in a conjunction.
	\caption{The syntax of normal-form formulae.}
	\label{fig:shmlnf-syn}
\end{figure}

Based on \Cref{def:normal-form}, in \Cref{fig:shmlnf-syn} we restrict the syntax of our \shml subset into \shmlwf. With this restricted syntax one can only define \emph{normalized} \shml formulae, \ie \shml formulae that adhere to \Cref{def:normal-form}. Concretely, in \shmlwf we introduce a syntactic restriction which \emph{combines} the conjunction operator $(\hAnd\hV_i)$ with the necessity operator (\hnec{\actS}\hV) into $\hAnd\hnec{\actS_i}\hV_i$ as shown in \Cref{fig:shmlnf-syn}. One can immediately notice that this restriction forbids from defining \shml formulae such as $\hV_2$, since the conjunct branches (\hnec{\actReq}\hnec{\actAns}\hVarX\,\hand\,\hnec{\actReq}\hnec{\actReq}\hfls) do \emph{not} define \emph{disjoint} events.

We aim to prove that despite the syntactic restrictions, \shmlwf is still as expressive as the unrestricted \shml subset\footnote{In \cite{Aceto2016Determinization} the authors prove that this result holds in relation to a version of the logic which only allows for defining \emph{concrete events}, we thus follow up on their proofs and extend them to the version that includes \emph{symbolic events}.}. To obtain this result, we devise a set of conversion algorithms and prove \Cref{thm:norm-equivalence}, \ie that \emph{any} formula $\hV{\in}\shml$ can be converted into a \emph{semantically equivalent} normalized formula $\hVV{\in}\shmlwf$. 
\begin{theorem}[Semantic Equivalence] \label{thm:norm-equivalence}
	$$ \forall \hV{\,\in\,}\shml, \exists \hVV{\,\in\,}\shmlwf \cdot \syn{\hV}=\syn{\hVV} $$ 
\end{theorem}
\noindent For instance, through the normalization algorithms we should be able to convert formula $\hV_2$ into (an unfolded version of) $\hV_0$, since these two formulae are \emph{semantically equivalent} to each other.

\subsection{Reconstructing \shml into \shmlwf \wrt Singleton Symbolic Events} \label{sec:normalization}
{
	\setcounter{secnumdepth}{3}
Inspired from \cite{Aceto2016Determinization}, we define the normalization algorithm for \emph{singleton \shmlsl formulae} in terms of the \emph{four constructions} given below; each construction is accompanied by a proof guaranteeing semantic preservation, \ie that the result of each translation is equivalent to its input. The construction sequence is as follows:
\begin{enumerate}[\bf\textsection 1.\!]
	\item \textbf{Standardization of \shml:} \label{step:form-to-sfform} This step serves to convert a given \shml formula into a semantically equivalent formula that is an intermediate form known as the \emph{Standard Form}; this is discussed in \Cref{sec:form-to-sfform}.
	\item \textbf{Equation Form Conversion:} \label{step:sfform-to-sfsys} As explained in \Cref{sec:sfform-to-sfsys}, The standard form formula is then reformulated into a \emph{system of equations} which makes it easier to manipulate in later stages.
	\item \textbf{Normalization of Equations:} \label{step:sfsys-to-detsys} The normalization procedure reviewed in \Cref{sec:sfsys-to-detsys}, restructures the obtained system of equations into an equivalent system that is in the required normal form.
	\item \textbf{\shmlsl Form Conversion:} \label{step:detsys-to-shml} Finally, the normal form system of equations is converted back into an \shml formula that is in normal form; this conversion is described in \Cref{sec:detsys-to-shml}.
\end{enumerate}

\subsubsection{Standardization of \shmlsl} \label{sec:form-to-sfform}
The first step towards achieving the required normal form requires converting the given \shml formula into a semantically equivalent \shmlsf formula, \ie an \shml formula that satisfies the structural constraints of \emph{standard formed formulae} as defined in \Cref{def:standard-form-formulas} below.

\begin{definition}[Standard Form Formulae] \label{def:standard-form-formulas}
	According to \cite{Aceto2016Determinization}, a formula $\hV\in\shml$ is in \emph{standard form} if \emph{all free and unguarded recursion variables} $\hVarX_i$ in $\hV$, are at the \emph{topmost level}, \ie if $\hV = \hVV\land\hAND{i\in Q}X_i$ where $\hVV$ does not contain any free and unguarded recursion
	variables and $Q$ is a finite set of indices.\bqed
\end{definition}

\begin{example}[Standard Form Formulae]
	Formula $(\hmax{\hVarY}{(\hnec{\recv{i}{3}}\hVarY\hand\hVarX)})\hand\hnec{\recv{i}{3}}\hfls$ is \emph{not} in standard form since the free logical variable \hVarX is not scoped under the topmost conjunction, and instead it is scoped under the maximal fixpoint $\hmax{\hVarY}{(\ldots)}$. This formula can thus be easily standardized by elevating \hVarX to the topmost conjunction and thus obtain $((\hmax{\hVarY}{(\hnec{\recv{i}{3}}\hVarY)})\hand\hnec{\recv{i}{3}}\hfls\hand\hVarX)\in\shmlsf$. \bqed
\end{example}

\begin{figure}[t]
	\noindent\textbf{Construction}
	\begin{align*}
		&\gensf{\hV} \defEquals
		\begin{xbrace}{ll}
		\hVV\Sub{\hmaxB{\hVarX_j}{\hVV}}{\hVarX_j}\hand\hAND{i\in\f{\hV'}\setminus\set{j}\hspace{-10mm}}\hVarX_i  & \qquad
		\begin{xbrace}{rll}
		\text{if } &\hV=\hmax{\hVarX_j\,}{\hV'} &\text{ and }\\ &\ \gensf{\hV'}=\hVV\hand\hAND{i\in\f{\hV'}\setminus\set{j}\hspace{-10mm}}\hVarX_i
		\end{xbrace}	 \\[8mm]
		(\hVV_1\hand\hVV_2)\hand\hAND{i\in\f{\hV_1}\hspace{-5mm}}\hVarX_i \hand\hAND{i\in\f{\hV_2}\hspace{-5mm}}\hVarX_i & \qquad
		\begin{xbrace}{rll}
		\text{if } &\hV=(\hV_1\hand\hV_2) &\text{ and }\\[1mm] &\gensf{\hV_1}=\hVV_1\hand\hAND{i\in\f{\hV_1}\hspace{-5mm}}\hVarX_i & \text{ and }\\ &\gensf{\hV_2}=\hVV_2\hand\hAND{i\in\f{\hV_2}\hspace{-5mm}}\hVarX_i
		\end{xbrace}	 \\[12mm]
		\hV & \qquad\text{otherwise} \\[3mm]
		\end{xbrace}\\[1mm]		
		&\qquad\text{where  } \f{\hV}=\big\{ i \,\big\vert\, \text{if } X_i \text{ occurs free and unguarded in }\hV \big\} 
	\end{align*}
	\caption{The Standardization Algorithm}
	\label{fig:standardization-alg}
\end{figure}

In \Cref{fig:standardization-alg} we present the construction algorithm $\gensf{-}::\shml\mapsto\shmlsf$. This construction reformulates a given formula \hV such that every occurrence of a free and unguarded logical (recursion) variable, $\hVarX_i$ (\ie $i\in\f{\hV}$), is elevated to the topmost conjunction to obtain the required standard form, and in the process, any bound logical variable $\hVarX_j\notin\f{\hV}$ is also \emph{unfolded}. The unfolding is required to ensure that the resultant conjunction branches are \emph{always guarded by a necessity operation}. For example, the free logical variable \hVarX in formula $(\hmaxB{\hVarY}{\hnec{\recv{i}{3}}\hVarY\hand\hVarX})\hand\hnec{\recv{i}{3}}\hfls$ is elevated to the topmost conjunction, while the bound variable \hVarY is unfolded so to obtain $(\hnec{\recv{i}{3}}\hmaxB{\hVarY}{\hnec{\recv{i}{3}}\hVarY}\hand\hnec{\recv{i}{3}}\hfls)\hand\hVarX$.

More specifically, when analysing a conjunction, \ie \gensf{\hV_1\hand\hV_2}, the construction is reapplied on the individual branches, \ie \gensf{\hV_1} and \gensf{\hV_2}, in order to obtain $\hVV_1\hand\hAND{i\in\f{\hV_1}\hspace{-6mm}}\hVarX_i$ and $\hVV_2\hand\hAND{i\in\f{\hV_2}\hspace{-6mm}}\hVarX_i$ \resp Conjunctions $\hAND{i\in\f{\hV_1}\hspace{-6mm}}\hVarX_i$ and $\hAND{i\in\f{\hV_2}\hspace{-6mm}}\hVarX_i$ represent every free and unguarded logical variable defined in $\hV_1$ and $\hV_2$ \resp These free variables are added at the topmost conjunction such that the result of $\gensf{\hV_1\hand\hV_2}$ is $(\hVV_1\hand\hVV_2)\hand\hAND{i\in\f{\hV_1}\hspace{-6mm}}\hVarX_i\hand\hAND{i\in\f{\hV_2}\hspace{-6mm}}\hVarX_i$.

When analysing a maximal fixpoint, \ie $\gensf{\hmax{\hVarX_j}{\hV'}}$, the construction is immediately reapplied on $\hV'$, such that \gensf{\hV'} returns $\hVV\hand\hAND{i\in\f{\hV'}\setminus\set{j}\hspace{-10mm}}\hVarX_i$; the construction then uses this result to construct $\hVV\Sub{\hmaxB{\hVarX_j}{\hVV}}{\hVarX_j}\hand\hAND{i\in\f{\hV'}\setminus\set{j}\hspace{-10mm}}\hVarX_i$. Notice that the first part of the reconstructed formula, \ie $\hVV\Sub{\hmaxB{\hVarX_j}{\hVV}}{\hVarX_j}$ defines a substitution which unfolds the formula. The second part, \ie $\hAND{i\in\f{\hV'}\setminus\set{j}\hspace{-10mm}}\hVarX_i$ then serves to ensure that all the free and unguarded variables defined in \hV, except $\hVarX_j$, are grouped and added to the topmost conjunction layer; variable $\hVarX_j$ is not included as this is now bound under the maximal fixpoint, \ie $\hmaxB{\hVarX_j}{\ldots}$.

\begin{example}[Standardization of \shmlsl]\label{ex-shml-to-shmlsf}\normalfont
	\noindent Recall \shml formula $\hV_2$ from \Cref{ex:enf-shml} (restated below). 
	\begin{align*}
		\hV_2 \defeq&\; \hmax{\hVarX_{0}}{\hV_2'} \\
	 	\hV_2' \defeq&\; (\hnec{\actReq}\hnec{\actAns}\hVarX_{0})\hand(\hnec{\actReq}\hnec{\actReq}\hfls)
	\end{align*}
	We recursively apply $\gensf{-}$ on $\hV_2$ to obtain the result from the following derivation:
	\begin{derivation}\normalfont
		Since $\hV_2=\hmax{\hVarX_0}{\hV_2'}$ we know
		$$\gensf{\hmax{\hVarX_0}{\hV_2'}} = \hVV\Sub{\hmaxB{\hVarX_0}{\;\hVV_2'}}{\hVarX_0}\hand\htru $$
		where $\gensf{\hV_2'}=\hVV_2'\hand\hAND{j\in\f{\hV_2'}\setminus\set{0}\hspace{-10mm}}\hVarX_j$ and since $\f{\hV_2'}\setminus\set{0}=\emptyset$, then $(\hAND{j\in\f{\hV_2'}\setminus\set{0}\hspace{-10mm}}\hVarX_i){\,=\,}\htru$. \\[2mm]
		Since $\hV_2'=(\hnec{\actReq}\hnec{\actAns}\hVarX_0\hand\hnec{\actReq}\hnec{\actReq}\hfls)$ we know
		$$\gensf{\hnec{\actReq}\hnec{\actAns}\hVarX_0\hand\hnec{\actReq}\hnec{\actReq}\hfls} = (\hVV_2''\hand\hVV_2''')\hand\htru\hand\htru$$ \vspace{-6mm}
		{
			$$\begin{array}{rrcl}
				\text{where } & \hVV_2''&=&\gensf{\hnec{\actReq}\hnec{\actAns}\hVarX_0}\!=\hnec{\actReq}\hnec{\actAns}\hVarX_0\!\hand\htru\\
				\text{ and } & \hVV_2'''&=&\gensf{\hnec{\actReq}\hnec{\actReq}\hfls}\!=\hnec{\actReq}\hnec{\actReq}\hfls\!\hand\!\htru 
			\end{array}$$
			and since $\f{\hV_2''}{\,=\,}\emptyset{\,=\,}\f{\hV_2'''}$ then $\hAND{j\in\f{\hV_2''}\hspace{-5mm}}\hVarX_j=\htru=\hAND{j\in\f{\hV_2'''}\hspace{-5mm}}\hVarX_j$.
		}\bigskip
		
		\noindent Therefore, the resultant formula is the following:		
		$$\begin{array}{@{\!\!\!}r@{\,}c@{\,}l}
		\gensf{\hV_2} &=&
		\begin{xbrackets}{rl}
		\hnec{\actReq}\hnec{\actAns}\hVarX_0&\hand\\[2mm]\hnec{\actReq}\hnec{\actReq}\hfls
		\end{xbrackets}\Sub{\hmaxB{\hVarX_0}{\hnec{\actReq}\hnec{\actAns}\hVarX_0\hand\hnec{\actReq}\hnec{\actReq}\hfls}}{\hVarX_0}\\ 
		&&\multicolumn{1}{c}{\big\{\text{By applying the substitution we obtain the following formula}\big\}} \\
		&=& (\hnec{\actReq}\hnec{\actAns}\hmaxB{\hVarX_0}{\hnec{\actReq}\hnec{\actAns}\hVarX_0\hand\hnec{\actReq}\hnec{\actReq}\hfls}) \hand (\hnec{\actReq}\hnec{\actReq}\hfls)\\ 
		&=& \hV^{\textsf{sf}}_2
		\end{array} $$ \\[-12mm]\bqed
	\end{derivation} 
\end{example}

\paragraph{Proving Semantic Preservation for \pmb{\gensf{-}}.} To prove that standardization construction \gensf{-} preserves the original semantics of the given \shml formula, we must prove that the following criterion holds:
	$$\forall\hV{\in}\shml\cdot\gensf{\hV}\equiv\hV \text{  where  } \hV\in\shmlsf $$
\begin{proof} We refer to \emph{Lemma 8} from \cite{Aceto2016Determinization} in order to prove that construction \gensf{-} preserves the semantics of the given formula \hV and thus creates a semantically equivalent standardized formula \hVSF. Although \emph{Lemma 8} is proven \wrt a version of \shml that only allows for defining concrete events, the proof of this lemma still applies to our setting.

In fact, \emph{Lemma 8} shows that semantics are preserved when moving the free and unguarded logical variables to the topmost conjunction and when unfolding the formula, \ie as done by construction \gensf{-}, and pays no regard to the type of events described in the necessities; adapting the proof for our setting thus only requires minor syntactic changes, \ie changing \acta into \actS. 

\end{proof}

\subsubsection{Equation Form Conversion} \label{sec:sfform-to-sfsys}
The second construction reformulates \shmlsf formulae into an equivalent \emph{system of equations}. As defined in \Cref{def:system-of-equations} (below), systems of equations provide an alternative way of defining \uhml formulae in terms of a set of equations between a logical variable and a formula.

\begin{definition}[System of Equations] \label{def:system-of-equations} A system of equations \sys is defined as a triple $\syseq{\eqq}{\hVarX}{\sysrecset}$, where \hVarX represents the \emph{principle logical variable} which identifies the starting equation, \sysrecset is a finite set of \emph{free logical variables}, and \eqq is an \emph{$n$-tuple of equations}, \ie $\Set{\hVarX_1=\hV_1, \hVarX_2=\hV_2, \ldots, \hVarX_n=\hV_n}$, where for $1\le i<j\le n$, $\hVarX_i$ is different from $\hVarX_j$, and each $\hV_i$ is a (possibly open) \shmlsl expression.
	
Maximal fixpoints within a system of equations are denoted by referring to a priorly defined recursion variable. We sometimes abuse the notation of \eqq and use it as a map from logical variables to the equated expression, \ie we denote $\eqq(\hVarX_i)=\hV_i$ in lieu of $\hVarX_i=\hV_i\in\eqq$. \bqed
\end{definition}
The equation form construction also needs to ensure that the constructed equations are also in standard form as defined by \Cref{def:standard-form-equations}.
\begin{definition}[Standard Form Equations] \label{def:standard-form-equations}
	Similar to \Cref{def:standard-form-formulas}, we say that an equation $\hVarX_i=\hV_i$ is in \emph{standard form} if $\hV_i\in\shmleq$, where \shmleq is defined as follows:
		$$ \hV\in\shmleq \bnfdef \hfls\;\; \bnfsepp \; \hAND{j\in\IndSet}\hnec{\actS_j}\hVarX_j \hand (\hAND{k\in\IndSet'}\hVarY_k)$$ 
	 for some finite sets of indices \IndSet and $\IndSet'$. A system \sysSF is in \emph{standard form} if \emph{every equation in the system is in standard form}, \ie $\sysSF=\syseq{\eqqSF}{\hVarX}{\sysrecset}$ such that $\forall \hVarX{=}\hV\in\eqq\cdot\hV\in\shmleq$.\bqed
\end{definition}

\begin{example}[System of Equations \vs Formulae]
	Intuitively, a recursive formula such as $\hV=\hmax{\hVarX_{0}}{\hnec{\recv{i}{3}}}(\hnec{\send{i}{4}}\hVarX_{0}\hand\hnec{\send{i}{4}}\hfls)$ can be represented in standard form via the following system of 3 equations:
	$$ \eqq \defeq \Set{\;\hVarX_{0}{=}\hnec{\recv{i}{3}}\hVarX_{1}, \quad \hVarX_{1}{=}\hnec{\send{i}{4}}\hVarX_{0}{\hand}\hnec{\send{i}{5}}\hVarX_{2}, \quad \hVarX_{2}=\hfls\;} $$
	Notice how recursion is represented by referring to $\hVarX_{0}$ in the second equation. Also, notice that since \hV starts with a maximal fixpoint defining $\hVarX_{0}$, we know that $\hVarX_{0}$ is also the principle logical variable of this system of equations. Moreover, we know that $\sysrecset{=}\sysrecsetE$ since all the logical variables defined in the system are bound, \ie variables $\hVarX_{0}$, $\hVarX_{1}$ and $\hVarX_{2}$ are all equated to some \shmleq formula. Hence, the resultant system of equations is $\syseq{\eqq}{\hVarX_{0}}{\sysrecsetE}$. \bqed
\end{example}

\begin{figure}[h]
\noindent\textbf{Construction}\\[-3mm]
	\begin{align*}
		&\gensys{\hV} \!\!\defEquals\!\!
		\begin{xbrace}{ll}
		\syseq{\Set{\hVarX=\htru}}{\hVarX}{\sysrecsetE} & \text{ if }\hV=\htru\hand\hAND{j\in Q}\hVarY_j \\[2mm]
		\syseq{\Set{\hVarX=\hfls}}{\hVarX}{\sysrecsetE} & \text{ if }\hV=\hfls\hand\hAND{j\in Q}\hVarY_j \\[2mm]
		\syseq{\Set{\hVarX=\hVarY}}{\hVarX}{\Set{\hVarY}} & \text{ if }\hV=\hVarY\\[4mm]
		\!\!\!\syseq{Eq_1 \;\cup\; Eq_2\; \cup \\\!\! \begin{xbraces}{c}\hVarX_0=
			\begin{xbrackets}{cc}
			Eq_1(\hVarX_1) &\!\!\hand\!\!\\ Eq_2(\hVarX_2)
			\end{xbrackets}	
			\end{xbraces}
		}{\hVarX_0}{ \sysrecset_1\!\cup\!\sysrecset_2\!\!\!} & \begin{array}{l}\!\text{ if }\hV=\hV_1\hand\hV_2 \\ \text{ and } \gensys{\hV_1}=\syseq{\!Eq_1}{\!\hVarX_1}{\!\sysrecset_1\!} \\ \text{ and } \gensys{\hV_2}=\syseq{\!Eq_2}{\!\hVarX_2}{\! \sysrecset_2\!} \end{array}\\[9mm]
		\syseq{Eq \cup \Set{\hVarY=Eq(\hVarX_1)}}{\hVarY}{ \sysrecset\setminus\Set{\hVarY}} & \begin{array}{l} \text{ if }\hV=\hmax{\hVarY}{\hV'}\\ \text{ and } \gensys{\hV'}=\syseq{Eq}{\hVarX_1}{ \sysrecset}\end{array}\\[5mm]
		\syseq{\Set{\hVarX_0=\hnec{\actS}\hVarX_1}\cup Eq}{\hVarX_0}{ \sysrecset } & \begin{array}{l} \text{ if }\hV=\hnec{\actS}\hV' \\ \text{ and } \gensys{\hV'}=\syseq{Eq}{\hVarX_1}{ \sysrecset }\end{array} 
		\end{xbrace} \\[-3mm]
	\end{align*}
	\caption{The conversion algorithm from a \shmlsf formula to a Standard Form System of equations.}
	\label{fig:sf-to-sys-alg}
\end{figure}

The construction $\gensys{-}::\shmlsf\mapsto(\eqqSF,\Var,\powerset{\Var})$, defined in \Cref{fig:sf-to-sys-alg}, compositionally inspects a given standard form formula \hV and translates it into an equivalent system of equations in standard form. For instance, \emph{truth}, \htru, and \emph{falsehood}, \hfls, are respectively translated into equations $\hVarX\!=\!\htru$ and $\hVarX\!=\!\hfls$, with \hVarX being the principle variable of the resultant system of equations. Similarly, a logical variable \hVarY is translated into equation $\hVarX\!=\!\hVarY$, with the addition that \hVarY is marked as being a free logical variable in \sysrecset to indicate that \hVarY is not currently bound to some fixpoint. 

Maximal fixpoints are converted into equation $\hVarY\!=\!\eqq(\hVarX_1)$, where $\hVarX_1$ is the principle variable of the system of equations obtained from the recursive application on the continuation $\hV'$ \ie $\gensys{\hV'}\!=\!\syseq{\eqq}{\hVarX}{\sysrecset}$, and $\eqq(\hVarX_1)$ refers to the formula $F_1$ equated to variable $\hVarX_1$ in \eqq, \ie if $\hVarX_1\!=\!F_1$ then $\eqq(\hVarX_1)\!=\!F_1$. Upon analysing a maximal fixpoint \hmax{\hVarY}{\hV'}, variable \hVarY is removed from \sysrecset, thus denoting that although \hVarY is free in $\hV'$, this is no longer the case in $\hV\!=\!\hmax{\hVarY}{\hV'}$.

In the case of conjunctions, $\hV_1\!\hand\!\hV_2$, these are reconstructed into a system of equations consisting in the systems of equations obtained from analysing $\hV_1$ and $\hV_2$, \ie $\eqq_1$ and $\eqq_2$, along with equation $\hVarX_0=\eqq_1(\hVarX_1)\!\hand\!\eqq_2(\hVarX_2)$, where $\hVarX_1$ and $\hVarX_2$ are the principle variables of $\hV_1$ and $\hV_2$ \resp 
Finally, necessity modalities, $\hnec{\actS}\hV'$, are translated into a system consisting in the equations, \eqq, obtained from \gensys{\hV'}, along with equation $\hVarX_0=\hnec{\actS}\hVarX_1$, where $\hVarX_1$ is the principle variable obtained from \gensys{\hV'}. 

\newcommand{\hVBSF}{\ensuremath{\hV^\texttt{sf}_2}\xspace}
\newcommand{\hVBSFA}{\ensuremath{\hV^\texttt{sf}_{2a}}\xspace}
\newcommand{\hVBSFB}{\ensuremath{\hV^\texttt{sf}_{2b}}\xspace}
\newcommand{\hVBSFC}{\ensuremath{\hV^\texttt{sf}_{2c}}\xspace}
\newcommand{\hVBSFD}{\ensuremath{\hV^\texttt{sf}_{2d}}\xspace}
\newcommand{\hVBSFE}{\ensuremath{\hV^\texttt{sf}_{2e}}\xspace}
\newcommand{\hVBSFF}{\ensuremath{\hV^\texttt{sf}_{2f}}\xspace}
\begin{example}\label{ex-shmlsf-to-syssf}\normalfont
	Recall the standard form, singleton \shml formula $\hV^{\textsf{sf}}_2$ obtained in \Cref{ex-shml-to-shmlsf}: 
	$$\begin{array}{r@{\;}c@{\;}l@{\qquad\qquad}r@{\;}c@{\;}l}
		\hVBSF &=& \hVBSFA \hand \hVBSFB & \hVBSFD &=& \hVBSFE\hand\hVBSFF   \\[2mm]
		\hVBSFA &=& \hnec{\actReq}\hnec{\actAns}\hVBSFC & \hVBSFE &=& \hnec{\actReq}\hnec{\actAns}\hVarX \\[2mm]
		\hVBSFB &=& \hnec{\actReq}\hnec{\actReq}\hfls & \hVBSFF &=& \hnec{\actReq}\hnec{\actReq}\hfls  \\[2mm]
		\hVBSFC &=& \hmaxB{\hVarX}{\hVBSFD}   \\		
	\end{array}	
	$$
	We recursively apply $\gensys{\hV}$ to obtain the resultant system of equations from the following derivation:
	\begin{derivation}\normalfont
		Since $\hVBSF = \hVBSFA \hand \hVBSFB$ we know 
		$$\gensys{\hVBSFA \hand \hVBSFB} = \syseq{\Set{\hVarX_{0}=\eqq_1(\hVarX_{1})\hand\eqq_1(\hVarX_{2})}\cup \eqq_1\cup \eqq_2}{\hVarX_{0}}{\sysrecsetE}$$
		where $\gensys{\hVBSFA}=\syseq{Eq_1}{\hVarX_{1}}{\sysrecsetE}$ and $\gensys{\hVBSFB}=\syseq{Eq_2}{\hVarX_{2}}{\sysrecsetE}$.\\ 
		
		\noindent We now consider $\gensys{\hVBSFA}$ (LHS) and $\gensys{\hVBSFB}$ (RHS) separately.\\
		
		\noindent\textbf{\underline{LHS}}
		\begin{context}
			Since $\hVBSFA=\hnec{\actReq}\hnec{\actAns}\hVBSFC$ we know
			$$\gensys{\hnec{\actReq}\hnec{\actAns}\hVBSFC} = \syseq{\Set{\hVarX_{1}=\hnec{\actReq}\hVarX_{3},\hVarX_{3}=\hnec{\actAns}\hVarX}\cup \eqq_1'}{\hVarX_{1}}{\sysrecsetE}$$ 
			\indent where $\gensys{\hVBSFC}=\syseq{Eq_1'}{\hVarX}{\sysrecsetE}$.\\
			
			\noindent Since $\hVBSFC=(\hmaxB{\hVarX}{\hVBSFD})$ we know
			$$\gensf{\hmaxB{\hVarX}{\hVBSFD}} = \syseq{\Set{\hVarX=\eqq_1''(\hVarX_{4})}\cup \eqq_1''}{\hVarX}{\sysrecsetE}$$ 
			\indent where $\gensys{\hVBSFD}=\syseq{Eq_1''}{\hVarX_4}{\sysrecsetE}$.\\
			
			\noindent Since $\hVBSFD=\hVBSFE\hand\hVBSFF $ we know
			$$\gensf{\hVBSFE\hand\hVBSFF} = \syseq{\Set{\hVarX_4=\eqq_3(\hVarX_{5})\hand\eqq_4(\hVarX_{6})}\cup \eqq_3\cup \eqq_4}{\hVarX_4}{\Set{\hVarX}}$$ 
			\indent where $\gensys{\hVBSFE}=\syseq{Eq_3}{\hVarX_5}{\Set{\hVarX}}$ and  $\gensys{\hVBSFF}=\syseq{Eq_4}{\hVarX_6}{\sysrecsetE}$.\\
			
			\noindent Since we have $\gensys{\hV_3}$ (LHS) and $\gensys{\hV_4}$ (RHS) we consider them separately.\\
			
			\noindent\textbf{\underline{LHS}}
			\begin{context}
				Since $\hVBSFE=\hnec{\actReq}\hnec{\actAns}\hVarX$ we know
				$$\gensys{\hnec{\actReq}\hnec{\actAns}\hVarX} = \syseq{\Set{\hVarX_{5}=\hnec{\actReq}\hVarX_{7},\hVarX_{7}=\hnec{\actAns}\hVarX_8,\hVarX_{8}=\hfls}}{\hVarX_{5}}{\Set{\hVarX}}$$
			\end{context}
		
			\noindent\textbf{\underline{RHS}}
			\begin{context}
				Since $\hVBSFF=\hnec{\actReq}\hnec{\actReq}\hfls$ we know
				$$\gensys{\hnec{\actReq}\hnec{\actReq}\hfls} = \syseq{\Set{\hVarX_{6}=\hnec{\actReq}\hVarX_{9},\hVarX_{9}=\hnec{\actReq}\hVarX_{10},\hVarX_{10}=\hfls}}{\hVarX_{6}}{\sysrecsetE}$$
			\end{context}
		\end{context}
	
		\noindent\textbf{\underline{RHS}}
		\begin{context}
			Since $\hVBSFB=\hnec{\actReq}\hnec{\actReq}\hfls$ we know
			$$\gensys{\hnec{\actReq}\hnec{\actReq}\hfls} = \syseq{\Set{\hVarX_{2}=\hnec{\actReq}\hVarX_{11},\hVarX_{11}=\hnec{\actReq}\hVarX_{12},\hVarX_{12}=\hfls}}{\hVarX_{2}}{\sysrecsetE}$$
		\end{context}	
		\noindent Hence, the result of this derivation is $\gensys{\hV}=\syseq{\eqq}{\hVarX_{0}}{\sysrecsetE}$ where 
		$$ \eqq=\begin{xbraces}{lll}
			\hVarX_{0}=\hnec{\actReq}\hVarX_{3}\hand\hnec{\actReq}\hVarX_{11},
			\greybox{\hVarX_{1}=\hnec{\actReq}\hVarX_{3}}, 
			\hVarX_{3}=\hnec{\actAns}\hVarX,\\ 
			\hVarX=\hnec{\actReq}\hVarX_{7}\hand\hnec{\actReq}\hVarX_{9}, \greybox{\hVarX_{4}=\hnec{\actReq}\hVarX_{7}\hand\hnec{\actReq}\hVarX_{9}},
			\greybox{\hVarX_{5}=\hnec{\actReq}\hVarX_{7}},\\ 
			\hVarX_{7}=\hnec{\actAns}\hVarX_{8}, \hVarX_{8}=\hVarX, 
			\greybox{\hVarX_{6}=\hnec{\actReq}\hVarX_{9}}, 
			\hVarX_{9}=\hnec{\actReq}\hVarX_{10},\\
			 \hVarX_{10}=\hfls, 
			 \greybox{\hVarX_{2}=\hnec{\actReq}\hVarX_{11}}, \hVarX_{11}=\hnec{\actReq}\hVarX_{12},
			\hVarX_{12}=\hfls
		\end{xbraces} $$
		Note that the greyed formulae were are redundant since they are not reachable from the principle logical variable $\hVarX_{0}$; for conciseness we will ignore them in forthcoming examples. \bqed
	\end{derivation} 
\end{example}

\paragraph{Proving Semantic Preservation for \pmb{\gensys{-}}.} To prove that construction \gensys{-} preserves the semantics of the given \shmlsf formula, we must prove that the following criterion holds:
$$\forall\hV{\in}\shmlsf\cdot\gensys{\hV}\equiv\sys \text{  where  } \sys \text{ is in \emph{Standard Form.}} $$ \vspace{-10mm}
\begin{proof}  The proof guaranteeing that the resultant system of equations \sys, constructed via \gensys{-}, is \emph{semantically equivalent} to the given standard form formula \hVSF follows from \emph{Lemma 10} given in \cite{Aceto2016Determinization}. Although this lemma is proven in relation to formulae that define concrete events, this lemma still applies for formulae defining symbolic events, since the construction is independent of the type of event described in the modal necessities.
	
\end{proof}

\subsubsection{Normalization of Equations} \label{sec:sfsys-to-detsys}
The third construction performs the actual normalization procedure as it converts systems of equations that are in standard form, into an equivalent \emph{systems of equations} that are in \emph{normal form} as defined in \Cref{def:normalized-eqns}.

\begin{definition}[Normalized System of Equations] \label{def:normalized-eqns} An equation $\hVarX_i{=}\hV_i$ is in \emph{normal form} if the subformulae of a conjunction are either \emph{free and unguarded logical variables}, or else \emph{guarded} by a \emph{disjoint necessity}. This ensures that \emph{at most only one necessity} guarding a branch in a conjunction can match a system action, \ie $\hV_i$ has the form \hfls or $\hAND{i\in \IndSet}\hnec{\actS_i}\hV_i\!\hand\!\hAND{j\in Q}\hVarY$ where $\bigdistinct{i\in\IndSet}{\actS_{i}}$. A system of equations \sysNF is in normal form when all of its equations, \eqqNF, are in normal form. \bqed
\end{definition}

\begin{figure}[h]
\noindent\textbf{Construction}
\begin{gather*}
	\gendet{\syseq{Eq}{\hVarX_i}{\sysrecset}} \defEquals \syseq{\eqqDet}{\hVarX_{\{i\}}}{\sysrecset} \\[2mm]
	\eqqDet \defeq
	 \Setdef{X_\IndSet=\!\!\!\!\hAND{\actS\in\detS{\IndSet}}\!\!\!\!\hnec{\actS}\hVarX_{\detD{\IndSet}{\acta}}\hand\!\!\!\!\hAND{j\in\detE{\IndSet}}\!\!\!\!\hVarY_j }{ \;\distinctUni{\detS{\IndSet}} \hfill\hfill \text{ and } \hfill\hfill \IndSet\subseteq \detI{\eqq} \hfill\hfill \text{ and }\\\; \hfls \text{ is \emph{not}} \text{ a subformula of }\eqq(\hVarX_\IndSet)} 
	 \\\quad  \cup \;\; \setdef{X_\IndSet=\hfls}{\IndSet\subseteq \detI{Eq} \text{ and } \hfls \text{ is a subformula of }Eq(\hVarX_\IndSet)}\\[5mm]	
	\begin{array}{crcl} 
		\text{where} & \detS{\IndSet} &\defEquals& \bigunion{i\in \IndSet}\setdef{\actS\!}{\!\hnec{\actS}\hVarX_j \text{ is a subformula in }\hV_i} \\ 
		&\detD{\IndSet}{\actS} &\defEquals& \bigunion{i\in \IndSet}\setdef{r\!}{\!\hnec{\actS}\hVarX_r \text{ is a subformula in }\hV_i} \\
		&\detE{\IndSet} &\defEquals& \bigunion{i\in \IndSet}\setdef{r\!}{\!\hVarY_r \text{ is unguarded in }\hV_i} \\
		&\detI{Eq} &\defEquals& \setdef{i}{\hVarX_i=\hV_i\in Eq}
	\end{array} 
\end{gather*}
\caption{The Normalization Algorithm for Systems of Equations}
\label{fig:sfsys-to-detsys-alg}
\end{figure}

\Cref{fig:sfsys-to-detsys-alg} presents the normalization algorithm in terms of the construction function $\gendet{-}::\syseq{\eqqSF}{\Var}{\powerset{\Var}}\mapsto\syseq{\eqqNF}{\Var}{\powerset{\Var}}$. This construction generates a new system of equations which contains the powerset combinations of the equations from the original system of equations. Intuitively, the construction takes two or more equations and combines the equated formulae with a conjunction.

\begin{example}Consider the system of equations $\syseq{\eqq}{\hVarX_{0}}{\sysrecsetE}$, where \eqq contains 3 equations $\hVarX_{0}{=}\hV_0$, $\hVarX_{1}{=}\hV_1$ and $\hVarX_{2}{=}\hV_2$. The construction thus takes all the combinations and creates a new system, namely, $\syseq{\eqqNF}{\hVarX_{\set{0}}}{\sysrecsetE}$ where 
$$ \eqq=\begin{xbraces}{c}
			\hVarX_{\set{0}}=\hV_0, \hVarX_{\set{0,1}}=\hV_0{\hand}\hV_1, \hVarX_{\set{0,1,2}}=\hV_0{\hand}\hV_1{\hand}\hV_2, \hVarX_{\set{1}}=\hV_1, \\ \hVarX_{\set{1,2}}=\hV_1\hand\hV_2, \hVarX_{\set{2}}=\hV_2, \hVarX_{\set{0,2}}=\hV_0{\hand}\hV_2
		\end{xbraces} $$
As in the original system of equations, the principle variable $\hVarX_{0}$ equates to $\hV_0$, the new principle variable for the reconstructed equation is therefore variable $\hVarX_{\set{0}}$, as this also equates to $\hV_0$. \bqed
\end{example}

While combining the equated formulae, the construction also normalizes the combined equated formulae. The first part of the construction, \ie $\hAND{\actS\in\detS{\IndSet}\hspace{-5mm}}\!\hnec{\actS}\hVarX_{\detD{\IndSet}{\acta}}$, thus makes sure that whenever the conjunction branches are guarded by necessities specifying a \emph{set of disjoint} events $\detS{\IndSet}$, then the conjunction is normalized by merging together the \emph{syntactically equal necessities} in the conjunction. This merger provides a resultant conjunction that has branches which are guarded by disjoint (syntactically different) necessities, \eg equation $\hVarX_0=\hnec{\actS}\hVarX_1\!\hand\!\hnec{\actS}\hVarX_2\!\hand\!\hnec{\actS'}\hVarX_3$ can be normalized into $\hVarX_{\set{0}}=\hnec{\actS}\hVarX_{\set{1,2}}\!\hand\!\hnec{\actS'}\hVarX_{\set{3}}$ by merging the first two branches, \ie $\hnec{\actS}\hVarX_1$ and $\hnec{\actS}\hVarX_2$ into $\hnec{\actS}\hVarX_\set{1,2}$.

Note how only the branches which guarded by \emph{syntactically equal} necessities are being merged together; this is done by taking a subset \IndSet of the powerset combinations of the indices provided by \detI{\eqq}, \ie $\IndSet\subseteq\detI{\eqq}$, where \detI{\eqq} returns all the indices specified in the set of equations \eqq. The merged indices are obtained using \detD{\IndSet}{\actS} which returns the index $r$ of \emph{every logical variable $\hVarX_r$ that is guarded by the same symbolic event \actS}.
		
Finally, the second part of the construction, \ie \hspace{-3mm}$\hAND{j\in\detE{\IndSet}}\!\!\!\!\hVarY_j$, ensures that every unguarded logical variable $\hVarY_j$ defined in the equation, is kept at the topmost level of the conjunction, therefore retaining the equation in \emph{standard form} as defined by \Cref{def:normalized-eqns}.	
	
\begin{example}\label{ex-syssf-to-sysdet}\normalfont
	Recall the standard form system of equations obtained in \Cref{ex-shmlsf-to-syssf}, \ie $\syseq{\eqqSF}{\hVarX_{0}}{\sysrecsetE}$ where	
	$$ \eqqSF=\begin{xbraces}{lll}
	\hVarX_{0}=\hnec{\actReq}\hVarX_{3}\hand\hnec{\actReq}\hVarX_{11}, \hVarX_{3}=\hnec{\actAns}\hVarX, \hVarX=\hnec{\actReq}\hVarX_{7}\hand\hnec{\actReq}\hVarX_{9},  \\\hVarX_{7}=\hnec{\actAns}\hVarX_{8}, \hVarX_{8}=\hVarX,  \hVarX_{9}=\hnec{\actReq}\hVarX_{10}, \hVarX_{10}=\hfls, \hVarX_{11}=\hnec{\actReq}\hVarX_{12},	\\\hVarX_{12}=\hfls
	\end{xbraces} $$
	\noindent When the construction rule is applied, it generates every possible combination and merges the modal necessities where necessary. By definition of $\gendet{\syseq{\eqq}{\hVarX_{0}}{\sysrecsetE}}$ we therefore obtain $\syseq{\eqqDet}{\hVarX_{\set{0}}}{\sysrecsetE}$ where
	$$ \eqqDet=\Set{\hVarX_{\set{0}}=\hnec{\actReq}\hVarX_{\set{3,11}}} \cup \eqqDet' $$
	For instance, note how this construction collapses $\hVarX_{0}=\hnec{\actReq}\hVarX_{3}\hand\hnec{\actReq}\hVarX_{11}$ into $\hVarX_{\set{0}}=\hnec{\actReq}\hVarX_{\set{3,11}}$, where continuations $\hVarX_{3}$ and $\hVarX_{11}$ were combined into a single conjunct continuation $\hVarX_{\set{3,11}}$, in this way every event specified by a necessity in the conjunction becomes \emph{disjoint}, \ie none of the  reconstructed singleton symbolic events can denote a set containing the same concrete event as another symbolic event residing in the same conjunction. Similarly, the algorithm also merges $\hVarX{=}\hnec{\actReq}\hVarX_{7}{\hand}\hnec{\actReq}\hVarX_{9}$ into $\hVarX{=}\hnec{\actReq}\hVarX_{7,9}$. The combined formulae for variables $\hVarX_{\set{3,11}}$ and $\hVarX_{\set{7,9}}$ are also constructed by the normalization algorithm as defined in $\eqqNF'$ below:
	$$ \eqqNF'=\begin{xbraces}{lll} \hVarX_{\set{3,11}}=\hnec{\actAns}\hVarX{\hand}\hnec{\actReq}\hVarX_{\set{12}},  \hVarX=\hnec{\actReq}\hVarX_{\set{7,9}}, \\ \hVarX_{\set{7,9}}=\hnec{\actAns}\hVarX_{\set{8}}{\hand}\hnec{\actReq}\hVarX_{\set{10}}, \hVarX_{\set{8}}=\hVarX, \\\hVarX_{\set{10}}=\hfls, \hVarX_{\set{11}}=\hnec{\actReq}\hVarX_{\set{12}}, \hVarX_{\set{12}}=\hfls, \quad\ldots \end{xbraces} $$
	For conciseness, some combinations have been ignored from the resultant set as they are \emph{not reachable} from the principle variable $\hVarX_{\set{0}}$ and are thus redundant. \bqed 
\end{example}

\paragraph{Proving Semantic Preservation for \pmb{\gendet{-}}.} To prove that construction \gendet{-} preserves the semantics of the given standardized system of equations, we must prove that the following holds:
$$\sys \text{ is in \emph{Standard Form} } \imp \gendet{\sys}\equiv\sys \text{  where  } \gendet{\sys} \text{ is in \emph{Normal Form.}} $$ \vspace{-10mm}
\begin{proof} In order to prove that construction \gendet{-} produces a normalized system of equations \sysNF that is \emph{semantically equivalent} to the given system of equations \sys, we refer to \emph{Lemma 11} in \cite{Aceto2016Determinization}.

However, \emph{Lemma 11} holds \wrt a version of our construction which requires that the necessities define concrete events; this is necessary since the construction must merge together the conjunct branches that are prefixed by syntactically equal modal necessities, \eg $\hVarX_{0}{=}\hnec{\recv{i}{3}}\hVarX_{1}\hand\hnec{\recv{i}{3}}\hVarX_{2}$ becomes $\hVarX_{0}{=}\hnec{\recv{i}{3}}\hVarX_{\set{1,2}}$ since both branches are prefixed by the same concrete necessity \hnec{\recv{i}{3}}.

\emph{Lemma 11}, however, still holds for our construction since \gendet{-} requires condition \distinctUni{\detS{\IndSet}} to hold. This condition states that all the symbolic events guarding a conjunction must be \emph{disjoint unless syntactically equal}. In this way, if two or more symbolic events are syntactically equal, they would be merged by the construction, while the other (non-syntactically equal) necessities are guaranteed to be disjoint. Hence, a concrete system event can only satisfy at most one symbolic modal necessity, \ie in the same way as per normalized conjunctions defining concrete events. 

\end{proof}
	
\subsubsection{\shmlsl Form Conversion} \label{sec:detsys-to-shml}
The final step for obtaining the final normalized formula only requires regenerating the normalized formula, $\hVV\in\shmlwf$, from a given normalized system of equations, $\sysNF{=}\syseq{\eqqNF}{\hVarX}{\sysrecset}$. 

\begin{figure}[h]
	\noindent\textbf{Construction}
	\begin{gather*}
		\genwf{\syseq{Eq}{\hVarX_i}{\sysrecset}} \defEquals \toshml{\hVarX_i}{\eqq} \\[2mm]
		\toshml{\hV}{\eqq} \defEquals 
			\begin{xbrace}{l@{\quad}l}\\[3mm]
				\hV & \text{if }\fv{\hV}{=}\sysrecsetE \\[2mm]
				\toshml{\hV\s}{\eqq} & \text{if }\fv{\hV}{=}\sysrecset \text{ then } \s{=}\Setdef{\subE{\hmax{\hVarX_{0}}{\hV_0}}{\hVarX_{0}}}{\hVarX_{0}{=}\hV_0{\,\in\,}\eqq\\[2mm]\text{ and }\,\hVarX_{0}{\in}\sysrecset}\\[3mm]
			\end{xbrace}
	\end{gather*}
	\caption{Converting a normalized system of equations into an \shmlwf formula.}
	\label{fig:detsys-to-shml-alg}
\end{figure}

\Cref{fig:detsys-to-shml-alg} presents the final construction, $\genwf{-}::\syseq{\eqqNF}{\Var}{\powerset{\Var}}\!\mapsto\shmlwf$, which converts a normalized system of equations into a semantically equivalent \shmlwf formula. The algorithm internally employs function $\toshmlSym::(\shmlwf\times\eqq)\mapsto\shmlwf$ to create a normalized \shmlwf formula that is semantically equivalent to the given system of equations. This function starts by taking as input the principle variable $\hVarX_{i}$ along with the set of equations \eqq, it then searches for the equation $\hVarX_i{=}\hV_i$ in \eqq and converts it into a substitution environment which substitutes variable $\hVarX_i$ with $\hmax{\hVarX_i}{\hV_i}$, \ie $\sub{\hmax{\hVarX_i}{\hV_i}}{\hVarX_i}$. 

This substitution is then applied to $\hVarX_i$ and the function recurses with the substituted value, \ie $\toshml{\hmax{\hVarX_i}{\hV_i}}{\eqq}$; recursion stops when the resultant formula \hV becomes closed, in which case it is returned as the normalized \shmlwf formula. 

\begin{example}\label{ex-sysdet-to-shmlwf}\normalfont
	Consider the following normalized system of equations obtained in \Cref{ex-syssf-to-sysdet}, $\syseq{\eqq}{\hVarX_{\set{0}}}{\sysrecsetE}$ where 
	$$ \eqqDet=\begin{xbraces}{lll} \hVarX_{\set{0}}=\hnec{\actReq}\hVarX_{\set{3,11}}, \hVarX_{\set{3,11}}=\hnec{\actAns}\hVarX\hand\hnec{\actReq}\hVarX_{\set{12}}, \\ \hVarX=\hnec{\actReq}\hVarX_{\set{7,9}}, \hVarX_{\set{7,9}}=\hnec{\actAns}\hVarX_{\set{8}}\!\hand\!\hnec{\actReq}\hVarX_{\set{10}}, \hVarX_{\set{8}}=\hVarX, \\\hVarX_{\set{10}}=\hfls,  \hVarX_{\set{11}}=\hnec{\actReq}\hVarX_{\set{12}}, \hVarX_{\set{12}}=\hfls \end{xbraces} $$
	
	\noindent By $\genwf{\syseq{\eqq}{\hVarX_{\set{0}}}{\sysrecsetE}}$ we obtain $\hV\in\shmlwf$ where
	$$\begin{array}{rcl}
		\hVV_2&\defeq& \toshml{\hVarX_{\set{0}}}{\eqqNF} \\[1mm]
		&\defeq&\hmaxBA{\hVarX_{\set{0}}}{\hnec{\actReq}\hmaxBA{\hVarX_{\set{3,11}}\\}{\hnec{\actAns}\hmaxBA{\hVarX}{\hnec{\actReq}\hmax{\hVarX_{\set{7,9}}}{\hnec{\actAns}\hmax{\hVarX_{\set{8}}}{\hVarX}\\\hand\hnec{\actReq}\hmax{\hVarX_{\set{12}}}{\hfls}\!\!}\!\!}\!\! \\ \hand \hnec{\actReq}\hmax{\hVarX_{\set{10}}}{\hfls}\!\!\!}\!\!\!} 
	\end{array}  $$ 
	The above formula can be optimized by removing redundant maximal fixpoint declarations such as \hmax{\hVarX_{\set{0}}}{} and \hmax{\hVarX_{\set{3,11}}}{}, \ie fixpoints that declare a variable which is never referenced to throughout the rest of the formula. Hence we can optimize our formula as follows:
	$$\hVV_2\defeq\hnec{\actReq}(\hnec{\actAns}\hmaxB{\hVarX}{\hnec{\actReq}(\hnec{\actAns}\hVarX\hand\hnec{\actReq}\hfls)}\hand \hnec{\actReq}\hfls)\in\shmlwf$$	
	Notice that the obtained optimized formula, \ie $\hVV_2$ is in fact an unfolded version of formula $\hV_0$, and that both $\hV_0$ and also $\hVV_2$ are in \emph{normal form} \ie $\hV_0,\hVV_2\in\shmlwf$. 
	$$\begin{array}{rcl}
	\hV_0 &\defeq& \hVdefConc\\
	\multicolumn{3}{c}{\text{\{ By unfolding \hmax{\hVarX_{0}}{} we obtain \}}}\\
	&=& \hnec{\actReq}(\hnec{\actAns}\hmaxB{\hVarX}{\hnec{\actReq}(\hnec{\actAns}\hVarX\hand\hnec{\actReq}\hfls)}\hand \hnec{\actReq}\hfls)\\
	&=&\hVV_2\in\shmlwf
	\end{array}  $$
	\\[-12mm]\bqed
\end{example}

\paragraph{Proving Semantic Preservation for \pmb{\genwf{-}}.} To prove that construction \genwf{-} preserves the semantics of the given normalized system of equations, we must prove that the following holds:
$$\sys \text{ is in \emph{Normal Form} } \imp \genwf{\sys}\equiv\sys \text{  where  } \genwf{\sys}{\in}\shmlwf $$ \vspace{-10mm}

\begin{proof} 
	Since construction \genwf{-} is independent of the type of events defined in the modal necessities of the given system of equations, we refer to \emph{Lemma 12} as proof that this construction preserves the semantics of the given system of equations \sysNF, such that it produces a semantically equivalent formula $\hV{\in}\shmlwf$.
	
\end{proof}

\subsection{Reconstructing \shml into \shmlwf \wrt \emph{any} Symbolic Event}\label{sec:new-determinization}

Up until now we have only considered normalizing \shml formulae defining singleton symbolic events, as these events are easy to statically differentiate from each other as required for merging conjunct branches in \Cref{step:sfsys-to-detsys} (see \Cref{sec:sfsys-to-detsys}). However, the necessities in the logic we consider can also describe symbolic events which denote wider sets of concrete events \eg  $\hnec{\symReqA}\hV$ where \syn{\symReqA} = \set{\ldots, \actReq, \recv{k}{\req}, \ldots}. 

One major difference between singleton and other symbolic events is that the former can \emph{easily be distinguished} from one another by a simple syntactic check \eg singleton events $\actSN{\recv{\dvV_1}{\dvVV_1}}{\dvV_1{=}i{\,\land\,}\dvVV_1{=}\req}$ and $\actSN{\recv{\dvV_2}{\dvVV_2}}{\dvV_2{=}i{\,\land\,}\dvVV_2{=}\cls}$ can be represented as \actReq and \actCls \resp and can thus be easily distinguished since $\actReq{\,\neq\,}\actCls$. In general, however, this distinction is not always possible with symbolic events. 

For instance, consider \recv{\dvV}{5} and \recv{i}{\dvVV}, these two events, although syntactically different, they define \emph{intersecting sets} of input events, \ie $\syn{\recv{\dvV}{5}}{\cap}\syn{\recv{i}{\dvVV}}$, meaning that both symbolic events can match the same concrete system event \recv{i}{5}. Hence, this makes it harder to statically differentiate and distinguish between symbolic events, which is crucial when applying the normalization construction step \Sref{step:sfsys-to-detsys}.

As shown in \Cref{ex:norm-symev}, normalizing a non-singleton symbolic formula using the algorithm we presented so far, may sometimes fail to produce an equivalent formula that is in normal form.

\newcommand{\hVSFC}{\ensuremath{\hVSF_{3}}}
\newcommand{\sysSFC}{\ensuremath{\sysSF_{3}}}
\newcommand{\eqqSFC}{\ensuremath{\eqq^{\textsf{sf}}_{3}}}
\newcommand{\eqqSFCP}{\ensuremath{\eqq^{\textsf{sf'}}_{3}}}
\newcommand{\eqqSFCPP}{\ensuremath{\eqq^{\textsf{sf''}}_{3}}}
\newcommand{\sysNFC}{\ensuremath{\sys^{\textsf{nf}}_{3}}}
\newcommand{\eqqNFC}{\ensuremath{\eqq^{\textsf{nf}}_{3}}}

\begin{example}[Normalizing Symbolic Formulae]\label{ex:norm-symev} Consider the non-singleton symbolic formula $\hV_3$ given below.
	$$ \hV_3 \defEquals \hVBdef$$
	By applying \Sref{step:form-to-sfform} on $\hV_3$ we obtain the following standard form formula,
	$$ \hVSFC= \gensf{\hV_3}= (\hnec{\symAnsC}\hnec{\symAnsC}(\hV_3)) \hand	(\hnec{\symReqD}\hnec{\symReqE}\hfls) $$
	By \Sref{step:sfform-to-sfsys} we then obtain the following system of equations which we can then normalize via \Sref{step:sfsys-to-detsys} as shown below.
	$$ \sysSFC=\gensf{\hVSFC} = \syseq{\eqqSFC}{\hVarX_0}{\sysrecsetE}  \qquad \text{ where}$$
	\[
	\eqqSFC = \begin{xbraces}{c}
				\hVarX_{0}=\hnec{\symReqC}\hVarX_3\hand\hnec{\symReqD}\hVarX_{11}, \hVarX_{3}=\hnec{\symAnsC}\hVarX,\\[3mm] \hVarX=\hnec{\symReqC}\hVarX_7\hand\hnec{\symReqD}\hVarX_9, \hVarX_{7}=\hnec{\symAnsC}\hVarX_8, \hVarX_{8}=\hVarX, \\[3mm]
				\hVarX_{9}=\hnec{\symReqE}\hVarX_{10}, \hVarX_{10}=\hfls, \hVarX_{11}=\hnec{\symReqE}\hVarX_{12}, \hVarX_{12}=\hfls
			\end{xbraces}
	\]
		
	\noindent However, when applying \Sref{step:sfsys-to-detsys} on $\sysSFC$, the algorithm fails to combine symbolic events \symReqC and \symReqD as despite not being disjoint, they are neither syntactically equal. Hence, equation $\hVarX_{0}=\hnec{\symReqC}\hVarX_3\hand\hnec{\symReqD}\hVarX_{11}$ is not merged into $\hVarX_{0}=\hnec{\actSN{\patReqA}{\dvV{\neq}h{\,\land\,}\dvV{\neq}j}}\hVarX_{\set{3,11}}$, but simply remains the same as shown below.
	$$ \sysNFC=\gensys{\sysSFC} = \syseq{\eqqNFC}{\hVarX_{\set{0}}}{\sysrecsetE}  \qquad\qquad \text{ where}$$
	\[
	\eqqNFC = \begin{xbraces}{c}
		\hVarX_{\set{0}}=\hnec{\symReqC}\hVarX_{\set{3}}\hand\hnec{\symReqD}\hVarX_{\set{11}}\\[3mm] \hVarX=\hnec{\symReqC}\hVarX_{\set{7}}\hand\hnec{\symReqD}\hVarX_{\set{9}}, \\[3mm]
		 \hVarX_{\set{7}}=\hnec{\symAnsC}\hVarX_{\set{8}}, \hVarX_{\set{8}}=\hVarX, , \hVarX_{\set{3}}=\hnec{\symAnsC}\hVarX,  \\[3mm]
		\hVarX_{\set{9}}=\hnec{\symReqE}\hVarX_{\set{10}}, \hVarX_{\set{10}}=\hfls,  \\[3mm] \hVarX_{\set{11}}=\hnec{\symReqE}\hVarX_{\set{12}}, \hVarX_{\set{12}}=\hfls
	\end{xbraces}
	\]	
	Due to this, when applying \Sref{step:detsys-to-shml}, we end up with $\hVV_3$ (stated below), which despite being semantically equivalent to the original formula $\hV_3$, it is still not in normal form (\ie $\hVV_3{\,\notin\,}\shmlwf$), since its conjunctions are not guarded by necessities defining disjoint events.
	$$ 	\begin{array}{rcl} 
		\hVV_3 &=& \gensf{\hV_3}= (\hnec{\symAnsC}\hnec{\symAnsC}(\hV_3)) \hand (\hnec{\symReqD}\hnec{\symReqE}\hfls) \\
			  &=& \hVSFC 
		\end{array}
	$$ 
	Even though formula $\hV_3$ is a generic version of $\hV_2$ (\ie $\hV_3$ is less restrictive than $\hV_2$), our normalization algorithm fails to produce an equivalent normalized formula since $\hVV_3\notin\shmlwf$. \bqed
\end{example}

Therefore, to be able to use the existing normalization procedure introduced in \cite{Aceto2016Determinization}, we must add extra steps to make sure that we can \emph{clearly distinguish} between symbolic events, \eg \recv{\dvV}{5} and \recv{i}{\dvVV} can be replaced by the following guarded patterns: 
\begin{itemize}
	\item \recv{\dvV}{5} can be encoded as $\actSN{\recv{\dvV}{\dvVV}}{\dvVV{=}5 \land \dvV{=}i}$ or $\actSN{\recv{\dvV}{\dvVV}}{\dvVV{=}5 \land x{\neq}i}$; while
	\item \recv{i}{\dvVV} can be encoded as $\actSN{\recv{\dvV}{\dvVV}}{\dvVV{=}5 \land \dvV{=}i}$ or $\actSN{\recv{\dvV}{\dvVV}}{\dvVV{\neq}5 \land x{=}i}$.
\end{itemize}
Using this technique we are able to statically tell whether one encoded symbolic event is equal to another event or not. For instance, the reconstructed symbolic events $\actSN{\recv{\dvV}{\dvVV}}{\dvV=5 \land x=i}$ and $\actSN{\recv{\dvV}{\dvVV}}{\dvVV\neq5 \land \dvV=i}$ can now be statically distinguished by using a simple syntactic check, since their contradicting conditions, \ie $\dvVV{=}5$ and $\dvVV{\neq}5$ \resp, guarantee that the reconstructed events are also disjoint, \ie $\syn{\actSN{\recv{\dvV}{\dvVV}}{\dvVV{=}5 \land \dvV{=}i}}\cap\syn{\actSN{\recv{\dvV}{\dvVV}}{\dvVV{\neq}5 \land \dvV{=}i}}{=}\emptyset$.

In the following section we thus formalize and present the additional steps that must be performed when working \wrt symbolic events in order to ensure that the obtained conjunct branches are unable to match the same concrete event.

{
	\renewcommand{\theparagraph}{\bf (\roman{paragraph})}
	\setcounter{secnumdepth}{4}
\subsubsection{Additional Steps for Normalizing Necessities defining Symbolic Events} 

We formally define three additional construction rules that must be applied between steps \Sref{step:sfform-to-sfsys} and \Sref{step:sfsys-to-detsys}. These new constructions convert conjunctions that are guarded by necessities defining non-disjoint symbolic events, into equivalent conjunctions guarded by \emph{disjoint necessities}, \ie necessities describing symbolic events that are both \emph{syntactically and semantically different} such that the sets of concrete events they denote do not intersect. The additional steps are the following:

\begin{enumerate}[\bf{\textsection}i.\!]
	\item \textbf{Uniformity of Symbolic Events:} \label{step:alpha-equiv-cons} In this step, we inspect conjunct modal necessities and substitute their data variables with the same fresh variable whenever they define pattern equivalent symbolic events, \eg we convert $\hnec{\actSN{\recv{\dvV_1}{\dvV_2}}{\predc(\dvV_1,\dvV_2)}}\hnec{\actSN{\recv{\dvV_3}{\dvV_4}}{\predc(\dvV_3,\dvV_4)}}\hfls\hand\hnec{\actSN{\recv{\dvVV_1}{\dvVV_2}}{\predc(\dvVV_1,\dvVV_2)}}\hnec{\actSN{\recv{\dvVV_3}{\dvV_4}}{\predc(\dvVV_3,\dvVV_4)}}\hfls$ into $\hnec{\actSN{\recv{\dvVVV_1}{\dvVVV_2}}{\predc(\dvVVV_1,\dvVVV_2)}}\hnec{\actSN{\recv{\dvVVV_3}{\dvVVV_4}}{\predc(\dvVVV_3,\dvVVV_4)}}\hfls\hand\hnec{\actSN{\recv{\dvVVV_1}{\dvVVV_2}}{\predc(\dvVVV_1,\dvVVV_2)}}\hnec{\actSN{\recv{\dvVVV_3}{\dvVVV_4}}{\predc(\dvVVV_3,\dvVVV_4)}}\hfls$; we discuss this in detail in \Cref{sec:alpha-equiv-cons}.
	\item \textbf{Condition Reformulation of Conjunct Symbolic Events:}  \label{step:truth-combos-cons} Once uniformed, the conjunctions are recomposed such that the reconstructed conjunctions are guaranteed to define necessities that define disjoint symbolic events; the details on how this is done are discussed in \Cref{sec:truth-combos-cons}.
\end{enumerate}

Internally, the constructions presented in \Cref{sec:alpha-equiv-cons,sec:truth-combos-cons} both make use of the \traverse function in order to process the given set of equations in a \emph{tree-like} manner as defined in \Cref{fig:traverse}.

\begin{figure}[t]
\textbf{Traversal Functions.}
\begin{align*}
	\traverseFun{\eqq}{\IndSet}{\projFunSym}{\acc} &\defeq\, 
	\begin{xbrace}{c@{\qquad}l}
		\traverseFun{\eqq'}{\IndSet'}{\projFunSym}{\acc'} & \text{if }\eqq{\neq}\emptyset\text{ and }\IndSet{\neq}\emptyset \\[2mm] & \text{then } \acc'{=}\projFun{\eqq}{\IndSet}{\acc}  \\[2mm] & \text{and } \eqq'{=}\eqq{\setminus}\eqqres{\eqq}{\IndSet}  \\[2mm] & \text{and } \IndSet'{=}\bigunion{j\in\IndSet}\childrenFun{\eqq}{j} \\[5mm]
		\acc & \text{otherwise}
	\end{xbrace}\\[5mm]
	\childrenFun{\eqq}{i} &\defeq\, \setdef{j}{\eqq(\hVarX_i){=}\hAND{j\in\IndSet}\hnec{\actS_j}\hVarX_j{\hand}\hV \text{ and } j{\neq}i \text{ and }\hVarX_j{\in}\dom{\eqq}}\\[3mm]
	\eqqres{\eqq}{\IndSet}  &\defeq\, \Setdef{\hVarX_i{=}\hV_i}{(\hVarX_i{=}\hV_i){\,\in\,}\eqq \text{ and } i{\in}\IndSet\!\!}
\end{align*}
\caption{The Breath First Traversal Algorithm.}
\label{fig:traverse}
\end{figure}

The function $\traverse::(\eqq\times\powerset{\Index}\times\ProjFun\times\Acc)\mapsto\Acc$ is a \emph{higher order function} which takes as input: a set of equations \eqq, a set of indices \IndSet an arbitrary projection function \projFunSym, and an accumulator argument \acc.

\newcommand{\XOEq}{\hVarX_{0}{=}\hnec{\actS_1}\hVarX_{1}{\,\hand\,}\hnec{\actS_2}\hVarX_{2}{\,\hand\,}\hnec{\actS_3}\hVarX_{3}}
\newcommand{\XAEq}{\hVarX_{1}{=}\hnec{\actS_4}\hVarX_{0}}
\newcommand{\XBEq}{\hVarX_{2}{=}\hnec{\actS_5}\hVarX_{5}}
\newcommand{\XCEq}{\hVarX_{3}{=}\hnec{\actS_6}\hVarX_{3}}
\newcommand{\XEEq}{\hVarX_{5}{=}\hfls}

\paragraph{Performing a Traversal.}
The \traverse function is generally used to conduct a \emph{breath first} traversal on the given equation set, starting from the equation that equates to the principle variable as the root of the tree traversal, \eg as shown in \Cref{fig:traversal-ex}, for system $\syseq{\eqq}{\hVarX_{0}}{\sysrecset}$, equation $\XOEq$ is the root of the traversal since $\hVarX_{0}$ is the principle variable.

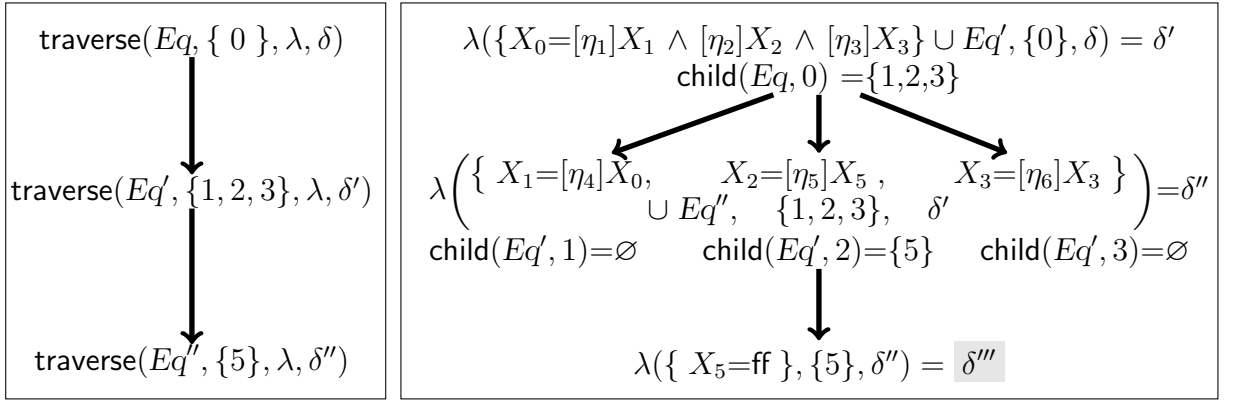
\begin{figure}[t]
	\begin{tikzpicture}
	\path [draw=black] (-5,1.75) rectangle (5.75,7);
	\path [draw=black] (-10.2,1.75) rectangle (-5.2,7);
	
	\begin{pgfonlayer}{nodelayer}
		\node [style=none] (0) at (-7.75, 6.5) {\traverseFun{\eqq}{\set{\;0\;}}{\projFunSym}{\acc}};
		\node [style=none] (1) at (-7.75, 4.5) {\traverseFun{\eqq'}{\set{1,2,3}}{\projFunSym}{\acc'}};
		\node [style=none] (2) at (-7.75, 2.25) {\traverseFun{\eqq''}{\set{5}}{\projFunSym}{\acc''}};
		\node [style=none] (3) at (0.5, 6.5) {$\projFun{\set{\XOEq}\cup\eqq'}{\set{0} }{\acc}=\acc'$};
		\node [style=none] (4) at (0.5, 6) {\childrenFun{\eqq}{0}=\set{1,2,3}};
		\node [style=none] (5) at (0.5, 4.5) {\projFunSym\!\!$\begin{xbrackets}{c}
				\Set{\;\XAEq,\qquad \XBEq\;,\qquad \XCEq\;}\\\cup\;\eqq'', \quad 
				\set{1,2,3}, \quad \acc'
			\end{xbrackets}\!{=}\acc''$};
		\node [style=none] (6) at (-3.25, 3.7) {$\childrenFun{\eqq'}{1}{=}\emptyset$};
		\node [style=none] (7) at (0.5, 3.7) {$\childrenFun{\eqq'}{2}{=}\set{5}$};
		\node [style=none] (8) at (4, 3.7) {$\childrenFun{\eqq'}{3}{=}\emptyset$};
		\node [style=none] (9) at (0.5, 2.25) {$\projFun{\set{\;\XEEq\;}}{\set{5} }{\acc''}=\greybox{\acc'''}$};
		\node [style=none] (10) at (-2.25, 5) {};
		\node [style=none] (11) at (3, 5) {};
	\end{pgfonlayer}
	\begin{pgfonlayer}{edgelayer}
		\draw [style=arrow] (0) to (1);
		\draw [style=arrow] (1) to (2);
		\draw [style=arrow] (4) to (5);
		\draw [style=arrow] (7) to (9);
		\draw [style=arrow] (4) to (10);
		\draw [style=arrow] (4) to (11);
	\end{pgfonlayer}
\end{tikzpicture}
	\vspace{-5mm}
	\caption{A pictorial view of an example equation set traversal.}
	\label{fig:traversal-ex}
\end{figure}

The children of the root are calculated via the $\children{\,::\,}(\eqq\times\Index){\mapsto}\powerset{\Index}$ function. This function takes as input a set of equations \eqq along with the index of the parent equation, \eg index $0$ for equation $\XOEq$. It then scans the equated formula and returns the set containing the indices of every branch, defined in the equated formula, which is prefixed by a modal necessity, \eg in \Cref{fig:traverse}, the children of $\XOEq$ are $\Set{1,2,3}$, such that branches $\hnec{\actS_1}\hVarX_{1}$, $\hnec{\actS_2}\hVarX_{2}$ and $\hnec{\actS_3}\hVarX_{3}$ are \emph{siblings} as they are defined at the \emph{same conjunction layer}.

Cycles in the traversal are avoided since the \children function is always executed \wrt a restricted set of equations, \ie one which \emph{does not} include the parent equation. For instance, while analysing equation $\XAEq$ (see \Cref{fig:traverse}), \traverse is evaluated \wrt $\eqq'$ which does not include the parent equation, \ie since $\eqq'{=}\eqq{\setminus}\eqqres{\eqq}{\set{0}}$ where $\eqqres{\eqq}{\set{0}}=\set{\XOEq}$. sIn this way, when computing the children of $\hVarX_{1}$ (via \childrenFun{\eqq'}{1}) index $0$ is not added to the resultant set of child indices, since $\hVarX_{0}\notin\dom{\eqq'}$'; this avoids cycling back to some (grand) parent equation.

Additionally, the \children function avoids cycling back to the (immediate) parent by removing the parent's index form the returned set of child indices, \eg when evaluating \childrenFun{\eqq'}{3} to retrieve the child indices of equation $\XCEq$, index $3$ is removed thus avoiding the creation of a loop in the traversal.

\paragraph{Applying the projection function during traversal.}
While traversing the equation set, the \traverse function can apply an arbitrary projection function \projFunSym. Despite being an arbitrary function, \projFunSym must adhere to the following type: $\projFunSym::(\eqq\times\powerset{\Index}\times\Acc){\mapsto}\Acc$, \ie \projFunSym must be a function which takes three inputs, including the current set of equations \eqq, a set of indices \IndSet and an accumulator value \acc; as a result \projFunSym must return an updated version of the accumulator, \ie $\acc'$.

\paragraph{Traversal Termination and Return Value.}
The traversal terminates when either all the equations in \eqq have been process such that the \traverse function is applied \wrt $\eqq{=}\emptyset$, or whenever no further children can be visited since none of the branches have any valid children, \ie for every branch $i$, $\childrenFun{\eqq}{i}{=}0$. The latter is an optimization which omits the redundant processing of equations that are not reachable from the principle equation.
\begin{example} Consider the system of equations \syseq{\eqq}{\hVarX_{0}}{\emptyset} where 
	$$\eqq{=}\Set{\hVarX_{0}{=}\hnec{\actS}\hVarX_{1},\, \hVarX_{1}{=}\hfls,\hVarX_{2}{=}\hfls}$$
	The traversal starts from equation $\hVarX_{0}{=}\hnec{\actS}\hVarX_{1}$ as the root, followed by $\hVarX_{1}{=}\hfls$ as its immediate child, however equation $\hVarX_{2}{=}\hfls$ is the child of neither equation and is thus ignored by the traversal. \bqed
\end{example}
The latest version of the accumulator value is returned once the traversal is complete.

\newcommand{\dvVVA}{\ensuremath{\dvVV^1}}
\newcommand{\dvVVB}{\ensuremath{\dvVV^2}}
\newcommand{\dvVVC}{\ensuremath{\dvVV^3}}
\newcommand{\dvVVD}{\ensuremath{\dvVV^4}}
\newcommand{\dvVVE}{\ensuremath{\dvVV^5}}
\newcommand{\dvVVF}{\ensuremath{\dvVV^6}}
\newcommand{\dvVVG}{\ensuremath{\dvVV^7}}
\newcommand{\dvVVH}{\ensuremath{\dvVV^8}}
\newcommand{\dvVVI}{\ensuremath{\dvVV^9}}
\newcommand{\dvVVJ}{\ensuremath{\dvVV^{10}}}
\newcommand{\dvVVK}{\ensuremath{\dvVV^{11}}}
\newcommand{\dvVVL}{\ensuremath{\dvVV^{12}}}
\newcommand{\dvVVM}{\ensuremath{\dvVV^{13}}}
\newcommand{\dvVVN}{\ensuremath{\dvVV^{14}}}

\subsubsection{Uniformity of Symbolic Events} \label{sec:alpha-equiv-cons}
Intuitively, this part of the normalization algorithm \emph{renames the data variables} of the \emph{pattern equivalent} symbolic events (\ie symbolic events defining equivalent patterns; see \Cref{sec:ptrn-match}) defined in necessity operations guarding branches within the same conjunction level, to the \emph{same variable names} to obtain a \emph{uniform system of equations} as defined in \Cref{def:generic-sys-eqs}.

\begin{definition}[Uniform System of Equations]\label{def:generic-sys-eqs}
	An equation is \emph{uniform} when it is in \emph{standard form} and when every \emph{pattern equivalent event} (see \Cref{sec:symevent-match}) defined \emph{by sibling necessities within a conjunction} defines the exact \emph{same} data variable names. A system of equations is uniform when all of its equations are uniform. \bqed
\end{definition}

\begin{example}
	The symbolic necessity events defined in $\hVarX_0\!=\!\hgnec{\recv{\dvV_1}{\dvV_2}}{c_1(\dvV_1,\dvV_2)}\hVarX_1\!\hand\!\hgnec{\recv{\dvVV_1}{\dvVV_2}}{c_2(\dvVV_1,\dvVV_2)}\hVarX_2$ are both pattern equivalent, however they are \emph{not uniform} since they \emph{do not} define the same variables. This formula can however be easily made uniform by applying \genuni{-} which renames $\dvV_1$ and $\dvVV_1$ to the same $\dvVVV_1$ and similarly $\dvV_2$ and $\dvVV_2$ to a fresh variable $\dvVVV_2$, so to obtain $\hVarX_0\!=\!\hgnec{\recv{\dvVVV_1}{\dvVVV_2}}{c_1(\dvVVV_1,\dvVVV_2)}\hVarX_1\!\hand\!\hgnec{\recv{\dvVVV_1}{\dvVVV_2}}{c_2(\dvVVV_1,\dvVVV_2)}\hVarX_2$  \bqed
\end{example}

\begin{figure}[t]
\noindent\textbf{Construction}

$$
\begin{array}{r@{\,}c@{\,}l}
	\genuni{\syseq{\eqq}{\hVarX_0}{\sysrecset}} &\defeq& \syseq{\uniFun{\eqq}{\subMap}}{\hVarX_0}{\sysrecset}\\
	&& \text{ where } \subMap{=}\traverseFun{\eqq}{\set{0}}{\partition}{\emptyset} \\[5mm]
	\uniFun{\eqq}{\subMap} &\defeq& \setdef{\hVarX_i{=}\hAND{j\in\IndSet}\hnec{\actS_j(\subMap(j))}\hVarX_j{\hand}\hV}{\hVarX_i{=}\hAND{j\in\IndSet}\hnec{\actS_j}\hVarX_j{\hand}\hV\in\eqq}\\[5mm]
	\partitionFun{\eqq}{\IndSet}{\subMap} &\defeq& \Setdef{\submapstoE{j}{\subMap(i)\cupplus\sub{\dvVVV^n}{\dvV^n}}\\[5mm]\submapstoE{k}{\subMap(l)\cupplus\sub{\dvVVV^n}{\dvVV^n}} }
				{\forall i,l\in\IndSet\cdot
					\eqq(i){=}\hAND{j\in\IndSet'}\hnec{\actS_j(\dvV^n)}\hVarX_j{\hand}\hV \\ 
					\text{ and } 
					\eqq(l){=}\hAND{k\in\IndSet''}\hnec{\actS_k(\dvVV^n)}\hVarX_k{\hand}\hV \textsl{ s.t.} \\[3mm]
					\text{ if }\actS_j(\dvV^n) \text{ is \emph{pattern equivalent} to }\\[2mm] \,\actS_k(\dvVV^n), \text{ then we assign the \emph{same} }\\[2mm]
					\text{ set of fresh variables }\dvVVV^n.
				}\cupplus\subMap
\end{array}
$$
\caption{The Uniformity Algorithm for Symbolic Events}
\label{fig:alpha-equiv-cons-alg}
\end{figure}

In \Cref{fig:alpha-equiv-cons-alg}, we present the construction $\genuni{-}::\syseq{\eqqSF}{\Var}{\powerset{\Var}}\!\mapsto\syseq{\eqqUni}{\Var}{\powerset{\Var}}$. This construction internally uses the \uni function to create the required uniform set of equations \eqqUni from the given standardized equation set \eqqSF. More specificallym \uni reconstructs the equation set by performing a linear scan during which it converts equations of the form $\hVarX_i{=}\hAND{j\in\IndSet}\hnec{\actS_j\subMap(j)}\hVarX_j{\hand}\hV$ where $\subMap::\Index{\mapsto}\s$ is a map that provides a substitution environment \s for a given index $i$.

Intuitively, a \emph{well-formed} \subMap map should provide substitutions that \emph{uniformly} rename the data variables of pattern equivalent modal necessities that are defined as \emph{siblings within a tree of conjunctions}, \ie the data variables defined in such necessities are renamed to the \emph{same fresh set of variable names}, such that the patterns of the renamed necessities become \emph{syntactically equal}. 
\begin{definition}[A Well-Formed \subMap Map]
	We say that \subMap is a \emph{well-formed map} for a set of equations \eqq, whenever it provides a set of mappings which allow for
	\begin{enumerate}[$(i)$]
		\item renaming the \emph{data variables} of each \emph{pattern equivalent sibling necessity}, defined in \eqq, to the \emph{same} set of fresh variables, and for
		\item renaming any \emph{reference} to a data variable that is bound by a \emph{renamed parent necessity} defined in \eqq. \bqed
	\end{enumerate} 
\end{definition}
We assume that when an index $i$ is not in the domain of the \subMap map (\ie $i\notin\dom{\subMap}$) then $\subMap(i)=\emptyset$.

\newcommand{\hnecA}{\ensuremath{\hgnec{\recv{\dvVA}{\dvVB}}{\dvVA{\neq}i}}\xspace}
\newcommand{\hnecB}{\ensuremath{\hgnec{\recv{\dvVC}{\dvVD}}{\dvVD{\neq}3}}\xspace}
\newcommand{\hnecC}{\ensuremath{\hgnec{\send{\dvVE}{\dvVF}}{\ctru}}\xspace}
\newcommand{\hnecD}{\ensuremath{\hgnec{\send{\dvVG}{\dvVH}}{\dvVG{=}\dvVC}}\xspace}

\newcommand{\hnecAN}{\ensuremath{\hgnec{\recv{\dvVVA}{\dvVVB}}{\dvVVA{\neq}i}}\xspace}
\newcommand{\hnecBN}{\ensuremath{\hgnec{\recv{\dvVVA}{\dvVVB}}{\dvVVB{\neq}3}}\xspace}
\newcommand{\hnecCN}{\ensuremath{\hgnec{\send{\dvVVC}{\dvVVD}}{\ctru}}\xspace}
\newcommand{\hnecDN}{\ensuremath{\hgnec{\send{\dvVVC}{\dvVVD}}{\dvVVC{=}\dvVVA}}\xspace}

\renewcommand{\XOEq}{\ensuremath{\hVarX_{0}{=}\hnecA\hVarX_{1}{\hand}\hnecB\hVarX_{2}}}
\renewcommand{\XAEq}{\ensuremath{\hVarX_{1}{=}\hnecC\hVarX_{3}}}
\renewcommand{\XBEq}{\ensuremath{\hVarX_{2}{=}\hnecD\hVarX_{4}}}
\renewcommand{\XCEq}{\ensuremath{\hVarX_{3}{=}\hfls}}
\newcommand{\XDEq}{\ensuremath{\hVarX_{4}{=}\hfls}}

\newcommand{\XOEqN}{\ensuremath{\hVarX_{0}{=}\hnecAN\hVarX_{1}{\hand}\hnecBN\hVarX_{2}}}
\newcommand{\XAEqN}{\ensuremath{\hVarX_{1}{=}\hnecCN\hVarX_{3}}}
\newcommand{\XBEqN}{\ensuremath{\hVarX_{2}{=}\hnecDN\hVarX_{4}}}
\newcommand{\XCEqN}{\ensuremath{\hVarX_{3}{=}\hfls}}
\newcommand{\XDEqN}{\ensuremath{\hVarX_{4}{=}\hfls}}

\begin{example}\label{ex:uniform-submap-gen} Consider the following system of equations \syseq{\eqq}{\hVarX_{0}}{\sysrecsetE} where
	$$ \eqq=\begin{xbraces}{c}
				\XOEq, \; \XAEq, \\ \XBEq, \quad \XCEq, \quad \XDEq
			\end{xbraces} 
	$$
	For convenience, we also represent these equations as a tree starting from equation $\XOEq$ as the root of the tree. We also assume the knowledge of a \emph{well-formed} \subMap map that has the following form, \ie 
	$$ 	\subMap=\begin{xbraces}{c}
					\submapstoEP{1}{\subE{\dvVA}{\dvVVA},\subE{\dvVB}{\dvVVB}}, \quad \submapstoEP{2}{\subE{\dvVC}{\dvVVA},\subE{\dvVD}{\dvVVB}},\\ 
					\submapsto{3}{\subMap(1)}{\subE{\dvVE}{\dvVVC},\subE{\dvVF}{\dvVVD}},\quad \submapsto{4}{\subMap(2)}{\subE{\dvVG}{\dvVVC},\subE{\dvVH}{\dvVVD}}
				\end{xbraces}. \medskip
	$$
%
	\begin{figure}[t]
		\centering

\newcommand{\hnecAS}{\ensuremath{\hnec{\actSN{\recv{\dvVA}{\dvVB}}{\dvVA{\neq}i}\subMap(1)}}\xspace}
\newcommand{\hnecBS}{\ensuremath{\hnec{\actSN{\recv{\dvVC}{\dvVD}}{\dvVD{\neq}3}\subMap(2)}}\xspace}
\newcommand{\hnecCS}{\ensuremath{\hnec{\actSN{\send{\dvVE}{\dvVF}}{\ctru}\subMap(3)}}\xspace}
\newcommand{\hnecDS}{\ensuremath{\hnec{\actSN{\send{\dvVG}{\dvVH}}{\dvVG{=}\dvVC}\subMap(4)}}\xspace}

\newcommand{\XOEqS}{\ensuremath{\hVarX_{0}{=}\hnecAS\hVarX_{1}{\hand}\hnecBS\hVarX_{2}}}
\newcommand{\XAEqS}{\ensuremath{\hVarX_{1}{=}\hnecCS\hVarX_{3}}}
\newcommand{\XBEqS}{\ensuremath{\hVarX_{2}{=}\hnecDS\hVarX_{4}}}
\newcommand{\XCEqS}{\ensuremath{\hVarX_{3}{=}\hfls}}
\newcommand{\XDEqS}{\ensuremath{\hVarX_{4}{=}\hfls}}

\begin{tikzpicture}
	\begin{pgfonlayer}{nodelayer}
		\node [style=shadowRect] (0) at (0.75, 4.25) {\color{black}\XOEqN};
		\node [style=none] (0Label) at (5.35, 4.95) {\uni};
		\node [style=shadowRect] (1) at (-3.25, 1.25) {\color{black}\XAEqN};
		\node [style=none] (1Label) at (-1.2, 2.2) {\uni};		
		\node [style=shadowRect] (2) at (4.75, 1.25) {\color{black}\XBEqN};
		\node [style=none] (2Label) at (7.25, 2.2) {\uni};
		
		\node [style=rect] (3) at (0, 5) {\XOEqS};
		\node [style=rect] (4) at (-4, 2) {\XAEqS};
		\node [style=rect] (5) at (4, 2) {\XBEqS};
		\node [style=rect] (6) at (-4, -1) {\XCEqS};
		\node [style=rect] (7) at (4, -1) {\XDEqS};
	\end{pgfonlayer}
	\begin{pgfonlayer}{edgelayer}
		\draw [style=arrow] (3) to (4);
		\draw [style=arrow] (3) to (5);
		\draw [style=arrow, bend left=80, out=90, looseness=1.75 ] ($(3.east)+(0,0.0)$) to ($(0.east)-(0.0,0)$);
		\draw [style=arrow, bend left=80, out=90, looseness=1.5 ] ($(4.east)+(0,0.2)$) to ($(1.north east)-(0.2,0)$);
		\draw [style=arrow, bend left=80, out=90, looseness=1.5 ] ($(5.east)+(0,0.2)$) to ($(2.north east)-(0.2,0)$);
		\draw [style=arrow] (4) to (6);
		\draw [style=arrow] (5) to (7);
	\end{pgfonlayer}
\end{tikzpicture}
		\caption{A Tree representation of the \uni traversal performed on \eqq.}
		\label{fig:traverse-uni}
	\end{figure}
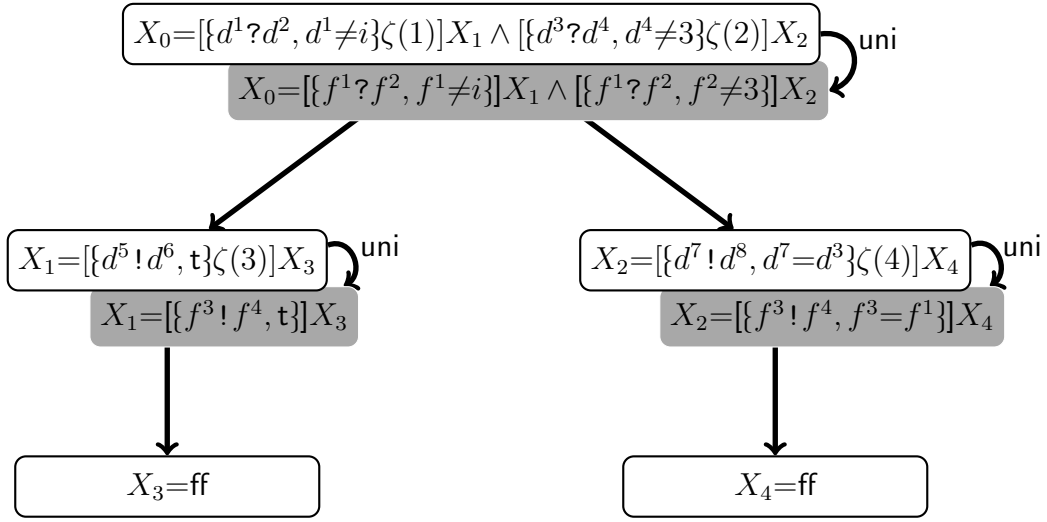
	As shown by the tree representation in \Cref{fig:traverse-uni}, necessities \hnecA and \hnecB are pattern equivalent siblings within the conjunction defined by equation $\hVarX_{0}$; in order to be uniformed, the substitution map \subMap projects indices $1$ and $2$ onto environments \subb{\subE{\dvVA}{\dvVVA},\subE{\dvVB}{\dvVVB}} and \subb{\subE{\dvVC}{\dvVVA},\subE{\dvVD}{\dvVVB}} \resp Once the substitution is applied to both necessities we obtain \hnecAN and \hnecBN, in lieu of \hnecA and \hnecB. Notice how the patterns in both of the resultant necessities are now \emph{syntactically equal}, meaning that the resultant equation \XOEqN is now \emph{uniform}.
	
	As illustrated in the equation tree above, necessities \hnecC and \hnecD are also \emph{pattern equivalent siblings} within the conjunction defined in $\hVarX_{0}$. In order to make them uniform, \subMap provides the following mappings, namely, $\submapsto{3}{\subMap(1)}{\subE{\dvVE}{\dvVVC},\subE{\dvVF}{\dvVVD}}$ and $\submapsto{4}{\subMap(2)}{\subE{\dvVG}{\dvVVC},\subE{\dvVH}{\dvVVD}}$, where enable for renaming the aforementioned necessities into \hnecCN and \hnecDN.
	
	Notice how the filtering condition $\dvVG{=}\dvVC$ in \hnecD was also renamed to $\dvVVC{=}\dvVVA$ as variable \dvVG is substituted by \dvVVC when its binding necessity \hnecB is uniformed into \hnecBN. This substitution was made possible since mapping $\subMap(4)$ includes the substitutions returned by the parent's index, \ie $\subMap(2)$; this allows applying the substitutions performed upon the parent, to its children, which is important to keep the equation closed \wrt the data variables. \bqed
\end{example}

\paragraph{Creating a Well-Formed \subMap Map.} 
Up until now we have been assuming the existence of a \emph{well-formed} \subMap map which provides all the necessary information, without having any knowledge as to how it is created. 

The \subMap map is created as a result of conducting a breath first traversal, via the \traverse function, on the given equation set, using the \partition function as the \projFunSym projection function required by \traverse. The function $\partition::(\eqq\times\powerset{\Index}\times\Acc){\mapsto}\Acc$ follows the format dictated by \projFunSym, \ie it takes as input a set of equations \eqq, a set of indices \IndSet and an accumulator, in this case \subMap; it returns an updated version of \subMap as a result.

In order to update \subMap, the \partition function inspects the sibling equations denoted by the indices in \IndSet and as a result it creates a \emph{substitution environment} which renames the variables of each pattern equivalent sibling necessity, to the same fresh set of variables. 

\begin{example} Recall the system of equations defined earlier in \Cref{ex:uniform-submap-gen}, \ie \syseq{\eqq}{\hVarX_{0}}{\sysrecsetE} where
	$$ \eqq=\begin{xbraces}{c}
	\XOEq, \;\; \XAEq, \\[3mm] \XBEq, \quad \XCEq, \quad \XDEq
	\end{xbraces}  \medskip
	$$	
	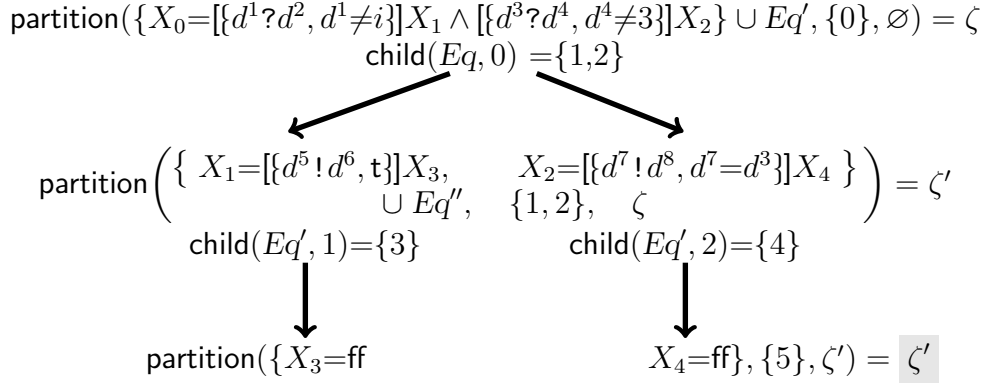
\begin{figure}[t]
		\centering

\begin{tikzpicture}
	\begin{pgfonlayer}{nodelayer}
	
		\node [style=none] (3) at (0.5, 6.5) {$\partitionFun{\set{\XOEq}\cup\eqq'}{\set{0} }{\emptyset}{\;=\;}\subMap$};
		\node [style=none] (4) at (0.5, 6) {\childrenFun{\eqq}{0}=\set{1,2}};
		\node [style=none] (5) at (0.5, 4.3) {\partition\!\!$\begin{xbrackets}{c}
				\Set{\;\XAEq,\qquad \XBEq\;}\\\cup\;\eqq'', \quad 
				\set{1,2}, \quad \subMap
			\end{xbrackets}\!{\;=\;}\subMap'$};
		\node [style=none] (6) at (-2, 3.5) {$\childrenFun{\eqq'}{1}{=}\set{3}$};
		\node [style=none] (8) at (3, 3.5) {$\childrenFun{\eqq'}{2}{=}\set{4}$};
		\node [style=none] (9) at (0.5, 2) {$\qquad\quad\partitionFun{\set{\XCEq\qquad\quad\qquad\qquad\qquad\XDEq}}{\set{5} }{\subMap'}{\;=\;}\greybox{\subMap'}$};
		\node [style=none] (10) at (-2.25, 5) {};
		\node [style=none] (11) at (3, 5) {};
		
		\node [style=none] (12) at (-2, 2.25) {};
		\node [style=none] (13) at (3, 2.25) {};
	\end{pgfonlayer}
	\begin{pgfonlayer}{edgelayer}
		\draw [style=arrow] (4) to (10);
		\draw [style=arrow] (4) to (11);
		
		\draw [style=arrow] (6) to (12);
		\draw [style=arrow] (8) to (13);
	\end{pgfonlayer}
\end{tikzpicture}
		\caption{A breath first traversal using \partition to obtain \subMap.}
		\label{fig:traverse-partition}
	\end{figure}
	\indent \Cref{fig:traverse-partition} depicts the breath first traversal performed by the \traverse function in which the projection function \partition was applied on each set of siblings. Notice that when \partition is applied on the root equation, the initially empty map gets extended by 2 entries, namely $\subMap{=}\emptyset{\cupplus}\Set{\submapsto{1}{\emptyset}{\subE{\dvVVA}{\dvVA},\subE{\dvVVB}{\dvVB}}, \submapsto{2}{\emptyset}{\subE{\dvVVA}{\dvVC},\subE{\dvVVB}{\dvVD}}}$; as shown in \Cref{ex:uniform-submap-gen}, this allows for the sibling necessities defined in $\hVarX_{0}$ to be uniformed.
	
	The \subMap map is further extended into $\subMap'{=}\subMap{\cupplus}\Set{\submapsto{3}{\subMap(1)}{\subE{\dvVVC}{\dvVE},\subE{\dvVVD}{\dvVF}}, \submapsto{4}{\subMap(2)}{\subE{\dvVVC}{\dvVG},\subE{\dvVVD}{\dvVH}}}$, since the partition function recognises that the sibling necessities \hnecC and \hnecD are also pattern equivalent; it therefore maps variables \dvVE and \dvVG to the same fresh variable \dvVVC, and \dvVF and \dvVH to \dvVVD. \bqed
\end{example}

\newcommand{\eqqUNIC}{\ensuremath{\eqq^\textsf{uni}_3}}
\newcommand{\eqqUNICP}{\ensuremath{\eqq^\textsf{uni'}_3}}
\newcommand{\eqqUNICPP}{\ensuremath{\eqq^\textsf{uni''}_3}}
\newcommand{\sysUNIC}{\ensuremath{\sys^\textsf{\;uni}_3}}
\newcommand{\dvVVVA}{\ensuremath{\dvVVV^1}}
\newcommand{\dvVVVB}{\ensuremath{\dvVVV^2}}
\newcommand{\dvVVVC}{\ensuremath{\dvVVV^3}}
\newcommand{\dvVVVD}{\ensuremath{\dvVVV^4}}
\newcommand{\dvVVVE}{\ensuremath{\dvVVV^5}}
\newcommand{\dvVVVF}{\ensuremath{\dvVVV^6}}
\newcommand{\dvVVVG}{\ensuremath{\dvVVV^7}}
\newcommand{\dvVVVH}{\ensuremath{\dvVVV^8}}
\newcommand{\dvVVVI}{\ensuremath{\dvVVV^9}}
\newcommand{\dvVVVJ}{\ensuremath{\dvVVV^{10}}}
\newcommand{\dvVVVK}{\ensuremath{\dvVVV^{11}}}
\newcommand{\dvVVVL}{\ensuremath{\dvVVV^{12}}}
\newcommand{\dvVVVM}{\ensuremath{\dvVVV^{13}}}
\newcommand{\dvVVVN}{\ensuremath{\dvVVV^{14}}}

\begin{example}\label{ex-generic}\normalfont
	We now recall the \emph{standardized} system of equations \sysSF obtained from \Cref{ex:norm-symev} (restated below); for convenience, we will stop using the shorthand notation and we will instead use guarded symbolic event notation in which the patterns are fully opened and guarded by a filtering equation, \eg we write $\actSN{\recv{\dvV}{\dvVA}}{\dvVA{=}\req\!\land\!\dvV{\neq}h}$ instead of \symReqC.
	$$\sysSFC=\syseq{\eqqSFC}{\hVarX_{0}}{\sysrecsetE} \qquad \text{ where}$$
	$$
	\begin{array}{rcl}
		\eqqSFC &=& \begin{xbraces}{c}
					\hVarX_0 = \hand \begin{array}{c}
					\hgnec{\recv{\dvV}{\dvVA}}{\dvVA{=}\req\!\land\!\dvV{\neq}h}\hVarX_{3} \\[3mm] 
					\hgnec{\recv{\dvVV}{\dvVVA}}{\dvVVA{=}\req \!\land\!\dvVV{\neq}j}\hVarX_{11}
			\end{array}
		\end{xbraces}\cup\eqqSFCP
		\\[7mm]
		\eqqSFCP&=&\begin{xbraces}{c}
			\hVarX_{3}=\hgnec{\send{\dvVB}{\dvVC}}{\dvVB{=}\dvV{\,\land\,}\dvVC{=}\ans}\hVarX_{13}, \\[3mm] \hVarX_{11}=\hgnec{\recv{\dvVVB}{\dvVVC}}{\dvVVB{=}\dvVV{\,\land\,}\dvVVC{=}\req}\hVarX_{12}
		\end{xbraces}\cup\eqqSFCPP
		\\[7mm]
		\eqqSFCPP&=&\begin{xbraces}{c}		
			\hVarX_{13}=\hand\begin{array}{c}
				\hgnec{\recv{\dvVD}{\dvVE}}{\dvVD{=}\dvV{\,\land\,}\dvVE{=}\req\!\land\!\dvVD{\neq}h}\hVarX_7 \\
				\hgnec{\recv{\dvVVD}{\dvVVE}}{\dvVVD{=}\dvVV{\,\land\,}\dvVVE{=}\req\!\land\!\dvVVD{\neq}j}\hVarX_9
				\end{array},  \\[6mm]
			\hVarX_{7}=\hgnec{\send{\dvVF}{\dvVG}}{\dvVF{=}\dvV{\,\land\,}\dvVG{=}\ans}\hVarX_8, \\[3mm] 
			\hVarX_{9}=\hgnec{\recv{\dvVVF}{\dvVVG}}{\dvVVF{=}\dvV{\,\land\,}\dvVVG{=}\req}\hVarX_{10},  \\[5mm] 
			\hVarX_{8}=\hVarX_{13}, \quad
			\hVarX_{10}=\hfls, \quad
			\hVarX_{12}=\hfls
		\end{xbraces}
	\end{array}
	$$	
	
	\noindent To attain the uniform equivalent of \sysSFC via construction \genuni{-} we must first create a \emph{well-formed} \subMap map using \traverseFun{\eqqSFC}{\set{0}}{\partition}{\emptyset}. The traversal is initiated with the initial set of indices \IndSet, as being equal to \set{0}, this is required since our formula starts from the principal logical variable $\hVarX_0$, \ie the \traverse function starts by inspecting equation $\hVarX_{0}{=}\hV_0$.
		
	Since the sibling necessities define equivalent patterns $\recv{\dvV}{\dvVA}$ and $\recv{\dvVV}{\dvVVA}$, once applied the \traverse function applies \partitionFun{\eqqSF}{\set{3,11}}{\emptyset} over these sibling necessities, and creates an initial \subMap map, where
	$$\subMap{\;=\;}\Set{\submapstoEP{3}{\subE{\dvVVVA}{\dvV},\;\subE{\dvVVVB}{\dvVA}},\submapstoEP{11}{\subE{\dvVVVA}{\dvVV},\;\subE{\dvVVVB}{\dvVVA}}}.$$
	Hence, this map will later on permit the \uni function to replace $\dvV$ and $\dvVV$ with the same fresh variable $\dvVVVA$ by applying \sub{\dvVVVA}{\dvV} and \sub{\dvVVVA}{\dvVV}, and similarly $\dvVV$ and $\dvVVA$ with $\dvVVVB$ via \sub{\dvVVVB}{\dvVA} and \sub{\dvVVVB}{\dvVVA} accordingly.
	
	The \traverse function is subsequently applied \wrt the children of $\hVarX_{0}$ \ie $\IndSet'{=}\set{3,11}$, during which it applies the \partition function on equations $\hVarX_{3}=\hgnec{\send{\dvVB}{\dvVC}}{\dvVB{=}\dvV{\,\land\,}\dvVC{=}\ans}\hVarX_{13}$, and $\hVarX_{11}=\hgnec{\recv{\dvVVB}{\dvVVC}}{\dvVVB{=}\dvVV{\,\land\,}\dvVVC{=}\req}\hVarX_{12}$. 
	Since patterns $\send{\dvVB}{\dvVC}$ and $\recv{\dvVVB}{\dvVVC}$ are \emph{not} equivalent, they do \emph{not} require renaming as they are already \emph{uniform}, hence \partitionFun{\eqqSFCP}{\set{13,12}}{\subMap} (where $\set{13,12}$ are the children of $\hVarX_{3}$ and $\hVarX_{11}$) returns $$\subMap'{\;=\;}\subMap\cupplus\Set{\submapstoE{13}{\subMap(3)},\submapstoE{12}{\subMap(11)}}.$$
	Although none of the variables declared in these patterns require renaming, the substitution map still includes mappings $\submapstoE{13}{\subMap(3)}$ and $\submapstoE{12}{\subMap(11)$  which allows for renaming variable references $\dvV$ and $\dvVV$ (which are bound by a parent necessity defined in $\hVarX_{0}$), into $\dvVVVA$, thus keeping the system of equations closed \wrt data variables.
		
	The \traverse function keeps on performing the breath first traversal until finally it creates the required map, \ie
	$$ 
	\begin{array}{rcl}
		\subMap'' &=& \subMap'\cupplus\set{\submapsto{7}{\subMap'(13)}{\subE{\dvVVVC}{\dvVD},\subE{\dvVVVD}{\dvVE}},
			\submapsto{9}{\subMap'(13)}{\subE{\dvVVVC}{\dvVVD},\subE{\dvVVVD}{\dvVVE}}} \\ 
		\subMap''' &=& \subMap''\cupplus\set{\submapstoE{8}{\subMap''(7)}, \submapstoE{10}{\subMap''(9)}}
	\end{array}	
	$$
	Finally, when we apply the \uni function using the mappings provided in $\subMap'''$, we obtain the following system of equations, \ie \sysUNIC{=}\syseq{\eqqUNIC}{\hVarX_{0}}{\sysrecsetE} where
	$$\eqqUNIC = \begin{xbraces}{c}
		\hVarX_0 = 
			\hgnec{\recv{\dvVVVA}{\dvVVVB}}{\dvVVVB{=}\req\!\land\!\dvVVVA{\neq}h}\hVarX_{3} \hand
			\hgnec{\recv{\dvVVVA}{\dvVVVB}}{\dvVVVB{=}\req \!\land\!\dvVVVA{\neq}j}\hVarX_{11}
		\\[3mm]
		\hVarX_{3}=\hgnec{\send{\dvVB}{\dvVC}}{\dvVB{=}\dvVVVA{\,\land\,}\dvVC{=}\ans}\hVarX_{13}, \\[3mm] \hVarX_{11}=\hgnec{\recv{\dvVVB}{\dvVVC}}{\dvVVB{=}\dvVVVA{\,\land\,}\dvVVC{=}\req}\hVarX_{12}\\[3mm]
		\hVarX_{13}=\hand\begin{array}{c}
			\hgnec{\recv{\dvVVVC}{\dvVVVD}}{\dvVVVC{=}\dvVVVA\land\dvVVVD{=}\req\!\land\!\dvVVVA{\neq}h}\hVarX_7 \\
			\hgnec{\recv{\dvVVVC}{\dvVVVD}}{\dvVVVC{=}\dvVVVA\land\dvVVVD{=}\req\!\land\!\dvVVVA{\neq}j}\hVarX_9
		\end{array},  \\[6mm]
		\hVarX_{7}=\hgnec{\send{\dvVF}{\dvVG}}{\dvVF{=}\dvVVVA{\,\land\,}\dvVG{=}\ans}\hVarX_8, \\[3mm] 
		\hVarX_{9}=\hgnec{\recv{\dvVVF}{\dvVVG}}{\dvVVF{=}\dvVVVA{\,\land\,}\dvVVG{=}\req}\hVarX_{10},  \\[3mm] 
		\hVarX_{8}=\hVarX_{13}, \qquad 
		\hVarX_{10}=\hfls, \qquad
		\hVarX_{12}=\hfls
	\end{xbraces}$$ \\[-15mm]\bqed
\end{example}

\paragraph{Proving Semantic Preservation for \pmb{\genuni{-}}.} To prove that construction \genuni{-} preserves the semantics of the given standardized system of equations, we must prove that the following criterion holds:
$$\sys \text{ is in \emph{Standard Form}} \!\!\imp\!\! \genuni{\sys}\equiv\sys \text{ where } \genuni{\sys} \text{ is \emph{Uniform}}.$$ 
In the proof given below, we make use of the following lemmas:
\begin{lemma}\label{lemma:norm-1a}
	\traverseFun{\eqq}{\set{0}}{\partition}{\emptyset}{=}\subMap \quad \textsl{implies} \\ \subMap is a \emph{well-formed} map for \eqq.
\end{lemma}
\begin{lemma}\label{lemma:norm-1b}
	$\forall(\hVarX_j{=}\hV_j){\,\in\,}\eqq\cdot$ equation $\hVarX_j{=}\hV_j$ is in \emph{Standard form}, and \subMap is a \emph{well-formed} map for \eqq \textsl{implies} $\uniFun{\eqq}{\subMap}{\equiv}\eqq$ and $\forall(\hVarX_k{=}\hVV_k){\,\in\,}\uniFun{\eqq}{\subMap}\cdot$ equation $(\hVarX_k{=}\hVV_k)$ is \emph{Uniform}.
\end{lemma}

\Cref{lemma:norm-1a} dictates that whenever a \subMap map is obtained by conducting a breath first traversal on an equation set \eqq using the \partition projection function, then the resultant map is \emph{well-formed} \wrt \eqq.

\Cref{lemma:norm-1b} builds on the result of the previous lemma by stating that upon obtaining a \emph{well-formed} map for \eqq and when all the equations in \eqq are in Standard Form, a \emph{semantically equivalent}, Uniform equation set can be obtained by applying the \uni function on  \eqq using the well-formed \subMap map to obtain the required uniformity. 

The proofs for both of these lemmas are provided in Appendix \Cref{sec:proof-uni-lemmas}.
\begin{proof}
	Initially we know 
	\begin{gather}
		\sys\text{ is in Standard Form} \label{proof:uni-main-1}
	\end{gather}
	By \pref{proof:uni-main-1} and definition of \sys we also know
	\begin{gather}
		\syseq{\eqq}{\hVarX_0}{\sysrecset}\text{ is in Standard Form} \label{proof:uni-main-2}
	\end{gather}
	because
	\begin{gather}
		\forall(\hVarX_j{=}\hV_j)\in\eqq\cdot \text{equation }\hVarX_j{=}\hV_j \text{ is in Standard Form} \label{proof:uni-main-3}
	\end{gather}
	We create a \subMap map for our equation set \eqq by using a breath first traversal that applies the \partition projection function, such that we know
	\begin{gather}
		\traverseFun{\eqq}{\set{0}}{\partition}{\emptyset} = \subMap \label{proof:uni-main-4}
	\end{gather}
	By \pref{proof:uni-main-4} and \Cref{lemma:norm-1a}, we know
	\begin{gather}
		\subMap\text{ is a \emph{well-formed} map for }\eqq  \label{proof:uni-main-5}
	\end{gather}
	By \pref{proof:uni-main-3}, \pref{proof:uni-main-5} and \Cref{lemma:norm-1b}, we know
	\begin{gather}
		\uniFun{\eqq}{\subMap}{\equiv}\eqq \label{proof:uni-main-6} \\
		\forall(\hVarX_k{=}\hVV_k)\in\uniFun{\eqq}{\subMap}\cdot\text{ equation }(\hVarX_k{=}\hVV_k) \text{ is \emph{Uniform}} \label{proof:uni-main-7}
	\end{gather}
	By \pref{proof:uni-main-6}, \pref{proof:uni-main-7} and the definition of \genuni{\sys} we can conclude 
	\begin{center}
		$\genuni{\sys}\equiv\sys$  where  $\genuni{\sys}$ is \emph{Uniform}.
	\end{center}
\end{proof}

\subsubsection{Condition Reformulation of Conjunct Symbolic Events} \label{sec:truth-combos-cons}
Reformulating the filtering conditions of conjunct symbolic events involves reconstructing a uniform system of equations into a semantically equivalent \emph{equi-disjoint} conjunction as defined in \Cref{def:equi-disjoint} below.

\begin{definition}[System of Equi-Disjoint Equations] \label{def:equi-disjoint}
	An equation is \emph{equi-disjoint} when it is \emph{uniform}, and when multiple necessities defined at the top-level of the same conjunction \ie $\hAND{j\in\IndSet}\hnec{\actS_j}\hVarX_j$, are \emph{unable to be matched by the same concrete system action} \acta, unless they are \emph{syntactically equal}; formally defined as:
	$$ \text{if }\hVarX_{i}=\hAND{j\in\IndSet}\hnec{\actS_j}\hVarX_j \text{ then }\forall k,l\in\IndSet\cdot\syn{\actS_k}{\cap}\syn{\actS_l}\neq\emptyset \implies \actS_k{\,=\,}\actS_l$$
	A system of equations is \emph{equi-disjoint} when all of its equations are \emph{equi-disjoint}. \bqed
\end{definition}

\begin{example} As defined by \Cref{def:equi-disjoint}, we can deduce that equation $\hVarX_0=(\hgnec{\recv{\dvV}{\dvVV}}{\dvVV{>}5}\hVarX_1)\hand(\hgnec{\recv{\dvV}{\dvVV}}{\dvVV{>}5}\hVarX_2)\hand(\hgnec{\recv{\dvV}{\dvVV}}{\dvVV{\leq}5}\hVarX_3)$ is \emph{equi-disjoint} since there does not exist a system action that is able to satisfy both $\hgnec{\recv{\dvV}{\dvVV}}{\dvVV{>}5}$ and $\hgnec{\recv{\dvV}{\dvVV}}{\dvVV{\leq}5}$, \ie $\syn{\actSN{\recv{\dvV}{\dvVV}}{\dvVV{\leq}5}}{\,\cap\,}\syn{\actSN{\recv{\dvV}{\dvVV}}{\dvVV{>}5}}{\,=\,}\emptyset$.

The only two branches that are satisfiable by the same system actions are $\hgnec{\recv{\dvV}{\dvVV}}{\dvVV{>}5}\hVarX_1$ and $\hgnec{\recv{\dvV}{\dvVV}}{\dvVV{>}5}\hVarX_2$ but they are both prefixed by \emph{syntactically equal} necessities \ie $\syn{\actSN{\recv{\dvV}{\dvVV}}{\dvVV{>}5}}{\,\cap\,}\syn{\actSN{\recv{\dvV}{\dvVV}}{\dvVV{>}5}}{\,\neq\,}\emptyset$ since $\actSN{\recv{\dvV}{\dvVV}}{\dvVV{>}5}{\,=\,}\actSN{\recv{\dvV}{\dvVV}}{\dvVV{>}5}$. 

However, equation $\hVarX_{1}{\,=\,}\hgnec{\recv{\dvV_1}{\dvVV_1}}{\ctru}\hVarX_4{\,\hand\,}\hgnec{\recv{\dvV_1}{\dvVV_1}}{\dvVV_1{\,\neq\,}5}\hVarX_5$ is \emph{not} equi-disjoint since $\syn{\actSN{\recv{\dvV_1}{\dvVV_1}}{\ctru}}{\,\cap\,}\syn{\actSN{\recv{\dvV_1}{\dvVV_1}}{\dvVV_1{\,\neq\,}5}}{\,\neq\,}\emptyset$ but $\actSN{\recv{\dvV_1}{\dvVV_1}}{\ctru}{\,\neq\,}\actSN{\recv{\dvV_1}{\dvVV_1}}{\dvVV_1{\,\neq\,}5}$. \bqed
\end{example}

\begin{remark}
	Note that normalization construction \Sref{step:sfsys-to-detsys} actually checks that the branches are \emph{Equi-Disjoint} by using \distinctUni{\detS{\IndSet}}. This is since \detS{\IndSet} returns the \emph{set} of symbolic events defined by the branches in the guarded conjunction that is being analysed, \ie $\bigunion{i\in\IndSet}\Setdef{\actS_j}{\hnec{\actS_j}\hVarX_j \text{ is a subformula in }\hV_i}$. Since sets \emph{do not} contain repeated (syntactically equal) values, the predicate $\distinctUni{\detS{\IndSet}}$ can only return false when two or more events in \detS{\IndSet} are syntactically different yet still non-disjoint.  \bqed
\end{remark}


\renewcommand{\eqqREC}{\ensuremath{\eqq^\textsf{comb}}\xspace}

\begin{figure}[t]
\noindent\textbf{Construction}

$$
\begin{array}{r@{\,}c@{\,}l}
	\gencond{\syseq{\eqq}{\hVarX_0}{\sysrecset}}  &\defeq& \syseq{\traverseFun{\eqq}{\set{0}}{\condcomb}{\emptyset}}{\hVarX}{\sysrecset} 
	\\[3mm]
	\condcombFun{\eqq}{\IndSet}{\recEqq} &\defeq& \Setdef{\!\! \hVarX_i{=}\hAND{\predc_k\in\truthcombs{j}{\IndSet'}\hspace{-10mm}}\hgnec{\pate}{\predc_k}\hVarX_j{\hand}\hV \!\!}
	{
		(\hVarX_i{=}\hAND{j\in\IndSet''\hspace{-3mm}}\hgnec{\pate}{\predc_j}\hVarX_j{\hand}\hV){\in}\eqqres{\eqq}{\IndSet} \\[3mm]
		 \text{  and } \IndSet'{=}\bigunion{l\in\IndSet}\childrenFun{\eqq}{l} \\[3mm] 
		\textsl{  such that } \IndSet''\subseteq\IndSet' \!\!\!
	}\cupplus\recEqq
	\\[10mm]
	\truthcombs{i}{Q} &\defeq& \Setdef{\underline{c_i}\!\land\!c_j\ldots\land\!c_k,\\[2mm] \underline{c_i}\!\land\!\lnot c_j\ldots\land\!c_k, \\[2mm] \underline{c_i}\!\land\!\lnot c_j\ldots\land\!\lnot c_k}{\forall j\ldots n\in\IndSet \text{  where  } i\neq j\neq\ldots\neq n\\[2mm]\quad \text{ such that }\pate_i=\pate_j=\ldots=\pate_k}
\end{array}
$$
\caption{The Conjunction Reformulation Algorithm.}
\label{fig:truth-combos-cons-alg}
\end{figure}

\Cref{fig:truth-combos-cons-alg} presents the construction function, $\gencond{-}::\syseq{\eqqUni}{\Var}{\powerset{\Var}}$ for recomposing a uniform system of equation into an equi-disjoint one. Internally, this construction uses the \traverse function to perform a breath first traversal on the given uniform equation set, \eqqUNI, starting from the equation that equates to the principle variable, \ie with \IndSet{=}\set{0}. While conducting the traversal, this construction applies the \condcomb function in order to reconstruct the uniform conjunctions, \ie $\hAND{j\in\IndSet}\hnec{\actSN{\pate}{\predc}_j}\hVarX_j$ defined in $(\hVarX_i{=}\hV_i)\in\eqqUNI$, into equi-disjoint ones, thereby producing an \emph{equi-disjoint} equation set \eqqREC at the end of the traversal.

The function $\condcomb::(\eqqUNI\times\powerset{\Index}\times\Acc){\mapsto}\Acc$ is a projection function that takes as input a uniform equation set \eqqUNI, a set of indices \IndSet, and an accumulator \recEqq. The accumulator \recEqq contains a partial equi-disjoint set of equations which is first initialized to $\emptyset$ and is constantly extended by repeated \condcomb applications until the traversal is complete, in which case \recEqq is returned as the resultant equi-disjoint equation set.

In order to update \recEqq, the \condcomb function inspects the sibling equations denoted by the indices in \IndSet, \ie $(\hVarX_i{=}\hV_i)\in\eqqres{\eqq}{\IndSet}$, and computes the \emph{truth combinations} of the \emph{filtering conditions} defined by the sibling symbolic necessities (specified in these equations) which define the \emph{same (syntactically equal) patterns}. 

To compute these truth combinations, the \condcomb function starts by computing the child indices of the current sibling equations, denoted by \IndSet, by using the \children function, \ie $\IndSet'{=}\bigunion{l\in\IndSet}\childrenFun{\eqq}{l}$. Following this, it inspects the conjunctions defined in the selected equations, \ie $\hAND{j\in\IndSet''}\hgnec{\pate_j}{\predc_j}\hVarX_j{\,\hand\,}\hV$, and reconstructs them into $\hAND{\predc_k\in\truthcombs{j}{\IndSet'}\hspace{-8mm}}\hgnec{\pate_j}{\predc_k}\hVarX_j{\,\hand\,}\hV$. Notice that $\predc_k$ is a \emph{truth combination} of all the filtering conditions that are defined by symbolic necessities that specify \emph{syntactically equal patterns} and which are defined by the branches identified by the indices in $\IndSet'$, \eg if $\IndSet'{=}\set{1,2,3}$, then one possible truth combination $\predc_k$ is $\predc_1{\land}{\lnot\predc_2}{\land}\predc_3$.

The truth combinations, such as $\predc_k$, are generated through the \emph{combinatorial function} $\truthcombsSym::(\Index\times\powerset{\Index})$. This function takes as input the index of the branch that is being analysed, \ie the one identified by index $j$, along with the indices of all the sibling branches specified in $\IndSet'$. As a result, \truthcombs{j}{\IndSet'} returns the truth combinations in the which filtering condition, $\predc_j$, of the branch that is currently being reconstructed, \ie $\hgnec{\pate_j}{\predc_{j}}\hVarX_j$, is \emph{true}, \ie $\truthcombs{1}{\set{1,2,3}}{=}\Set{(\predc_1{\land}\predc_2{\land}\predc_3),\,(\predc_1{\land}\predc_2{\land}\lnot\predc_3),\,(\predc_1{\land}\lnot\predc_2{\land}\predc_3),\,(\predc_1{\land}\lnot\predc_2{\land}\lnot\predc_3)}$. These truth combinations are then used to reconstruct the existing branch into a collection of equi-disjoint branches. 
\begin{example} \label{ex:equi-disjoint-truthcombs} 
	Consider equation $\hVarX_{0}=\hgnec{\pate}{\predc_1}\hVarX_{1}\hand\hgnec{\pate}{\predc_2}\hVarX_{2}\hand\hgnec{\pate}{\predc_3}\hVarX_{3}$, using the truth combinations provided by \truthcombs{1}{\set{1,2,3}} we can reconstruct branch $\hgnec{\pate}{\predc_1}\hVarX_{1}$ into: $$
	\hgnec{\pate}{\underline{\predc_1}{\land}\predc_2{\land}\predc_3}\hVarX_{1}\hand
	\hgnec{\pate}{\underline{\predc_1}{\land}\predc_2{\land}\lnot\predc_3}\hVarX_{1}\hand
	\hgnec{\pate}{\underline{\predc_1}{\land}\lnot\predc_2{\land}\predc_3}\hVarX_{1}\hand
	\hgnec{\pate}{\underline{\predc_1}{\land}\lnot\predc_2{\land}\lnot\predc_3}\hVarX_{1}.$$
 	Similarly, with \truthcombs{2}{\set{1,2,3}} and \truthcombs{3}{\set{1,2,3}}, we can reconstruct branches $\hgnec{\pate}{\predc_2}\hVarX_{2}$ and $\hgnec{\pate}{\predc_3}\hVarX_{3}$ in the same way such that the resultant equation is:
 	$$
 	\hVarX_{0}{=} \begin{xbrackets}{c}
 		\hgnec{\pate}{{\predc_1}{\land}\predc_2{\land}\predc_3}\hVarX_{1}\hand
 		\hgnec{\pate}{{\predc_1}{\land}\predc_2{\land}\lnot\predc_3}\hVarX_{1}\hand \\[3mm]
 		\hgnec{\pate}{{\predc_1}{\land}\lnot\predc_2{\land}\predc_3}\hVarX_{1}\hand
 		\hgnec{\pate}{{\predc_1}{\land}\lnot\predc_2{\land}\lnot\predc_3}\hVarX_{1}\hand\\[4mm]
 		\hgnec{\pate}{\predc_1{\land}{\predc_2}{\land}\predc_3}\hVarX_{2}\hand 		
 		\hgnec{\pate}{\predc_1{\land}{\predc_2}{\land}\lnot\predc_3}\hVarX_{2}\hand \\[3mm]
 		\hgnec{\pate}{\lnot\predc_1{\land}{\predc_2}{\land}\predc_3}\hVarX_{2}\hand
 		\hgnec{\pate}{\lnot\predc_1{\land}{\predc_2}{\land}\lnot\predc_3}\hVarX_{2}\hand\\[4mm]
 		\hgnec{\pate}{\predc_1{\land}\predc_2{\land}{\predc_3}}\hVarX_{3}\hand 		
 		\hgnec{\pate}{\lnot\predc_1{\land}\predc_2{\land}{\predc_3}}\hVarX_{3}\hand \\[3mm]
 		\hgnec{\pate}{\predc_1{\land}\lnot\predc_2{\land}{\predc_3}}\hVarX_{3}\hand
 		\hgnec{\pate}{\lnot\predc_1{\land}\lnot\predc_2{\land}{\predc_3}}\hVarX_{3}
 	\end{xbrackets} 	
 	$$ \\[-12mm]\bqed
\end{example}
Note that the resultant reconstructed equations (case in point $\hVarX_{0}$ in \Cref{ex:equi-disjoint-truthcombs}) are \emph{equi-disjoint} as the truth combination conditions ensure that a concrete system event \acta can \emph{never} satisfy multiple symbolic necessities in the reconstructed branches, unless these are \emph{syntactically equal}.

Moreover, note that the truth combinations generated by function \truthcombs{j}{\IndSet'} \emph{do not include the cases where $\predc_j$ is false}. This is essential to ensure that none of the reconstructed branches can be satisfied when the original condition $\predc_j$ is true, thereby preserving the semantics of the original branch.
\begin{example} Recall equation $\hVarX_0=\hgnec{\pate}{\predc_1}\hVarX_{1}\hand\hgnec{\pate}{\predc_2}\hVarX_{2}\hand\hgnec{\pate}{\predc_3}\hVarX_{3}$ from \Cref{ex:equi-disjoint-truthcombs}. Note that logical variables $\hVarX_{1}$, $\hVarX_{2}$ and $\hVarX_{3}$ can only be evaluated when their prefixing necessities are satisfied by some concrete system event, meaning that continuation $\hVarX_{1}$ is reachable when $\predc_1$ is true, and \resp $\hVarX_{2}$ and $\hVarX_{3}$ when $\predc_2$ and $\predc_3$ are true. Hence, in the reconstructed equation, the conditions are \emph{never negated} when prefixing \resp logical variable as highlighted below:
	$$
	\hVarX_{0}{=} \begin{xbrackets}{c}
	\hgnec{\pate}{\underline{\predc_1}{\land}\predc_2{\land}\predc_3}\underline{\hVarX_{1}}\hand
	\hgnec{\pate}{\underline{\predc_1}{\land}\predc_2{\land}\lnot\predc_3}\underline{\hVarX_{1}}\hand \\[3mm]
	\hgnec{\pate}{\underline{\predc_1}{\land}\lnot\predc_2{\land}\predc_3}\underline{\hVarX_{1}}\hand
	\hgnec{\pate}{\underline{\predc_1}{\land}\lnot\predc_2{\land}\lnot\predc_3}\underline{\hVarX_{1}}\hand\\[6mm]
	\hgnec{\pate}{\predc_1{\land}\underline{\predc_2}{\land}\predc_3}\underline{\hVarX_{2}}\hand 		
	\hgnec{\pate}{\predc_1{\land}\underline{\predc_2}{\land}\lnot\predc_3}\underline{\hVarX_{2}}\hand \\[3mm]
	\hgnec{\pate}{\lnot\predc_1{\land}\underline{\predc_2}{\land}\predc_3}\underline{\hVarX_{2}}\hand
	\hgnec{\pate}{\lnot\predc_1{\land}\underline{\predc_2}{\land}\lnot\predc_3}\underline{\hVarX_{2}}\hand\\[6mm]
	\hgnec{\pate}{\predc_1{\land}\predc_2{\land}\underline{\predc_3}}\underline{\hVarX_{3}}\hand 		
	\hgnec{\pate}{\lnot\predc_1{\land}\predc_2{\land}\underline{\predc_3}}\underline{\hVarX_{3}}\hand \\[3mm]
	\hgnec{\pate}{\predc_1{\land}\lnot\predc_2{\land}\underline{\predc_3}}\underline{\hVarX_{3}}\hand
	\hgnec{\pate}{\lnot\predc_1{\land}\lnot\predc_2{\land}\underline{\predc_3}}\underline{\hVarX_{3}}
	\end{xbrackets} 	
	$$ \\[-12mm]\bqed
\end{example}
Once the traversal completes, the construction outputs the final accumulator value \recEqq containing the required equi-disjoint equation set.

\newcommand{\predA}{\dvVVVB{=}\req{\,\land\,}\dvVVVA{\neq}h}
\newcommand{\predB}{\dvVVVB{=}\req{\,\land\,}\dvVVVA{\neq}j}
\newcommand{\predC}{\dvVB{=}\dvVVVA{\,\land\,}\dvVC{=}\ans}
\newcommand{\predD}{\dvVVB{=}\dvVVVA{\,\land\,}\dvVVC{=}\req}
\newcommand{\predE}{\dvVVVC{=}\dvVVVA\land\dvVVVD{=}\req{\,\land\,}\dvVVVA{\neq}h}
\newcommand{\predF}{\dvVVVC{=}\dvVVVA\land\dvVVVD{=}\req{\,\land\,}\dvVVVA{\neq}j}
\newcommand{\predG}{\dvVF{=}\dvVVVA{\,\land\,}\dvVG{=}\ans}
\newcommand{\predH}{\dvVVF{=}\dvVVVA{\,\land\,}\dvVVG{=}\req}

\begin{example}[Applying the Construction]\normalfont
	Recall the uniform system of equations \sysUNIC obtained from \Cref{ex-generic} (restated below):
	$$\sysUNIC=\syseq{\eqqUNIC}{\hVarX_{0}}{\sysrecsetE} \qquad\qquad\qquad\text{ where}$$
	$$
	\begin{array}{rcl}
		\eqqUNIC &=& \begin{xbraces}{c}
					\hVarX_0 = \hand \begin{array}{c}
					\hgnec{\recv{\dvVVVA}{\dvVVVB}}{\predA}\hVarX_{3} \\[3mm]
					\hgnec{\recv{\dvVVVA}{\dvVVVB}}{\predB}\hVarX_{11}
					\end{array}
				\end{xbraces}\cupplus\eqqUNICP
		\\[5mm]
		\eqqUNICP&=&\begin{xbraces}{c}
					\hVarX_{3}=\hgnec{\send{\dvVB}{\dvVC}}{\predC}\hVarX_{13}, \\[3mm] \hVarX_{11}=\hgnec{\recv{\dvVVB}{\dvVVC}}{\predD}\hVarX_{12}
				\end{xbraces}\cupplus\eqqUNICPP
		\\[5mm]
		\eqqUNICPP&=&\begin{xbraces}{c}		
					\hVarX_{13}=\hand\begin{array}{c}
					\hgnec{\recv{\dvVVVC}{\dvVVVD}}{\predE}\hVarX_7 \\
					\hgnec{\recv{\dvVVVC}{\dvVVVD}}{\predF}\hVarX_9
					\end{array},  \\[6mm]
					\hVarX_{7}=\hgnec{\send{\dvVF}{\dvVG}}{\predG}\hVarX_8, \\[3mm] 
					\hVarX_{9}=\hgnec{\recv{\dvVVF}{\dvVVG}}{\predH}\hVarX_{10}, 
					\\[3mm] 
					\hVarX_{8}=\hVarX_{13},
					\hVarX_{10}=\hfls,
					\hVarX_{12}=\hfls
				\end{xbraces}
	\end{array}
	$$	
	
	The construction initiates the traversal from the principle equation, \ie $\hVarX_{0}{=}\hV_0$, by invoking \traverseFun{\eqqUni}{\set{0}}{\condcomb}{\emptyset}, which inspects the conjunct necessities defined in the root equation. Since events $\actSN{\recv{\dvVVVA}{\dvVVVB}}{\predA}$ and $\actSN{\recv{\dvVVVA}{\dvVVVB}}{\predB}$ define the same pattern $\recv{\dvVVVA}{\dvVVVB}$, by applying the \condcomb projection function this construction generates all the truth combinations of the filtering conditions defined in the aforementioned necessities of the formula, \ie $(\predA)$ and $(\predB)$, by using: 
	$$
	\begin{array}{rcl}
		\truthcombs{3}{\set{3,11}}&=&
		\begin{xbrackets}{c}
			((\predA)\land(\predB)),\\[2mm]
			((\predA)\land\lnot(\predB))
		\end{xbrackets}
		\\
		\text{ and }\\ 
		\truthcombs{11}{\set{3,11}}&=&
		\begin{xbrackets}{c}
			((\predA)\land(\predB)),\\[2mm]
			(\lnot(\predA)\land(\predB))
		\end{xbrackets} \\[6mm] 
	\end{array}
	$$
		
	Notice that the results of both $\truthcombs{3}{\set{3,11}}$ and $\truthcombs{11}{\set{3,11}}$ do \emph{not} include combinations in which the original condition (\ie $\predc_3=(\predA)$ and $\predc_{11}=(\predB)$ \resp) is negated, \ie truth combinations such as $\lnot(\predA)\land(\predB)$ and $\lnot(\predA)\land\lnot(\predB)$ are not included in the resultant set returned by $\truthcombs{3}{\set{3,11}}$, since they negate parts of the original condition \ie $\predc_3$.
	
	A branch is thus added to the reconstructed formula whenever the guard of the original necessity is \emph{not negated} in the generated truth combination, \eg branch $\hgnec{\recv{\dvVVVA}{\dvVVVB}}{\predA}\hVarX_{3}$ must only be replaced by
	$$
	\begin{array}{rl}
		\hgnec{\recv{\dvVVVA}{\dvVVVB}}{(\predA)\land(\predB)}\hVarX_{3} &\qquad \text{ and } \\
		\hgnec{\recv{\dvVVVA}{\dvVVVB}}{(\predA)\land\lnot(\predB)}\hVarX_{3}
	\end{array}
	$$ 
	but not 
	$$
	\begin{array}{rl}
		\hgnec{\recv{\dvVVVA}{\dvVVVB}}{\lnot(\predA)\land(\predB)}\hVarX_{3} &\qquad \text{ and }\\
		\hgnec{\recv{\dvVVVA}{\dvVVVB}}{\lnot(\predA)\land\lnot(\predB)}\hVarX_{3}
	\end{array}
	$$
	since the original condition is negated in these last two necessities, and hence, function \condcombFun{\eqqUNI}{\set{0}}{\emptyset} evaluates to \recEqq, where \vspace{-5mm}
	
	\newcommand{\sysRECC}{\ensuremath{\sys^\textsf{comb}_3}}
	\newcommand{\eqqRECC}{\ensuremath{\eqqREC_3}}
	\newcommand{\eqqRECCP}{\ensuremath{\eqq^\textsf{comb'}_3}}
	\newcommand{\eqqRECCPP}{\ensuremath{\eqq^\textsf{comb''}_3}}
	$$
	\begin{array}{rcl}
	\recEqq &=& \begin{xbraces}{c}
			\hVarX_0 =  \begin{array}{rc}
							& \hgnec{\recv{\dvVVVA}{\dvVVVB}}{(\predA)\land(\predB)}\hVarX_{3} \\[3mm]
							\hand &\hgnec{\recv{\dvVVVA}{\dvVVVB}}{(\predA)\land\lnot(\predB)}\hVarX_{3} \\[3mm]
							\hand &\hgnec{\recv{\dvVVVA}{\dvVVVB}}{(\predA)\land(\predB)}\hVarX_{11} \\[3mm]
							\hand &\hgnec{\recv{\dvVVVA}{\dvVVVB}}{\lnot(\predA)\land(\predB)}\hVarX_{11}
						\end{array}
		\end{xbraces}\\[12mm]
	\end{array}
	$$
	
	The traversal proceeds by computing the children of $\hVarX_0$ via $\bigunion{l\in\set{0}\hspace{-5mm}}\childrenFun{\eqqUNI}{l}$ which returns $\set{3,11}$ as the set of child indices; it then it recurses \wrt these indices, \ie via \traverseFun{\eqqUNIP}{\set{3,11}}{\condcomb}{\recEqq}. Once again the traversal applies \condcomb which attempts to generate the new truth combinations for \actSN{\send{\dvVB}{\dvVC}}{\predC} and \actSN{\recv{\dvVVB}{\dvVVC}}{\predD}, defined in $\hVarX_3$ and $\hVarX_{11}$ \resp via $\truthcombs{13}{\IndSet''}$ and $\truthcombs{12}{\IndSet''}$ where $\IndSet''=\bigunion{l\in\set{3,11}\hspace{-7mm}}\childrenFun{\eqqUNIP}{l}{=}\set{12,13}$. However, since these two symbolic events \emph{do not} define syntactically equal patterns, \ie $\send{\dvVB}{\dvVC}{\neq}\recv{\dvVVB}{\dvVVC}$, their filtering condition remains intact, \ie $\truthcombs{13}{\IndSet''}=\set{\dvVB{=}\dvVVVA\land\dvVC{=}\ans}$ and $\truthcombs{12}{\IndSet''}=\set{\dvVVB{=}\dvVVVA\land\dvVVC{=}\req}$. Hence, \condcombFun{\eqqUNIP}{\set{3,11}}{\recEqq} evaluates to \recEqq', where \vspace{-5mm}
	
	$$
		\begin{array}{rcl}
		\recEqq' &=& \recEqq\cupplus
			\begin{xbraces}{c}
				\hVarX_{3}=\hgnec{\send{\dvVB}{\dvVC}}{\dvVB{=}\dvVVVA\land\dvVC{=}\ans}\hVarX_{13}, \\[3mm] \hVarX_{11}=\hgnec{\recv{\dvVVB}{\dvVVC}}{\dvVVB{=}\dvVVVA\land\dvVVC{=}\req}\hVarX_{12}
			\end{xbraces}
		\end{array}
	$$
	
	The breath first traversal continues until all the remaining equations have been analysed and reconstructed into:
	\begin{flalign*}
		&\recEqq''\;=\;\recEqq'\,\cupplus\\
		&\begin{xbraces}{c}		
			\hVarX_{13}{=}\begin{xbrackets}{rc}
			&\hgnec{\recv{\dvVVVC}{\dvVVVD}}{(\predE)\land(\predF)}\hVarX_7 \\[3mm]
			\hand &\hgnec{\recv{\dvVVVC}{\dvVVVD}}{(\predE)\land\lnot(\predF)}\hVarX_7 \\[3mm]
			\hand &\hgnec{\recv{\dvVVVC}{\dvVVVD}}{(\predE)\land(\predF)}\hVarX_9 \\[3mm]
			\hand &\hgnec{\recv{\dvVVVC}{\dvVVVD}}{\lnot(\predE)\land(\predF)}\hVarX_9
			\end{xbrackets},  \\[14mm]
			\hVarX_{7}=\hgnec{\send{\dvVF}{\dvVG}}{\dvVF{=}\dvVVVA\land\dvVG{=}\ans}\hVarX_8, \\[3mm] 
			\hVarX_{9}=\hgnec{\recv{\dvVVF}{\dvVVG}}{\dvVVF{=}\dvVVVA\land\dvVVG{=}\req}\hVarX_{10},  \\[3mm] 
			\hVarX_{8}=\hVarX_{13},\quad
			\hVarX_{10}=\hfls,\quad
			\hVarX_{12}=\hfls
		\end{xbraces}
	\end{flalign*}
			
	\noindent Hence, the resultant system of equations is $ \sysREC=\syseq{\eqqRECC}{\hVarX_{0}}{\sysrecsetE} $ where $\eqqRECC=\recEqq$.
	
	Since all the necessities guarding conjunctions are now \emph{equi-disjoint}, \ie they can only be satisfied by the same action iff they are syntactically equal, the resultant equation can now be converted into \emph{normal form} by continuing from step \Sref{step:sfsys-to-detsys} of the normalization algorithm presented \Cref{sec:normalization}. Hence, by applying \Sref{step:sfsys-to-detsys} on \sysREC, we obtain $\sysNFC{=}\syseq{\eqqNFC}{\hVarX_{\set{0}}}{\sysrecsetE}$ where \eqqNFC\, is defined as:
		
		$$\begin{xbraces}{c}
			\hVarX_{\set{0}} =  \begin{xbrackets}{rc}
							& \hgnec{\recv{\dvVVVA}{\dvVVVB}}{(\predA)\land(\predB)}\hVarX_{\set{3,11}} \\[3mm]
							\hand &\hgnec{\recv{\dvVVVA}{\dvVVVB}}{(\predA)\land\lnot(\predB)}\hVarX_{\set{3}} \\[3mm]
							\hand &\hgnec{\recv{\dvVVVA}{\dvVVVB}}{\lnot(\predA)\land(\predB)}\hVarX_{\set{11}}
						\end{xbrackets}\\[12mm]
			\hVarX_{\set{3,11}}=(\hgnec{\send{\dvVB}{\dvVC}}{\predC}\hVarX_{\set{13}} \hand
								  \hgnec{\recv{\dvVVB}{\dvVVC}}{\predD}\hVarX_{\set{12}}
								), \\[6mm]
			\hVarX_{\set{3}}=\hgnec{\send{\dvVB}{\dvVC}}{\predC}\hVarX_{\set{13}}, \qquad \hVarX_{\set{11}}=\hgnec{\recv{\dvVVB}{\dvVVC}}{\predD}\hVarX_{\set{12}}, \\[5mm]
			\hVarX_{\set{13}}=\begin{xbrackets}{rc}
							&\hgnec{\recv{\dvVVVC}{\dvVVVD}}{(\predE)\land(\predF)}\hVarX_{\set{7,9}} \\[3mm]
							\hand &\hgnec{\recv{\dvVVVC}{\dvVVVD}}{(\predE)\land\lnot(\predF)}\hVarX_{\set{7}} \\[3mm]
							\hand &\hgnec{\recv{\dvVVVC}{\dvVVVD}}{\lnot(\predE)\land(\predF)}\hVarX_{\set{9}}
						\end{xbrackets},  \\[12mm]
			\hVarX_{\set{7,9}}=	(\hgnec{\send{\dvVF}{\dvVG}}{\predG}\hVarX_{\set{8}}
								\hand\hgnec{\recv{\dvVVF}{\dvVVG}}{\predH}\hVarX_{\set{10}}
								), \\[8mm] 			
			\hVarX_{\set{7}}=\hgnec{\send{\dvVF}{\dvVG}}{\predG}\hVarX_{\set{8}}, \qquad 
			\hVarX_{\set{9}}=\hgnec{\recv{\dvVVF}{\dvVVG}}{\predH}\hVarX_{\set{10}},  \\[5mm] 
			\hVarX_{\set{8}}=\hVarX_{\set{13}}, \qquad
			\hVarX_{\set{10}}=\hfls, \qquad
			\hVarX_{\set{12}}=\hfls
		\end{xbraces}
		$$
	such that by \Sref{step:detsys-to-shml} we finally obtain the required normalized formula $\hVV_3\in\shmlwf$.
	$$
		\pushQED{\bqed}
		\begin{array}{rcl}
				\hVV_3&=&
				\begin{xbrackets}{cl}
				& \begin{xbrackets}{l}
					\hgnec{\recv{\dvVVVA}{\dvVVVB}}{(\predA)\land(\predB)}\\
					\hgnec{\send{\dvVB}{\dvVC}}{\dvVB{=}\dvVVVA\land\dvVC{=}\ans}\\	
					(\hgnec{\recv{\dvVVB}{\dvVVC}}{\dvVVB{=}\dvVVVA\land\dvVVC{=}\req}\hfls \hand \hVV_3')
				\end{xbrackets}	
				\\[3mm]
				\hand 
				\\[3mm]&
				\begin{xbrackets}{l}
					\hgnec{\recv{\dvVVVA}{\dvVVVB}}{(\predA)\land\lnot(\predB)}\\
					\hgnec{\send{\dvVB}{\dvVC}}{\dvVB{=}\dvVVVA\land\dvVC{=}\ans}\\
					\hgnec{\recv{\dvVVB}{\dvVVC}}{\dvVVB{=}\dvVVVA\land\dvVVC{=}\req}\hfls \hand \hVV_3'
				\end{xbrackets}	 
				\\[3mm]
				\hand 
				\\[3mm]&
					\begin{xbrackets}{l}
						\hgnec{\recv{\dvVVVA}{\dvVVVB}}{\lnot(\predA)\land(\predB)}\\
						\hgnec{\recv{\dvVVB}{\dvVVC}}{\dvVVB{=}\dvVVVA\land\dvVVC{=}\req}\hfls
					\end{xbrackets}
				\end{xbrackets}			
			\\[25mm]
			\hVV_3'&=& \hmax{\hVarX_{13}}{}\\ &&
			\begin{xbrackets}{l}
			\quad\hgnec{\recv{\dvVVVC}{\dvVVVD}}{(\predE)\land(\predF)}\\
			\quad
			\Big(			
			\hgnec{\send{\dvVF}{\dvVG}}{\dvVF{=}\dvVVVA\land\dvVG{=}\ans}\hVarX_{13} \hand \hgnec{\recv{\dvVVF}{\dvVVG}}{\dvVVF{=}\dvVVVA\land\dvVVG{=}\req}\hfls  
			\Big)
			\\[3mm]
			\hand 
			\\[3mm]
			\quad\hgnec{\recv{\dvVVVC}{\dvVVVD}}{(\predE)\land\lnot(\predF)}\\[2mm]
			\quad\hgnec{\send{\dvVF}{\dvVG}}{\dvVF{=}\dvVVVA\land\dvVG{=}\ans}\hVarX_{13} 
			\\[3mm]
			\hand 
			\\[3mm] \quad\hgnec{\recv{\dvVVVC}{\dvVVVD}}{\lnot(\predE)\land(\predF)}\\[2mm]
			\quad\hgnec{\recv{\dvVVF}{\dvVVG}}{\dvVVF{=}\dvVVVA\land\dvVVG{=}\req}\hfls
			\end{xbrackets}
			\end{array}
	$$ \\[-6mm]\bqed
\end{example} 
}
\setcounter{secnumdepth}{3}

\paragraph{Proving Semantic Preservation for \pmb{\gencond{-}}.} To prove that construction \gencond{-} preserves the semantics of the given uniform system of equations, we must prove that the following criterion holds:
$$\sys \text{ is \emph{Uniform}} \!\!\imp\!\! \gencond{\sys}\equiv\sys \text{ where } \gencond{\sys} \text{ is \emph{Equi-Disjoint}}.$$ 
In the proof given below, we make use of the following lemma:
\begin{lemma}\label{lemma:norm-2a}
	\Cref{lemma:norm-2a}. $\forall(\hVarX_j{=}\hV_j)\in\eqq\cdot$ equation $\hVarX_j{=}\hV_j$ is \emph{Uniform} \quad \textsl{implies} \\ $\eqq{\,\equiv\,}\traverseFun{\eqq}{\set{0}}{\condcomb}{\emptyset}$ \qquad \textsl{and} \\ $\forall(\hVarX_k{=}\hVV_k)\in\traverseFun{\eqq}{\set{0}}{\condcomb}{\emptyset}\cdot$ equation $(\hVarX_k{=}\hVV_k)$ is \emph{Equi-Disjoint}.
\end{lemma}
This lemma states that whenever the equations in an equation set \eqq are \emph{all} Uniform, then a \emph{semantically equivalent}, \emph{equi-disjoint} equation set can be obtained by performing a breath first traversal upon \eqq using the \condcomb projection function. The proof for this lemma is provided in Appendix \Cref{sec:proof-comb-lemmas}.
\begin{proof}
	Initially we know 
	\begin{gather}
		\sys\text{ is Uniform} \label{proof:disjoint-main-1}
	\end{gather}
	By \pref{proof:disjoint-main-1} and definition of \sys we also know
	\begin{gather}
		\syseq{\eqq}{\hVarX_0}{\sysrecset}\text{ is Uniform} \label{proof:disjoint-main-2}
	\end{gather}
	because
	\begin{gather}
	\forall(\hVarX_j{=}\hV_j)\in\eqq\cdot \text{equation }\hVarX_j{=}\hV_j \text{ is Uniform} \label{proof:disjoint-main-3}
	\end{gather}
	By applying \gencond{-} on $\syseq{\eqq}{\hVarX_0}{\sysrecset}$ we know
	\begin{gather}
		\gencond{\syseq{\eqq}{\hVarX_0}{\sysrecset}} = \syseq{\traverseFun{\eqq}{\set{0}}{\condcomb}{\emptyset}}{\hVarX_{0}}{\sysrecset} \label{proof:disjoint-main-4}
	\end{gather}
	By \pref{proof:disjoint-main-3} and \Cref{lemma:norm-2a}, we know
	\begin{gather}
		\eqq{\,\equiv\,}\traverseFun{\eqq}{\set{0}}{\condcomb}{\emptyset} \label{proof:disjoint-main-5} \\
		\forall(\hVarX_k{=}\hVV_k){\,\in\,}\traverseFun{\eqq}{\set{0}}{\condcomb}{\emptyset}\cdot \textsf{ equation }(\hVarX_k{=}\hVV_k)\text{ is \emph{Equi-Disjoint}}  \label{proof:disjoint-main-6}
	\end{gather}
	Hence, by \pref{proof:disjoint-main-4}, \pref{proof:disjoint-main-5} and \pref{proof:disjoint-main-6} we can conclude
	\begin{center}
		$\gencond{\sys}\equiv\sys$ where $\gencond{\sys}$ is \emph{Equi-Disjoint}.
	\end{center}
\end{proof}

\section{Synthesising Deterministic Enforcers} \label{sec:enf-syn}
In the previous section we have presented an algorithm which demonstrates that for every formula $\hV\in\shml$ we can find a semantically equivalent normalized formula $\hVV\in\shmlwf$. By working on normalized formulae we are able to provide a more intuitive synthesis function which maps formulae to enforcement monitors in a similar way as was done in \cite{Francalanza2015Mon} for detection monitors (see \Cref{fig:mon-synthesis}). 

\begin{figure}[h]
	\noindent\textbf{Optimization Function}
	\begin{align*}
		\opt{\hV} &\defEquals 
			\begin{xbrace}{l@{\quad}l}
				\hV & \text{if }\hV{\in}\set{\hfls,\htru} \\[3mm]
				\opt{\hV'} & \text{if }\hV{=}\hmax{\hVarX_{0}}{\hV'}\text{ and }\hVarX_{0}{\notin}\fv{\hV'} \\[3mm]
				\hmax{\hVarX_{0}}{\opt{\hV'}} & \text{if }\hV{=}\hmax{\hVarX_{0}}{\hV'}\text{ and }\hVarX_{0}{\in}\fv{\hV'} \\[3mm]
				\hAND{i\in\IndSet}\hnec{\actS_i}\opt{\hV_{i}} & \text{if }\hV{=}\hAND{i\in\IndSet}\hnec{\actS_i}\hV_{i}
			\end{xbrace}
	\end{align*}
	\medskip
	
	\noindent\textbf{Synthesis Function}
	$$ \g{\hV}{} \defeq \g{\opt{\hV}}{\bot} \qquad\text{ where } \qquad\qquad\qquad\quad$$ 
	\begin{align*}
	\g{\!\hVarX}{\rho} &\defEquals \mx \\		
	\g{\hfls}{\rho} &\defEquals \my \qquad \text{when }\rho\!=\!\my \\
	\g{\htru}{\rho} &\defEquals \miden \\
	\g{\hmax{\hVarX}{\varphi}}{\rho} &\defEquals \mrec{\mx}{\g{\varphi}{\rho}}\\		
	\g{\hAnd\hnec{\actSN{\pate}{\predc}_i}{\varphi_{i}}}{\rhoRB} &\defEquals 
	\mrec{\my}{\mCh\begin{xbrace}{ll}
		\mact{\actSTN{\pate}{\predc}{\actt}_i}{\g{\varphi_i}{\my}} & \qquad \text{if } \hV_i{=}\hfls\\ 
		\mact{\actSTN{\pate}{\predc}{\pate}_i}{\g{\varphi_i}{\my}} & \qquad \text{otherwise}
		\end{xbrace}}
	\end{align*}
	\caption{Synthesis of Enforcement Monitors from \shmlwf formulae.}
	\label{fig:enf-synthesis}
\end{figure}

The synthesis function, $\g{\hV}{}::\shmlwf{\,\mapsto\,}\Enf$, presented in \Cref{fig:enf-synthesis} compositionally analyses a given formula $\hV{\,\in\,}\shmlwf$ in order to produce the required enforcement monitor. Before initiating the synthesis, function $\g{\hV}{}$ optimizes the given normalized formula \hV via the function $\optSym{\,::\,}\shmlwf{\,\mapsto\,}\shmlwf$. This function compositionally inspects a given formula \hV and removes maximal fixpoint declarations whenever their variable is never referenced in \hV. This ensures the removal of any redundant fixpoint declarations that might have been added during normalization (see \Cref{sec:detenf-norm}).

Once the formula is optimized, function $\g{\hV}{}$ initiates the synthesis, for obtaining the required enforcer, by invoking function $\g{-}{}::(\shmlwf\times\textsc{State}){\,\mapsto\,}\Enf$. This recursive function produces the required enforcer by compositionally analysing the given formula $\hV{\in}\shmlwf$. During analysis, it also maintains a state $\rho$ which can either be $\bot$, or may contain some recursive variable, $\my$; initially, $\rho$ is set to $\bot$, but is then modified accordingly during the synthesis derivation. 

The synthesis converts truth formulae, $\htru{\,\in\,}\shmlwf$, into the identity enforcer \miden, and logical variables, $\hVarX{\,\in\,}\shmlwf$, into the corresponding recursion enforcement variable, whereas maximal fixpoints, $\hmax{\hVarX}{\hV}{\,\in\,}\shmlwf$, are converted into a recursive enforcer; these three conversions are performed irrespective of the contents of $\rho$.  

Normalized conjunctions, $\hAnd\hgnec{\pate_i}{c_i}{\varphi_{i}}$, are synthesised into a \emph{recursive summation} of enforcers, \ie $\mrec{\my}{\mCh\prf{\eV}{\g{\hV_i}{\my}}}$, where $\eV$ can be either an \emph{identity transformation}, $\actSTN{\pate}{\predc}{\pate}_i$, or a \emph{suppression transformation}, $\actSTN{\pate}{\predc}{\actt}_i$ whenever the continuation $\hV_i$ is a falsehood, \ie $\hV_i{=}\hfls$. The latter ensures that the synthesised enforcer is capable of \emph{suppressing} any system action \acta that satisfies necessities that lead to a falsehood (\eg $\mtchS{\actSN{\pate}{\predc}}{\acta}{\,=\,}\s$ such that \acta satisfies necessity $\hgnec{\pate}{\predc}\hfls$), thereby preventing the system from violating the property. 

Also, notice how the fresh recursive variable, $\my$, introduced by the synthesised recursive construct, \mrec{\my}{}, is passed on to the subsequent applications of the synthesis function, \ie $\g{\hV_i}{\my}$. This is required to enable converting falsehood, \hfls, into the newly introduced recursive variable. In this way, an action is only suppressed whenever it satisfies a necessity that is followed by a \emph{falsehood}, \eg \hgnec{\recv{i}{x}}{x\!=\!5}{\hfls} becomes \mrec{\my}{\mact{\actSTD{\recv{i}{x}}{x\!=\!5}}{\my}}. This ensures that any action leading to a violation is \emph{continuously suppressed} until a non-matching action occurs, thereby preventing the violation from occurring.

\begin{example}\normalfont
	Recall formula $\hV_1{\,\in\,}\shmlwf$ from \Cref{ex:uhml-formula}: 
		$$ \hV_1=\hmax{\hVarX}{\hnec{\actReq}(\hnec{\actAns}\hVarX\,\hand\,\hnec{\actReq}\hfls)} $$ 
	Using the synthesis function defined in \Cref{fig:enf-synthesis} we can automatically generate enforcer $\eV_2$ (defined in \Cref{ex:enf-instrum}) as shown by the derivation below.
	\begin{align*}
		& \quad \g{\hmax{\hVarX}{\hnec{\actReq}(\hnec{\actAns}\hVarX\,\hand\,\hnec{\actReq}\hfls)}}{\bot} \\ 
		= & \quad \mrec{\mx}{\g{\hnec{\actReq}(\hnec{\actAns}\hVarX\,\hand\,\hnec{\actReq}\hfls)}{\bot} } \\
		= & \quad \mrec{\mx}{\mrec{\mz}{\mact{\actReq}{\g{(\hnec{\actAns}\hVarX\,\hand\,\hnec{\actReq}\hfls)}{\mz} }}} \\[-3mm]
		\intertext{Since $\hnec{\actReq}$ is followed by \hfls, we need to suppress every \actReq action.} \vspace{-3mm}
		= & \quad \mrec{\mx}{\mrec{\mz}{\mact{\actSID{\actReq}{\ctru}}{\mrec{\my}{(\mch{\mact{\actSID{\actAns}{\ctru}}{\g{\hVarX}{\my}}\,}{\,\mact{\actSTD{\actReq}{\ctru}}{\g{\hfls}{\my}}})}}}} \\
		= & \quad
		\mrec{\mx}{\mrec{\mz}{\mact{\actSID{\actReq}{\ctru}}{\mrec{\my}{(\mch{\mact{\actSID{\actAns}{\ctru}}{\hVarX}\,}{\,\mact{\actSTD{\actReq}{\ctru}}{\my}})}}}}
	\end{align*}
	\noindent In practice, the resultant monitor can be further optimized by removing any redundant recursive constructs such as \mrec{\mz}{}, thereby obtaining:
	$$ 
		\mrec{\mx}{\mact{\actSID{\actReq}{\ctru}}{\mrec{\my}{(\mch{\mact{\actSID{\actAns}{\ctru}}{\hVarX}\,}{\,\mact{\actSTD{\actReq}{\ctru}}{\my}})}}} 
	$$ \\[-17mm]\bqed
\end{example}

\subsection{Providing Formal Guarantees}
Now that we have established a synthesis function for normalised formulas, we proceed to prove that our synthesis actually generates \emph{deterministic monitors} (as defined by \Cref{def:det-enf}) that \emph{strongly enforce} \shmlwf formulas (as defined by \Cref{def:strong-enf}). 

\subsubsection{Ensuring Deterministic Behaviour of the Synthesized Enforcers}
We ensure the deterministic behaviour (as defined by \Cref{def:det-enf}) of any enforcer that can be synthesised by the synthesis algorithm provided in \Cref{fig:enf-synthesis}, by proving \Cref{thm:det-enf}.
\begin{theorem}[Deterministic Behaviour]\label{thm:det-enf}
	For every enforcer, \g{\hV}{}, that can be synthesised from a normalized formula $\hV{\in}\shmlwf$, 
	$$ \g{\hV}{}\wtraS{\ioact{\tr}{\tr'}}\eV' \text{  and  } \g{\hV}{}\wtraS{\ioact{\tr}{\tr''}}\eV'' \imp \eV'{=}\eV'' \text{  and  }\tr'{=}\tr'' $$
\end{theorem}
In order to prove \Cref{thm:det-enf}, we make use of \Cref{lemma:wf-enf} which states that our synthesis function, $\g{\hV}{}{=}\eV$, always produces \emph{well-formed enforcers}, \ie $\eV{\in}\Enfwf$.
\begin{lemma}[Enforcer Well-formedness]\label{lemma:wf-enf}
	$$\forall\hV{\in}\shmlwf\cdot\g{\hV}{}{=}\eV \imp \eV{\in}\Enfwf \qquad\quad \text{where}$$
	$$\eV{\in}\Enfwf \bnfdef \miden\bnfsepp\mx\;\bnfsepp\;\mrec{\mx}{\eV}\;\bnfsepp\;\mCh\mact{\actSTN{\pate_i}{\predc_i}{\pate'_i}}{\eV'_i} \quad \text{where }\bigdistinct{i{\in}\IndSet}\actSN{\pate_i}{\predc_i}$$
\end{lemma}
Enforcer well-formedness is a syntactic restriction that requires that every branch in a summation is prefixed by a \emph{disjoint} symbolic transducer\footnote{One can notice the similarities between normalized formulae $\hV{\in}\shmlwf$, which requires all conjunctions to be prefixed by disjoint necessities, and well-formed enforcers $\eV{\in}\Enfwf$.}.

Proving \Cref{thm:det-enf} also requires proving \emph{Single Step Determinism} (\Cref{lemma:enf-one-det}), which states that whenever a \emph{well-formed} enforcer \eV is able to perform two different reductions  for the \emph{same input event} \acta, \ie $\g{\hV}{}\traS{\ioact{\tr}{\tr'}}\eV'$ and $\g{\hV}{}\traS{\ioact{\tr}{\tr''}}\eV''$, then both reductions should process and modify \acta in the same way such that the output actions, $\actu'$ and $\actu''$, and the resultant continuation enforcers, $\eV'$ and $\eV''$, are both \emph{well-formed} and \emph{syntactically equal}.
\begin{lemma}[Single Step Determinism]\label{lemma:enf-one-det}
	For every \emph{well-formed} enforcer, $\eV{\in}\Enfwf$, 
	$$ \eV\traS{\ioact{\acta}{\actu'}}\eV' \text{  and  } \eV\traS{\ioact{\acta}{\actu''}}\eV'' \imp \eV'{=}\eV'' \text{  and  }\actu'{=}\actu'' $$
\end{lemma}
The proofs for \Cref{lemma:wf-enf} and \Cref{lemma:enf-one-det} are provided in Appendix \Cref{sec:app:deterministic-mon}.

\paragraph*{To Prove \Cref{thm:det-enf}}
\begin{align*}
	&\forall\hV\in\shmlwf\cdot\g{\hV}{}\wtraS{\ioact{\tr}{\tr'}}\eV' \text{  and  } \g{\hV}{}\wtraS{\ioact{\tr}{\tr''}}\eV'' \imp \eV'{=}\eV'' \text{  and  }\tr'{=}\tr''
	\intertext{By \Cref{lemma:wf-enf} we can instead prove}
	\implies &\forall\eV\in\Enfwf\cdot\eV\wtraS{\ioact{\tr}{\tr'}}\eV' \text{  and  } \eV\wtraS{\ioact{\tr}{\tr''}}\eV'' \imp \eV'{=}\eV'' \text{  and  }\tr'{=}\tr'' 
\end{align*}

\begin{proof}[by induction on the structure of the input trace \tr]
	\begin{case}[\tr{=}\varepsilon] We know 
		\begin{gather}
			\eV\in\Enfwf  \label{proof:det-enf-bc-1} \\
			\eV\wtraS{\wioact{\varepsilon}{\tr'}}\eV'  \label{proof:det-enf-bc-2} \\
			\eV\wtraS{\wioact{\varepsilon}{\tr''}}\eV''  \label{proof:det-enf-bc-3}
		\end{gather}
		Since $\tr{=}\varepsilon$, we know that \emph{no transitions} can be made by enforcer \eV in \eqref{proof:det-enf-bc-2} and \eqref{proof:det-enf-bc-3} since the input trace is \emph{empty}, and so no output trace can be produced in both cases. Hence, we conclude
		\begin{gather}
			\eV'=\eV=\eV''\label{proof:det-enf-bc-4} \\
			\tr'=\varepsilon=\tr''\label{proof:det-enf-bc-5} 
		\end{gather}
	\end{case}
	
	\begin{case}[\tr{=}\acta;\trr] We know 
		\begin{gather}
			\eV\in\Enfwf  \label{proof:det-enf-ic-1} \\
			\eV\wtraS{\wioact{\acta;\trr}{\tr'}}\eV'  \label{proof:det-enf-ic-2} \\
			\eV\wtraS{\wioact{\acta;\trr}{\tr''}}\eV''  \label{proof:det-enf-ic-3}
		\end{gather}
		By \eqref{proof:det-enf-ic-2} and the definition of $\wtraS{\wioact{\acta;\trr}{\tr'}}$, we know
		\begin{gather}
			\eV\traS{\ioact{\acta}{\actu'}}\eV'''  \label{proof:det-enf-ic-4} \\
			\eV'''\wtraS{\wioact{\trr}{\trr'}}\eV'  \label{proof:det-enf-ic-5} \\
			\tr'=\actu';\trr' \label{proof:det-enf-ic-6}
		\end{gather}
		Similarly, by \eqref{proof:det-enf-ic-3} and the definition of $\wtraS{\wioact{\acta;\trr}{\tr'}}$, we know
		\begin{gather}
			\eV\traS{\ioact{\acta}{\actu''}}\eV''''  \label{proof:det-enf-ic-7} \\
			\eV''''\wtraS{\wioact{\trr}{\trr''}}\eV'  \label{proof:det-enf-ic-8} \\
			\tr''=\actu'';\trr'' \label{proof:det-enf-ic-9}
		\end{gather}
		By \eqref{proof:det-enf-ic-1}, \eqref{proof:det-enf-ic-4}, \eqref{proof:det-enf-ic-7} and \emph{Single Step Determinism} (\Cref{lemma:enf-one-det}), we know
		\begin{gather}
			\actu'=\actu'' \label{proof:det-enf-ic-10} \\
			\eV''''=\eV''' \label{proof:det-enf-ic-11} \\
			\eV''',\eV''''\in\Enfwf \label{proof:det-enf-ic-12}
		\end{gather}
		By \eqref{proof:det-enf-ic-5}, \eqref{proof:det-enf-ic-8}, \eqref{proof:det-enf-ic-11}, \eqref{proof:det-enf-ic-12} and IH we know 
		\begin{gather}
			\eV'=\eV'' \label{proof:det-enf-ic-13} \\
			\tr'=\tr'' \label{proof:det-enf-ic-14} 
		\end{gather}
		Hence, from \eqref{proof:det-enf-ic-6}, \eqref{proof:det-enf-ic-9}, \eqref{proof:det-enf-ic-10} and \eqref{proof:det-enf-ic-14} we conclude
		\begin{gather}
			\tr'=(\actu';\trr')=(\actu'';\trr'')=\tr'' \label{proof:det-enf-ic-15} 
		\end{gather}
		$\therefore$ Case holds by \eqref{proof:det-enf-ic-13} and \eqref{proof:det-enf-ic-15}.
	\end{case}

\end{proof}

\subsubsection{Proving Strong Enforceability by the Synthesized Enforcers}
In order to prove that our synthesised deterministic enforcers are capable of \emph{strongly enforcing} the formula they were derived from, we prove \Cref{thm:strong-enf}.

\begin{theorem}[Strong Enforcement] \label{thm:strong-enf} For every \emph{satisfiable} formula, $\hV{\in}\Sat$, that is also in \emph{normal form}, \ie $\hV{\in}\shmlwf$,
	$$ \g{\hV}{}\;\textsl{ strongly enforces }\; \hV $$
\end{theorem}
\noindent In order to prove \Cref{thm:strong-enf}, by the definition of \emph{Strong Enforcement} (\Cref{def:strong-enf}) we must prove the following lemmas:
\begin{lemma}[Enforcement Soundness] \label{lemma:soundness}
	$$\forall\pV{\,\in\,}\Proc\cdot \i{\g{\hV}{}}{\pV}{\,\in\,}\syn{\hV} $$
\end{lemma}
\begin{lemma}[Enforcement Transparency] \label{lemma:transparency}
	$$\forall\pV{\,\in\,}\Proc\cdot \pV{\,\in\,}\syn{\hV} \; \imp \; \i{\g{\hV}{}}{\pV}\bisim\pV $$	
\end{lemma}
\noindent Proving these two lemmas requires making use of \Cref{lemma:opt-equiv}, thus allowing us to work up to \emph{optimized formulae}, \ie $\hVV{\,\in\,}\shmlwfopt$.
\begin{lemma} \label{lemma:opt-equiv} For every normalized formula $\hV{\in}\shmlwf$
	\begin{gather*}
		\opt{\hV}=\hVV \; \imp \; \hVV{\equiv}\hV \, \text{ and }\, \hVV{\in}\shmlwfopt \qquad\qquad \text{where} \\
		\hV,\hVV\in\shmlwfopt \bnfdef \hVarX\;\bnfsepp\;\hfls\;\bnfsepp\;\htru\;\bnfsepp\hAND{i\in\IndSet}\hnec{\actS_i}\hV_i \text{ where }\bigdistinct{i\in\IndSet}\actS_i \;\bnfsepp \hmax{\hVarX}{\hV} \text{ where } \hVarX{\in}\fv{\hV}
	\end{gather*}
\end{lemma}
\noindent \Cref{lemma:opt-equiv} states that \emph{semantics are preserved}  when a normalized formula $\hV{\in}\shmlwf$ is optimized into $\hVV{\in}\shmlwfopt$, where \hVV is said to be optimized when every fixpoint variable \hVarX, that is bound to a maximal fixpoint \hmax{\hVarX}{\hV}, is used \emph{at least once} in the continuation formula \hV. We prove this lemma by \emph{structural induction} on \hV in Appendix \Cref{sec:proof-opt-equiv}.

Moreover, in order to facilitate our proofs we use an alternative satisfaction semantics for \shml as explained below.

\paragraph{Alternative \shml Semantics} An alternative semantics for \shml was presented by Aceto \etal in \cite{Aceto1999TestingHML,Aceto2007Book} in terms of a \emph{satisfaction relation}, \vSatS. When restricted to \shml, \vSatS is the \emph{largest relation} satisfying the implications defined in \Cref{fig:uhml-sat}. 
\begin{figure}[h]
	\begin{displaymath}
	\begin{array}{r@{\;\,}c@{\;\,}ll}
	\pV&\vSatS&\htru & \textsl{  always} \\[.5mm]
	\pV&\vSatS&\hfls & \textsl{  never} \\[.5mm]
	\pV&\vSatS&\bigwedge_{i\in\IndSet}\!\hV_i & \textsl  {  whenever } \pV\vSatS\hV_i \textsl{ for all } i{\,\in\,}\IndSet \\[.5mm]
	\pV&\vSatS&\hnec{\actS}\hV & \textsl{  whenever } (\forall\pVV\cdot\pV\traS{\acta}\pVV \textsl{ and } \mtchS{\actS}{\acta}=\s)\,\imp\, \pVV\vSatS\hV \\[.5mm]
	\pV&\vSatS&\hmax{\hVarX}{\hV} & \textsl{  whenever } \pV\vSatS\hV\Sub{\hmax{\hVarX}{\hV}}{\hVarX} \\[.5mm]
	\end{array}
	\end{displaymath}
	\caption{A Satisfaction Relation for \shml formulae}
	\label{fig:uhml-sat}
\end{figure}

The satisfaction relation states that truth, \htru, is \emph{always satisfied}, while falsehood, \hfls, can \emph{never be satisfied}. Conjunctions, $\hAnd\hV_i$ are satisfied when \emph{all branches} are satisfied (\ie $\forall i{\,\in\,}\IndSet \text{ such that } \pV\vSatS\hV_i$), while 
necessities, $\hnec{\actS}\hV$, are satisfied by a process \pV when \emph{all derivatives} \pVV that are reachable over an action \acta where $\mtchS{\actS}{\acta}\!=\!\s$ (possibly none), also satisfy $\hV\s$, \ie $\pVV{\,\vSatS\,}\hV\s$. 
Finally, a process \pV satisfies a maximal fixpoint \hmax{\hVarX}{\hV} when it is also able to satisfy an \emph{unfolded version} of \hV, \ie $\pV\vSatS\hV\sub{\hmax{\hVarX}{\hV}}{\hVarX}$. 

The authors proved that these satisfaction semantics, $\pV\vSatS\hV$, correspond to the denotational semantics of the \shml subset of \uhml, \syn{\hV}, presented in \Cref{fig:recHML}, such that $\pV\vSatS\hV$ can be used in lieu of $\pV\in\syn{\hV}$ (see \cite{Aceto1999TestingHML,Aceto2007Book} for more detail).

\paragraph{Proving Enforcement Soundness} In order to prove that our synthesised enforcers are \emph{sound} we must show that \Cref{lemma:soundness} holds, \ie that 
	$$\forall\pV{\,\in\,}\Proc,\hV\in\shmlwf\text{ when }\hV{\in}\Sat \cdot \i{\g{\hV}{}}{\pV}{\,\in\,}\syn{\hV} $$
Since $\pV{\,\in\,}\syn{\hV} \equiv \pV{\,\vSatS\,}\hV$, and by the definition of \g{-}{}, we get
	$$ \forall\pV{\,\in\,}\Proc,\hV\in\shmlwf\text{ when }\hV{\in}\Sat \cdot \i{\g{\opt{\hV}}{\bot}}{\pV}{\,\vSatS\,}\hV $$
Since by \Cref{lemma:opt-equiv} we know that \opt{\hV} implies $\hV{\equiv}\hVV$ and $\hVV{\in}\shmlwfopt$, we can thus prove
	$$ \forall\pV{\,\in\,}\Proc,\hVV\in\shmlwfopt\text{ when }\hVV{\in}\Sat \cdot \i{\g{\hVV}{\bot}}{\pV}{\,\vSatS\,}\hVV $$
Since $\g{\hfls}{\bot}$ does not yield a result, by definition of \g{-}{\bot} we know that $\bot$ cannot be added to the resultant enforcer and hence substitution \sub{\eV}{\bot} is negligible, meaning that $\g{\hV}{\bot}\sub{\eV}{\bot}$ is equivalent to $\g{\hV}{\bot}$. Hence, we can instead prove
$$ \forall,\pV{\,\in\,}\Proc,\eV{\,\in\,}\Enfwf,\hVV\in\shmlwfopt\text{ when }\hVV{\in}\Sat\; \cdot\; \i{\g{\hVV}{\bot}\sub{\eV}{\bot}}{\pV}{\,\vSatS\,}\hVV $$
This, however, allows us to prove a stronger result ranging over all possible $\rho$ (\ie where $\rho{\,=\,}\bot$ is a specific instance), as follows
$$ \forall\rho,\pV{\,\in\,}\Proc,\eV{\,\in\,}\Enfwf,\hVV\in\shmlwfopt\text{ when }\hVV{\in}\Sat\; \cdot\; \g{\hVV}{\rho}\sub{\eV}{\rho}=\eV'\!\! \imp\!\! \i{\eV'}{\pV}{\,\vSatS\,}\hVV $$
We prove this result coinductively by showing that there exists a \emph{satisfaction relation} \R (\ie a relation that conforms to the rules given in \Cref{fig:uhml-sat}) such that for every optimized formula $\hV{\,\in\,}\shmlwfopt$ we cab show that the enforced process $\i{\eV'}{\pV}$ and formula \hVV are related by \R (defined below), meaning that process $\i{\eV'}{\pV}$ satisfies \hVV.
$$ \R\;\defeq\;\setdef{(\i{\eV'}{\pV},\hVV)}{\hVV{\in}\Sat\;\text{ and }\;\forall\eV\in\Enfwf\cdot\g{\hVV}{\rho}\sub{\eV}{\rho}=\eV'} $$
The proof for this Soundness result is provided in Appendix \Cref{sec:proof-soundness}.

\paragraph{Proving Enforcement Transparency} To prove that the synthesised enforcers are also \emph{transparent} we must prove that \Cref{lemma:transparency} holds, \ie that 
	$$\forall\pV{\,\in\,}\Proc,\hV{\,\in\,}\shmlwf\cdot \pV{\,\in\,}\syn{\hV} \; \imp \; \i{\g{\hV}{}}{\pV}\bisim\pV $$	
Since $\pV{\,\in\,}\syn{\hV} \equiv \pV{\,\vSatS\,}\hV$, and by the definition of \g{-}{}, we get
	$$\forall\pV{\,\in\,}\Proc,\hV{\,\in\,}\shmlwf\cdot \pV{\,\vSatS\,}\hV \; \imp \; \i{\g{\opt{\hV}}{\bot}}{\pV}\bisim\pV $$
Since by \Cref{lemma:opt-equiv} we know that \opt{\hV} implies $\hV{\equiv}\hVV$ and $\hVV{\in}\shmlwfopt$, we can instead prove
	$$\forall\pV{\,\in\,}\Proc,\hVV{\,\in\,}\shmlwfopt\cdot \pV{\,\vSatS\,}\hVV \; \imp \; \i{\g{\hVV}{\bot}}{\pV}\bisim\pV $$
For all $\eV{\,\in\,}\Enfwf$ we know that $\g{\hV}{\bot}\sub{\eV}{\bot}\equiv\g{\hV}{\bot}$, since by definition of \g{-}{\bot} we know that $\bot$ cannot be added to the resultant enforcer since $\g{\hfls}{\bot}$ does not yield a result, and hence substitution \sub{\eV}{\bot} is negligible, such that, we can instead prove
	$$\forall\pV{\,\in\,}\Proc,\eV{\,\in\,}\Enfwf,\hVV{\,\in\,}\shmlwfopt\cdot \pV{\,\vSatS\,}\hVV \; \imp \; \i{\g{\hVV}{\bot}\sub{\eV}{\bot}}{\pV}\bisim\pV $$	
This, however, allows us to prove a stronger result ranging over all possible $\rho$ (\ie where $\rho{\,=\,}\bot$ is a specific instance), as follows
 	$$\forall\rho,\pV{\,\in\,}\Proc,\eV{\,\in\,}\Enfwf,\hVV{\,\in\,}\shmlwfopt\cdot \pV{\,\vSatS\,}\hVV\, \text{ and }\,\g{\hVV}{\rho}\sub{\eV}{\rho}=\eV' \; \imp \; \i{\eV'}{\pV}\bisim\pV $$
We prove this result in a coinductive manner by showing that there exists a \emph{bisimulation relation} \R (defined below), such that for every optimized formula $\hV{\in}\shmlwfopt$, we can show that processes 
\i{\eV'}{\pV} and \pV are related by relation \R, and are thus \emph{bisimilar}.
$$ \R\;\defeq\;\setdef{(\i{\eV'}{\pV},\pV)}{\pV{\,\vSatS\,}\hVV\;\text{ and }\;\forall\eV\in\Enfwf\cdot\g{\hVV}{\rho}\sub{\eV}{\rho}=\eV'} $$
The proof for this Transparency result is provided in Appendix \Cref{sec:proof-transparency}.

\section{Concluding Remarks} In this chapter we have investigated the enforceability of safety properties via suppressions. Since the \shml subset was proven to be maximally expressive \wrt safety properties \cite{Francalanza2015Mon}, we have focussed on defining a novel mapping between the \shml subset and suppression enforcers in the form of a synthesis function. To reduce the complexity of synthesising deterministic enforcers, we assume a syntactic restriction that limits the domain of our synthesis function to normalized formulae. Despite this restriction, we have also produced a novel algorithm (based on \cite{Aceto2016Determinization}) capable of converting any formula $\hV\in\shml$ into a semantically equivalent formula \hVV which is also in \emph{normal form}, \ie $\hVV\in\shmlwf$. This result advocates that our restricted syntax, \shmlwf, is still as expressive as the full (unrestricted) syntax of \shml.


\chapter{Related Work}
\label{sec:background}
In this section we review the state-of-the-art related to runtime enforcement and enforceability and compare it to our work.

\section{Enforceability} \label{sec:related-work-enf}
In general, the term \emph{enforceability} refers to the \emph{relationship} between the meaning of a property and its ability to be \emph{forcibly imposed} upon a system at runtime. In our work, we investigate this relationship \wrt properties that are expressed as \uhml formulae, and as a result we identify a syntactic subset of the logic that allows for defining properties that can be enforced at runtime. Through our synthesis function, we also maintain a clear separation between the declarative specification, \ie the \uhml logic, and the operational enforcement model, \ie our enforcers.
 
Various works \cite{schneider2000,Ligatti2005,Falcone2012,Martinelli2005} have presented a wide variety of definitions for enforceability. However, unlike our work, most of these definitions do not distinguish between the specification and the enforcer, thereby requiring the property to be expressed in terms of the enforcement mechanism itself.

For instance, when Schneider first explored enforceability in \cite{schneider2000}, he stated that a property is enforceable if its \emph{violation} can be \emph{detected} by a \emph{truncation automaton} which in turn prevents it by terminating the system. In this setting, properties thus require to be specified in terms of the enforcement model itself, \ie as a truncation automaton. Furthermore, since this enforcement model can only prevent the occurrence of misbehaviour via system termination, the set of enforceable properties is thus limited to just safety properties. 

As a continuation of Schneider's work, Ligatti \etal in \cite{Ligatti2005,Ligatti2006PhD,Ligatti2009}, sought to widen the set of enforceable properties by introducing \emph{edit automata} \ie an enforcement mechanism capable of \emph{suppressing} and \emph{inserting} system actions. Based on this enforcement model, Ligatti \etal introduced a new notion of enforceability stating that a property is enforceable if it can be expressed as an edit automaton that is able to \emph{transform} an invalid system execution into a valid one. 

Edit automata were shown to be capable of enforcing instances of safety and liveness properties, along with other properties such as infinite renewal properties, \ie properties that are satisfied by infinite executions that have an infinite number of valid prefixes, and violated when an infinite execution has only a finite number of valid prefixes.

Based on these enforcement mechanisms, the authors define different notions of enforceability based on the criteria of 
\begin{enumerate}[$(i)$]
	\item \emph{enforcement soundness}, \ie an enforcer is \emph{sound} when it always converts invalid executions into valid ones, and
	\item \emph{transparency}, \ie a \emph{strongly transparent} enforcer \emph{does not modify} valid executions at all, while a \emph{weakly transparent} one can only modify valid traces into semantically equivalent ones (for some notion of equivalence).
\end{enumerate}

Similar to Schneider \cite{schneider2000}, in this setting, their exists no clear separation between the specification and the enforcement mechanism, such that the properties are required to be expressed in terms of the enforcement model itself, \ie as edit automata. The authors thus state that a property is \emph{strongly enforceable} if it can be expressed as an edit-automaton which enforces the said property in a \emph{sound} and \emph{strongly transparent} manner; a weaker notion for enforceability is similarly defined in terms of soundness and \emph{weak transparency}. 

The fundamental difference between these definitions for enforceability and ours is that, we investigate enforceability \wrt a logic, \ie we identify which subsets of our logic, \ie \uhml, can be enforced \emph{soundly} and \emph{transparently}. Furthermore, since we define enforceability \wrt the entire process, and not \wrt just a single execution trace, we are able to prove stronger results for soundness and transparency. For instance, in the case of transparency, we can prove that in certain cases, when a process \pV already satisfies formula \hV, then the enforcer \eV that is synthesised from \hV, does not change the behaviour of \pV in any way. We prove this result by showing that the enforced process \i{\eV}{\pV} is \emph{strongly bisimilar} to the original process \pV. This result is thus stricter than trace equivalence constraint imposed by Ligatti \etal's strong transparency criteria.

Other researchers such as Fong \cite{fong2004} and Talhi \etal \cite{talhi2008}, investigated the enforceability of properties \wrt enforcement automata with limited resources such as a bounded history of system events. This research shows that although some properties are enforceable in theory via unbounded enforcement automata, in practice it would require an infeasible amount of memory. They thus showed that, in practice, the level of enforceability of an enforcement automaton is relative to the bounds imposed by the available resources. This implies that although some properties may be theoretically enforceable, they might not be so when limited by practical constraints.

In their work, Bielova \etal \cite{Bielova2011PhD,Bielova2011,Bielova2011Predictability} remarked that the transparency constraint, previously introduced by Ligatti \etal, only dictates how the monitor should react when analysing a \emph{valid} execution prefix, \ie the meaning of valid executions should remain intact. However, neither soundness nor transparency specify the extent of the modifications that an enforcer should be able to apply over an invalid prefix. 

The authors thus propose a notion of \emph{predictability} that restricts the enforcers from transforming the invalid executions in an arbitrary way. More precisely, an enforcer is \emph{predictable} if one can predict the number of transformations that the enforcer will apply in order to transform an invalid execution into a valid one.

Predictability thus prevents enforcers from applying unnecessary transformations upon an invalid execution. Based on this notion they thus devise a more stringent notion of enforceability that states that a property is enforceable if there exists a \emph{sound}, \emph{transparent} and \emph{predictable} enforcer.

By synthesising enforcers from a logic, we always produce enforcers that are, in some sense, predictable since the synthesised enforcers are designed to enforce properties of the same type, in the same way. For example, our syntheses converts safety property $\hnec{\acta}\hnec{\actb}\hfls$ into a \emph{deterministic} enforcer which enforces the property by continuously suppressing events that \emph{directly lead to a violation}, \ie in this case every occurrence of event \actb that occurs after event \acta is thus suppressed. A similar approach is applied to other properties of the same type (\ie safety) such as $\hmax{\hVarX}{\hnec{\acta}(\hnec{\actb}\hVarX\hand\hnec{\actg}\hfls)}$, where only the events preceding the violation, \ie \actg, are required to be suppressed when preceded by an \acta action and zero or more \actb actions. 

\section{Separating the Property from the Enforcer}
As remarked by Bielova in \cite{Bielova2011PhD}, most of the definitions for enforceability do not make a clear separation between the logic behind the property and the \resp enforcement mechanisms. In fact, the enforcement definitions we have seen so far in \Cref{sec:related-work-enf}, do not make a distinction between the enforcement mechanism and the logic for specifying correctness properties, as they assume that properties are specified in terms of the \resp enforcement mechanism, \eg as truncation automata, edit automata \etc.

Various research \cite{Bielova2011PhD,Falcone2012,Pinisetty2012,Martinelli2005} has been conducted to introduce a \emph{synthesis phase} which allows for expressing properties in terms of a more abstract model from which the required enforcer is then derived. This allows the specifier to focus on defining \emph{what} should be enforced and \emph{not} on \emph{how} this should be done.

Most of this research, however, was conducted \wrt automata based specifications, such as Policy automata \cite{Bielova2011PhD} and Streett automata \cite{Falcone2012}, unlike in our case where we study the enforceability of branching time logic formulae. Mapping automata-based specifications to enforcement automata is arguably easier to attain compared to mapping a logic to an enforcer, mainly since both the specification and the enforcement models are defined as variants of automata. 

Bielova in \cite{Bielova2011PhD} thus proposes an algorithm for automatically constructing enforcement automata from policy automata, \ie an automata based representation that combines the acceptance states of B\:{u}chi Automata and Finite State Automata. This therefore contrasts with our synthesis function which derives the required enforcers directly from branching-time logic formulae. 

Similarly, Falcone \etal in \cite{Falcone2010,Falcone2011,Falcone2012}, propose a way how a variety of properties, defined in terms of Streett automata, can be mapped onto an enforcement automaton that enforces the \resp property. More precisely, Falcone \etal showed that most of the property classes defined within the \emph{Safety-Progress hierarchy} \cite{Pnueli1990} are enforceable, as they can be encoded as Streett automata and subsequently converted into enforcement automata. 

Hence, as opposed to Ligatti \etal, Bielova and Falcone \etal separate the specification of the property from the enforcement mechanism. This was done by providing construction rules that convert properties expressed as Policy and Streett automata into the \resp enforcement automata.

In relation to the work by Falcone \etal, Pinisetty \etal \cite{Pinisetty2012,Pinisetty2014} studied the enforceability of safety and co-safety \emph{Timed Properties} expressed as a variant of Timed Automata. Similar to our work, non-deterministic specifications must first be converted into deterministic ones. 

Despite this step being conceptually similar to our normalization phase, since the properties considered are expressed as Timed Automata, the determinization of such non-deterministic properties was thus attainable via the conventional determinization construction rules that are standard in automata theory. Conversely, the normalization of \uhml specifications was only recently explored in \cite{Aceto2016Determinization} and required being heavily adapted to our setting. 

Following the determinization step, the resultant deterministic timed automata are then projected onto the \resp enforcement mechanisms accordingly. Their work was also implemented as a prototype tool, using Python, which allowed for assessing the feasibility of their proposed timed enforcement.

More similar to our work, in \cite{Martinelli2005,Martinelli2006Wits,Matteucci2006Tool}, Martinelli \etal performed an initial study of the enforceability of a logic. Hence, they partially addressed this issue by devising a synthesis algorithm that converts formulae expressed in modal $\mu$-calculus \cite{Kozen1983MuCalc} (a reformulation of \uhml), into either a truncation, suppression, insertion or an edit automaton, as decided by the specifier.

This synthesis algorithm thus burdens the specifier with still having to manually deduce the appropriate enforcement automaton (if any) capable of enforcing the said formula, and with having to manually define a function which specifies the points where the synthesised automaton should perform the required enforcement. By contrast, we identify the subsets of \uhml properties that can be enforced \wrt insertions and suppressions such that the specifier is completely relieved from additional manual work related to the enforcement mechanisms required for enforcing the specified property.

The presented synthesis algorithm is defined in terms of two techniques, namely, \emph{partial model checking} \cite{Andersen1995} and \emph{satisfiability} \cite{Streett1989}. The former is used to simplify the formula and modify it according to the function defined by the specifier, while the latter is the main mechanism used for obtaining the required enforcer; this is achieved by finding a process which satisfies the modified formula obtained from partially model checking the original formula; this contrasts with the way we generate the required enforcers via a compositional analysis of the specified \uhml formula.

In a multi-pronged verification approach, our synthesis may, however, benefit from their former technique, \ie partial model checking, to statically verify the non-enforceable parts of the given formula and reduce it into an enforceable one which can then be synthesised using our synthesis algorithm.

\section{Monitorability} 
\emph{Monitorability} amounts to the relationship between the meaning of a property and its ability to be \emph{recognized} at runtime \cite{Cassar2017RV,Francalanza2017FMSD}. As both runtime monitoring and enforcement are dynamic analysis techniques capable of observing the behaviour of a system, the differences between these two techniques is sometimes blurred such as in \cite{Colombo2012PolyLarva,Colombo2009,Chen2005,Rosu2012JavaMop}. 

However, in contrast to enforceability which requires imposing property compliance, a property is said to be \emph{monitorable} when there exists a \emph{detection monitor} that can issue a \emph{definitive verdict} determining whether the system under scrutiny satisfies or violates the property.

Along the years, several variants for monitorability have been defined \wrt various criteria. Viswanathan \etal in \cite{Viswanathan2005} defined monitorable properties as a strict subset of \emph{safety properties} defined over \emph{infinite execution traces}. This restrictive definition came as the result of limiting the monitor's detection capabilities to only detecting misbehaviour after observing a finite prefix of a potentially infinite execution trace. As a by-product, this monitoring limitation restricts the domain of monitorable properties to just safety properties. 

Pnueli \etal in \cite{Pnueli2006} defined a wider notion of monitorability which states that properties are monitorable only when a \emph{positive} or a \emph{negative} verdict can be issued by the monitor after analysing a \emph{finite} execution trace. Specifically, a monitor should be able to issue a negative (\resp positive) verdict, for a safety (\resp co-safety) property $\varphi$, only in cases when a finite prefix of an execution trace \emph{provides enough information} from which the monitor can deduce that any possible continuation of the prefix still violates (\resp satisfies) property $\varphi$. The authors also determined that once a verdict is issued by the monitor \wrt a finite prefix of the complete execution trace, this implicitly applies to the complete execution, and hence, the monitoring becomes redundant and can therefore be halted.

Inspired by \cite{Pnueli2006}, Bauer \etal in \cite{Bauer2006,Bauer2011} formulated another definition for monitorability in which they adopted the notion of good and bad prefix from the field of model checking \cite{Kupferman2001}. The authors conjecture that it is possible to detect a violation or satisfaction for properties describing infinite behaviour, by only observing a finite prefix of a potentially infinite execution. More specifically, a finite execution prefix is said to be a bad prefix (\resp good prefix) \wrt a property $\varphi$ if it provides enough information to conclude that property $\varphi$ was violated (\resp satisfied). A finite execution prefix is, however, said to be \emph{ugly} if it \emph{does not} provide the necessary information to draw either of these conclusions. 

Based on these three types of prefixes, the authors define new semantics for Linear Time Logic (LTL) which reasons about finite traces. This variant of the logic was called $\text{LTL}_3$, for its ability of producing three verdicts, namely, \emph{true} ($\top$), \emph{false} ($\bot$) \resp indicating property satisfaction and violation, along with an inconclusive verdict ($?$). In this way, an $\text{LTL}_3$ property $\varphi$ is monitorable \wrt an execution trace $t$, if there exists a bad (or good) prefix of $t$ that violates (or satisfies) $\varphi$.

Bauer \etal further extend their work in \cite{Bauer2007,Bauer2010} by extending the monitoring semantics of $\text{LTL}_3$ in order to refine the inconclusive verdict (\ie $?$), into a more informative verdict. Hence, the authors proposed new LTL semantics defining a 4-valued truth-domain which allows for producing two concrete verdicts, namely \emph{true} and \emph{false}, along with approximation results denoting \emph{presumably true} and \resp \emph{presumably false}. Despite still being inconclusive, the latter two approximation verdicts provide more information compared to the single inconclusive verdict defined in $\text{LTL}_3$. This is therefore ideal for providing approximate positive (\resp negative) verdicts for properties that are satisfied (\resp violated) by infinite executions. 

For instance, a monitor can easily deduce the violation of a safety property after observing bad prefix of an infinite execution trace, however, determining satisfaction of the same property requires analysing the entire infinite sequence. Using the new truth domain, monitors are thus able to issue a presumably true verdict for execution prefixes that do not violate a safety property, and conversely, a presumably false for finite execution prefixes that do not satisfy a co-safety property; in $\text{LTL}_3$ the same inconclusive verdict (\ie $?$) is issued in both cases. 

Falcone \etal in \cite{Falcone2012} generalized the definition by Bauer \etal by \emph{parameterizing} it with various truth-domains. Using this definition they are able to identify cases when a truth domain is not expressive or fine-grained enough to monitor for some \emph{specific classes of properties} pertaining to the \emph{safety-progress hierarchy} \cite{Pnueli1990,chang1992,Falcone2009EM}. Their new definition of monitorability is therefore based on the \emph{distinguishability} of good and bad execution sequences. This alternative definition is able to better distinguish between inconclusive situations that a monitor might encounter while analysing a finite trace. 

The definitions we have seen so far have always been explored \wrt linear-time logics or automata-based specifications, and execution traces. Francalanza \etal in \cite{Francalanza2015Mon,Francalanza2017FMSD}, studied the monitorability of branching-time properties expressed in terms of Hennessy-Milner Logic with recursion (\uhml) \wrt the \emph{entire computational structure} of the process (\ie not just traces).

Francalanza \etal thus defined their own definition of monitorability which states that a \uhml property is monitorable, if for every process capable of executing a particular trace, there exists a monitor that can issue a \emph{positive} or \emph{negative} verdict by only analysing the executed trace. Similar to previous work by Pnueli \etal \cite{Pnueli2006}, once a verdict is issued by a monitor, this becomes \emph{irrevocable} and hence the monitor must consistently provide the \emph{same} verdict.

In relation to this notion of monitorability, the authors identify the maximally expressive of \uhml that syntactically characterises the monitorable \uhml formulae, \ie they identify a syntactic restriction for \uhml that allows for defining all the possible \uhml formulae that are monitorable according to their notion of monitorability. The identified subset was thus termed as Monitorable HML (\mhml) and consists in the union of Safe-HML (\shml) and Co-Safe-HML (\chml). As the name implies, Safe-HML can only be used for specifying safety properties, while Co-Safe-HML only allows for specifying co-safety (liveness) properties. 

In this approach, the authors thus preserve the original semantics of the chosen logic, \ie \uhml, and instead identify the monitorable subset of this logic thus keeping the meaning of the logic independent from the employed verification technique. This is therefore different from the body of work proposed by Bauer \etal \cite{Bauer2006,Bauer2007,Bauer2010,Bauer2011} where they redefined the semantics of their chosen logic, \ie LTL, \wrt finite traces in order to appease the selected verification technique, \ie runtime monitoring.

The authors also prove the existence of a relationship between the logic and the monitor's verdicts. More precisely, the authors prove that for any process \pV which satisfies (\resp violates) an arbitrary \mhml property \hV, there exists a detection monitor \mV that is able to detect the property satisfaction (\resp violation) by analysing a witness trace executed by \pV, and thus issue an irrevocable positive (\resp negative) verdict as a result. In addition, the authors prove that the converse also holds, \ie that for any process \pV which can be monitored by a monitor \mV capable of issuing a positive or a negative verdict, there exists a \mhml formula that can be either satisfied or violated by process \pV. 

On top of this, the authors also provide a synthesis function capable of analysing and monitorable \mhml formula and deriving a monitor that is able to detect the violation or satisfaction of the property it was derived from, by issuing the appropriate verdict. Their theory is also supported by a runtime verification tool for monitoring Erlang programs called \detecter \cite{FraSey2015,Attard2016,Cassar2015Fesca,Cassar2014Foclasa,Cassar2017Betty}. This tool is able to synthesise Erlang detection monitors from the monitorable subset of \uhml identified in \cite{Francalanza2015Mon}.

\section{Supervisory-Control} 
\emph{Supervisory-Control} \cite{Ramadge1987,VanHulst2017} is a static analysis technique that ensures that a system always adheres to a specific correctness property by \emph{modifying} its behaviour in a \emph{pre-deployment phase}. More precisely, a synthesis function in the sense of supervisory-control, has the task of taking an existing program and reconstructing it in a way which conforms to the given correctness property \cite{VanHulst2017}. In general, the modifications made to the system's internal behaviour ensure that the system always conforms to the property by \emph{completely removing} invalid execution sequences from the resultant modified system.

This differs from runtime enforcement since, instead of modifying the internal system behaviour \emph{pre-deployment}, enforcers ensure that the system conforms to the property by performing \emph{on-the-fly} modifications (\ie event suppression or insertion) while the system executes, and not during synthesis. Hence, the enforcer ensures that the system behaves as specified by the property by \emph{wrapping around} the system and acting as a \emph{proxy} for its inputs and outputs. In fact, unlike supervisory-control, in enforcement the system under scrutiny is barely modified during synthesis and is generally treated as being a \emph{black box}; this makes enforcement ideal in cases where the internal behaviour of the system is unknown pre-deployment.

Van Hulst \etal in \cite{VanHulst2017} studied a variant of \uhml \wrt supervisory-control by establishing a synthesis algorithm that produces a controlled system that complies to the specified property. Their synthesis is therefore designed to produce the controlled system by statically analysing a given system and reformulating its behaviour to keep it in line with some correctness property expressed as a formula of the logic variant. As an evaluation of their work, the authors prove that their synthesis always terminates, and that it adheres to a number of constraints, such as \emph{validity}, \emph{maximality} and \emph{controllability}. 

Validity can be seen as the equivalent of soundness in RE, as this constraint serves to ensure that the synthesised controlled system must \emph{always} satisfy the \resp property. On the other hand, maximality is, in some sense, equivalent to transparency, since this criterion dictates that the synthesis should remove the least possible behaviour from the original system, such that any valid behaviour is left intact. In addition, controllability is used to ensure that the synthesis affect other behaviour that was not specified by the property.

Similar work on supervisory-control was also conducted in relation to other logics such as the modal \mucalc \cite{Basu2007,Pinchinat2005}, LTL \cite{Wolff2013} and CTL* \cite{Ehlers2014} amongst others.

\chapter{Conclusion}
\label{sec:conc}
In this report we outlined preliminary results towards developing theoretical foundations for understanding the enforceability of properties that are expressed in terms of a highly expressive logic. As stated in \Cref{sec:intro-aims}, in order to achieve this aim, we intend to address the following objectives:
\begin{enumerate}[\it O1.]
	\item Developing an abstract model for runtime enforcers and define enforceability \wrt \uhml.
	\item Assess the correctness of our abstract enforcers.
	\item Assessing the implementability of our enforcers by redefining them into a more implementable version.
	\item Developing a prototype implementation for our enforcers and evaluating their feasibility.
\end{enumerate} 
Until now, we have conducted an initial investigation of our first two objectives by studying the ability to enforce \uhml properties via event suppression; this allowed us to identify \shml as being the subset of \uhml which is enforceable via sound and transparent suppression enforcers. We summarize our main novel contributions as follows:
\begin{itemize} \setlength{\itemindent}{-1.2em}
	\item In \Cref{sec:enf-model} we have mainly addressed our first objective by defining:
	\begin{itemize}
		\item \emph{Symbolic Transformations}, that formally define a mapping mechanism for transforming a concrete system event into a (possibly) different one as specified by the transformation pattern;
		\item \emph{An LTS semantics for Enforcers}, which formalise the structure and dynamic behaviour of enforcers, along with the interaction between the enforcer, \eV and the process under scrutiny, \pV, in the form of the instrumented LTS, \i{\eV}{\pV}; and 
		\item \emph{A Formal Definition for Enforceability}, defining the relationship between the meaning of a \uhml property and its ability to be adequately enforced at runtime by an enforcer. With this definition we establish that an enforcer \eV \emph{strongly enforces} a formula \hV whenever it is able to do it in a \emph{sound} and \emph{transparent} manner. By defining the enforced system as an LTS, \i{\eV}{\pV}, we were able to provide novel definitions Soundness and Transparency which are \emph{stronger} than the classic definitions (see \Cref{sec:enforceability}).
	\end{itemize}
	\item In \Cref{sec:enf-synthesis} we continued addressing objective \emph{O1} by:
	\begin{itemize}
		\item Identifying \emph{\shmlsl as being Suppression Enforceable} thus denoting that every safety property expressible in \uhml can be enforced via action suppression.
		\item Defining a \emph{Normalization algorithm} for reducing a symbolic \shml formula into a semantically equivalent formula that is in \emph{normal form}, from which we can then synthesise deterministic enforcers; this algorithm was heavily inspired from the work in \cite{Aceto2016Determinization}; and by
		\item Developing a \emph{Synthesis algorithm} for converting normalized symbolic \shml formulae into suppression enforcers.
	\end{itemize}
	\item In \Cref{sec:enf-synthesis} we also contributed towards objective \emph{O2} by:
		\begin{itemize}
			\item Guaranteeing \emph{semantic preservation} for our \emph{normalization algorithm} by proving that each normalization step preserves the original semantics of the given formula (or system of equations); in Appendix \Cref{sec:app:proofs-new-determinization} we provide the proofs of the lemmas used when proving semantic equivalence; and by
			\item Proving that the synthesised enforcers always \emph{behave deterministically} and can always \emph{strongly enforce} the formula they were synthesised from; in Appendix \Cref{sec:app:correctness-proofs} we provide the proofs for the supporting lemmas used when proving the aforementioned results.
		\end{itemize}
\end{itemize}


\section{Future Work}
We plan to extend this work along two different avenues, namely, $(i)$ by expanding the work on objectives \emph{O1} and \emph{O2} by investigating a wider notion of enforceability, and $(ii)$ by addressing objectives \emph{O3} and \emph{O4} by exploring the implementability and feasibility of our enforcers. To tackle $(i)$ would require enlarging the subset of enforceable specifications by investigating the enforceability of \uhml formulae \wrt action insertions. This will potentially require extending the formal model, proofs and other results obtained so far according to the new notion of enforceability. 

Meanwhile, in $(ii)$, we envision the development of another synthesis function that converts the subset of enforceable \uhml specifications into enforcers that are defined in terms of a more implementable model. However, to ensure that these implementable enforcers are also well-behaved (\ie deterministic, sound and transparent), it would require proving correspondence to the enforcers defined by the abstract enforcement model we have so far. Based on this new synthesis we aim to develop a runtime enforcement tool implementation that checks whether a given specification is enforceable or not, when possible, converts it into an equivalent enforceable property, and finally generates the required enforcer automatically. At the time of writing we are unaware of the existence of such a tool.

\bibliographystyle{abbrv}
\bibliography{refs}

\begin{thebibliography}{10}

\bibitem{ibm2005}
An architectural blueprint for autonomic computing.
\newblock Technical report, IBM, 2005.

\bibitem{Achilleos2018FSTTCS}
L.~Aceto, A.~Achilleos, A.~Francalanza, and A.~Ing{\'o}lfsd{\'o}ttir.
\newblock {Monitoring for Silent Actions}.
\newblock In S.~Lokam and R.~Ramanujam, editors, {\em 37th IARCS Annual
  Conference on Foundations of Software Technology and Theoretical Computer
  Science (FSTTCS 2017)}, volume~93 of {\em Leibniz International Proceedings
  in Informatics (LIPIcs)}, pages 7:1--7:14, Dagstuhl, Germany, 2018. Schloss
  Dagstuhl--Leibniz-Zentrum fuer Informatik.

\bibitem{Aceto2016Determinization}
L.~Aceto, A.~Achilleos, A.~Francalanza, A.~Ing{\'o}lfsd{\'o}ttir, and S.~{\"O}.
  Kjartansson.
\newblock Determinizing monitors for hml with recursion.
\newblock {\em arXiv preprint}, 2016.

\bibitem{Aceto1999TestingHML}
L.~Aceto and A.~Ing\'{o}lfsd\'{o}ttir.
\newblock Testing hennessy-milner logic with recursion.
\newblock In {\em FoSSaCS}, volume 1578 of {\em LNCS}, pages 41--55. Springer,
  1999.

\bibitem{Aceto2007Book}
L.~Aceto, A.~Ing\'{o}lfsd\'{o}ttir, K.~G. Larsen, and J.~Srba.
\newblock {\em Reactive Systems: Modelling, Specification and Verification}.
\newblock Cambridge University Press, New York, NY, USA, 2007.

\bibitem{Achilleos2018Fossacs}
A.~Achilleos, A.~Francalanza, L.~Aceto, and A.~Ing{\'{o}}lfsd{\'{o}}ttir.
\newblock A framework for parametrized monitorability.
\newblock In {\em FOSSACS}, 2018.
\newblock (to appear).

\bibitem{Andersen1995}
H.~R. Andersen.
\newblock Partial model checking.
\newblock In {\em Logic in Computer Science, 1995. LICS'95. Proceedings., Tenth
  Annual IEEE Symposium on}, pages 398--407. IEEE, 1995.

\bibitem{Cassar2017Betty}
D.~P. Attard, I.~Cassar, A.~Francalanza, L.~Aceto, and A.~Ingolfsdottir.
\newblock A runtime monitoring tool for actor-based systems.
\newblock {\em Behavioural Types: from Theory to Tools.}, 2017.

\bibitem{Attard2016}
D.~P. Attard and A.~Francalanza.
\newblock {\em A Monitoring Tool for a Branching-Time Logic}, pages 473--481.
\newblock Springer International Publishing, Cham, 2016.

\bibitem{Azzopardi2016}
S.~Azzopardi, C.~Colombo, and G.~J. Pace.
\newblock A model-based approach to combining static and dynamic verification
  techniques.
\newblock In {\em ISoLA {(1)}}, volume 9952 of {\em Lecture Notes in Computer
  Science}, pages 416--430, 2016.

\bibitem{Basu2007}
S.~Basu and R.~Kumar.
\newblock Quotient-based control synthesis for partially observed
  non-deterministic plants with mu-calculus specifications.
\newblock In {\em CDC}, pages 5294--5299. IEEE, 2007.

\bibitem{Bauer2006}
A.~Bauer, M.~Leucker, and C.~Schallhart.
\newblock Monitoring of real-time properties.
\newblock In S.~Arun-Kumar and N.~Garg, editors, {\em FSTTCS 2006: Foundations
  of Software Technology and Theoretical Computer Science}, pages 260--272,
  Berlin, Heidelberg, 2006. Springer Berlin Heidelberg.

\bibitem{Bauer2007}
A.~Bauer, M.~Leucker, and C.~Schallhart.
\newblock The good, the bad, and the ugly, but how ugly is ugly?
\newblock In {\em International Workshop on Runtime Verification}, pages
  126--138. Springer, 2007.

\bibitem{Bauer2010}
A.~Bauer, M.~Leucker, and C.~Schallhart.
\newblock Comparing ltl semantics for runtime verification.
\newblock {\em Journal of Logic and Computation}, 20(3):651--674, 2010.

\bibitem{Bauer2011}
A.~Bauer, M.~Leucker, and C.~Schallhart.
\newblock Runtime verification for ltl and tltl.
\newblock {\em ACM Trans. Softw. Eng. Methodol.}, 20(4):14:1--14:64, Sept.
  2011.

\bibitem{Ligatti2002}
L.~Bauer, J.~Ligatti, and D.~Walker.
\newblock More enforceable security policies.
\newblock In {\em Proceedings of the Workshop on Foundations of Computer
  Security (FCS’02), Copenhagen, Denmark}. Citeseer, 2002.

\bibitem{BenAri1983}
M.~Ben-Ari, A.~Pnueli, and Z.~Manna.
\newblock The temporal logic of branching time.
\newblock {\em Acta Informatica}, 20(3):207--226, 1983.

\bibitem{Bielova2011PhD}
N.~Bielova.
\newblock {\em A theory of constructive and predictable runtime enforcement
  mechanisms}.
\newblock PhD thesis, University of Trento, 2011.

\bibitem{Bielova2008}
N.~Bielova and F.~Massacci.
\newblock Do you really mean what you actually enforced?
\newblock In {\em International Workshop on Formal Aspects in Security and
  Trust}, pages 287--301. Springer, 2008.

\bibitem{Bielova2011}
N.~Bielova and F.~Massacci.
\newblock Do you really mean what you actually enforced?-edited automata
  revisited.
\newblock {\em International Journal of Information Security}, 10(4):239--254,
  2011.

\bibitem{Bielova2011Predictability}
N.~Bielova and F.~Massacci.
\newblock Predictability of enforcement.
\newblock In {\em International Symposium on Engineering Secure Software and
  Systems}, pages 73--86. Springer, 2011.

\bibitem{Bruening2011}
D.~Bruening and Q.~Zhao.
\newblock Practical memory checking with dr. memory.
\newblock In {\em Proceedings of the 9th Annual IEEE/ACM International
  Symposium on Code Generation and Optimization}, pages 213--223. IEEE Computer
  Society, 2011.

\bibitem{Bruening2004PhdDynamoRIO}
D.~L. Bruening.
\newblock {\em Efficient, Transparent, and Comprehensive Runtime Code
  Manipulation}.
\newblock PhD thesis, Cambridge, MA, USA, 2004.
\newblock AAI0807735.

\bibitem{Cassar2014Foclasa}
I.~Cassar and A.~Francalanza.
\newblock On synchronous and asynchronous monitor instrumentation for
  actor-based systems.
\newblock In {\em Proceedings 13th International Workshop on Foundations of
  Coordination Languages and Self-Adaptive Systems, {FOCLASA} 2014, Rome,
  Italy, 6th September 2014.}, pages 54--68, 2014.

\bibitem{Cassar2015}
I.~Cassar and A.~Francalanza.
\newblock Runtime adaptation for actor systems.
\newblock In {\em Runtime Verification}, pages 38--54. Springer, 2015.

\bibitem{Cassar2016IFM}
I.~Cassar and A.~Francalanza.
\newblock On implementing a monitor-oriented programming framework for actor
  systems.
\newblock In {\em International Conference on Integrated Formal Methods}, pages
  176--192. Springer, 2016.

\bibitem{Cassar2015Fesca}
I.~Cassar, A.~Francalanza, and S.~Said.
\newblock Improving runtime overheads for detecter.
\newblock In {\em Proceedings 12th International Workshop on Formal Engineering
  approaches to Software Components and Architectures, {FESCA} 2015, London,
  United Kingdom, April 12th, 2015.}, pages 1--8, 2015.

\bibitem{chang1992}
E.~Chang, Z.~Manna, and A.~Pnueli.
\newblock Characterization of temporal property classes.
\newblock In {\em International Colloquium on Automata, Languages, and
  Programming}, pages 474--486. Springer, 1992.

\bibitem{Charafeddine2014}
H.~Charafeddine, K.~El{-}Harake, Y.~Falcone, and M.~Jaber.
\newblock Runtime enforcement for component-based systems.
\newblock {\em CoRR}, abs/1406.5708, 2014.

\bibitem{Chen2005}
F.~Chen and G.~Ro{\c{s}}u.
\newblock {\em Java-MOP: A Monitoring Oriented Programming Environment for
  Java}, pages 546--550.
\newblock Springer Berlin Heidelberg, Berlin, Heidelberg, 2005.

\bibitem{FraCini2015}
C.~Cini and A.~Francalanza.
\newblock An ltl proof system for runtime verification.
\newblock In {\em International Conference on Tools and Algorithms for the
  Construction and Analysis of Systems}, pages 581--595. Springer, 2015.

\bibitem{Clarke1989}
E.~M. Clarke and I.~A. Draghicescu.
\newblock {\em Linear Time, Branching Time and Partial Order in Logics and
  Models for Concurrency: School/Workshop, Noordwijkerhout, The Netherlands May
  30 -- June 3, 1988}, chapter Expressibility results for linear-time and
  branching-time logics, pages 428--437.
\newblock Springer, 1989.

\bibitem{Clarke2008}
E.~M. Clarke and E.~A. Emerson.
\newblock Design and synthesis of synchronization skeletons using branching
  time temporal logic.
\newblock In {\em 25 Years of Model Checking}, pages 196--215. Springer, 2008.

\bibitem{Clarke1999Book}
E.~M. Clarke, O.~Grumberg, and D.~Peled.
\newblock {\em Model checking}.
\newblock MIT press, 1999.

\bibitem{Colombo2012PolyLarva}
C.~Colombo, A.~Francalanza, R.~Mizzi, and G.~J. Pace.
\newblock polylarva: Runtime verification with configurable resource-aware
  monitoring boundaries.
\newblock In {\em SEFM}, pages 218--232, 2012.

\bibitem{Colombo2009}
C.~Colombo, G.~Pace, and G.~Schneider.
\newblock Larva --- {S}afer monitoring of real-time java programs ({T}ool
  paper).
\newblock In {\em SEFM}, pages 33--37, 2009.

\bibitem{Ehlers2014}
R.~Ehlers, S.~Lafortune, S.~Tripakis, and M.~Vardi.
\newblock Bridging the gap between supervisory control and reactive synthesis:
  Case of full observation and centralized control.
\newblock {\em IFAC Proceedings Volumes}, 47(2):222 -- 227, 2014.
\newblock 12th IFAC International Workshop on Discrete Event Systems (2014).

\bibitem{Emerson1980}
E.~Emerson and E.~Clarke.
\newblock Characterizing correctness properties of parallel programs using
  fixpoints.
\newblock In {\em Automata, Languages and Programming}, volume 85/1980 of {\em
  Lecture Notes in Computer Science}, pages 169--181. Springer, Berlin /
  Heidelberg, 1980.

\bibitem{Emerson1986}
E.~A. Emerson and J.~Y. Halpern.
\newblock \&ldquo;sometimes\&rdquo; and \&ldquo;not never\&rdquo; revisited: On
  branching versus linear time temporal logic.
\newblock {\em J. ACM}, 33(1):151--178, Jan. 1986.

\bibitem{Emerson1985}
E.~A. Emerson and C.-L. Lei.
\newblock Modalities for model checking (extended abstract): Branching time
  strikes back.
\newblock In {\em Proceedings of the 12th ACM SIGACT-SIGPLAN Symposium on
  Principles of Programming Languages}, POPL '85, pages 84--96, New York, NY,
  USA, 1985. ACM.

\bibitem{Schneider1999}
U.~Erlingsson and F.~B. Schneider.
\newblock Sasi enforcement of security policies: A retrospective.
\newblock In {\em Proceedings of the 1999 Workshop on New Security Paradigms},
  NSPW '99, pages 87--95, New York, NY, USA, 1999. ACM.

\bibitem{Falcone2010}
Y.~Falcone.
\newblock You should better enforce than verify.
\newblock In {\em Runtime Verification}. Springer, Jan. 2010.

\bibitem{Falcone2009EM}
Y.~Falcone, J.-C. Fernandez, and L.~Mounier.
\newblock Enforcement monitoring wrt. the safety-progress classification of
  properties.
\newblock In {\em Proceedings of the 2009 ACM Symposium on Applied Computing},
  SAC '09, pages 593--600, New York, NY, USA, 2009. ACM.

\bibitem{Falcone2012}
Y.~Falcone, J.-C. Fernandez, and L.~Mounier.
\newblock What can you verify and enforce at runtime?
\newblock {\em International Journal on Software Tools for Technology
  Transfer}, 14(3):349, June 2012.

\bibitem{Falcone2011}
Y.~Falcone, L.~Mounier, J.-C. Fernandez, and J.-L. Richier.
\newblock Runtime enforcement monitors: composition, synthesis, and enforcement
  abilities.
\newblock {\em Formal Methods in System Design}, 38(3):223, June 2011.

\bibitem{fong2004}
P.~W. Fong.
\newblock Access control by tracking shallow execution history.
\newblock In {\em Security and Privacy, 2004. Proceedings. 2004 IEEE Symposium
  on}, pages 43--55. IEEE, 2004.

\bibitem{Francalanza2016TheoMon}
A.~Francalanza.
\newblock {\em A Theory of Monitors}, pages 145--161.
\newblock Springer, 2016.

\bibitem{Cassar2017RV}
A.~Francalanza, L.~Aceto, A.~Achilleos, D.~P. Attard, I.~Cassar, D.~D. Monica,
  and A.~Ing{\'{o}}lfsd{\'{o}}ttir.
\newblock A foundation for runtime monitoring.
\newblock In {\em RV}, pages 8--29, 2017.

\bibitem{Francalanza2015Mon}
A.~Francalanza, L.~Aceto, and A.~Ing{\'{o}}lfsd{\'{o}}ttir.
\newblock On verifying hennessy-milner logic with recursion at runtime.
\newblock In {\em Runtime Verification - 6th International Conference, {RV}
  2015 Vienna, Austria, September 22-25, 2015. Proceedings}, pages 71--86,
  2015.

\bibitem{Francalanza2017FMSD}
A.~Francalanza, L.~Aceto, and A.~Ing{\'{o}}lfsd{\'{o}}ttir.
\newblock Monitorability for the hennessy-milner logic with recursion.
\newblock {\em Formal Methods in System Design}, 51(1):87--116, 2017.

\bibitem{FraSey2015}
A.~Francalanza and A.~Seychell.
\newblock Synthesising correct concurrent runtime monitors.
\newblock {\em Formal Methods in System Design}, 46(3):226--261, 2015.

\bibitem{Hennessy1985}
M.~Hennessy and R.~Milner.
\newblock Algebraic laws for nondeterminism and concurrency.
\newblock {\em J. ACM}, 32(1):137--161, Jan. 1985.

\bibitem{JacquesSilva2012}
G.~Jacques-Silva, B.~Gedik, R.~Wagle, K.-L. Wu, and V.~Kumar.
\newblock Building user-defined runtime adaptation routines for stream
  processing applications.
\newblock {\em Proc. VLDB Endow.}, 5(12):1826--1837, Aug. 2012.

\bibitem{Ligatti2005}
D.~W. Jay~Ligatti, Lujo~Bauer.
\newblock Edit automata: Enforcement mechanisms for run-time security policies,
  2005.

\bibitem{Rosu2012JavaMop}
D.~Jin, P.~O. Meredith, C.~Lee, and G.~Rosu.
\newblock Javamop: Efficient parametric runtime monitoring framework.
\newblock In M.~Glinz, G.~C. Murphy, and M.~Pezzè, editors, {\em ICSE}, pages
  1427--1430. IEEE Computer Society, 2012.

\bibitem{Kell2008Survey}
S.~Kell.
\newblock A survey of practical software adaptation techniques.
\newblock {\em J.UCS}, 14(13):2110--2157, 2008.

\bibitem{Kozen1983MuCalc}
D.~C. Kozen.
\newblock Results on the propositional $\mu$-calculus.
\newblock {\em Theoretical Computer Science}, 27:333--354, 1983.

\bibitem{Kupferman2001}
O.~Kupferman and M.~Y. Vardi.
\newblock Model checking of safety properties.
\newblock {\em Formal Methods in System Design}, 19(3):291--314, 2001.

\bibitem{Lamport1980}
L.~Lamport.
\newblock "sometime" is sometimes "not never": On the temporal logic of
  programs.
\newblock In {\em Proceedings of the 7th ACM SIGPLAN-SIGACT Symposium on
  Principles of Programming Languages}, POPL '80, pages 174--185. ACM, 1980.

\bibitem{Lang2012TACAS}
F.~Lang and R.~Mateescu.
\newblock Partial model checking using networks of labelled transition systems
  and boolean equation systems.
\newblock In C.~Flanagan and B.~K{\"o}nig, editors, {\em TACAS}, pages
  141--156, Berlin, Heidelberg, 2012. Springer Berlin Heidelberg.

\bibitem{Larsen1990}
K.~G. Larsen.
\newblock Proof systems for satisfiability in hennessy-milner logic with
  recursion.
\newblock {\em Theoretical Computer Science}, 72(2):265--288, 1990.

\bibitem{Leucker2009}
M.~Leucker and C.~Schallhart.
\newblock A brief account of runtime verification.
\newblock {\em The Journal of Logic and Algebraic Programming}, 78(5):293--303,
  2009.

\bibitem{Ligatti2009}
J.~Ligatti, L.~Bauer, and D.~Walker.
\newblock Run-time enforcement of nonsafety policies.
\newblock {\em ACM Trans. Inf. Syst. Secur.}, 12(3):19:1--19:41, Jan. 2009.

\bibitem{Ligatti2010}
J.~Ligatti and S.~Reddy.
\newblock {\em A Theory of Runtime Enforcement, with Results}, pages 87--100.
\newblock Springer Berlin Heidelberg, Berlin, Heidelberg, 2010.

\bibitem{Ligatti2006PhD}
J.~A. Ligatti.
\newblock {\em Policy Enforcement via Program Monitoring}.
\newblock PhD thesis, Princeton, NJ, USA, 2006.
\newblock AAI3214569.

\bibitem{Rinard2012}
F.~Long, V.~Ganesh, M.~Carbin, S.~Sidiroglou, and M.~Rinard.
\newblock Automatic input rectification.
\newblock In {\em ICSE}, pages 80--90, Piscataway, NJ, USA, 2012. IEEE Press.

\bibitem{Martinelli2005}
F.~Martinelli and I.~Matteucci.
\newblock Partial model checking, process algebra operators and satisfiability
  procedures for (automatically) enforcing security properties.
\newblock In {\em Foundations of Computer Security}, page 133. Citeseer, 2005.

\bibitem{Martinelli2006Wits}
F.~Martinelli and I.~Matteucci.
\newblock Modeling security automata with process algebras and related results.
\newblock In {\em Informal proceedings Presented at the 6th International
  Workshop on Issues in the Theory of Security (WITS’06)(March 2006)}.
  Citeseer, 2006.

\bibitem{Matteucci2006Tool}
I.~Matteucci.
\newblock A tool for the synthesis of controller programs.
\newblock In {\em International Workshop on Formal Aspects in Security and
  Trust}, pages 112--126. Springer, 2006.

\bibitem{Milner1992CCS}
R.~Milner, J.~Parrow, and D.~Walker.
\newblock A calculus of mobile processes, i.
\newblock {\em Information and computation}, 100(1):1--40, 1992.

\bibitem{Nain2007}
S.~Nain and M.~Y. Vardi.
\newblock {\em ATVA 2007 Tokyo}, chapter Branching vs. Linear Time: Semantical
  Perspective, pages 19--34.
\newblock Springer, Berlin, Heidelberg, 2007.

\bibitem{Pinchinat2005}
S.~Pinchinat and S.~Riedweg.
\newblock You can always compute maximally permissive controllers under partial
  observation when they exist.
\newblock In {\em Proceedings of the 2005, American Control Conference, 2005.},
  pages 2287--2292. IEEE, 2005.

\bibitem{Pinisetty2014}
S.~Pinisetty, Y.~Falcone, T.~J{\'e}ron, and H.~Marchand.
\newblock {Runtime Enforcement of Parametric Timed Properties with Practical
  Applications}.
\newblock In {\em {IEEE International Workshop on Discrete Event Systems}},
  pages 46--53, cachan, France, May 2014.

\bibitem{Pinisetty2012}
S.~Pinisetty, Y.~Falcone, T.~J{\'e}ron, H.~Marchand, A.~Rollet, and O.~L.
  Nguena~Timo.
\newblock Runtime enforcement of timed properties.
\newblock In S.~Qadeer and S.~Tasiran, editors, {\em Runtime Verification},
  pages 229--244, Berlin, Heidelberg, 2013. Springer Berlin Heidelberg.

\bibitem{Pinisetty2017}
S.~Pinisetty, P.~S. Roop, S.~Smyth, N.~Allen, S.~Tripakis, and R.~V. Hanxleden.
\newblock Runtime enforcement of cyber-physical systems.
\newblock {\em ACM Trans. Embed. Comput. Syst.}, 16(5s):178:1--178:25, Sept.
  2017.

\bibitem{Pinisetty2016}
S.~Pinisetty, P.~S. Roop, S.~Smyth, S.~Tripakis, and R.~von Hanxleden.
\newblock Runtime enforcement of reactive systems using synchronous enforcers.
\newblock {\em CoRR}, abs/1612.05030, 2016.

\bibitem{Pnueli1977}
A.~Pnueli.
\newblock The temporal logic of programs.
\newblock In {\em Foundations of Computer Science, 1977., 18th Annual Symposium
  on}, pages 46--57. IEEE, 1977.

\bibitem{Pnueli2006}
A.~Pnueli and A.~Zaks.
\newblock Psl model checking and run-time verification via testers.
\newblock In {\em International Symposium on Formal Methods}, pages 573--586.
  Springer, 2006.

\bibitem{Pnueli1990}
Z.~M.~A. Pnueli.
\newblock A hierarchy of temporal properties.
\newblock {\em Proc. of the 2th symph. ACM of principle of distributed
  computer}, 1990.

\bibitem{Ramadge1987}
P.~J. Ramadge and W.~M. Wonham.
\newblock Supervisory control of a class of discrete event processes.
\newblock {\em SIAM J. Control Optim.}, 25(1):206--230, Jan. 1987.

\bibitem{schneider2000}
F.~B. Schneider.
\newblock Enforceable security policies.
\newblock {\em ACM Transactions on Information and System Security (TISSEC)},
  3(1):30--50, 2000.

\bibitem{Stirling1996}
C.~Stirling.
\newblock Model checking and other games.
\newblock In {\em Notes for Mathfit Workshop on finite model theory, University
  of Wales, Swansea}, 1996.

\bibitem{Streett1989}
R.~S. Streett and E.~A. Emerson.
\newblock An automata theoretic decision procedure for the propositional
  mu-calculus.
\newblock {\em Inf. Comput.}, 81(3):249--264, June 1989.

\bibitem{talhi2008}
C.~Talhi, N.~Tawbi, and M.~Debbabi.
\newblock Execution monitoring enforcement under memory-limitation constraints.
\newblock {\em Information and Computation}, 206(2):158--184, 2008.

\bibitem{Tarski1955}
A.~Tarski.
\newblock A lattice-theoretical fixpoint theorem and its applications.
\newblock {\em Pacific J. Math.}, 5(2):285--309, 1955.

\bibitem{VanHulst2017}
A.~C. van Hulst, M.~A. Reniers, and W.~J. Fokkink.
\newblock Maximally permissive controlled system synthesis for non-determinism
  and modal logic.
\newblock {\em Discrete Event Dynamic Systems}, 27(1):109--142, Mar 2017.

\bibitem{Viswanathan2005}
M.~Viswanathan and M.~Kim.
\newblock Foundations for the run-time monitoring of reactive systems --
  fundamentals of the mac language.
\newblock In Z.~Liu and K.~Araki, editors, {\em Theoretical Aspects of
  Computing - ICTAC 2004}, pages 543--556, Berlin, Heidelberg, 2005. Springer
  Berlin Heidelberg.

\bibitem{Wolff2013}
E.~M. Wolff, U.~Topcu, and R.~M. Murray.
\newblock Efficient reactive controller synthesis for a fragment of linear
  temporal logic.
\newblock In {\em 2013 IEEE International Conference on Robotics and
  Automation}, pages 5033--5040, May 2013.

\end{thebibliography}

\newpage
\appendix

\chapter{Supporting Material}
\label{sec:app:supporting-material}
\newcommand{\map}{[\hVarX\!\mapsto\!S]}

This Appendix Chapter contains extra supporting information and examples that may help in better understanding our contributions. Within this section we refer to the following processes which are also pictorially represented in \Cref{fig:ex:lts}.\medskip
\begin{align*}
	\pV_1=&\rec{\mx}{\bigl(\esel{\prf{\actReq} \prf{\actAns}\mx}{\prf{\actCls}{\nil}}\bigr)} \\
	\pVV_1=&\rec{\mx}{\bigl(\esel{\prf{\actReq}{\prf{\actAns}\mx}{\esel{\,}{\,{\prf{\actReq}\mx}}}}{\prf{\actCls}{\nil}}\bigr)}\\
	\pVVV_1=&\;\rec{\mx}{\bigl(\esel{\prf{\actReq}{ \prf{\actAns}{\bigl({\esel{\prf{\actReq}{\prf{\actAns}{\mx}}}
						{\prf{\actCls}{\nil}}}\bigr) }}}{\prf{\actCls}{\nil}}\bigr)} \\
	\pVVVV_1=&\;\rec{\mx}{\bigl(\esel{\prf{\actt}{ \prf{\actReq}{ \prf{\actAns}\mx}}}{\prf{\actCls}{\nil}}\bigr)}
\end{align*}

\section{Bisimulation Game Characterisation}
\label{sec:app:bisim-games}

A \emph{Strong (}\resp \emph{Weak) bisimulation game} for LTS processes \pV and \pVV, is a turn-based game between two players, namely an \emph{attacker} which aims to disprove the Strong (\resp Weak) bisimulation, and a \emph{defender} which aims to prove the Strong (\resp Weak) bisimulation. The game is played in rounds, each of which considers a pair (\aka configuration) of LTS processes $(\pV_n,\pVV_n)$. The game starts its first round with the initial pair being $(\pV,\pVV)$, and in each round the players change the current pair according to the following rules:
\begin{enumerate}
	\item Given a pair $(\pV,\pVV)$, the attacker chooses either the first or second element, \ie \pV or \pVV, along with a suitable action \actu. Based on the choice, the attacker must perform a transition over the selected action \actu, \ie $\pV\traS{\actu}\pV'$ if \pV is chosen, and  $\pVV\traS{\actu}\pVV'$ when \pVV is chosen.
	\item To counter the attack, the defender must reply by making a transition over the same action \actu, by using the process which was not selected by the attacker, \ie if the attacker chose \pV and attacks with $\pV\traS{\actu}\pV'$, the defender must reply with a strong \actu-reduction $\pVV\traS{\actu}\pVV'$, when proving a Strong Bisimulation, and with a weak \actu-reduction $\pVV\wtraS{\hat{\actu}}\pVV'$, when proving a Weak one.
	\item After successfully defending an attack $\pV\traS{\actu}\pV'$, with a defensive Strong (\resp Weak) transition, the new pair becomes $(\pV',\pVV')$. The game continues for another round using the same rules.
\end{enumerate}
A \emph{play} of the game consists in a \emph{maximal sequence} of pairs constructed by the players according to the given rules, meaning that during a play an attacker must try every possible attack with the attempt to disprove the bisimulation, while the defender must defend for these attacks to prevent this. Hence, an \emph{attacker wins a finite play} whenever it is able to issue an attack, \eg $\pV\traS{\actu}\pV'$, for which the defender is unable to find a countering defensive move, \ie $\pV\ntraS{\actu}$ in a Strong Bisimulation Game, and $\pV\nwtraS{\hat{\actu}}$ in a Weak Game; otherwise the defender wins. In an \emph{infinite} play (\ie when comparing infinite LTSs), the defender wins if the play is infinite, \ie the attacker is bound to keep attacking infinitely with attacks that can be countered by the defender.

Whenever, one of the players, either the attacker or the defender, is able to provide a strategy that can \emph{always win the game}, regardless of how the opposing player chooses its moves, then we say that the player has a \emph{universal winning strategy}. Hence, two LTS processes \pV and \pVV are said to be \emph{Strong} (\resp \emph{Weak}) \emph{bisimilar} iff the \emph{defender} has a \emph{universal winning strategy} in the Strong (\resp Weak) bisimulation game which starts with the initial pair $(\pV,\pVV)$, otherwise the processes are not bisimilar iff the \emph{attacker} has a \emph{universal winning strategy}. 

The bisimulation games thus allow us to prove whether two processes are Strong (\resp Weak) bisimilar to each other or not. Using a strong bisimulation game we are able to find a strong bisimulation relation \R proving that processes $\pV_1$ and $\pVVV_1$ produce the \emph{same} runtime behaviour, \ie they both provide an answer for every request with the possibility of closing while waiting for requests. 

\begin{example}
	Recall example \Cref{ex:strong-bisim-p-r}, in this example we give an alternative way how to show that processes $\pV_1$ and $\pVVV_1$ from \Cref{fig:ex:lts} are Strongly Bisimilar via the \emph{strong bisimulation game characterization}. With this game we must therefore obtain a Strong bisimulation relation \R that proves that these two processes exhibit the same behaviour.
	\begin{proof}[\;for\;$\pmb{\pV_1\!\bisim\!\pVVV_1}$]
		To prove that $\pV_1\bisim\pVVV_1$ we use the \emph{Strong bisimulation game}, starting with the initial pair being $(\pV_1,\pVVV_1)$, to find a relation \R that observes \Cref{def:strong-bisim}. We denote an attacker's and defender's move as $A:$ and $D:$ \resp 
			
		\begin{block}[\emphbf{Round 1:}]
			We show $(\pV_1,\pVVV_1)\in\R$ as,
			\[\begin{array}{ccccc}
			\text{  A:  } &  \pV_1 \traS{\actReq}{\color{red}\pV_2}  & \qquad & \text{  D:  } & \pVVV_1 \traS{\actReq}{\color{red}\pVVV_2} \\
			\text{  A:  } &  \pV_1 \traS{\quad\actCls\quad}{\color{blue}\pV_3}  & \qquad & \text{  D:  } & \pVVV_2 \traS{\quad\actCls\quad}{\color{blue}\pVVV_5} \\
			\text{  A:  } & \pVVV_1 \traS{\actReq}{\color{red}\pVVV_2}  & \qquad &  \text{  D:  } & \pV_1 \traS{\actReq}{\color{red}\pV_2}\\
			\text{  A:  } & \pVVV_1 \traS{\quad\actCls\quad}{\color{blue}\pVVV_5}  & \qquad &  \text{  D:  } & \pV_1 \traS{\quad\actCls\quad}{\color{blue}\pV_3}
			\end{array}\]
			In the next round we must show $({\color{red}\pV_2},{\color{red}\pVVV_2})\in\R$ and $({\color{blue}\pV_3},{\color{blue}\pVVV_5})\in\R$.
		\end{block}
		
		\begin{block}[\emphbf{Round 2:}]
			We show $(\pV_2,\pVVV_2)\in\R$ as,
			\[\begin{array}{ccccc}
			\text{  A:  } &  \pV_2 \traS{\actAns}{\color{black}\pV_1}  & \qquad & \text{  D:  } & \pVVV_2 \traS{\actAns}{\color{black}\pVVV_3} \\
			\text{  A:  } &  \pVVV_2 \traS{\actAns}{\color{black}\pVVV_3}  & \qquad & \text{  D:  } & \pV_2 \traS{\actAns}{\color{black}\pV_1}
			\end{array}\]
			The case for $(\pV_3,\pVVV_5)\in\R$ holds immediately as neither $\pV_3$, nor $\pVVV_5$ can perform any further reductions. 
			In the next round we must show $(\pV_1,\pVVV_3)\in\R$.
		\end{block}
		
		\begin{block}[\emphbf{Round 3:}]
			We show $(\pV_1,\pVVV_3)\in\R$ as,
			\[\begin{array}{ccccc}
			\text{  A:  } &  \pV_1 \traS{\actReq}{\color{red}\pV_2}  & \qquad & \text{  D:  } & \pVVV_3 \traS{\actReq}{\color{red}\pVVV_4} \\
			\text{  A:  } &  \pV_1 \traS{\actCls}{\color{blue}\pV_3}  & \qquad & \text{  D:  } & \pVVV_3 \traS{\actCls}{\color{blue}\pVVV_5} \\
			\text{  A:  } & \pVVV_3 \traS{\actReq}{\color{red}\pVVV_4}  & \qquad &  \text{  D:  } & \pV_1 \traS{\actReq}{\color{red}\pV_2}\\
			\text{  A:  } & \pVVV_3 \traS{\actCls}{\color{blue}\pVVV_4}  & \qquad &  \text{  D:  } & \pV_1 \traS{\actCls}{\color{blue}\pV_3}
			\end{array}\]
			In the next round we must show that $(\pV_2,\pVVV_4)\in\R$.
		\end{block}
		
		\begin{block}[\emphbf{Round 4:}]
			We show $(\pV_2,\pVVV_4)\in\R$ as,
			\[\begin{array}{ccccc}
			\text{  A:  } &  \pV_2 \traS{\actAns}{\color{darkgreen}\pV_1}  & \qquad & \text{  D:  } & \pVVV_4 \wtraS{\actAns}{\color{darkgreen}\pVVV_1} \\
			\text{  A:  } & \pVVV_4 \traS{\actAns}{\color{darkgreen}\pVVV_1}  & \qquad &  \text{  D:  } & \pV_2 \wtraS{\actAns}{\color{darkgreen}\pV_1}\\
			\end{array}\]
			No further rounds are required for this game since we already know $(\pV_1,\pVVV_1)\in\R$ from Round 1.
		\end{block}
		Hence, given that the defender was always able to defend with a strong transition for every possible attack, we can conclude that there exists the following strong bisimulation relation \R, such that by the definition of strong bisimilarity we can conclude that $\pV_1\bisim\pVVV_1$, where:
		\[\R\defeq\Set{(\pV_1,\pVVV_1),(\pV_2,\pVVV_2),(\pV_1,\pVVV_3), (\pV_2,\pVVV_4), (\pV_3,\pVVV_5)}\]
	\end{proof}
\end{example}

\begin{example} 
	In \Cref{sec:lts} we had argued that showing $\pV_1{\not\bisim}\pVV_1$ is in general very hard to deduce since we need to prove that \emph{all} binary relations \R that can relate $\pV_1$ and $\pVV_1$, are \emph{not} a Strong Bisimulation relation. A more practical alternative is to resort to the bisimulation games. In the following proof we will use the \emph{Weak Bisimulation Games} to show that $\pV_1{\not\wbisim}\pVV_1$, which inherently implies that $\pV_1{\not\bisim}\pVV_1$.
	
	\begin{proof}[\;that\;$\pmb{\pV_1{\not\wbisim}\pVV_1}$]
		To prove that ${\not\bisim}$ we use the \emph{Weak Bisimulation Game}, starting with the initial pair being $(\pV_1,\pVV_1)$. 
		
		\begin{block}[\emphbf{Round 1:}]
			We show $(\pV_1,\pVV_1)\in\R$ as,
			\[\begin{array}{ccccc}
			\text{  A:  } &  \pV_1 \traS{\actReq}{\color{red}\pV_2}  & \qquad & \text{  D:  } & \pVV_1 \wtraS{\actReq}{\color{red}\pVV_2} \\
			\text{  A:  } &  \pV_1 \traS{\actCls}{\color{blue}\pV_3}  & \qquad & \text{  D:  } & \pVV_1 \wtraS{\actCls}{\color{blue}\pVV_3} \\
			\text{  A:  } & \pVV_1 \traS{\actReq}{\color{red}\pVV_2}  & \qquad &  \text{  D:  } & \pV_1 \wtraS{\actReq}{\color{red}\pV_2}\\
			\text{  A:  } & \pVV_1 \traS{\actReq}{\color{darkgreen}\pVV_1}  & \qquad &  \text{  D:  } & \pV_1 \wtraS{\actReq}{\color{darkgreen}\pV_2}\\
			\text{  A:  } & \pVV_1 \traS{\actCls}{\color{blue}\pVV_3}  & \qquad &  \text{  D:  } & \pV_1 \wtraS{\actCls}{\color{blue}\pV_3}
			\end{array}\]
			In the next round we must show $({\color{red}\pV_2},{\color{red}\pVV_2})\in\R$, $({\color{darkgreen}\pV_2},{\color{darkgreen}\pVV_1})\in\R$ and $({\color{blue}\pV_3},{\color{blue}\pVV_3})\in\R$.
		\end{block}
		
	\begin{block}[\emphbf{Round 2:}]
		We show $(\pV_3,\pVV_3)\in\R$ holds immediately as neither $\pV_3$ nor $\pVV_3$ can perform any further reductions. We also show $(\pV_2,\pVV_2)\in\R$ as,
		\[\begin{array}{ccccc}
			\text{  A:  } &  \pV_2 \traS{\actAns}{\color{black}\pV_1}  & \qquad & \text{  D:  } & \pVV_2 \wtraS{\actAns}{\color{black}\pVV_1} \\
			\text{  A:  } &  \pVV_2 \traS{\actAns}{\color{black}\pVV_1}  & \qquad & \text{  D:  } & \pV_2 \wtraS{\actAns}{\color{black}\pV_1}
			\end{array}\]
			However, the defender fails to counter the attacker's attack when showing that $(\pV_2,\pVV_1)\in\R$, since
			\[\begin{array}{ccccc}
			\text{  A:  } &  \pV_2 \traS{\actAns}{\color{black}\pV_1}  & \qquad & \text{  D:  } & \pVV_1 {\color{red}\pmb{\nwtraS{\actAns}}} 
			\end{array}\]			
		\end{block}
	Hence, given that the defender was unable to defend for one of the attacks, by \Cref{def:weak-bisim} we can conclude that $\pV_1{\not\wbisim}\pV_2$, which inherently implies that $\pV_1{\not\bisim}\pV_2$. 
	\end{proof}
\end{example}

\section{An Iterative Fixpoint Derivation Example}
\label{sec:app:tarski-proof}
Recall formula $\hV_{1}$ from \Cref{ex:uhml-formula} (restated below),
\[
\begin{array}{rcl@{\qquad\qquad}rcl}
	\hV_1 &=&\hmax{\hVarX}{\hV_1'} &
	\hV_1' &=& \hnec{\patReqA}{\hV_1''}\\	
	\hV_1'' &=&{\hV_1'''}\hand{\hV_1''''} &	
	\hV_1''' &=&\hnec{\patAns}{\hVarX}\\
	\hV_1'''' &=&\hnec{\patReqB}{\hfls}
\end{array}
\]
\noindent We can thus prove that $\pV_1\in\syn{\hV_1}$ as follows:
\begin{align*}
\syn{\hV_1} &= \sem{\hmax{\hVarX}{\hV_1'}}{\emptyset}\\
&=  S \subseteq \sem{\hV_1'}{\map}
\end{align*}
To establish the maximal set $S$ of processes which satisfy $\hV_1'$, we must iteratively evaluate $\sem{\hV_1'}{\map}$ starting with $S=\Proc=\set{\pV_1,\pV_2,\pV_3,\pVV_1,\pVV_2,\pVV_3}$ until the resultant set $S$ stops changing with each iteration, therefore denoting that the maximal fixpoint has been reached.

\newcommand{\sVD}{\sub{\vV}{\dvV}}
\newcommand{\sVF}{\sub{\vVV}{\dvVV}}
\newcommand{\symAnsA}{\actSN{\actAnsV}{\ctru}}

\paragraph{First Iteration} \pmb{$(S=\Proc)$} 

\[
\begin{array}{r@{}c@{}l}
\sem{\hV_1'}{\map} &=& \Setdef{\pV}{ \begin{xbrackets}{c}
		\forall\vV\in\Val\cdot \pV\wtraS{\actReqV}\pV' \textsl{   and}  \\[3mm] \mtchS{\symReqA}{\actReqV}\!=\!\sVD
	\end{xbrackets}\! \! \! \imp\! \pV'\in\sem{\hV_1''\sVD}{\map}} \\[6mm]
\multicolumn{3}{l}
{
	\begin{array}{|rcl}
	\sem{\hV_1''\sVD}{\map}  &=& \sem{\hV_1'''\sVD}{\map}\cap\sem{\hV_1''''\sVD}{\map}\\[2mm]
	\multicolumn{3}{|@{\quad}l}
	{
		\begin{array}{|rcl}
		\sem{\hV_1'''\sVD}{\map}  &=& \sem{\hnec{\patAns}{\hVarX}\sVD}{\map} \;\;=\;\; \sem{\hnec{\actAnsV}{\hVarX}}{\map} \\[2mm]
		\multicolumn{3}{|@{\qquad\quad}l}
		{
			\begin{array}{rcl}
				&=& \Setdef{\pV}{\begin{xbrackets}{c}\pV\wtraS{\actAnsV}\pV' \qquad\textsl{      and } \\[3mm]\mtchS{\symAnsA}{\actAnsV}\!=\!\sE\end{xbrackets}\! \! \! \imp\! \pV'\in\sem{\hVarX\sE}{\map}} \\[7mm]
				&=& \Setdef{\pV}{\begin{xbrackets}{c}\pV\wtraS{\actAnsV}\pV' \qquad\textsl{      and } \\[3mm]\mtchS{\symAnsA}{\actAnsV}\!=\!\sE\end{xbrackets}\! \! \! \imp\! \pV'\in S} \\
				&=& S=\Proc
			\end{array}
		}
		\end{array}
	}
	\end{array}
}
\end{array}
\]
\[
\begin{array}{r@{}c@{}l}
\multicolumn{3}{l}
{
	\begin{array}{|rcl}	
	\multicolumn{3}{|@{\quad}l}
	{
		\begin{array}{|rcl}
		\sem{\hV_1''''\sVD}{\map}  &=& \sem{\hnec{\patReqB}{\hfls}\sVD}{\map} \;\;=\;\; \sem{\hnec{\patReqB}{\hfls}}{\map} \\[2mm]
		\multicolumn{3}{|@{\qquad\quad}l}
		{
			\begin{array}{rcl}
			&=& \Setdef{\pV}{\begin{xbrackets}{c}\forall\vVV\in\Val\cdot \pV\wtraS{\actReqVV}\pV' \textsl{      and } \\[3mm]\mtchS{\symReqB}{\actReqVV}\!=\!\sVF\end{xbrackets}\! \! \! \imp\! \pV'\in\sem{\hfls\sVF}{\map}} \\[7mm]
			&=& \Setdef{\pV}{\begin{xbrackets}{c}\forall\vVV\in\Val\cdot \pV\wtraS{\actReqVV}\pV' \textsl{      and } \\[3mm]\mtchS{\symReqB}{\actReqVV}\!=\!\sVF\end{xbrackets}\! \! \! \imp\! \pV'\in\emptyset} \\
			&=& \set{\pV_2,\pV_3,\pVV_2,\pVV_3} 
			\end{array}
		}
		\end{array}
	}\\[2mm]
	\sem{\hV_1''\sVD}{\map} &=& \Proc\cap\set{\pV_3,\pV_4,\pVV_3,\pVV_4}\\
	&=& \set{\pV_3,\pV_4,\pVV_3,\pVV_4}	
	\end{array}
}\\[2mm]
\sem{\hV_1'}{\map} &=&   \Setdef{\pV}{ \begin{xbrackets}{c}
	\forall\vV\in\Val\cdot \pV\wtraS{\actReqV}\pV' \textsl{   and}  \\[3mm] \mtchS{\symReqA}{\actReqV}\!=\!\sVD
	\end{xbrackets}\! \! \! \imp\! \pV'\in\set{\pV_3,\pV_4,\pVV_3,\pVV_4}}\\
S &=& \set{\pV_1,\pV_2,\pV_3,\pVV_2,\pVV_3} 
\end{array}
\]
\\

\paragraph{Second Iteration} \pmb{$(S=\set{\pV_1,\pV_2,\pV_3,\pVV_2,\pVV_3})$} 

\[
\begin{array}{r@{}c@{}l}
\sem{\hV_1'}{\map} &=& \Setdef{\pV}{ \begin{xbrackets}{c}
	\forall\vV\in\Val\cdot \pV\wtraS{\actReqV}\pV' \textsl{   and}  \\[3mm] \mtchS{\symReqA}{\actReqV}\!=\!\sVD
	\end{xbrackets}\! \! \! \imp\! \pV'\in\sem{\hV_1''\sVD}{\map}} \\[6mm]
\multicolumn{3}{l}
{
	\begin{array}{|rcl}
	\sem{\hV_1''\sVD}{\map}  &=& \sem{\hV_1'''\sVD}{\map}\cap\sem{\hV_1''''\sVD}{\map}\\[2mm]
	\multicolumn{3}{|@{\quad}l}
	{
		\begin{array}{|rcl}
		\sem{\hV_1'''\sVD}{\map}  &=& \sem{\hnec{\patAns}{\hVarX}\sVD}{\map} \;\;=\;\; \sem{\hnec{\actAnsV}{\hVarX}}{\map} \\[2mm]
		\multicolumn{3}{|@{\qquad\quad}l}
		{
			\begin{array}{rcl}
			&=& \Setdef{\pV}{\begin{xbrackets}{c}\pV\wtraS{\actAnsV}\pV' \qquad\textsl{      and } \\[3mm]\mtchS{\symAnsA}{\actAnsV}\!=\!\sE\end{xbrackets}\! \! \! \imp\! \pV'\in\sem{\hVarX\sE}{\map}} \\[7mm]
			&=& \Setdef{\pV}{\begin{xbrackets}{c}\pV\wtraS{\actAnsV}\pV' \qquad\textsl{      and } \\[3mm]\mtchS{\symAnsA}{\actAnsV}\!=\!\sE\end{xbrackets}\! \! \! \imp\! \pV'\in S} \\
			&=& S=\set{\pV_1,\pV_2,\pV_3,\pVV_2,\pVV_3}
			\end{array}
		}
		\end{array}
	}
	\end{array}
}
\end{array}
\]
\[
\begin{array}{r@{}c@{}l}
\multicolumn{3}{l}
{
	\begin{array}{|rcl}	
	\multicolumn{3}{|@{\quad}l}
	{
		\begin{array}{|rcl}
		\sem{\hV_1''''\sVD}{\map}  &=& \sem{\hnec{\patReqB}{\hfls}\sVD}{\map} \;\;=\;\; \sem{\hnec{\patReqB}{\hfls}}{\map} \\[2mm]
		\multicolumn{3}{|@{\qquad\quad}l}
		{
			\begin{array}{rcl}
			&=& \Setdef{\pV}{\begin{xbrackets}{c}\forall\vVV\in\Val\cdot \pV\wtraS{\actReqVV}\pV' \textsl{      and } \\[3mm]\mtchS{\symReqB}{\actReqVV}\!=\!\sVF\end{xbrackets}\! \! \! \imp\! \pV'\in\sem{\hfls\sVF}{\map}} \\[7mm]
			&=& \Setdef{\pV}{\begin{xbrackets}{c}\forall\vVV\in\Val\cdot \pV\wtraS{\actReqVV}\pV' \textsl{      and } \\[3mm]\mtchS{\symReqB}{\actReqVV}\!=\!\sVF\end{xbrackets}\! \! \! \imp\! \pV'\in\emptyset} \\
			&=& \set{\pV_2,\pV_3,\pVV_2,\pVV_3} 
			\end{array}
		}
		\end{array}
	}\\[2mm]
	\sem{\hV_1''\sVD}{\map} &=& \set{\pV_1,\pV_2,\pV_3,\pVV_2,\pVV_3}\cap\set{\pV_3,\pV_4,\pVV_3,\pVV_4}\\
	&=& \set{\pV_3,\pV_4,\pVV_3,\pVV_4}	
	\end{array}
}\\[2mm]
\sem{\hV_1'}{\map} &=&   \Setdef{\pV}{ \begin{xbrackets}{c}
	\forall\vV\in\Val\cdot \pV\wtraS{\actReqV}\pV' \textsl{   and}  \\[3mm] \mtchS{\symReqA}{\actReqV}\!=\!\sVD
	\end{xbrackets}\! \! \! \imp\! \pV'\in\set{\pV_3,\pV_4,\pVV_3,\pVV_4}}\\
S &=& \set{\pV_1,\pV_2,\pV_3,\pVV_2,\pVV_3} 
\end{array}
\]

\bigskip

\noindent Notice how so far in our derivation, the set of processes $S$ has changed from \Proc to $\Proc\setminus\set{\pVV_1}$ after the first iteration, but has remained the same (\ie $\Proc\setminus\set{\pVV_1}$) after the second one. Hence, by the definition of $\sem{\hmax{\hVarX}{\hV}}{\emptyset}$, we can conclude:
\[ 
\begin{array}{rcl}
	\syn{\hV_1} &\defeq& \sem{\hmax{\hVarX}{\hnec{\actReq}(\hnec{\actReq}\hfls\,\hand\,\hnec{\actAns}\hVarX)}}{\emptyset} \\
			&=& \set{\pV_1,\pV_2,\pV_3,\pVV_2,\pVV_3}
\end{array}
\]

\chapter{Proving Semantic Preservation for Normalization}	
\label{sec:app:proofs-new-determinization}
As explained in \Cref{sec:normalization}, the algorithm for normalizing \emph{any} \shml formula (\ie defining any type of symbolic event and not just singleton events) amounts to the sequential application of the following 6 steps:
\begin{enumerate}[Step 1.]
	\item \label{norm-step:1} For each \shml formula \hV a \emph{semantically equivalent} formula $\hVSF$ that is in \emph{Standard Form} can be obtained using $\gensf{\hV}$.
	\begin{itemize}
		\item This construction was reviewed and explained \wrt symbolic events in \Cref{sec:form-to-sfform} along with a proof advocating for its semantic preservation.
	\end{itemize}
	\item \label{norm-step:2} For each \shml formula \hV that is in \emph{Standard Form}, an \emph{equivalent} System of Equations \sysSF that is in \emph{Standard Form} can be obtained using $\gensys{\hVSF}$.
	\begin{itemize}
		\item We have discussed this construction \wrt symbolic events in \Cref{sec:sfform-to-sfsys} accompanied by a proof denoting that the resultant system of equations \sysSF is semantically equivalent to the input formula \hVSF.
	\end{itemize}	
	\item \label{norm-step:4} For each System of Equations \sysSF that is in \emph{Standard Form}, an \emph{equivalent} System of Equations $\sysUNI$ that is \emph{Uniform} can be obtained using $\genuni{\sysSF}$.
	\begin{itemize}
		\item This construction was introduced and explained in \Cref{sec:alpha-equiv-cons}, for which we proved that the obtained uniform system of equations \sysUNI is semantically equivalent to the given standard form system of equations.
		\item The proof provided makes use of two lemmas, namely, \Cref{lemma:norm-1a} and \Cref{lemma:norm-1b}; we prove these lemmas in \Cref{sec:proof-uni-lemmas}.
	\end{itemize}	
	\item \label{norm-step:5} For each System of Equations \sysUNI that is \emph{Uniform}, an \emph{equivalent} System of Equations $\sysREC$ that is \emph{Equi-Disjoint}, can be obtained using $\gencond{\sysUNI}$.
	\begin{itemize}
		\item An introduction and explanation of this construction was given in \Cref{sec:truth-combos-cons}. In the same section we also prove that our construction is guaranteed to preserve the semantics of the given Uniform system of equations.
		\item The proof guaranteeing Semantic Preservation for this construction step makes use of \Cref{lemma:norm-2a} which we now prove in \Cref{sec:proof-comb-lemmas}.
	\end{itemize}
	\item \label{norm-step:6} For each System of Equations \sys that is \emph{Equi-Disjoint} and also in \emph{Standard Form}, an \emph{equivalent} System of Equations $\sysNF$ that is in \emph{Normal Form}, can be obtained using $\gendet{\sysREC}$.
	\begin{itemize}
		\item This construction step was presented and explained \wrt symbolic events in \Cref{sec:sfsys-to-detsys} accompanied by a proof sketch showing that the semantics of the original system of equations are preserved by the construction.
	\end{itemize}		
	\item \label{norm-step:7} For each System of Equations \sys that is \emph{Normal Form}, an \emph{equivalent} formula $\hVNF$ that is also in \emph{Normal Form}, can be obtained using $\genwf{\sysNF}$.
	\begin{itemize}
		\item This construction step was explained in relation to symbolic events in \Cref{sec:detsys-to-shml}, and a proof guaranteeing semantic preservation was also provided.
	\end{itemize}
\end{enumerate}

\section{Auxiliary Lemmas for Proving Semantic Preservation of Construction \pmb{\genuni{-}}} \label{sec:proof-uni-lemmas}
In this section we prove the following lemmas that are required for guaranteeing the semantic preservation of construction \genuni{-}.
\begin{description}
	\item[\Cref{lemma:norm-1a}.] \traverseFun{\eqq}{\set{0}}{\partition}{\emptyset}{=}\subMap \quad \textsl{implies} \\ \subMap is a \emph{well-formed} map for \eqq.
	\item[\Cref{lemma:norm-1b}.] $\forall(\hVarX_j{=}\hV_j){\,\in\,}\eqq\cdot$ equation $\hVarX_j{=}\hV_j$ is in \emph{Standard form}, and \subMap is a \emph{well-formed} map for \eqq \quad \textsl{implies} \quad $\uniFun{\eqq}{\subMap}{\equiv}\eqq$ and $\forall(\hVarX_k{=}\hVV_k){\,\in\,}\uniFun{\eqq}{\subMap}\cdot$ equation $(\hVarX_k{=}\hVV_k)$ is \emph{Uniform}.
\end{description}

\subsection{Proving \Cref{lemma:norm-1a}.}
In order to prove \Cref{lemma:norm-1a}, we use \Cref{lemma:norm-1c}. This new lemma states that a \emph{well-formed} $\subMap'$ map for \eqq is obtained upon performing a \partition traversal on a subset, $\eqq'$, of the given equation set \eqq, using an arbitrary \subMap map that is \emph{well-formed} \wrt a \emph{subset} of \eqq that is restricted to the indices defined by the domain of \subMap, \ie \eqqres{\eqq}{\dom{\subMap}}.
\begin{lemma}\label{lemma:norm-1c}
	$\forall\IndSet,\subMap\cdot \eqq'\subseteq\eqq\text{ and } \traverseFun{\eqq'}{\IndSet}{\partition}{\subMap}{=}\subMap' \text{ and } \subMap$ is a \emph{well-formed} map for \eqqres{\eqq}{\dom{\subMap}} \quad \textsl{implies} \quad 
	$\subMap'$ is a \emph{well-formed} map for \eqq.
\end{lemma}
The proof for this lemma is provided at the end of this section. 

\paragraph{To Prove \Cref{lemma:norm-1a}.}
\begin{center}
	\traverseFun{\eqq}{\set{0}}{\partition}{\emptyset}{=}\subMap \quad  \textsl{implies} \quad \subMap is a \emph{well-formed} map for \eqq.
\end{center}
\begin{proof}
	Initially we know 
	\begin{gather}
		\traverseFun{\eqq}{\set{0}}{\partition}{\emptyset}{=}\subMap \label{proof:lemma1a-1}
	\end{gather}
	By the definition of \eqqres{\eqq}{\IndSet} we know
	\begin{gather}
		\eqqres{\eqq}{\dom{\emptyset}}=\emptyset \label{proof:lemma1a-2}
	\end{gather}
	By \eqref{proof:lemma1a-2} and the definition of a \emph{well-formed} map we know
	\begin{gather}
		\emptyset \text{ is a \emph{Well-formed} map for }\eqqres{\eqq}{\dom{\emptyset}} \label{proof:lemma1a-3}
	\end{gather}
	By \eqref{proof:lemma1a-1}, \eqref{proof:lemma1a-3} and \Cref{lemma:norm-1c} we know
	\begin{gather*}
		\subMap\text{ is a \emph{well-formed} map for }\eqq
	\end{gather*}
\end{proof}

\paragraph{To Prove \Cref{lemma:norm-1c}.}
\begin{center}
	$\forall\IndSet,\subMap\cdot \eqq'\subseteq\eqq\text{ and } \traverseFun{\eqq'}{\IndSet}{\partition}{\subMap}{=}\subMap' \text{ and } \subMap$ is a \emph{well-formed} map for \eqqres{\eqq}{\dom{\subMap}} \quad \textsl{implies} \quad 
	$\subMap'$ is a \emph{well-formed} map for \eqq.
\end{center}
\begin{proof}[. \textbf{By induction on the structure of }$\pmb{\eqq'}$]
		
	\begin{case}[\eqq'=\emptyset]
		Initially we know
		\begin{gather}
			\traverseFun{\emptyset}{\IndSet}{\partition}{\subMap}{=}\subMap' \label{proof:lemma1c-bc-1} \\
			\subMap \text{ is a \emph{well-formed} map for } \eqqres{\eqq}{\dom{\subMap}} \label{proof:lemma1c-bc-2} \\
			\emptyset\subseteq\eqq
		\end{gather}
		Since $\eqq'{=}\emptyset$, by \eqref{proof:lemma1c-bc-1} and the definition of \traverse we know
		\begin{gather}
			\subMap=\subMap' \label{proof:lemma1c-bc-3}
		\end{gather}
		By \eqref{proof:lemma1c-bc-2} and \eqref{proof:lemma1c-bc-3} we know
		\begin{gather}
			\subMap' \text{ is a \emph{well-formed} map for } \eqqres{\eqq}{\dom{\subMap'}}  \label{proof:lemma1c-bc-4}
		\end{gather}
		By \eqref{proof:lemma1c-bc-1} and the definition of \traverse, we know that the traversal starts from the full equation set, \ie $\eqq'{\,=\,}\eqq$, using an empty \subMap map. With every recursive application of \traverse, the equation set $\eqq'$ 
		becomes smaller since when \traverse recurses it does so \wrt $\eqq''$, \ie a smaller version of the current $\eqq'$ which is computed via $\eqq''{=}\eqq'\setminus\eqqres{\eqq'}{\IndSet}$. By contrast, with every recursive application of \traverse, the \subMap accumulator becomes larger as it is updated with new mappings for each index specified by the set of indices \IndSet \ie with the indices of the equations that are removed from $\eqq'$ when creating $\eqq''$. Hence, when the \traverse function is recursively applied \wrt some $\eqq'''{=}\emptyset$, it means that all the equations specified in \eqq have been analysed by the traversal and their indices were thus added as maps in the resultant $\subMap'$. Hence, we can deduce
		\begin{gather}
			\eqqres{\eqq}{\dom{\subMap'}}=\eqq  \label{proof:lemma1c-bc-5}
		\end{gather}
		Finally, by \eqref{proof:lemma1c-bc-4} and \eqref{proof:lemma1c-bc-5} we conclude
		\begin{gather}
			\subMap' \text{ is a \emph{well-formed} map for } \eqq.
		\end{gather}
	\end{case}
	
	\begin{case-noreset}[\eqq'\neq\emptyset]
		Initially we know
		\begin{gather}
		\traverseFun{\eqq'}{\IndSet}{\partition}{\subMap}{=}\subMap' \label{proof:lemma1c-ic-1} \\
		\subMap \text{ is a \emph{well-formed} map for } \eqqres{\eqq}{\dom{\subMap}} \label{proof:lemma1c-ic-2} \\
		\eqq'\subseteq\eqq \label{proof:lemma1c-ic-2.5}
		\end{gather}
		We consider two subcases:
		\begin{itemize}
			\item[\pmb{--}] $\pmb{\IndSet=\emptyset:}$
				Since $\IndSet{=}\emptyset$, by \eqref{proof:lemma1c-ic-1} and the definition of \traverse we know
				\begin{gather}
				\subMap=\subMap' \label{proof:lemma1c-ic-4}
				\end{gather}
				By \eqref{proof:lemma1c-ic-2} and \eqref{proof:lemma1c-ic-4} we know
				\begin{gather}
				\subMap' \text{ is a \emph{well-formed} map for } \eqqres{\eqq}{\dom{\subMap'}}  \label{proof:lemma1c-ic-5}
				\end{gather}
					Since $\IndSet{=}\emptyset$, this means that the traversal has reached a point where no more children can be computed, which means that all the \emph{relevant equations} (\ie those reachable from the principle variable) have been analysed. This means that any other equation in \eqq (that is not in \eqqres{\eqq}{\dom{\subMap'}}, if any) is \emph{redundant} and \emph{irrelevant}. Hence, since from \eqref{proof:lemma1c-ic-5} we know that $\subMap''$ is a \emph{well-formed} map for the \emph{relevant subset} of equations in \eqq, \ie \eqqres{\eqq}{\dom{\subMap'}}, then it is also \emph{well-formed} for the full blown subset of equations \eqq (\ie including any unreachable, redundant equations). Therefore, we can conclude
				\begin{gather*}
				\subMap' \text{ is a \emph{well-formed} map for } \eqq. 
				\end{gather*}				
			\item[\pmb{--}] $\pmb{\IndSet{\neq}\emptyset:}$
				By \eqref{proof:lemma1c-ic-1} and the definition of \traverse we know
				\begin{gather}
					\subMap''=\partitionFun{\eqq'}{\IndSet}{\subMap} \label{proof:lemma1c-ic-6}\\
					\eqq''=\eqq'\setminus\eqqres{\eqq'}{\IndSet} \label{proof:lemma1c-ic-7}\\
					\IndSet'=\bigunion{j\in\IndSet}\childrenFun{\eqq'}{j} \label{proof:lemma1c-ic-8}\\
					\traverseFun{\eqq''}{\IndSet'}{\partition}{\subMap''}{=}\subMap' \label{proof:lemma1c-ic-9}
				\end{gather}
				From \eqref{proof:lemma1c-ic-7} and \eqref{proof:lemma1c-ic-2.5} we can deduce
				\begin{gather}
					\eqq''\subseteq\eqq \label{proof:lemma1c-ic-10}
				\end{gather}
				By \eqref{proof:lemma1c-ic-2} and the definition of a \emph{well-formed} map we know that \subMap provides a set of mappings which allow for:
				\begin{flalign}
					&\begin{array}{l}
						\bullet\quad \text{renaming the \emph{data variables} of each \emph{pattern equivalent sibling necessity},}\\ \qquad \text{defined in \eqqres{\eqq}{\dom{\subMap}}, to the \emph{same} set of fresh variables.} \vspace{-9mm}
					\end{array}
					\label{proof:lemma1c-ic-11} \\[3mm]
					&\begin{array}{l}
						\bullet\quad \text{renaming any \emph{reference} to a data variable that is bound by a \emph{renamed }}\\\qquad \text{\emph{parent necessity} defined in }\eqqres{\eqq}{\dom{\subMap}}.\vspace{-9mm}
					\end{array}					 
					\label{proof:lemma1c-ic-12}
				\end{flalign}
				By \eqref{proof:lemma1c-ic-6} and the definition of \partition we know
				\begin{gather}
					\subMap'' = \subMap \cupplus \Setdef{\submapstoE{j}{\subMap(i)\cupplus\sub{\dvVVV^n}{\dvV^n}}\\[5mm]\submapstoE{k}{\subMap(l)\cupplus\sub{\dvVVV^n}{\dvVV^n}} }
					{\forall i,l\in\IndSet\cdot
						\eqq'(i){=}\hAND{j\in\IndSet''}\hnec{\actS_j(\dvV^n)}\hVarX_j{\hand}\hV \\ 
						\text{ and } 
						\eqq'(l){=}\hAND{k\in\IndSet'''}\hnec{\actS_k(\dvVV^n)}\hVarX_k{\hand}\hV \textsl{ s.t.} \\[3mm]
						\text{ if }\actS_j(\dvV^n) \text{ is \emph{pattern equivalent} to }\\[2mm] \,\actS_k(\dvVV^n), \text{ then we assign the \emph{same} }\\[2mm]
						\text{ set of fresh variables }\dvVVV^n.
					} \label{proof:lemma1c-ic-13}
				\end{gather} 
				From \eqref{proof:lemma1c-ic-13} we know that $\subMap''$ includes a mapping for each sibling branch that defines a pattern equivalent necessity. The added mappings map the child indices (\ie $j,k{\in}\IndSet'$ since by \eqref{proof:lemma1c-ic-8} we know that $\IndSet''$ and $\IndSet'''$ are subsets of $\IndSet'$) of the conjunction branches, defined by equations identified by the parent indices (\ie $i{\in}\IndSet$) specified in \IndSet, to a substitution environment which renames the \resp variable names of these conjunct pattern equivalent sibling necessities, to the same fresh set of variable names, thereby making the equivalent sibling patterns, syntactically equal. Hence, by \eqref{proof:lemma1c-ic-11} we can deduce that $\subMap''$ provides a set of mappings which allow for \vspace{-12mm}
				\begin{flalign}
				&\begin{array}{l}
				\bullet\quad\text{renaming the \emph{data variables} of each \emph{pattern equivalent sibling necessity},}\\\qquad\text{defined in \eqqres{\eqq}{\dom{\subMap}\cup\IndSet'}, to the \emph{same} set of fresh variables.} \vspace{-30mm}
				\end{array} 
				\label{proof:lemma1c-ic-14} 
				\end{flalign}
				Similarly, from \eqref{proof:lemma1c-ic-13} we also know that the mappings in $\subMap''$ include the substitutions performed upon the parent necessities, \ie in each mapping $j{\,\mapsto\,}\s_j$, the mapped substitution environment $\s_j$ also includes $\subMap(i)$ where $i\in\IndSet$ is the parent index of $j{\,\in\,}\IndSet'$. Hence, by  \eqref{proof:lemma1c-ic-12} we can deduce that the mappings provided by $\subMap''$ also allow for  \vspace{-2mm}
				\begin{flalign}
				&\begin{array}{l}
					\bullet\quad \text{renaming any \emph{reference} to a data variable that is bound by a \emph{renamed }}\\\qquad \text{\emph{parent necessity} defined in }\eqqres{\eqq}{\dom{\subMap}\cup\IndSet'}. \vspace{-9mm}
				\end{array} 
				\label{proof:lemma1c-ic-15} 
				\end{flalign}
				Hence, by \eqref{proof:lemma1c-ic-14}, \eqref{proof:lemma1c-ic-15} and the definition of a \emph{well-formed} map we know
				\begin{gather}
					\subMap'' \text{ is a \emph{well-formed} map for } \eqqres{\eqq}{\dom{\subMap}\cup\IndSet'}. \label{proof:lemma1c-ic-16} 
				\end{gather}
				From \eqref{proof:lemma1c-ic-13} we know that $\subMap''$ includes a mapping for each child branch, identified by $j\in\IndSet''$ and $k\in\IndSet'''$ (where $\IndSet''$ and $\IndSet'''$ are both subsets of $\IndSet'$), that is defined in the equation identified by index $i\in\IndSet$ and which defines a pattern equivalent necessity. Hence, we know that the domain of $\subMap''$ is an extension of the domain of \subMap which additionally contains the child indices defined in $\IndSet'$, such that we can deduce
				\begin{gather}
					\dom{\subMap''}=\dom{\subMap}\cup\IndSet'\label{proof:lemma1c-ic-17} 
				\end{gather}
				Therefore, from \eqref{proof:lemma1c-ic-16} and \eqref{proof:lemma1c-ic-17} we can infer
				\begin{gather}
					\subMap'' \text{ is a \emph{well-formed} map for } \eqqres{\eqq}{\dom{\subMap''}}. \label{proof:lemma1c-ic-18} 
				\end{gather}
				Finally, by \eqref{proof:lemma1c-ic-9}, \eqref{proof:lemma1c-ic-10}, \eqref{proof:lemma1c-ic-18} and IH we can conclude
				\begin{gather*}
					\subMap' \text{ is a \emph{well-formed} map for } \eqq.
				\end{gather*}
 		\end{itemize}
	\end{case-noreset}
\end{proof}

\subsection{Proving \Cref{lemma:norm-1b}.}
\begin{center}
	$\forall(\hVarX_j{=}\hV_j){\in}\eqq\cdot$ equation $\hVarX_j{=}\hV_j$ is in \emph{Standard form}, \text{and} \subMap is a \emph{well-formed} map for \eqq \quad \textsl{implies} \quad  $\forall(\hVarX_k{=}\hVV_k){\in}\uniFun{\eqq}{\subMap}\cdot$ equation $(\hVarX_k{=}\hVV_k)$ is \emph{Uniform} \text{and} $\uniFun{\eqq}{\subMap}{\equiv}\eqq$.
\end{center}
\begin{proof}[. \textbf{By induction on the structure of \eqq}]
	
	\begin{case}[\eqq=\emptyset]
		This case holds trivially since $\eqq=\emptyset=\uniFun{\emptyset}{\subMap}$.
	\end{case}
	
	\newcommand{\subeqn}{\hVarX_i{=}\hAND{j\in\IndSet}\hnec{\actSN{\pate_j(\dvVVV^n_j)}{\predc_j(\dvVVV^m_{<i})}}\hV_j{\,\hand\,}\hV}
	\newcommand{\cureqn}[1]{\hVarX_i{=}\hAND{j\in\IndSet}\hnec{\actSN{\pate_j(\dvV^n_j)}{\predc_j(\dvVV^m_{<i})}#1}\hV_j{\,\hand\,}\hV}
	\newcommand{\cureqnSet}{\Set{\cureqn{}}\cupplus\eqq'}
	\newcommand{\unieqnSet}{\Set{\cureqn{\;\underline{\subMap(j)}\;}}\cupplus\uniFun{\eqq'}{\subMap}}
	
	\begin{case}[\eqq=\cureqnSet]
		Initially we know
		\begin{gather}
			\begin{array}{l}
				\forall(\hVarX_k{=}\hV_k)\in\cureqnSet\cdot\\\qquad
				\text{ equation }\hVarX_k{=}\hV_k \text{ is in \emph{Standard Form}.} 
			\end{array}\label{proof:lemma1b-1}\\[3mm]
			\subMap \text{ is a \emph{well-formed} map for } \eqq \label{proof:lemma1b-2}
		\end{gather}
		Since $\eqq'\subset\eqq$ from \eqref{proof:lemma1b-1} we know
		\begin{gather}
			\text{Equation }\cureqn{} \text{ is in \emph{Standard Form}}.\label{proof:lemma1b-3}\\
			\forall(\hVarX_k{=}\hV_k)\in\eqq'\cdot\text{ equation }\hVarX_k{=}\hV_k \text{ is in \emph{Standard Form}.} \label{proof:lemma1b-4}
		\end{gather}
		Since $\eqq'\subset\eqq$ from \eqref{proof:lemma1b-2} we know
		\begin{gather}
			\subMap \text{ is a \emph{well-formed} map for } \eqq' \label{proof:lemma1b-7}
		\end{gather}
		Hence by \eqref{proof:lemma1b-4}, \eqref{proof:lemma1b-7} and IH we know
		\begin{gather}
			\forall(\hVarX_k{=}\hVV_k)\in\uniFun{\eqq'}{\subMap}\cdot\text{ equation } (\hVarX_k{=}\hVV_k) \text{ is \emph{Uniform}}.  \label{proof:lemma1b-8}\\
			\uniFun{\eqq'}{\subMap}{\equiv}\eqq'  \label{proof:lemma1b-9}
		\end{gather}
		By applying the \uni function on \eqq and \subMap we obtain
		\begin{gather}
			\begin{array}{cl}
				&\uniFun{\cureqnSet}{\subMap} \\
				=&\unieqnSet
			\end{array}
			\label{proof:lemma1b-10}
		\end{gather}
		By \eqref{proof:lemma1b-2} and the definition of a \emph{well-formed} map we know that \subMap provides a set of mappings which allow for \vspace{-8mm}
		\begin{flalign}
			&\begin{array}{l}
				\bullet\quad \text{renaming the \emph{data variables} of each \emph{pattern equivalent sibling necessity},}\\ \qquad \text{defined in \eqq, to the \emph{same} set of fresh variables.} \vspace{-18mm}
			\end{array}
			\label{proof:lemma1b-5} \\[0mm]
			&\begin{array}{l}
				\bullet\quad \text{renaming any \emph{reference} to a data variable that is bound by a \emph{renamed }}\\\qquad \text{\emph{parent necessity} defined in }\eqq.\vspace{-9mm}
			\end{array}					 
			\label{proof:lemma1b-6}
		\end{flalign}
		Hence, \eqref{proof:lemma1b-5} and \eqref{proof:lemma1b-6} allow us to deduce that mapping $\subMap(j)$ in \eqref{proof:lemma1b-10} produces a substitution environment which renames the data variables $\dvV^n_j$ (defined by pattern $\pate(\dvV^n_j)$) to some set of fresh variables $\dvVVV^n_j$, which is the \emph{same} for all the other conjunct sibling necessities that are pattern equivalent to $\hnec{\actSN{\pate_j(\dvV^n_j)}{\predc_j(\dvVV^m_{<i})}}$. Hence, by the definition of a \emph{Uniform Equation}, we can deduce
		\begin{gather}
			\text{Equation }\cureqn{} \text{ is \emph{Uniform}}.\label{proof:lemma1b-11}
		\end{gather}
		By \eqref{proof:lemma1b-8}, \eqref{proof:lemma1b-10} and \eqref{proof:lemma1b-11} we can thus conclude
		\begin{gather}
			\forall(\hVarX_k{=}\hV_k)\in\uniFun{\eqq}{\subMap}\cdot\text{ equation }\hVarX_k{=}\hV_k \text{ is \emph{Uniform}.} \label{proof:lemma1b-12}
		\end{gather}
		By \eqref{proof:lemma1b-5} and \eqref{proof:lemma1b-6} we can deduce that equation $\cureqn{}$ is \emph{semantically equivalent} to the equation reconstructed by the \uni function in \eqref{proof:lemma1b-10}, \ie $\cureqn{\subMap(j)}$. This holds since when the substitution environment, returned by $\subMap(j)$, is applied on the equated formula, it substitutes symbolic event $\actSN{\pate_j(\dvV^n_j)}{\predc_j(\dvVV^m_{<i})}$ by $\actSN{\pate_j(\dvVVV^n_j)}{\predc_j(\dvVVV^m_{<i})}$. Notice that pattern $\pate_j(\dvVVV^n_j)$ is \emph{equivalent} to the original pattern $\pate_j(\dvV^n_j)$ since it only varies by the name of the data variables it defines, while condition $\predc_j(\dvVVV^m_{<i})$ is also equivalent to $\predc_j(\dvVV^m_{<i})$ since by \eqref{proof:lemma1b-6} we know that $\subMap(j)$ (where $\subMap(j)$ also contains $\subMap(i)$ where $i$ is the parent of $j$) substitutes accordingly the references to variables defined by renamed parent necessities that are being made by the filtering condition $\predc_j(\dvVV^m_{<i})$, \ie $\subMap(j)$ renames $\dvVV^m_{<i}$ to the variable names, $\dvVVV^m_{<i}$, assigned to the renamed parent necessities; this preserves the semantics of the equation by keeping it closed \wrt data variables. Hence, we can deduce
		\begin{gather}
			\begin{array}{cl}
					&\cureqn{} \\
				\equiv&\subeqn \\
				\equiv&\cureqn{\subMap(j)}
			\end{array} \label{proof:lemma1b-13}
		\end{gather}
		Finally, by \eqref{proof:lemma1b-9}, \eqref{proof:lemma1b-10} and \eqref{proof:lemma1b-13} we can conclude
		\begin{gather}
		\begin{array}{l}
			\qquad\cureqnSet \\ \equiv \; \unieqnSet \\[5mm]
			\ie\quad \eqq \equiv\uniFun{\eqq}{\subMap}
			\end{array} \label{proof:lemma1b-14}
		\end{gather}
		$\therefore$ This case holds by \eqref{proof:lemma1b-12} and \eqref{proof:lemma1b-14}.
	\end{case}
\end{proof}

\section{Auxiliary Lemmas for Proving Semantic Preservation of Construction \pmb{\gencond{-}}} \label{sec:proof-comb-lemmas}
In this section we provide the proof for \Cref{lemma:norm-2a} (restated below) which is are required for ensuring the semantic preservation of construction \gencond{-}.
\begin{description}
	\item[\Cref{lemma:norm-2a}] $\forall(\hVarX_j{=}\hV_j)\in\eqq\cdot$ equation $\hVarX_j{=}\hV_j$ is \emph{Uniform} \quad \textsl{implies} \\ $\eqq{\,\equiv\,}\traverseFun{\eqq}{\set{0}}{\condcomb}{\emptyset}$ \qquad \textsl{and} \\ $\forall(\hVarX_k{=}\hVV_k){\,\in\,}\traverseFun{\eqq}{\set{0}}{\condcomb}{\emptyset}\cdot$ equation $(\hVarX_k{=}\hVV_k)$ is \emph{Equi-Disjoint}.
\end{description}

\subsection{Proving \Cref{lemma:norm-2a}.}
In order to prove \Cref{lemma:norm-2a}, we use \Cref{lemma:norm-2b}. This new lemma states that one can obtain an \emph{Equi-disjoint} equation set, $\recEqq'$, that is \emph{semantically equivalent} to the original equation set \eqq, by conducting a traversal upon a \emph{Uniform} subset of $\eqq$ (\ie $\eqq'$), using an \emph{Equi-disjoint} accumulator equation set $\recEqq$, where $\recEqq$ must be \emph{semantically equivalent} to a subset of \eqq that is restricted to the indices associated to the logical variables specified by the domain of \recEqq, \ie $\recEqq\equiv\eqqres{\eqq}{\domInd{\recEqq}}$, where 
$$\domInd{\recEqq}\defeq\setdef{i}{\hVarX_i\in\dom{\recEqq}}$$

\begin{lemma}\label{lemma:norm-2b}
	$\forall\IndSet,\recEqq\cdot \eqq'\subseteq\eqq$ \textsl{ and }  $\traverseFun{\eqq'}{\IndSet}{\condcomb}{\recEqq}{=}\recEqq'$ \textsl{ and }
	$\eqqres{\eqq}{\domInd{\recEqq}}{\equiv}\recEqq$  \textsl{ and }
	$\forall(\hVarX_j{=}\hV_j){\,\in\,}\eqq'\cdot$ equation $\hVarX_j{=}\hV_j$ is \emph{Uniform} \textsl{ and } 
	$\forall(\hVarX_k{=}\hVV_k){\,\in\,}\recEqq\cdot$ equation $(\hVarX_k{=}\hVV_k)$ is \emph{Equi-Disjoint} \quad \textsl{implies} \quad $\forall(\hVarX_k{=}\hVV_k){\,\in\,}\recEqq'\cdot$ equation $(\hVarX_k{=}\hVV_k)$ is \emph{Equi-Disjoint} \textsl{ and } $\eqq{\,\equiv\,}\recEqq'$ 
\end{lemma}
The proof for this lemma is provided at the end of this section. 

\paragraph{To Prove \Cref{lemma:norm-2a}.}
\begin{center}
	$\forall(\hVarX_j{=}\hV_j)\in\eqq\cdot$ equation $\hVarX_j{=}\hV_j$ is \emph{Uniform} \quad \textsl{implies} \\ $\eqq{\,\equiv\,}\traverseFun{\eqq}{\set{0}}{\condcomb}{\emptyset}$ \qquad \textsl{and} \\ $\forall(\hVarX_k{=}\hVV_k){\,\in\,}\traverseFun{\eqq}{\set{0}}{\condcomb}{\emptyset}\cdot$ equation $(\hVarX_k{=}\hVV_k)$ is \emph{Equi-Disjoint}.
\end{center}
\begin{proof}
	Initially we know 
	\begin{gather}
		\forall(\hVarX_j{=}\hV_j)\in\eqq\cdot\text{ equation }\hVarX_j{=}\hV_j\text{ is \emph{Uniform}} \label{proof:lemma2a-1}
	\end{gather}
	By applying the \traverse function on \eqq starting from $\IndSet{=}\set{0}$ and $\recEqq{=}\emptyset$ we know
	\begin{gather}
		\traverseFun{\eqq}{\set{0}}{\condcomb}{\recEqq}=\recEqq' \label{proof:lemma2a-2}\\
		\recEqq=\emptyset \label{proof:lemma2a-3}
	\end{gather}	
	By \eqref{proof:lemma2a-3} and the definition of a \eqqres{\eqq}{\IndSet} we know
	\begin{gather}
		\eqqres{\eqq}{\dom{\emptyset}}=\emptyset=\recEqq \label{proof:lemma2a-4}
	\end{gather}
	From \eqref{proof:lemma2a-3} we can also deduce 
	\begin{gather}
		\forall(\hVarX_k{=}\hVV_k)\in\recEqq\cdot\text{ equation }\hVarX_k{=}\hVV_k\text{ is \emph{Equi-Disjoint}} \label{proof:lemma2a-5}
	\end{gather}	
	By \eqref{proof:lemma2a-1}, \eqref{proof:lemma2a-2}, \eqref{proof:lemma2a-4}, \eqref{proof:lemma2a-5} and \Cref{lemma:norm-2b} we know
	\begin{gather}
		\eqq\equiv\recEqq' \label{proof:lemma2a-6}\\
		\forall(\hVarX_k{=}\hVV_k)\in\recEqq'\cdot\text{ equation }\hVarX_k{=}\hVV_k\text{ is \emph{Equi-Disjoint}} \label{proof:lemma2a-7}
	\end{gather}
	$\therefore$ This Lemma holds by \eqref{proof:lemma2a-6} and \eqref{proof:lemma2a-7}.
	
\end{proof}

\paragraph{To Prove \Cref{lemma:norm-2b}.}
\begin{center}
	$\forall\IndSet,\recEqq\cdot \eqq'\subseteq\eqq$ \textsl{ and }  $\traverseFun{\eqq'}{\IndSet}{\condcomb}{\recEqq}{=}\recEqq'$ \textsl{ and }
	$\eqqres{\eqq}{\domInd{\recEqq}}{\equiv}\recEqq$  \textsl{ and }
	$\forall(\hVarX_j{=}\hV_j){\in}\eqq'\cdot$ equation $\hVarX_j{=}\hV_j$ is \emph{Uniform} \textsl{ and } 
	$\forall(\hVarX_k{=}\hVV_k){\in}\recEqq\cdot$ equation $(\hVarX_k{=}\hVV_k)$ is \emph{Equi-Disjoint} \quad \textsl{implies} \quad $\forall(\hVarX_k{=}\hVV_k)\in\recEqq'\cdot$ equation $(\hVarX_k{=}\hVV_k)$ is \emph{Equi-Disjoint} \textsl{ and } $\eqq{\,\equiv\,}\recEqq'$ 
\end{center}
\begin{proof}[. By induction on the structure of \IndSet]
	
	\begin{case}[\IndSet{\,=\,}\emptyset]
		Initially we know 
		\begin{gather}
			\eqq'\subseteq\eqq \label{proof:lemma2b-bc-1}\\
			\traverseFun{\eqq'}{\emptyset}{\condcomb}{\recEqq}{=}\recEqq'  \label{proof:lemma2b-bc-2}\\
			\eqqres{\eqq}{\domInd{\recEqq}}{\equiv}\recEqq  \label{proof:lemma2b-bc-3}\\
			\forall(\hVarX_j{=}\hV_j)\in\eqq'\cdot\text{ equation }\hVarX_j{=}\hV_j\text{ is \emph{Uniform}} \label{proof:lemma2b-bc-4}\\
			\forall(\hVarX_k{=}\hVV_k)\in\recEqq\cdot\text{ equation }(\hVarX_k{=}\hVV_k)\text{ is \emph{Equi-Disjoint}} \label{proof:lemma2b-bc-5}
		\end{gather}
		By \eqref{proof:lemma2b-bc-2} and the definition of \traverse we know
		\begin{gather}
			\recEqq=\recEqq' \label{proof:lemma2b-bc-6}
		\end{gather}
		From \eqref{proof:lemma2b-bc-5} and \eqref{proof:lemma2b-bc-6} we can deduce
		\begin{gather}
			\forall(\hVarX_k{=}\hVV_k)\in\recEqq'\cdot\text{ equation }(\hVarX_k{=}\hVV_k)\text{ is \emph{Equi-Disjoint}} \label{proof:lemma2b-bc-7}
		\end{gather}
		From \eqref{proof:lemma2b-bc-3} and \eqref{proof:lemma2b-bc-6} we also know
		\begin{gather}
			\eqqres{\eqq}{\domInd{\recEqq'}}{\equiv}\recEqq'  \label{proof:lemma2b-bc-8}
		\end{gather} 
		Since $\IndSet{=}\emptyset$, by \eqref{proof:lemma2b-bc-2} and the definition of \traverse we know the traversal has reached a point where no more children can be computed, which means that all the \emph{relevant equations} (\ie those reachable from the principle variable) have been analysed. This implies that any other equation in \eqq (if any) is \emph{redundant} and \emph{irrelevant}. Hence, since from \eqref{proof:lemma2b-bc-8} we know that the equations in $\recEqq'$ are \emph{equivalent to the relevant subset of equations in \eqq}, \ie \eqqres{\eqq}{\domInd{\recEqq'}}, and hence we can conclude
		\begin{gather}
			\recEqq'\equiv\eqq \label{proof:lemma2b-bc-9}
		\end{gather}
		$\therefore$ This subcase holds by \eqref{proof:lemma2b-bc-7} and \eqref{proof:lemma2b-bc-9}.
	\end{case}

	\newcommand{\eqnSet}{\Setdef
		{\!\! \hVarX_i{=}\hAND{\predc_k\in\truthcombs{j}{\IndSet'}\hspace{-10mm}}\hgnec{\pate}{\predc_k}\hVarX_j{\hand}\hV (=\hVV_i) 
			\!\!}
		{
			(\hVarX_i{=}\hAND{j\in\IndSet''\hspace{-3mm}}\hgnec{\pate}{\predc_j}\hVarX_j{\hand}\hV){\in}\eqqres{\eqq}{\IndSet} \\[3mm]
			\text{  and } \IndSet'{=}\bigunion{l\in\IndSet}\childrenFun{\eqq}{l} \\[3mm] 
			\textsl{  such that } \IndSet''\subseteq\IndSet' \!\!\!
		}}
	\begin{case}[\IndSet{\,\neq\,}\emptyset]
		Initially we know 
		\begin{gather}
			\eqq'\subseteq\eqq \label{proof:lemma2b-ic-1}\\
			\traverseFun{\eqq'}{\IndSet}{\condcomb}{\recEqq}{=}\recEqq'  \label{proof:lemma2b-ic-2}\\
			\eqqres{\eqq}{\domInd{\recEqq}}{\equiv}\recEqq  \label{proof:lemma2b-ic-3}\\
			\forall(\hVarX_j{=}\hV_j)\in\eqq'\cdot\text{ equation }\hVarX_j{=}\hV_j\text{ is \emph{Uniform}} \label{proof:lemma2b-ic-4.5}\\
			\forall(\hVarX_k{=}\hVV_k)\in\recEqq\cdot\text{ equation }(\hVarX_k{=}\hVV_k)\text{ is \emph{Equi-Disjoint}} \label{proof:lemma2b-ic-4}
		\end{gather}
		We consider two subcases:
		\begin{itemize}
				\item[\pmb{--}] $\pmb{\eqq'=\emptyset:}$ Since $\eqq'=\emptyset$, by \eqref{proof:lemma2b-ic-2} and the definition of \traverse we know 
				\begin{gather}
					\recEqq=\recEqq'  \label{proof:lemma2b-ic-5}
				\end{gather}
				By \eqref{proof:lemma2b-ic-3}, \eqref{proof:lemma2b-ic-4} and \eqref{proof:lemma2b-ic-5} we know
				\begin{gather}
					\eqqres{\eqq}{\domInd{\recEqq'}}{\equiv}\recEqq' \label{proof:lemma2b-ic-6}\\
					\forall(\hVarX_k{=}\hVV_k)\in\recEqq'\cdot\text{ equation }(\hVarX_k{=}\hVV_k)\text{ is \emph{Equi-Disjoint}} \label{proof:lemma2b-ic-7}
				\end{gather}
				By \eqref{proof:lemma2b-ic-2} and the definition of \traverse we know that the traversal starts from the full equation set, \ie $\eqq'=\eqq$, using an empty accumulator, \ie $\recEqq{=}\emptyset$, that would eventually contain the resultant Equi-Disjoint equation set. Every recursive application of the \traverse function is then performed \wrt: a \emph{smaller} version \eqq, \ie $\eqq'{=}\eqq{\setminus}\eqqres{\eqq}{\IndSet}$, and a \emph{larger} accumulator $\recEqq'$ containing the reformulated, Equi-Disjoint equations whose indices are defined in \IndSet (and which where removed from $\eqq'$). Hence, when $\eqq'$ becomes $\emptyset$ it means that 
				\begin{gather}
					\domInd{\recEqq'}=\domInd{\eqq}  \label{proof:lemma2b-ic-7.5}
				\end{gather}
				Hence, from \eqref{proof:lemma2b-ic-7.5} and by the definition of \eqqres{\eqq}{\IndSet} we can deduce
				\begin{gather}
					\eqqres{\eqq}{\domInd{\recEqq}}=\eqqres{\eqq}{\domInd{\eqq}}=\eqq  \label{proof:lemma2b-ic-8}
				\end{gather}
				Therefore by \eqref{proof:lemma2b-ic-6} and \eqref{proof:lemma2b-ic-8} we conclude
				\begin{gather}
					\eqq\equiv\recEqq'  \label{proof:lemma2b-ic-9}
				\end{gather}
				$\therefore$ This subcase holds by \eqref{proof:lemma2b-ic-7} and \eqref{proof:lemma2b-ic-9}.
			\item[\pmb{--}] $\pmb{\eqq'\neq\emptyset:}$ 
				By \eqref{proof:lemma2b-ic-2} and the definition of \traverse we know 
				\begin{gather}
					\condcombFun{\eqq'}{\IndSet}{\recEqq}{=}\recEqq''  \label{proof:lemma2b-ic-10} \\
					\eqq''=\eqq'\setminus\eqqres{\eqq'}{\IndSet}  \label{proof:lemma2b-ic-11} \\
					\IndSet'=\bigunion{l\in\IndSet}\childrenFun{\eqq}{l} \label{proof:lemma2b-ic-12} \\
					\traverseFun{\eqq''}{\IndSet'}{\condcomb}{\recEqq''}=\recEqq' \label{proof:lemma2b-ic-13}
				\end{gather}
				From \eqref{proof:lemma2b-ic-1} and \eqref{proof:lemma2b-ic-11} we know 
				\begin{gather}
					\eqq''\subseteq\eqq \label{proof:lemma2b-ic-13.5}
				\end{gather}
				By \eqref{proof:lemma2b-ic-10} and the definition of \condcomb we know 
				\begin{gather}
				\recEqq'' = \recEqq{\,\cupplus\,}\eqnSet
					\label{proof:lemma2b-ic-14}
				\end{gather}
				By \eqref{proof:lemma2b-ic-14} and the definition of \truthcombs{j}{\IndSet'}, we know that the conjunctions in the reformulated equations (\ie in every $\hVV_i$) now include an additional branch for each condition $\predc_k\in\truthcombs{j}{\IndSet'}$ where $\predc_k$ is a \emph{compound condition} \eg $\predc_0\land\predc_1\land\ldots\land\predc_n$ or $\predc_0\land\lnot\predc_1\land\ldots\land\lnot\predc_n$. These compound conditions consist in a \emph{truth combination} of the filtering conditions of the sibling symbolic events which specify \emph{syntactically equal patterns}; this is guaranteed since by \eqref{proof:lemma2b-ic-4.5} we know that the equations in $\eqq'$ are \emph{uniform}, meaning that all sibling pattern equivalent events are guaranteed to be syntactically equal as well.
				
				\qquad Hence, the reconstructed symbolic events in these additional guarded branches are \emph{unable} to match the same concrete event \acta unless they are syntactically equal (\ie define the same pattern and condition), since despite their pattern being syntactically equal, \emph{only one} compound filtering condition can at most be satisfied by the matching concrete event \acta; a case in point is when equation $\hVarX_0{=}\hgnec{\pate}{\predc_1}\hVarX_{1}{\land}\hgnec{\pate}{\predc_2}\hVarX_{2}$ is reconstructed into $\hVarX_0{=}\hgnec{\pate}{\predc_1{\land}\predc_2}\hVarX_{1}{\hand} \hgnec{\pate}{\predc_1{\land}\lnot\predc_2}\hVarX_{2} {\hand}\hgnec{\pate}{\predc_2}\hVarX_{2}{\hand}\hgnec{\pate}{\predc_2}\hVarX_{2}$. Therefore, by \eqref{proof:lemma2b-ic-14} and the definition of \emph{Equi-Disjoint}, we can deduce that				
				\begin{gather}
					\begin{array}{c}
						\forall(\hVarX_k{=}\hVV_k)\in\eqnSet\cdot\\\text{ equation }(\hVarX_k{=}\hVV_k)\text{ is \emph{Equi-Disjoint}} \label{proof:lemma2b-ic-15}
					\end{array}
				\end{gather}  
				Hence, by \eqref{proof:lemma2b-ic-5}, \eqref{proof:lemma2b-ic-14} and \eqref{proof:lemma2b-ic-15} we can conclude
				\begin{gather}
					\forall(\hVarX_k{=}\hVV_k)\in\recEqq''\cdot\text{ equation }(\hVarX_k{=}\hVV_k)\text{ is \emph{Equi-Disjoint}} \label{proof:lemma2b-ic-16}
				\end{gather}
				Finally, we argue that the reconstructed equations in \eqref{proof:lemma2b-ic-14} (\ie $\hVarX_i{=}\hVV_i$) \emph{semantically equivalent} to the original ones (\ie $(\hVarX_i{=}\hV_i){\in}\eqqres{\eqq}{\IndSet}$) since whenever a guarded branch, $\hnec{\actSN{\pate}{\predc_i}}\hVarX_i$, is reconstructed into (possibly) multiple branches, $\hnec{\actSN{\pate}{\predc_i\!\land\!\predc_j\!\ldots\!\predc_k}}\hVarX_i\hand\hnec{\actSN{\pate}{\predc_i\!\land\!\lnot\predc_j\!\ldots\!\predc_k}}\hVarX_i\hand\ldots\hand\hnec{\actSN{\pate}{\predc_i\!\land\!\lnot\predc_j\!\ldots\!\lnot\predc_k}}\hVarX_i$, via the truth combination function \truthcombs{i}{\IndSet'}, the condition, $\predc_i$, of the original branch is \emph{never negated}. This guarantees that continuation $\hVarX_i$ can only be reached when the original condition $\predc_i$ is \emph{true}, and thus preserves the original semantics of the branch. Therefore, we conclude
				\begin{gather}
					\eqnSet\equiv\eqqres{\eqq}{\IndSet}  \label{proof:lemma2b-ic-17}
				\end{gather} 
				By \eqref{proof:lemma2b-ic-3}, \eqref{proof:lemma2b-ic-14} and \eqref{proof:lemma2b-ic-17} we know
				\begin{gather}
					\eqqres{\eqq}{\domInd{\recEqq''}}\equiv\recEqq''  \label{proof:lemma2b-ic-18}
				\end{gather}
				By \eqref{proof:lemma2b-ic-4.5} and \eqref{proof:lemma2b-ic-13.5} we know
				\begin{gather}
					\forall(\hVarX_j{=}\hV_j)\in\eqq''\cdot\text{ equation }\hVarX_j{=}\hV_j\text{ is \emph{Uniform}} \label{proof:lemma2b-ic-19}
				\end{gather}
				Finally, by \eqref{proof:lemma2b-ic-13}, \eqref{proof:lemma2b-ic-13.5}, \eqref{proof:lemma2b-ic-16}, \eqref{proof:lemma2b-ic-18}, \eqref{proof:lemma2b-ic-19} and IH we know
				\begin{gather}
					\eqq\equiv\recEqq' \label{proof:lemma2b-ic-20} \\
					\forall(\hVarX_k{=}\hVV_k)\in\recEqq'\cdot\text{ equation }(\hVarX_k{=}\hVV_k)\text{ is \emph{Equi-Disjoint}} \label{proof:lemma2b-ic-21}
				\end{gather}
				$\therefore$ This case holds by \eqref{proof:lemma2b-ic-20} and \eqref{proof:lemma2b-ic-21}.
		\end{itemize}
	\end{case}
\end{proof}

\chapter{Proving Enforcement Correctness}
\label{sec:app:correctness-proofs}
In this section we present proofs ascertaining the correctness of our enforcers. Particularly, in \Cref{sec:app:deterministic-mon} we prove auxiliary lemmas required for proving \Cref{thm:det-enf}, \ie the synthesised enforcers are \emph{Deterministic}, while in \Cref{sec:app:sound-trans}, we show that these enforcers can also \emph{Strongly enforce} the \shml formula they were synthesised from.

\section{Proving Determinism for the Synthesised Enforcers}
\label{sec:app:deterministic-mon}
To address the issue of determinism we prove \Cref{lemma:wf-enf} which states that the synthesis function always produces \emph{well-formed enforcers} from a normalized formula, and \Cref{lemma:enf-one-det}, which states that an enforcer always processes an input action and thus reduces in the same way, and thus behaves deterministically with every reduction. \medskip

\subsection{Proving Enforcer Well-formedness (\Cref{lemma:wf-enf})} \label{sec:app:proof:wf-enf}
\begin{rtp}
	$$\forall\hV{\in}\shmlwf\cdot\g{\hV}{}{=}\eV \imp \eV{\in}\Enfwf$$
	By the definition of \g{-}{} we can instead prove
	$$\forall\hV{\in}\shmlwf\cdot \g{\opt{\hV}}{\bot}{=}\eV \imp \eV{\in}\Enfwf$$
	We therefore quantify over all possible $\rho$ and prove a stronger result, \ie 
	$$\forall\rho,\hV{\in}\shmlwf\cdot \opt{\hV}{=}\hVV \text{  and  }\g{\hVV}{\rho}{=}\eV \imp \eV{\in}\Enfwf$$ \\[-18mm]
\end{rtp}
\begin{proof}[ by induction on the structure of \pmb{\hV}]
	
	\begin{case}[\hV=\htru]
		This case holds trivially since $\opt{\htru}{\,=\,}\htru$ and $\g{\htru}{\rho}{\,=\,}\miden$ where $\miden{\,\in\,}\Enfwf$.
	\end{case}
	
	\begin{case}[\hV=\hfls] We know
		\begin{gather}
			\opt{\hfls}{\,=\,}\hfls \label{proof:wf-enf-ff-1} \\
			\g{\hfls}{\rho} = \eV \label{proof:wf-enf-ff-2}
		\end{gather}
		We must consider two subcases for $\rho$.
		\begin{itemize}
			\item[$\rho=\bot$:] Case does not apply since $\g{\hfls}{\bot}$ does \emph{not} produce an enforcer \eV.
			\item[$\rho=\my$:] From \eqref{proof:wf-enf-ff-2} and since $\rho=\my$ we know 
			\begin{gather}
				\g{\hfls}{\my} = \my \label{proof:wf-enf-ff-3}
			\end{gather}
			$\therefore$ Case holds by \eqref{proof:wf-enf-ff-3} since $\my\in\Enfwf$.
		\end{itemize}
	\end{case}
	
	\newcommand{\curform}{\hAND{i\in\IndSet}\hgnec{\pate_i}{\predc_i}\hV_i}
	\newcommand{\curformOpt}{\hAND{i\in\IndSet}\hgnec{\pate_i}{\predc_i}\opt{\hV_i}}
	\newcommand{\curEnf}{\mrec{\my}{\mCH{i\in\IndSet}
			\begin{xbrace}{rc}
				\mact{\actSTD{\pate_i}{\predc_i}}{\eV_i} \quad& \text{(if $\g{\hfls_i}{\my}=\eV_i$)} \\
				\mact{\actSID{\pate_i}{\predc_i}}{\eV_i} \quad& \text{(otherwise)}
			\end{xbrace} }}
	\begin{case}[\hV=\curform] We know
		\begin{gather}
			\opt{\curform}{\,=\,}\curformOpt \label{proof:wf-enf-and-1} \\
			\g{\curformOpt}{\rho} = \curEnf \label{proof:wf-enf-and-2}
		\end{gather}
		because by the definition of \g{-}{\my} we know
		\begin{gather}
			\forall i\in\IndSet\cdot \g{\opt{\hV_i}}{\my}{=}\eV_i  \label{proof:wf-enf-and-3}
		\end{gather}
		By applying the IH on \eqref{proof:wf-enf-and-3} we know
		\begin{gather}
			\forall i\in\IndSet\cdot \eV_i\in\Enfwf  \label{proof:wf-enf-and-4}
		\end{gather}
		Hence, from \eqref{proof:wf-enf-and-2}, \eqref{proof:wf-enf-and-3}, \eqref{proof:wf-enf-and-4} and the definition of \Enfwf we can conclude 
		\begin{gather}
			\begin{xbrackets}{c}
			\curEnf
			\end{xbrackets} \in \Enfwf
		\end{gather}
	\end{case}
	
	\renewcommand{\curform}{\hmax{\hVarX}{\hV}}
	\renewcommand{\curformOpt}{\hmax{\hVarX}{\opt{\hV}}}
	\renewcommand{\curEnf}{\mrec{\mx}{\eV}}
	\begin{case}[\curform \text{ and }\hVarX{\,\in\,}\fv{\hV}]
		\begin{gather}
			\opt{\curform}{\,=\,}\curformOpt \label{proof:wf-enf-max-a-1} \\
			\g{\curformOpt}{\rho} = \curEnf \label{proof:wf-enf-max-a-2}
		\end{gather}
		because by the definition of \g{-}{\my} we know
		\begin{gather}
			 \g{\opt{\hV}}{\rho}{=}\eV  \label{proof:wf-enf-max-a-3}
		\end{gather}
		By applying the IH on \eqref{proof:wf-enf-max-a-3} we know
		\begin{gather}
			\eV\in\Enfwf  \label{proof:wf-enf-max-a-4}
		\end{gather}
		Hence, from \eqref{proof:wf-enf-max-a-2}, \eqref{proof:wf-enf-max-a-4} and the definition of \Enfwf we can conclude 
		\begin{gather}
			\curEnf\in\Enfwf
		\end{gather}
	\end{case}

	\renewcommand{\curformOpt}{\opt{\hV}}
	\renewcommand{\curEnf}{\eV}
	\begin{case}[\curform \text{ and }\hVarX{\,\notin\,}\fv{\hV}]
		\begin{gather}
			\opt{\curform}{\,=\,}\curformOpt \label{proof:wf-enf-max-b-1} \\
			\g{\curformOpt}{\rho} = \curEnf \label{proof:wf-enf-max-b-2}
		\end{gather}
		By applying the IH on \eqref{proof:wf-enf-max-b-2} we know
		\begin{gather}
			\curEnf\in\Enfwf  \label{proof:wf-enf-max-b-3}
		\end{gather}
	\end{case}

\end{proof}

\subsection{Proving Single Step Determinism (\Cref{lemma:enf-one-det})} \label{sec:app:proof:enf-one-det}
\begin{rtp}	 
	$$ \forall\eV{\in}\Enfwf\cdot \eV\traS{\ioact{\acta}{\actu'}}\eV' \text{  and  } \eV\traS{\ioact{\acta}{\actu''}}\eV'' \imp \eV'{=}\eV'' \text{  and  }\actu'{=}\actu''$$
\end{rtp}\\[-12mm]
\begin{proof}[ by rule induction on \pmb{$\eV\traS{\ioact{\acta}{\actu'}}\eV'$}] 
	
	\begin{case}[\rtit{eId}] Initially we know 
		\begin{gather}
			\eV\traS{\ioact{\acta}{\actu'}}\eV' \label{proof:enf-one-det-id-1}
		\end{gather} 
		where
		\begin{gather}
			\eV=\miden \label{proof:enf-one-det-id-2}\\
			\eV'=\miden \label{proof:enf-one-det-id-3}\\
			\actu'=\acta \label{proof:enf-one-det-id-4}\\
			\eV\in\Enfwf \label{proof:enf-one-det-id-4.5}
		\end{gather} 
		and
		\begin{gather}
			\eV\traS{\ioact{\acta}{\actu''}}\eV'' \label{proof:enf-one-det-id-5}
		\end{gather} 
		By \eqref{proof:enf-one-det-id-2}, \eqref{proof:enf-one-det-id-5} and \rtit{eId} we know 
		\begin{gather}
			\eV''=\miden \label{proof:enf-one-det-id-6}\\
			\actu''=\acta \label{proof:enf-one-det-id-7}
		\end{gather} 
		Hence, from \eqref{proof:enf-one-det-id-3}, \eqref{proof:enf-one-det-id-6} and \eqref{proof:enf-one-det-id-4} and \eqref{proof:enf-one-det-id-7} we conclude
		\begin{gather}
			\eV'=\eV''=\miden \label{proof:enf-one-det-id-8}\\
			\actu'=\actu'' \label{proof:enf-one-det-id-9}
		\end{gather}
		From \eqref{proof:enf-one-det-id-2}, \eqref{proof:enf-one-det-id-4.5} and \eqref{proof:enf-one-det-id-8} we also know 
		\begin{gather}
			\eV',\eV''\in\Enfwf \label{proof:enf-one-det-id-10}
		\end{gather} 
		$\therefore$ Case holds by \eqref{proof:enf-one-det-id-8}, \eqref{proof:enf-one-det-id-9} and \eqref{proof:enf-one-det-id-10}.
	\end{case} 
	
	\newcommand{\cureV}{\mrec{\mx}{\eV}}
	\begin{case}[\rtit{eRec}] Initially we know 
		\begin{gather}
			\cureV\traS{\ioact{\acta}{\actu'}}\eV' \label{proof:enf-one-det-rec-1}
		\end{gather} 
		because
		\begin{gather}
			\eV\sub{\cureV}{\mx}\traS{\ioact{\acta}{\actu'}}\eV' \label{proof:enf-one-det-rec-2}
		\end{gather} 
		and
		\begin{gather}
			\cureV\traS{\ioact{\acta}{\actu''}}\eV'' \label{proof:enf-one-det-rec-3} \\
			\cureV\in\Enfwf  \label{proof:enf-one-det-rec-4}
		\end{gather} 
		By \eqref{proof:enf-one-det-rec-3} and \rtit{eRec} we know
		\begin{gather}
			\eV\sub{\cureV}{\mx}\traS{\ioact{\acta}{\actu''}}\eV'' \label{proof:enf-one-det-rec-5}
		\end{gather} 
		Since $\eV\sub{\cureV}{\mx}$ is the unfolded equivalent of $\cureV$, from \eqref{proof:enf-one-det-rec-4} we can deduce
		\begin{gather}
			\eV\sub{\cureV}{\mx}\in\Enfwf \label{proof:enf-one-det-rec-6}
		\end{gather} 
		Hence, by \eqref{proof:enf-one-det-rec-2}, \eqref{proof:enf-one-det-rec-5}, \eqref{proof:enf-one-det-rec-6} and IH we know
		\begin{gather}
			\eV'=\eV'' \label{proof:enf-one-det-rec-7}\\
			\actu'=\actu'' \label{proof:enf-one-det-rec-8}\\
			\eV',\eV''\in\Enfwf \label{proof:enf-one-det-rec-9}
		\end{gather}
 		$\therefore$ Case holds by \eqref{proof:enf-one-det-rec-7}, \eqref{proof:enf-one-det-rec-8} and \eqref{proof:enf-one-det-rec-9}.
	\end{case} 

	\renewcommand{\cureV}{\mCH{i\in\IndSet}\eV_i}
	\begin{case}[\rtit{eSel}] Initially we know 
		\begin{gather}
			\cureV\traS{\ioact{\acta}{\actu'}}\eV_j' \label{proof:enf-one-det-sel-1}
		\end{gather} 
		because
		\begin{gather}
			\exists j\in\IndSet\cdot\eV_j\traS{\ioact{\acta}{\actu'}}\eV_j' \label{proof:enf-one-det-sel-2}
		\end{gather} 
		and
		\begin{gather}
			\cureV\traS{\ioact{\acta}{\actu''}}\eV_k' \label{proof:enf-one-det-sel-3} \\
			\cureV\in\Enfwf  \label{proof:enf-one-det-sel-4}
		\end{gather} 
		By \eqref{proof:enf-one-det-sel-3} and \rtit{eSel} we know
		\begin{gather}
			\exists k\in\IndSet\cdot\eV_k\traS{\ioact{\acta}{\actu''}}\eV_k' \label{proof:enf-one-det-sel-4.5}
		\end{gather} 
		By \eqref{proof:enf-one-det-sel-4} and the definition of \Enfwf we know that every branch $\eV_i$ is prefixed by \emph{disjoint} symbolic transformations, such that we know
		\begin{gather}
			\forall i\in\IndSet\cdot\eV_i=\mact{\actSTN{\pate_i}{\predc_i}{\pate_i'}}{\eV_i'}
			 \label{proof:enf-one-det-sel-5} \\
			\bigdistinct{i\in\IndSet}\actSN{\pate_i}{\predc_i} \label{proof:enf-one-det-sel-6}
		\end{gather} 
		By \eqref{proof:enf-one-det-sel-5} and \eqref{proof:enf-one-det-sel-6} we know that only one of the branches in \eqref{proof:enf-one-det-sel-4} can match the concrete system event \acta, and hence no matter how many times \eV is executed \wrt \acta, the same branch is always chosen. Therefore, the same branch is chosen in both reductions \eqref{proof:enf-one-det-sel-2} and \eqref{proof:enf-one-det-sel-4.5}, such that we can conclude
		\begin{gather}
			j=k \label{proof:enf-one-det-sel-7}
		\end{gather}
		Hence, by \eqref{proof:enf-one-det-sel-2}, \eqref{proof:enf-one-det-sel-4.5} and \eqref{proof:enf-one-det-sel-7} we know
		\begin{gather}
			\eV_j\traS{\ioact{\acta}{\actu'}}\eV_j'  \label{proof:enf-one-det-sel-8} \\
			\eV_j\traS{\ioact{\acta}{\actu''}}\eV_k' \label{proof:enf-one-det-sel-9} 
		\end{gather}
		Since $j\in\IndSet$, from \eqref{proof:enf-one-det-sel-4} we know
		\begin{gather}
			\eV_j\in\Enfwf  \label{proof:enf-one-det-sel-10}
		\end{gather}
		Hence, by \eqref{proof:enf-one-det-sel-8}, \eqref{proof:enf-one-det-sel-9}, \eqref{proof:enf-one-det-sel-10} and IH we know
		\begin{gather}
			\eV_j'=\eV_k' \label{proof:enf-one-det-sel-11}\\
			\actu'=\actu'' \label{proof:enf-one-det-sel-12}\\
			\eV_j',\eV_k'\in\Enfwf \label{proof:enf-one-det-sel-13}
		\end{gather}
		$\therefore$ Case holds by \eqref{proof:enf-one-det-sel-11}, \eqref{proof:enf-one-det-sel-12} and \eqref{proof:enf-one-det-sel-13}.
	\end{case}

	\renewcommand{\cureV}{\mact{\actSTN{\pate}{\predc}{\pate'}}{\eV}}
	\begin{case}[\rtit{eTrns}] Initially we know 
		\begin{gather}
			\cureV\traS{\ioact{\acta}{\actu'}}\eV\s \label{proof:enf-one-det-trns-1}
		\end{gather} 
		because
		\begin{gather}
			\actSTN{\pate}{\predc}{\pate'}(\acta)=(\actu',\s) \label{proof:enf-one-det-trns-2}
		\end{gather} 
		and
		\begin{gather}
			\cureV\traS{\ioact{\acta}{\actu''}}\eV'' \label{proof:enf-one-det-trns-3} \\
			\cureV\in\Enfwf  \label{proof:enf-one-det-trns-4}
		\end{gather} 
		By \eqref{proof:enf-one-det-trns-2}, \eqref{proof:enf-one-det-trns-3} and \rtit{eTrns} we know
		\begin{gather}
			\eV''=\eV\s \label{proof:enf-one-det-trns-6}\\
			\actu''=\actu' \label{proof:enf-one-det-trns-7}
		\end{gather} 
		By \eqref{proof:enf-one-det-trns-4} and the definition of \Enfwf, we know that $\cureV$ is a special case for $\mCH{i\in\IndSet}\mact{\actSTN{\pate_i}{\predc_i}{\pate'_i}}{\eV_i}$, \ie where \IndSet contains only one index, and hence we know
		\begin{gather}
			\eV\in\Enfwf \label{proof:enf-one-det-trns-5}
		\end{gather} 
		Moreover, when \s is applied on an enforcer \eV, this does not modify the structure of \eV, but just binds data variables defined in a prefixing symbolic event to the data defined in the matching concrete system event. Hence, from \eqref{proof:enf-one-det-trns-5} we deduce
		\begin{gather}
			\eV\s\in\Enfwf \label{proof:enf-one-det-trns-8}
		\end{gather} 
		$\therefore$ Case holds by \eqref{proof:enf-one-det-trns-6}, \eqref{proof:enf-one-det-trns-7} and \eqref{proof:enf-one-det-trns-8}.
	\end{case} 

\end{proof}

\section{Proving Strong Enforceability for the Synthesised Enforcers}
\label{sec:app:sound-trans}
We prove \Cref{thm:strong-enf} (restated below), by proving that the enforcers synthesised by our synthesis function are \emph{sound} and \emph{transparent}. We prove these two criteria in \Cref{sec:proof-soundness,sec:proof-transparency} \resp Finally, we prove the supporting lemma, \Cref{lemma:opt-equiv}, in \Cref{sec:proof-opt-equiv}.\\

\subsection{Proving Soundness} \label{sec:proof-soundness}
	$$ \forall\rho,\pV{\,\in\,}\Proc,\eV{\,\in\,}\Enfwf,\hV\in\shmlwfopt\text{ when }\hV{\in}\Sat\; \cdot\; \g{\hV}{\rho}\sub{\eV}{\rho}=\eV'\!\! \imp\!\! \i{\eV'}{\pV}{\,\vSatS\,}\hV $$
	To prove this theorem we must show that relation \R (below) is a \emph{satisfaction relation} (\vSatS) as defined by the rules in \Cref{fig:uhml-sat}. 
	$$ \R\;\defeq\;\setdef{(\i{\eV'}{\pV},\hV)}{\hV{\in}\Sat\;\text{ and }\;\forall\eV\in\Enfwf\cdot\g{\hV}{\rho}\sub{\eV}{\rho}=\eV'} $$
 \\[-10mm]
\begin{proof} By coinduction on the structure of $\hV$. 
	
	\begin{case}[\hV=\hVarX] Does not apply since $\hVarX$ is an open formula and thus $\hVarX\notin\Sat$ \end{case} 	
	
	\begin{case}[\hV=\hfls] Does not apply since $\hfls\notin\Sat$\end{case} 
	
	\begin{case}[\hV=\htru] Holds trivially since \emph{any process} satisfies \htru, which confirms that  $(\i{\miden}{\pV},\htru)\in\R$, since $\g{\htru}{\rho}\sub{\eV}{\rho}=\g{\htru}{\rho}=\miden$. \end{case} 
	
	\begin{case}[\hV=\hmax{\hVarX}{\hV} \text{ and } \hVarX{\in}\fv{\hV}] We know
		\newcommand{\formula}{\hmax{\hVarX}{\hV}}
		\newcommand{\maxSub}{\hV\,\sub{\formula}{\hVarX}}
		\begin{gather}
			(\i{\eV'}{\pV},\formula)\in\R \label{proof:str-soundness-max-1}
		\end{gather}
		because		
		\begin{gather}
			\formula\in\Sat \label{proof:str-soundness-max-2} \\
			\g{\formula}{\rho}\sub{\eV}{\rho}=\eV' \label{proof:str-soundness-max-3}
		\end{gather}
		By \eqref{proof:str-soundness-max-3} and the definition of $\g{-}{\rho}$, we know
		\begin{gather}
			\g{\formula}{\rho}\sub{\eV}{\rho}=\mrec{\mx}{\eV''}=\eV' \label{proof:str-soundness-max-4} \\
			\g{\hV}{\rho}\sub{\eV}{\rho}=\eV'' \label{proof:str-soundness-max-5}
		\end{gather}
		\begin{proofRemark}[
			To prove that \R is a satisfaction relation, we must prove that when $\g{\maxSub}{\rho}\sub{\eV}{\rho}=\eV'''$ the following holds:
			$$ (\i{\eV'''}{\pV},\maxSub)\in\R$$	]
		\end{proofRemark}
		By \eqref{proof:str-soundness-max-2} and the definition of \Sat we can deduce
		\begin{gather}
			\exists \pVV\cdot \pVV\vSatS\formula \label{proof:str-soundness-max-6} 
		\end{gather}
		By \eqref{proof:str-soundness-max-6} and the definition of \vSatS we can know
		\begin{gather}
			\exists \pVV\cdot \pVV\vSatS\maxSub \label{proof:str-soundness-max-7} 
		\end{gather}
		By \eqref{proof:str-soundness-max-7} and the definition of \Sat we know 
		\begin{gather}
			\maxSub\in\Sat \label{proof:str-soundness-max-8} 
		\end{gather}
		Since $\maxSub$ is the unfolded equivalent of $\formula$, by \eqref{proof:str-soundness-max-4}, \eqref{proof:str-soundness-max-5} and the definition of $\g{-}{\rho}$, we know
		\begin{gather}
			\g{\maxSub}{\rho}\sub{\eV}{\rho}=\eV''\sub{\mrec{\mx}{\eV''}}{\mx} \label{proof:str-soundness-max-9}
		\end{gather}
		Hence, by \eqref{proof:str-soundness-max-8}, \eqref{proof:str-soundness-max-9} and the definition of \R we know 
		\begin{gather}
			(\i{\eV''\sub{\mrec{\mx}{\eV''}}{\mx}}{\pV},\maxSub)\in\R 
		\end{gather}		
	\end{case}
		
	\begin{case}[\hV=\hAnd\hgnec{\pate_i}{\predc_i}{\hV_i} \text{ where } \bigdistinct{i\in\IndSet}\actSN{\pate_i}{\predc_i}] We know
		\newcommand{\formula}{\hAnd\hgnec{\pate_i}{c_i}{\hV_i}}
		\newcommand{\dropForm}{
			\begin{xbrackets}{c}
				\mrec{\my}{\mCh\begin{xbrace}{ll}
					\mact{\actSTD{\pate_i}{\predc_i}}{\eV''_i} & \, (\text{if }\g{\hfls_i}{\my}=\eV''_i)\\
					\mact{\actSID{\pate_i}{\predc_i}}{\eV''_i} & \, (\text{otherwise})
				\end{xbrace}}
			\end{xbrackets}
		}	
		\newcommand{\dropFormUnfolded}{
			\begin{xbrackets}{c}
				\mCh\begin{xbrace}{ll}
						\mact{\actSTD{\pate_i}{\predc_i}}{\eV''_i\sub{\eV'}{\my}} & \, (\text{if }\g{\hfls_i}{\my}=\eV''_i)\\
						\mact{\actSID{\pate_i}{\predc_i}}{\eV''_i\sub{\eV'}{\my}} & \, (\text{otherwise})
				\end{xbrace}
			\end{xbrackets}
		}
		\begin{gather}
			(\i{\eV'}{\pV},\formula)\in\R \label{proof:str-soundness-nec-1}
		\end{gather}
		because		
		\begin{gather}
			\formula\in\Sat \label{proof:str-soundness-nec-2} \\
			\bigdistinct{i\in\IndSet}\actSN{\pate_i}{\predc_i} \label{proof:str-soundness-nec-3}\\
			\g{\formula}{\rhoRB}\sub{\eV}{\rho}=\eV' \label{proof:str-soundness-nec-4}
		\end{gather}
		By \eqref{proof:str-soundness-nec-4} and the definition of $\g{-}{\rho}$, we know
		\begin{gather}
			\g{\formula}{\rhoRB}\sub{\eV}{\rho}=\dropForm=\eV' \label{proof:str-soundness-nec-5} \\
			\forall i\in\IndSet\cdot\eV_i''=\g{\hV}{\my}\sub{\eV}{\rho}=\g{\hV}{\my}
			 \label{proof:str-soundness-nec-6}
		\end{gather}
		\begin{proofRemark}[
			To prove that \R is a satisfaction relation, we must prove that whenever $\g{\hgnec{\pate_i}{\predc_i}\hV_i}{\rho}\sub{\eV}{\rho}=\eV_i$ the following condition holds:
			$$\forall i{\,\in\,}\IndSet\cdot(\i{\eV_i}{\pV},\hnec{\actSN{\pate_i}{\predc_i}}\hV_i)\in\R$$
			But for \R to be a satisfaction relation, by definition of \vSatS we must prove
			$$\forall i{\,\in\,}\IndSet\cdot (\forall\pVV'\cdot\i{\eV_i}{\pV}\traS{\acta}\pVV' \text{ and } \actSN{\pate_i}{\predc_i}(\acta)=\s) \imp (\pVV',\hV_i\s)\in\R$$
			Also, since by \eqref{proof:str-soundness-nec-3} we know that guarded conjunctions in \shmlwf are \emph{disjoint}, we know that the \emph{same} event \acta can match and satisfy the condition of \emph{at most one} necessity guarding a specific branch. Since the case where \emph{none of the branches match is satisfied trivially}, we can simply prove the case where only \emph{one} branch matches \acta, \ie we must show that whenever $\g{\formula}{\rhoRB}\sub{\eV}{\rho}=\eV'$ then
			$$\exists j\!\in\!\IndSet \cdot (\i{\eV'}{\pV} \tra{\acta} \pVV' \textsl{ and } \actSN{\pate_j}{\predc_j}(\acta)=\s) \imp (\pVV',\hV_j\s)\in\R $$ ]
		\end{proofRemark}	
		Hence we start by assuming the knowledge of
		\begin{gather}
			\i{\eV'}{\pV}\traS{\acta}\pVV' \label{proof:str-soundness-nec-7} \\ 
			\exists j\!\in\!\IndSet \cdot \actSN{\pate_j}{\predc_j}(\acta)=\s  \label{proof:str-soundness-nec-8} 
		\end{gather}
		By \eqref{proof:str-soundness-nec-5} and \eqref{proof:str-soundness-nec-7} we know 
		\begin{gather}
			\i{\dropForm}{\pV} \tra{\acta} \pVV' \label{proof:str-soundness-nec-9} 
		\end{gather}
		By \eqref{proof:str-soundness-nec-9} and \rtit{iEnf} we know
		\begin{gather}
			\pVV' = \i{\eV'''}{\pV'}  \label{proof:str-soundness-nec-10} \\
			\dropForm\traS{\actaa}\eV''' \label{proof:str-soundness-nec-11} \\
			\pV\traS{\acta}\pV' \label{proof:str-soundness-nec-12}
		\end{gather}
		By \eqref{proof:str-soundness-nec-11} and \rtit{eRec} we know
		\begin{gather}
			\dropFormUnfolded\traS{\actaa}\eV''' \label{proof:str-soundness-nec-13}
		\end{gather}
		Since by \eqref{proof:str-soundness-nec-3} we know that the summands in \eqref{proof:str-soundness-nec-13} are prefixed by a \emph{disjoint} transducer, we thus know that only \emph{one} branch may be satisfied by action \acta. Hence, by applying rule \rtit{eSel} on \eqref{proof:str-soundness-nec-13} we know
		\begin{gather}
			\exists j\in\IndSet\cdot\mact{\actSTN{\pate_j}{\predc_j}{\actg}}{\eV_j''\sub{\eV'}{\my}}\traS{\actaa}\eV''' \qquad \text{ where }\actg\in\set{\actt,\pate_j} \label{proof:str-soundness-nec-14}
		\end{gather}
		Since the output action of the \actaa-reduction in \eqref{proof:str-soundness-nec-14} is $\acta\equiv\pate_j\s$, we know that the selected branch \emph{cannot be a suppression operation} (otherwise the output would have been \actt), hence we know
		\begin{gather}
			\exists j\in\IndSet\cdot\mact{\actSID{\pate_j}{\predc_j}}{\g{\hV_i}{\my}\sub{\eV}{\my}}\traS{\actaa}\eV'''	 \label{proof:str-soundness-nec-15}
		\end{gather}
		By \eqref{proof:str-soundness-nec-8} and the definition of \syn{\actSTN{\pate}{\predc}{\pate'}}, we know
		\begin{gather}
			\exists j\in\IndSet\cdot\actSID{\pate_j}{\predc_j}(\acta)=(\acta,\s) \label{proof:str-soundness-nec-16}
		\end{gather}
		By \eqref{proof:str-soundness-nec-15}, \eqref{proof:str-soundness-nec-16} and \rtit{eTrns} we know
		\begin{gather}
			\eV'''=\eV_j''\s\sub{\eV'}{\my}  \label{proof:str-soundness-nec-17}
		\end{gather}
		Since $j\in\IndSet$, from \eqref{proof:str-soundness-nec-6} we can deduce
		\begin{gather}
			\eV_j''=\g{\hV_j}{\my}  \label{proof:str-soundness-nec-18}
		\end{gather}
		Hence, by \eqref{proof:str-soundness-nec-17}, \eqref{proof:str-soundness-nec-18} and the definition of \g{-}{\rho}, we can conclude
		\begin{gather}
			\g{\hV_j\s}{\my}\sub{\eV'}{\my}=\eV_j''\s\sub{\eV'}{\my}=\eV'''  \label{proof:str-soundness-nec-19}
		\end{gather}
		Now, by \eqref{proof:str-soundness-nec-2} and the definition of \Sat, we know
		\begin{gather}
			\exists \pVV\cdot \pVV\vSatS\formula \label{proof:str-soundness-nec-20}
		\end{gather}
		By \eqref{proof:str-soundness-nec-20} and definition of \vSatS we know
		\begin{gather}
			\exists \pVV, \forall \pVV',i\in\IndSet\cdot (\pVV\traS{\acta}\pVV' \text{ and } \actSN{\pate_i}{\predc_i}(\acta)=\s) \imp \pVV'\vSatS\hV_j\s \label{proof:str-soundness-nec-21}
		\end{gather}
		Since by \eqref{proof:str-soundness-nec-3} we know that all branches are \emph{disjoint} from each other, we can deduce that \eqref{proof:str-soundness-nec-21} can be satisfied if there is \emph{one} branch that matches \acta (or none at all), such that we know
		\begin{gather}
			\forall \pVV',\exists\pVV, j\in\IndSet\cdot (\pVV\traS{\acta}\pVV' \text{ and } \actSN{\pate_j}{\predc_j}(\acta)=\s) \imp \pVV'\vSatS\hV_j\s \label{proof:str-soundness-nec-22}
		\end{gather}
		Therefore, by \eqref{proof:str-soundness-nec-8}, \eqref{proof:str-soundness-nec-12} and \eqref{proof:str-soundness-nec-22} we know
		\begin{gather}
			\exists \pVV'\cdot \pVV'\vSatS\hV_j\s \label{proof:str-soundness-nec-23}
		\end{gather}
		Hence, by \eqref{proof:str-soundness-nec-23}, the definition of \Sat we know
		\begin{gather}
			 \hV_j\s\in\Sat \label{proof:str-soundness-nec-24}
		\end{gather}
		Finally, by \eqref{proof:str-soundness-nec-19}, \eqref{proof:str-soundness-nec-24} and the definition of \R we know
		\begin{gather}
			\exists j\in\IndSet\cdot (\i{\eV'''}{\pV'},\hV_j\s)\in\R \label{proof:str-soundness-nec-25}
		\end{gather}
		Hence by assumptions \eqref{proof:str-soundness-nec-7}, \eqref{proof:str-soundness-nec-8}, \eqref{proof:str-soundness-nec-10} and deduction \eqref{proof:str-soundness-nec-25} we conclude
		\begin{gather*}
			\begin{array}{c}
				\exists j\in\IndSet\cdot (\i{\eV'}{\pV}\traS{\acta}\i{\eV_j''\s\sub{\eV'}{\my}}{\pV'} \text{ and } \actSN{\pate_i}{\predc_i}(\acta)=\s) \\ \imp (\i{\eV_j''\s\sub{\eV'}{\my}}{\pV'},\hV_j\s)\in\R
			\end{array}
		\end{gather*}	
	\end{case} 	
	$\; $	
\end{proof}
\subsection{Proving Transparency} \label{sec:proof-transparency}
\begin{rtp}
	$$\forall\rho,\pV{\,\in\,}\Proc,\eV{\,\in\,}\Enfwf,\hV{\,\in\,}\shmlwfopt\cdot \pV{\,\vSatS\,}\hV\, \text{ and }\,\g{\hV}{\rho}\sub{\eV}{\rho}=\eV' \; \imp \; \i{\eV'}{\pV}\bisim\pV $$
	To Prove this lemma we must show that relation \R (below) is a \emph{strong bisimulation relation}.
	$$ \R\;\defeq\;\setdef{(\i{\eV'}{\pV},\pV)}{\pV{\,\vSatS\,}\hV\;\text{ and }\;\forall\eV\in\Enfwf\cdot\g{\hV}{\rho}\sub{\eV}{\rho}=\eV'} $$
	Hence we must show that \R satisfies the following conditions:
	\begin{enumerate}[\quad(a)]
		\item if $\pV\traS{\actu}\pV'$ then $\i{\eV'}{\pV}\traS{\actu}S'$ and $(\pV',S')\in\R$
		\item if $\i{\eV'}{\pV}\traS{\actu}S'$ then $\pV\traS{\actu}\pV'$ and $(\pV',S')\in\R$
	\end{enumerate} $\\[-20mm]\;$	
\end{rtp}

\begin{proof} By coinduction on the structure of $\hV$. 
	
	\begin{case}[\hV=\hfls]
		Case does not apply since $\nexists \pV\cdot \pV\vSatS\hfls$.
	\end{case}

	\begin{case}[\hV=\hVarX]
		Case does not apply since $\hVarX$ is an open-formula and $\nexists \pV\cdot \pV\vSatS\hVarX$.
	\end{case}

	\begin{case}[\hV=\htru]
		Initially we know 
		\begin{gather}
			(\pV,\i{\eV'}{\pV})\in\R \label{proof:trans-tt-1}
		\end{gather}
		because
		\begin{gather}
			\pV\vSatS\htru \label{proof:trans-tt-2}\\
			\g{\htru}{\rho}\sub{\eV}{\rho}=\eV'  \label{proof:trans-tt-3}
		\end{gather}
		By \eqref{proof:trans-tt-3} and the definition of \g{-}{\rho} we know that function \g{-}{\rho} replaces every occurrence of \hfls with $\rho$, in this case we have no falsehood declarations, and hence we know
		\begin{gather}
			\forall\eV\cdot\g{\htru}{\rho}\sub{\eV}{\rho}=\g{\htru}{\rho}=\miden=\eV' \label{proof:trans-tt-4}
		\end{gather} 
		By the definition of \vSatS also we know that \htru is satisfied by \emph{any} process, hence we deduce
		\begin{gather}
			\pV'\vSatS\htru \label{proof:trans-tt-5}
		\end{gather}
		Hence by \eqref{proof:trans-tt-3}, \eqref{proof:trans-tt-5} and the definition of \R we know.
		\begin{gather}
			(\pV',\i{\eV'}{\pV'})\in\R \label{proof:trans-tt-6}
		\end{gather}
		\begin{itemize}
			\item[\pmb{--}] \textbf{To Prove (a):} Since $\actu\in\{\acta,\actt\}$, we must consider the following two subcases.
			\begin{itemize}
				\item[\pmb{--}] \pmb{\actu=\actt:} We start by assuming 
					\begin{gather}
						\pV\tra{\actt}\pV' \label{proof:trans-tt-7}
					\end{gather}
					By \eqref{proof:trans-tt-7} and \rtit{iAsyP} we know 
					\begin{gather}
						\i{\eV'}{\pV}\traS{\actt}\i{\eV'}{\pV'} \label{proof:trans-tt-8}
					\end{gather}
					Hence by assumption \eqref{proof:trans-tt-7} and deductions \eqref{proof:trans-tt-6} and \eqref{proof:trans-tt-8} we know
					\begin{gather*}
						\pV\tra{\actt}\pV' \imp \i{\eV'}{\pV}\traS{\actt}\i{\eV'}{\pV'} \text{ and } (\pV',\i{\eV'}{\pV'})\in\R
					\end{gather*}
				\item[\pmb{--}] \pmb{\actu=\acta:} We start by assuming 
					\begin{gather}
						\pV\tra{\acta}\pV' \label{proof:trans-tt-9}
					\end{gather}
					By \rtit{eId} we know
					\begin{gather}
						\miden\tra{\ioact{\acta}{\acta}}\miden \label{proof:trans-tt-10}
					\end{gather}
					By \eqref{proof:trans-tt-9}, \eqref{proof:trans-tt-10} and \rtit{iEnf} we know 
					\begin{gather}
						\i{\miden}{\pV}\tra{\acta}\i{\miden}{\pV'} \label{proof:trans-tt-11}
					\end{gather}
					By \eqref{proof:trans-tt-4} and \eqref{proof:trans-tt-11} we can thus deduce
 					\begin{gather}
 						\i{\eV'}{\pV}\tra{\acta}\i{\eV'}{\pV'} \label{proof:trans-tt-12}
 					\end{gather}
					Hence by assumption \eqref{proof:trans-tt-9} and deductions \eqref{proof:trans-tt-6} and \eqref{proof:trans-tt-12} we know
					\begin{gather*}
						\pV\tra{\acta}\pV' \imp \i{\eV'}{\pV}\tra{\acta}\i{\eV'}{\pV'} \text{ and } (\pV',\i{\eV'}{\pV'})\in\R
					\end{gather*}
			\end{itemize}
			\item[\pmb{--}] \textbf{To Prove (b):} Since $\actu\in\{\acta,\actt\}$, we must consider the following two subcases.
			\begin{itemize}
				\item[\pmb{--}] \pmb{\actu=\actt:} We start by assuming 
				\begin{gather}
					\i{\eV'}{\pV}\tra{\actt}\pVV \label{proof:trans-tt-13}
				\end{gather}
				By \eqref{proof:trans-tt-13} and \rtit{iEnf} we know 
				\begin{gather}
					\pVV=\i{\eV''}{\pV'} \label{proof:trans-tt-14}\\
					\eV'\tra{\actat}\eV'' \label{proof:trans-tt-15}\\
					\pV\traS{\acta}\pV' \label{proof:trans-tt-16}
				\end{gather}
				From \eqref{proof:trans-tt-4} and \eqref{proof:trans-tt-15} we can deduce
				\begin{gather}
					\miden\tra{\actat}\eV'' \label{proof:trans-tt-17}
				\end{gather}
				Since there does not exist a rule in our model that allows a monitor of the form $\miden$ to reduce using a \actat-transition, this means that assumption \eqref{proof:trans-tt-17} \emph{can never occur}, which implies that this case does not apply.
				\item[\pmb{--}] \pmb{\actu=\acta:} We start by assuming 
				\begin{gather}
					\i{\eV'}{\pV}\tra{\acta}\pVV \label{proof:trans-tt-18}
				\end{gather}
				By \eqref{proof:trans-tt-4} and \eqref{proof:trans-tt-18} we know 
				\begin{gather}
					\i{\miden}{\pV}\tra{\acta}\pVV \label{proof:trans-tt-19}
				\end{gather}
				By \eqref{proof:trans-tt-19} \rtit{iEnf} and \rtit{eId} we know
				\begin{gather}
					\pV\tra{\acta}\pV'   \label{proof:trans-tt-20} \\
					\pVV=\i{\miden}{\pV'} \label{proof:trans-tt-21}
				\end{gather}
				By \eqref{proof:trans-tt-4} and \eqref{proof:trans-tt-21} we know
				\begin{gather}
					\pVV=\i{\eV'}{\pV'} \label{proof:trans-tt-22}
				\end{gather}				
				Hence by assumptions \eqref{proof:trans-tt-18}, \eqref{proof:trans-tt-22} and deductions \eqref{proof:trans-tt-6} and \eqref{proof:trans-tt-20} we know
				\begin{gather*}
					\i{\eV'}{\pV}\traS{\acta}\i{\eV'}{\pV'} \imp \pV\traS{\acta}\pV' \text{ and } (\pV',\i{\eV'}{\pV'})\in\R
				\end{gather*}
			\end{itemize}
		\end{itemize}
	\end{case}

	\begin{case}[\hV=\hmax{\hVarX}{\hV} \text{ where }\hVarX{\in}\fv{\hV}]
		\newcommand{\formula}{\hmax{\hVarX}{\hV}}
		\newcommand{\maxsub}{\hV\sub{\formula}{\hVarX}}
		Initially we know 
		\begin{gather}
			(\pV,\i{\eV'}{\pV})\in\R \label{proof:trans-max-1}
		\end{gather}
		because	
		\begin{gather}
			\pV\vSatS\formula \label{proof:trans-max-2}\\
			\g{\formula}{\rho}\sub{\eV}{\rho}=\eV' \label{proof:trans-max-3}
		\end{gather}
		By \eqref{proof:trans-max-3} and the definition of \g{-}{\rho} we know
		\begin{gather}
			\g{\formula}{\rho}\sub{\eV}{\rho}=\mrec{\mx}{\eV''}=\eV' \label{proof:trans-max-4}\\
			\g{\hV}{\rho}\sub{\eV}{\rho}=\eV'' \label{proof:trans-max-5}
		\end{gather}
		Since $\hVarX{\in}\fv{\hV}$, we know that the fixpoint variable \hVarX is defined in the continuation formula \hV, hence if we apply the synthesis function on the \emph{unfolded version} of \formula, \ie \maxsub, from \eqref{proof:trans-max-4} we know that the synthesis produces an \emph{unfolded version} of $\eV'$, such that we can deduce
		\begin{gather}
			\g{\maxsub}{\rho}\sub{\eV}{\rho}=\eV''\sub{\mrec{\mx}{\eV''}}{\mx} \label{proof:trans-max-6}
		\end{gather}		
		By \eqref{proof:trans-max-2} and the definition of \vSatS we know
		\begin{gather}
			\pV\vSatS\maxsub  \label{proof:trans-max-7}
		\end{gather}
		Hence, by \eqref{proof:trans-max-6}, \eqref{proof:trans-max-7} and the definition of \R we know
		\begin{gather}
			(\pV, \i{\eV''\sub{\mrec{\mx}{\eV''}}{\mx}}{\pV})\in\R  \label{proof:trans-max-8}
		\end{gather}
		\begin{itemize}
			\item[\pmb{--}] \textbf{To Prove (a):} We start by assuming
				\begin{gather}
					\pV\tra{\actu}\pV' \label{proof:trans-max-9}
				\end{gather}
				From \eqref{proof:trans-max-8} and IH we know 
				\begin{gather}
					\pV\tra{\actu}\pV' \imp \i{\eV''\sub{\mrec{\mx}{\eV''}}{\mx}}{\pV}\traS{\actu}\pVV' \text{ and } (\pV',\pVV')\in\R  \label{proof:trans-max-10}
				\end{gather}
				By \eqref{proof:trans-max-9} and \eqref{proof:trans-max-10} we deduce
				\begin{gather}
					\i{\eV''\sub{\mrec{\mx}{\eV''}}{\mx}}{\pV}\traS{\actu}\pVV' \label{proof:trans-max-11} \\
					(\pV',\pVV')\in\R  \label{proof:trans-max-12}
				\end{gather}
				By \eqref{proof:trans-max-9}, \eqref{proof:trans-max-11} and \rtit{iEnf} we know
				\begin{gather}
					\eV''\sub{\mrec{\mx}{\eV''}}{\mx}\traS{\actau}\eV''' \label{proof:trans-max-13} \\
					\pVV'=\i{\eV'''}{\pV'}  \label{proof:trans-max-14}
				\end{gather}
				By applying rule \rtit{eRec} on \eqref{proof:trans-max-13} we know
				\begin{gather}
					\mrec{\mx}{\eV''}\tra{\actau}\eV''' \label{proof:trans-max-15} 
				\end{gather}
				By \eqref{proof:trans-max-9}, \eqref{proof:trans-max-14}, \eqref{proof:trans-max-15} and \rtit{iEnf} we know
				\begin{gather}
					\i{\mrec{\mx}{\eV''}}{\pV}\tra{\actu}\pVV' \label{proof:trans-max-16} 
				\end{gather}
				By \eqref{proof:trans-max-4} and \eqref{proof:trans-max-16} we deduce
				\begin{gather}
					\i{\eV'}{\pV}\tra{\actu}\pVV' \label{proof:trans-max-17} 
				\end{gather}				
				Hence by assumption \eqref{proof:trans-max-9} and deductions \eqref{proof:trans-max-12} and \eqref{proof:trans-max-17} we conclude
				\begin{gather*}
					\pV\traS{\actu}\pV' \imp \i{\eV'}{\pV}\traS{\actu}\pVV'  \text{ and }  (\pV',\pVV')\in\R
				\end{gather*}
			\item[\pmb{--}] \textbf{To Prove (b):} Since $\actu\in\{\acta,\actt\}$, we must consider the following two subcases.
			\begin{itemize}
				\item[\pmb{--}] \pmb{\actu=\acta:} We start by assuming 
				\begin{gather}
					\i{\eV'}{\pV}\tra{\actu}\pVV' \label{proof:trans-max-18}
				\end{gather}
				By \eqref{proof:trans-max-18} and \rtit{iEnf} we know 
				\begin{gather}
					\pVV'=\i{\eV'''}{\pV'}  \label{proof:trans-max-19} \\
					\eV'\traS{\actau}\eV''' \label{proof:trans-max-20} \\
					\pV\traS{\acta}\pV' \label{proof:trans-max-21}
				\end{gather}
				By \eqref{proof:trans-max-4} and \eqref{proof:trans-max-20} we know
				\begin{gather}
					\mrec{\mx}{\eV''}\traS{\actau}\eV'''  \label{proof:trans-max-22}
				\end{gather}
				By applying rule \rtit{eRec} on \eqref{proof:trans-max-22} we know
				\begin{gather}
					\eV''\sub{\mrec{\mx}{\eV''}}{\mx}\traS{\actau}\eV'''  \label{proof:trans-max-23}
				\end{gather}
				By \eqref{proof:trans-max-19}, \eqref{proof:trans-max-21}, \eqref{proof:trans-max-23} and \rtit{iEnf} we know
				\begin{gather}
					\i{\eV''\sub{\mrec{\mx}{\eV''}}{\mx}}{\pV}\traS{\actu}\pVV' \label{proof:trans-max-24}
				\end{gather}
				By \eqref{proof:trans-max-8} and IH we know 
				\begin{gather}
					\i{\eV''\sub{\mrec{\mx}{\eV''}}{\mx}}{\pV}\traS{\actu}\pVV' \imp \pV\traS{\actu}\pV' \text{ and } (\pV',\pVV')\in\R \label{proof:trans-max-25}
				\end{gather}
				From \eqref{proof:trans-max-24} and \eqref{proof:trans-max-25} we thus deduce
				\begin{gather}
					\pV\traS{\actu}\pV' \label{proof:trans-max-26} \\
					(\pV',\pVV')\in\R \label{proof:trans-max-27}
				\end{gather}
				Hence by assumptions \eqref{proof:trans-max-18} and deductions \eqref{proof:trans-max-26} and \eqref{proof:trans-max-27} we conclude
				\begin{align*}
					\i{\eV'}{\pV}\tra{\actu}\pVV' \imp \pV\traS{\actu}\pV'  \text{ and }  (\pV',\pVV')\in\R
				\end{align*}
			\end{itemize}
		\end{itemize}
	\end{case}
	
	\begin{case}[\hV=\hAnd\hnec{\actSN{\pate_i}{\predc_i}}{\hV_i}]
		\newcommand{\formula}{\hAnd\hnec{\actSN{\pate_i}{\predc_i}}{\hV_i}}
		\newcommand{\dropForm}{
				\mrec{\my}{
					\begin{xbrackets}{c}
						\sum_{i\in\IndSet}\begin{xbrace}{ll}
							\mact{\actSTD{\pate_i}{\predc_i}}{\eV''_i} & \, (\text{if }\g{\hfls_i}{\my}{=}\eV_i'')\\
							\mact{\actSID{\pate_i}{\predc_i}}{\eV''_i} & \, (\text{otherwise})
						\end{xbrace}
				\end{xbrackets}}
		}	
		\newcommand{\dropFormWithJ}{
			\begin{xbrackets}{c}
				\mrec{\my}{
					\begin{xbrackets}{c}
						\sum_{i\in\IndSet\setminus\set{j}}\begin{xbrace}{ll}
							\mact{\actSTD{\pate_i}{\predc_i}}{\eV''} & \, (\text{if }\g{\hfls}{\my}{=}\eV'')\\
							\mact{\actSID{\pate_i}{\predc_i}}{\eV''} & \, (\text{otherwise})
						\end{xbrace}\\
						\mch{}{\quad\mact{\actSTN{\pate_j}{\predc_j}{\pate'}}{\g{\hV_j}{\eV}}}
					\end{xbrackets}}
			\end{xbrackets}
		}	
		\newcommand{\dropFormSel}{
			\begin{xbrackets}{c}
			\sum_{i\in\IndSet\setminus\set{j}}\begin{xbrace}{ll}
				\mact{\actSTD{\pate_i}{\predc_i}}{\eV_j''\sub{\eV'}{\my}} & \, (\text{if }\g{\hfls}{\my}{=}\eV'')\\
				\mact{\actSID{\pate_i}{\predc_i}}{\eV_j''\sub{\eV'}{\my}} & \, (\text{otherwise})
			\end{xbrace}\\
			\mch{}{\mact{\quad\actSTN{\pate_j}{\predc_j}{\pate'}}{\eV''\sub{\eV'}{\my}}}
			\end{xbrackets}
		}	
		\newcommand{\dropFormB}{
		\begin{xbrackets}{c}
			\mrec{\my}{\dropFormSelB}
		\end{xbrackets}
		}	
		\newcommand{\dropFormSelB}{
			\mCh\begin{xbrace}{ll}
				\mact{\actSTD{\pate_i}{\predc_i}}{\eV''\sub{\eV'}{\my}} & \, (\text{if }\g{\hfls}{\my}{=}\eV'')\\
				\mact{\actSID{\pate_i}{\predc_i}}{\eV''\sub{\eV'}{\my}} & \, (\text{otherwise})
			\end{xbrace}
		}	
		Initially we know 
		\begin{gather}
			(\pV,\i{\eV'}{\pV})\in\R \label{proof:trans-nec-1}
		\end{gather}
		because
		\begin{gather}
			\pV\vSatS\formula \label{proof:trans-nec-2} \\
			\bigdistinct{i\in\IndSet}{\actSN{\pate_i}{\predc_i}} \label{proof:trans-nec-3} \\
			\eV'=\g{\formula}{\rhoRB}\sub{\eV}{\rho} \label{proof:trans-nec-4}
		\end{gather}
		By \eqref{proof:trans-nec-4} and the definition of \g{-}{\rho} we know
		\begin{gather}
			\g{\formula}{\rhoRB}\sub{\eV}{\rho}=\dropForm=\eV' \label{proof:trans-nec-5}\\
			\forall i\in\IndSet\cdot\eV_i''=\g{\hV}{\my}\sub{\eV}{\rho}=\g{\hV}{\my} \label{proof:trans-nec-5.1}
		\end{gather}
		By \eqref{proof:trans-nec-2} and the definition of \vSatS we know
		\begin{gather}
			\forall i\in\IndSet\cdot\pV\vSatS\hnec{\actSN{\pate_i}{\predc_i}}\hV_i  \label{proof:trans-nec-6}
		\end{gather}
		By \eqref{proof:trans-nec-6} and the definition of \vSatS we know
		\begin{gather}
			\forall i{\,\in\,}\IndSet \cdot(\forall\pV'\cdot\pV\traS{\acta}\pV'\text{ and }\actSN{\pate_j}{\predc_j}(\acta)=\s)\imp\pV'\vSatS\hV_i\s  \label{proof:trans-nec-7}
		\end{gather}
		Since by \eqref{proof:trans-nec-3} we know that a concrete event \acta can match \emph{at most one} symbolic event defined in the guarding necessities, then we know that \emph{at most one branch} can be selected at runtime, hence from \eqref{proof:trans-nec-7} we can deduce
		\begin{gather}
			\exists j\in\IndSet\cdot(\forall\pV'\cdot\pV\traS{\acta}\pV'\text{ and }\actSN{\pate_i}{\predc_i}(\acta)=\s)\imp\pV'\vSatS\hV_j\s \label{proof:trans-nec-8}
		\end{gather}
		\begin{itemize}
			\item[\pmb{--}] \textbf{To Prove (a):} Since $\actu\!\in\!\{\acta,\actt\}$, we must consider the following two subcases.
			\begin{itemize}
				\item[\pmb{--}] \pmb{\actu=\actt:} To prove this subcase we assume
				\begin{gather}
					\pV\tra{\actt}\pV' \label{proof:trans-nec-9}
				\end{gather}
					This case holds trivially since \eqref{proof:trans-nec-9} contradicts the assumption of \eqref{proof:trans-nec-8}, thus trivially satisfying the implication in \eqref{proof:trans-nec-8}.
					\item[\pmb{--}] \pmb{\actu=\acta:} We start by assuming 
					\begin{gather}
						\pV\tra{\acta}\pV' \label{proof:trans-nec-10}
					\end{gather}
					We further investigate the following cases:
					\begin{itemize}
						\item[\pmb{--}] $\pmb{\forall j\!\in\!\IndSet\cdot\mtch{\pate_j}{\acta}=\sundef \;(\ie\text{ no matching branches}):}$ 
						\\This case is trivially satisfied by \eqref{proof:trans-nec-8} since $\nexists j\in\IndSet\cdot\actSN{\pate_i}{\predc_i}(\acta)=\s$.
						\item[\pmb{--}] $\pmb{\exists j\!\in\!\IndSet\cdot\actSN{\pate_j}{\predc_j}(\acta)=\s \; (\ie\text{ 1 matching branch}):}$ We know
						\begin{gather}
							\exists j\!\in\!\IndSet\cdot\actSN{\pate_j}{\predc_j}(\acta)=\s \label{proof:trans-nec-11}
						\end{gather}
						By \eqref{proof:trans-nec-8}, \eqref{proof:trans-nec-10} and \eqref{proof:trans-nec-11} we know
						\begin{gather}
							\exists j\!\in\!\IndSet\cdot \pV'\vSatS\hV_j\s \label{proof:trans-nec-12}
						\end{gather}
						By \eqref{proof:trans-nec-9} and the definition of \syn{\actSTN{\pate}{\predc}{\pate'}} we know
						\begin{gather}
							\exists j\!\in\!\IndSet\cdot \actSTN{\pate}{\predc}{\pate'}(\acta)=(\pate'\s,\s) \label{proof:trans-nec-13}							
						\end{gather}
						By \eqref{proof:trans-nec-13} and rule \rtit{eTrns} we know
						\begin{gather}
							\exists j\!\in\!\IndSet\cdot \mact{\actSTN{\pate_j}{\predc_j}{\pate'}}{\eV_j''\sub{\eV'}{\my}} \traS{\ioact{\acta}{\pate'\s}}\eV_j''\s\sub{\eV'}{\my} \label{proof:trans-nec-14} 
						\end{gather}
						By \eqref{proof:trans-nec-10}, \eqref{proof:trans-nec-14} and rule \rtit{eSel} we know
						\begin{gather}
							\begin{array}{r}
							\exists j\!\in\!\IndSet\cdot \dropFormSel \\[5mm] \traS{\;\ioact{\acta}{\pate'\s}\;}\eV_j''\s\sub{\eV'}{\my}
							\end{array}
							\label{proof:trans-nec-15} 
						\end{gather}
						By \eqref{proof:trans-nec-10}, \eqref{proof:trans-nec-15} and rules \rtit{eRec + iEnf} we know
						\begin{gather}
							\begin{array}{r}
							\exists j\!\in\!\IndSet\cdot \i{\dropForm}{\pV} \\[5mm]  \traS{\;\pate'\s\;} \i{\eV_j''\s\sub{\eV'}{\my}}{\pV'} 
							\end{array}
							\label{proof:trans-nec-16} 
						\end{gather}
						Since $j{\,\in\,}\IndSet$, from \eqref{proof:trans-nec-5.1} we can deduce
						\begin{gather}
							\g{\hV_j}{\my}\sub{\eV'}{\my}=\eV_j''\sub{\eV'}{\my} \label{proof:trans-nec-17}	
						\end{gather}
						As we consider \emph{optimized} formulae, we cannot have a case where \hfls is embedded within a maximal fixpoint, \eg \hmax{\hVarX}{\hfls} would have been optimized into \hfls. Hence, since $\nexists\pV\cdot\pV{\,\vSatS\,}\hfls$, from \eqref{proof:trans-nec-11} we can deduce 
						\begin{gather}
							\hV_j\s\neq\hfls  \label{proof:trans-nec-18}
						\end{gather}
						By \eqref{proof:trans-nec-18} we know that actions satisfying $\hnec{\actSN{\pate_j}{\predc_j}}\hV_j$ in \eqref{proof:trans-nec-16} \emph{will not be suppressed} since $\hV_j\s\neq\hfls$, which means that $\pate'\!=\!\pate_j\!\neq\!\actt$. Hence, by the definition of \g{-}{\rho}, we know
						\begin{gather}
							\exists j\in\IndSet\cdot\i{\eV'}{\pV}\traS{\pate_j\s}\i{\eV_j''\s\sub{\eV'}{\my}}{\pV'} \label{proof:trans-nec-19} \\
							\pate'=\pate_j \text{  such that }\pate_j\s=\acta \label{proof:trans-nec-20}
						\end{gather}
						Hence, by \eqref{proof:trans-nec-19} and \eqref{proof:trans-nec-20} we know
						\begin{gather}
							\exists j\in\IndSet\cdot\i{\eV'}{\pV}\traS{\acta}\i{\eV_j''\s\sub{\eV'}{\my}}{\pV'} \label{proof:trans-nec-21}
						\end{gather}
						Since $\hV_j\s$ is the closed equivalent of $\hV_i$ (\wrt data variables), from \eqref{proof:trans-nec-17} we can deduce
						\begin{gather}
							\g{\hV_j\s}{\my}\sub{\eV'}{\my}=\eV_j''\s\sub{\eV'}{\my} \label{proof:trans-nec-22}
						\end{gather}
						By \eqref{proof:trans-nec-12}, \eqref{proof:trans-nec-22} and the definition of \R we know
						\begin{gather}
							\exists j\in\IndSet\cdot(\pV',\i{\eV_j''\s\sub{\eV'}{\my}}{\pV'}){\,\in\,}\R \label{proof:trans-nec-23}
						\end{gather}
						Hence, by assumption \eqref{proof:trans-nec-10} and deductions \eqref{proof:trans-nec-21} and \eqref{proof:trans-nec-23} we can conclude
						\begin{gather*}
						\begin{array}{c}
							\exists j\in\IndSet\cdot \pV\traS{\acta}\pV' \imp \i{\eV'}{\pV}\traS{\acta}\i{\eV_j''\s\sub{\eV'}{\my}}{\pV'}  \\  \text{ and } (\pV',\i{\eV_j''\s\sub{\eV'}{\my}}{\pV'}){\,\in\,}\R
						\end{array}
						\end{gather*}
					\end{itemize}
			\item[\pmb{--}] \textbf{To Prove (b):} Since $\actu\!\in\!\{\acta,\actt\}$, we must consider the following two subcases.
				\begin{itemize}
					\item[\pmb{--}] \pmb{\actu=\actt:} We start by assuming 
						\begin{gather}
							\i{\eV'}{\pV}\traS{\actt}\pVV \label{proof:trans-nec-24}
						\end{gather}
						By \eqref{proof:trans-nec-5} and \eqref{proof:trans-nec-24} we know
						\begin{gather}
							\i{\dropForm}{\pV}\traS{\actt}\pVV \label{proof:trans-nec-25}
						\end{gather}
						By \eqref{proof:trans-nec-25} and \rtit{iEnf + eRec} we know
						\begin{gather}
							\dropFormSelB\traS{\actat}\eV''' \label{proof:trans-nec-26} \\
							\pV\traS{\acta}\pV'	\label{proof:trans-nec-27} \\
							\pVV=\i{\eV'''}{\pV'} \label{proof:trans-nec-28}
						\end{gather}
						By \eqref{proof:trans-nec-26} and \rtit{eSel} we know
						\begin{gather}
							\exists j\in\IndSet\cdot \mact{\actSTN{\pate_i}{\predc_i}{\pate'}}{\eV''_j\sub{\eV'}{\my}} \traS{\actat} \eV''' \label{proof:trans-nec-29}
						\end{gather}
						Hence, by \eqref{proof:trans-nec-29} and \rtit{eTrns} we know
						\begin{gather}
							\actSTN{\pate_j}{\predc_j}{\pate'}(\acta)=(\pate'\s,\s) \label{proof:trans-nec-30}
						\end{gather}
						By \eqref{proof:trans-nec-30} and the definition of \syn{\actSTN{\pate}{\predc}{\pate'}} we know
						\begin{gather}
							\actSN{\pate_j}{\predc_j}(\acta)=\s  \label{proof:trans-nec-31}
						\end{gather}
						Hence, by \eqref{proof:trans-nec-8}, \eqref{proof:trans-nec-27} and \eqref{proof:trans-nec-31} we know
						\begin{gather}
							\exists j\in\IndSet\cdot\pV'\vSatS\hV_j\s  \label{proof:trans-nec-32}
						\end{gather}
						However, as the reduction in \eqref{proof:trans-nec-29} is performed over action \actat, this can only be achieved when the matched branch is prefixed by a \emph{suppression transducer}, \ie where $\pate'{=}\actt$, hence from \eqref{proof:trans-nec-26} we know that $\eV''_j\sub{\eV'}{\my}$ performs a suppression operation when $\g{\hfls}{\my}=\eV_j''\sub{\eV'}{\my}$, such that we know
						\begin{gather}
							\hV_j=\hfls  \label{proof:trans-nec-33}
						\end{gather}
						Hence, this case does not apply (and is thus satisfied trivially) since by definition of \vSatS we know that $\nexists\pV\cdot\pV{\,\vSatS\,}\hfls$, which contradicts with \eqref{proof:trans-nec-32} and \eqref{proof:trans-nec-33}.
					\item[\pmb{--}] \pmb{\actu=\acta:} We start by assuming 
						\begin{gather}
							\i{\eV'}{\pV}\traS{\acta}\pVV \label{proof:trans-nec-34}
						\end{gather}
						By \eqref{proof:trans-nec-5} and \eqref{proof:trans-nec-34} we know
						\begin{gather}
							\i{\dropForm}{\pV}\traS{\acta}\pVV \label{proof:trans-nec-35}
						\end{gather}
						By \eqref{proof:trans-nec-35} and \rtit{iEnf + eRec} we know
						\begin{gather}
							\dropFormSelB\traS{\actaa}\eV''' \label{proof:trans-nec-36} \\
							\pV\traS{\acta}\pV'	\label{proof:trans-nec-37} \\
							\pVV=\i{\eV'''}{\pV'} \label{proof:trans-nec-38}
						\end{gather}
						By \eqref{proof:trans-nec-3}, \eqref{proof:trans-nec-36} and \rtit{eSel} we know
						\begin{gather}
							\exists j\in\IndSet\cdot \mact{\actSTN{\pate_i}{\predc_i}{\pate'}}{\eV_j''\sub{\eV}{\my}} \traS{\actaa} \eV''' \label{proof:trans-nec-39}
						\end{gather}
						Since the reduction in \eqref{proof:trans-nec-39} is performed over an \actaa\!\!\!\! action, this can only be achieved when the matched branch is guarded by an identity transducer, such that $\pate'=\pate_j$. Hence, we can infer
						\begin{gather}
							\exists j\in\IndSet\cdot \mact{\actSID{\pate_j}{\predc_j}}{\eV''_j\sub{\eV}{\my}}\traS{\actaa}\eV''' \label{proof:trans-nec-40} 
						\end{gather}
						By \eqref{proof:trans-nec-40} and \rtit{eTrns} we know 
						\begin{gather}
							\eV'''=\eV_j''\s\sub{\eV}{\my} \label{proof:trans-nec-41} \\
							\mtchS{\actSID{\pate_j}{\predc_j}}{\acta}=(\pate_j\s,\s) \label{proof:trans-nec-42} 
						\end{gather}
						From \eqref{proof:trans-nec-42} and the definition of $\syn{\actSTN{\pate}{\predc}{\pate'}}$ we know
						\begin{gather}
							\mtchS{\actSN{\pate_j}{\predc_j}}{\acta}=\s \label{proof:trans-nec-43} 
						\end{gather}
						By \eqref{proof:trans-nec-8}, \eqref{proof:trans-nec-37} and \eqref{proof:trans-nec-43} we can deduce
						\begin{gather}
							\pV'\vSatS\hV_j\s \label{proof:trans-nec-44} 
						\end{gather}
						Since $j{\,\in\,}\IndSet$ from \eqref{proof:trans-nec-5.1} we know
						\begin{gather}
							\g{\hV_j}{\my}=\eV_j'' \label{proof:trans-nec-45}
						\end{gather}
						Hence, from \eqref{proof:trans-nec-41}, \eqref{proof:trans-nec-45} and the definition of \g{-}{\rho} we can deduce
						\begin{gather}
							\g{\hV_j\s}{\my}\sub{\eV}{\my}=\eV_j''\s\sub{\eV}{\my}=\eV''' \label{proof:trans-nec-46}
						\end{gather}
						Therefore, by \eqref{proof:trans-nec-44}, \eqref{proof:trans-nec-46} and the definition of \R, we know
						\begin{gather}
							(\pV',\i{\eV'''}{\pV'})\in\R \label{proof:trans-nec-47} 
						\end{gather}
						Hence by assumptions  \eqref{proof:trans-nec-34}, \eqref{proof:trans-nec-38} and deductions  \eqref{proof:trans-nec-37} and  \eqref{proof:trans-nec-47} we can finally conclude
						\begin{gather*}
							\i{\eV'}{\pV}\traS{\acta}\i{\eV'''}{\pV'} \imp \pV\traS{\acta}\pV' \text{ and } (\pV',\i{\eV'''}{\pV'})\in\R 
						\end{gather*}
				\end{itemize}
			\end{itemize}
		\end{itemize}
	\end{case}
		
\end{proof}
\subsection{Proving \Cref{lemma:opt-equiv}} \label{sec:proof-opt-equiv} \vspace{-5mm}
\begin{rtp}
	$ \opt{\hV}{\,=\,}\hVV \imp \hV{\,\equiv\,}\hVV \text{  and  } \hVV{\,\in\,}\shmlwfopt $
\end{rtp}\vspace{-5mm}

\begin{proof} By structural induction on $\hV$.
	
	\begin{Cases}[\hV=\hVV\;\text{ where }\;\hVV{\,\in\,}\Set{\htru,\hfls,\hVarX}]
		\\\indent Holds trivially since $\opt{\hVV}{=}\hVV$ and $\hVV{\,\in\,}\shmlwfopt$.
	\end{Cases}

	\begin{case}[\hV=\hmax{\hVarX}{\hV'}]
		We know
		\begin{gather}
			\opt{\hmax{\hVarX}{\hVV'}}=\hVV \label{proof:opt-equiv-max-1}
		\end{gather}
		By definition of \optSym, we must consider two subcases
		\begin{itemize}
			\item[\pmb{--}] $\pmb{\hVarX{\,\in\,}\fv{\hV'}:}$ We know 
			\begin{gather}
				\hVV=\hmax{\hVarX}{\hVV'} \label{proof:opt-equiv-max-2} \\
				\hVV'=\opt{\hV'} \label{proof:opt-equiv-max-3}
			\end{gather}
			By \eqref{proof:opt-equiv-max-3} and IH we know
			\begin{gather}
				\hVV'\equiv\hV'  \label{proof:opt-equiv-max-4} \\
				\hVV'\in\shmlwfopt  \label{proof:opt-equiv-max-5}
			\end{gather}
			By \eqref{proof:opt-equiv-max-1}, \eqref{proof:opt-equiv-max-2} and \eqref{proof:opt-equiv-max-4} we can deduce
			\begin{gather}
				\hmax{\hVarX}{\hVV'}\equiv\hmax{\hVarX}{\hV'}  \label{proof:opt-equiv-max-6} 
			\end{gather}
			By \eqref{proof:opt-equiv-max-5} and the definition of \shmlwfopt we know
			\begin{gather}
				\hmax{\hVarX}{\hVV'}{\,\in\,}\shmlwfopt  \label{proof:opt-equiv-max-7} 
			\end{gather}
			$\therefore$ This subcase holds by \eqref{proof:opt-equiv-max-6} and \eqref{proof:opt-equiv-max-7}.
			\item[\pmb{--}] $\pmb{\hVarX{\,\notin\,}\fv{\hV'}:}$  We know 
			\begin{gather}
				\hVV=\opt{\hV'} \label{proof:opt-equiv-max-8}
			\end{gather}
			By \eqref{proof:opt-equiv-max-8} and IH we know
			\begin{gather}
				\hVV\equiv\hV'  \label{proof:opt-equiv-max-9} \\
				\hVV\in\shmlwfopt  \label{proof:opt-equiv-max-10}
			\end{gather}
			Since $\hVarX{\,\notin\,}\fv{\hV'}$, we know that \hVarX is never referenced in $\hV'$, thus making the maximal fixpoint declaration \hmax{\hVarX}{} redundant, hence from \eqref{proof:opt-equiv-max-9} we can deduce 
			\begin{gather}
				\hmax{\hVarX}{\hVV'}\equiv\hV'\equiv\hVV  \label{proof:opt-equiv-max-11} 
			\end{gather}
			$\therefore$ This subcase holds by \eqref{proof:opt-equiv-max-10} and \eqref{proof:opt-equiv-max-11}.
		\end{itemize}
	\end{case}
	
	\begin{case}[\hV=\hAnd\hnec{\actS_i}\hV_i]
		We know
		\begin{gather}
		\opt{\hAnd\hnec{\actS_i}\hV_i}=\hAnd\hnec{\actS_i}\hVV_i \label{proof:opt-equiv-nec-1}
		\end{gather}
		because
		\begin{gather}
			\forall i{\in}\IndSet\cdot\opt{\hV_i}=\hVV_i \label{proof:opt-equiv-nec-2}
		\end{gather}
		By \eqref{proof:opt-equiv-nec-2} and IH we know
		\begin{gather}
			\forall i{\in}\IndSet\cdot \hVV_i\equiv\hV_i  \label{proof:opt-equiv-nec-3} \\
			\forall i{\in}\IndSet\cdot \hVV_i\in\shmlwfopt  \label{proof:opt-equiv-nec-4}
		\end{gather}
		From \eqref{proof:opt-equiv-nec-1}, \eqref{proof:opt-equiv-nec-2} and \eqref{proof:opt-equiv-nec-3} we can deduce
		\begin{gather}
			\hAnd\hnec{\actS_i}\hV_i\equiv\hAnd\hnec{\actS_i}\hVV_i  \label{proof:opt-equiv-nec-5} 
		\end{gather}
		From  \eqref{proof:opt-equiv-nec-1}, \eqref{proof:opt-equiv-nec-2}, \eqref{proof:opt-equiv-nec-4} and the definition of \shmlwfopt we know
		\begin{gather}
			\hAnd\hnec{\actS_i}\hVV_i{\,\in\,}\shmlwfopt  \label{proof:opt-equiv-nec-6} 
		\end{gather}
		$\therefore$ This subcase holds by \eqref{proof:opt-equiv-nec-5} and \eqref{proof:opt-equiv-nec-6}.
	\end{case}
	
\end{proof}


\end{document}